\documentclass{aa}
\usepackage{siunitx}
\usepackage{graphicx}
\usepackage{txfonts}
\usepackage{subfigure}

\newcommand\blfootnote[1]{%
        \begingroup
        \renewcommand\thefootnote{}\footnote{#1}%
        \addtocounter{footnote}{-1}%
        \endgroup
}

\graphicspath{{figs/}}
\usepackage[colorlinks=true, allcolors=blue]{hyperref}
\def\hii{\hbox{{\rm H {\scriptsize II}}}~}
\newcommand{\U}{\mathrm}
\makeatletter
\renewcommand*\aa@pageof{, page \thepage{} of \pageref*{LastPage}}
\makeatother
\newcolumntype{L}[1]{>{\raggedright\arraybackslash}p{#1}}
\newcolumntype{C}[1]{>{\centering\arraybackslash}p{#1}}
\newcolumntype{R}[1]{>{\raggedleft\arraybackslash}p{#1}}
\begin{document} 
        
        \title{The effects of stellar feedback on molecular clumps in the Lagoon Nebula (M8)\footnotemark}
        
        \author{K. Angelique Kahle\inst{1,2}
                \and
                Friedrich Wyrowski\inst{1}
                \and
                Carsten K\"{o}nig\inst{1}
                \and
                Ivalu Barlach Christensen\inst{1}
                \and
                Maitraiyee Tiwari\inst{1}
                \and
                Karl M. Menten\inst{1}   
        }
        
        \institute{\inst{1} Max-Planck-Institut f\"{u}r Radioastronomie, Auf dem H\"{u}gel 69, 53121 Bonn, Germany\\
                \inst{2} Max-Planck-Institut f\"{u}r Astronomie, K\"{o}nigstuhl 17, 69117 Heidelberg, Germany\\
                \email{kahle@mpia.de}
        }
        \titlerunning{Stellar feedback in M8}
        \authorrunning{K. A. Kahle et al.}
        \date{Received Dec 19, 2023; accepted April 04, 2024}
        
        \abstract
        {The Lagoon Nebula (M8) is host to multiple regions with recent and ongoing massive star formation, due to which it appears as one of the brightest \hii regions in the sky. M8-Main and M8 East, two prominent regions of massive star formation, have been studied in detail over the past few years, while large parts of the nebula and its surroundings have received little attention. These largely unexplored regions comprise a large sample of molecular clumps that are affected by the presence of massive O- and B-type stars. Thus, exploring the dynamics and chemical composition of these clumps will improve our understanding of the feedback from massive stars on star-forming regions in their vicinity.}
        {We established an inventory of species observed towards 37 known molecular clumps in M8 and investigated their physical structure. We compared our findings for these clumps with the galaxy-wide sample of massive dense clumps observed as part of the APEX Telescope Large Area Survey of the Galaxy (ATLASGAL). Furthermore, we investigated the region for signs of star formation and stellar feedback.}
        {To obtain an overview of the kinematics and chemical abundances across the sample of molecular clumps in the M8 region, we conducted an unbiased line survey for each clump. We used the Atacama Pathfinder EXperiment (APEX) 12\,m submillimetre telescope and the 30m telescope of the Institut de Radioastronomie Millim\'{e}trique (IRAM) to conduct pointed on-off observations of 37 clumps in M8. These observations cover bandwidths of 53\,GHz and 40\,GHz in frequency ranges from 210\,GHz to 280\,GHz and from 70\,GHz to 117\,GHz, respectively. Temperatures were derived from rotational transitions of acetonitrile, methyl acetylene, and para-formaldehyde. Additional archival data from the \textit{Spitzer}, \textit{Herschel}, MSX, APEX, WISE, JCMT, and AKARI telescopes were used to investigate the morphology of the region and to derive the physical parameters of the dust emission by fitting spectral energy distributions to the observed flux densities.}
        {Across the observed M8 region, we identify 346 transitions from 70 different molecular species, including isotopologues. While many species and fainter transitions are detected exclusively towards M8 East, we also observe a large chemical variety in many other molecular clumps. We detect tracers of photo-dissociation regions (PDRs) across all the clumps, and 38\% of these clumps show signs of star formation. In our sample of clumps with extinctions between 1 and 60\,mag, we find that PDR tracers are most abundant in clumps with relatively low H$_2$ column densities. When comparing M8 clumps to ATLASGAL sources at similar distances, we find them to be slightly less massive (median $10\,\si{M_\odot}$) and have compatible luminosities (median $200\,\si{L_\odot}$) and radii (median $0.16$\,pc). In contrast, dust temperatures of the clumps in M8 are found to be increased by approximately 5\,K (25\%), indicating substantial external heating of the clumps by radiation of the present O- and B-type stars.}
        {This work finds clear and widespread effects of stellar feedback on the molecular clumps in the Lagoon Nebula. While the radiation from the O- and B-type stars possibly causes fragmentation of the remnant gas and heats the molecular clumps externally, it also gives rise to extended PDRs on the clump surfaces. Despite this fragmentation, the dense cores within 38\% of the observed clumps in M8 are forming a new generation of stars.}
        
        \maketitle
        %
        %
        \section{Introduction}\label{sec:intro} 
        To understand the origins of our own Solar System and the distribution of the stars around us, it is important to have a broad understanding of the formation of stellar objects and their subsequent feedback on the molecular clouds where they were born. While models explaining the formation of individual low-mass stars are well established, the processes involved in high-mass star formation are more complicated due to their formation in clusters~\citep{yorke2007,krumholz2014clusters,motte2018}. Usually, star-forming sites are located in close proximity to one another inside large-scale molecular clouds, and these sites are affected by the radiation of already existing massive O- and B-type stars.\blfootnote{$^\star$ The full versions of Tables~\ref{tab:app:idtransitions},~\ref{tab:app:line_properties}, and~\ref{tab:app:cd} are available at the CDS via anonymous ftp to cdsarc.u-strasbg.fr (130.79.128.5) or via http://cdsweb.u-strasbg.fr/cgi-bin/qcat?J/A+A/.}
            
        Studying the effects of stellar feedback on star formation is crucial \citep[e.g.][]{schneider2020sofia}. On one hand, the transferred momentum from stars can lead to a compression of the surrounding molecular gas, inducing star formation in dense regions. On the other hand, stellar radiation can contribute to the disruption of molecular clumps, preventing star formation.
        
        \begin{figure}[tbp]
                \centering
                \includegraphics[width=0.49\textwidth]{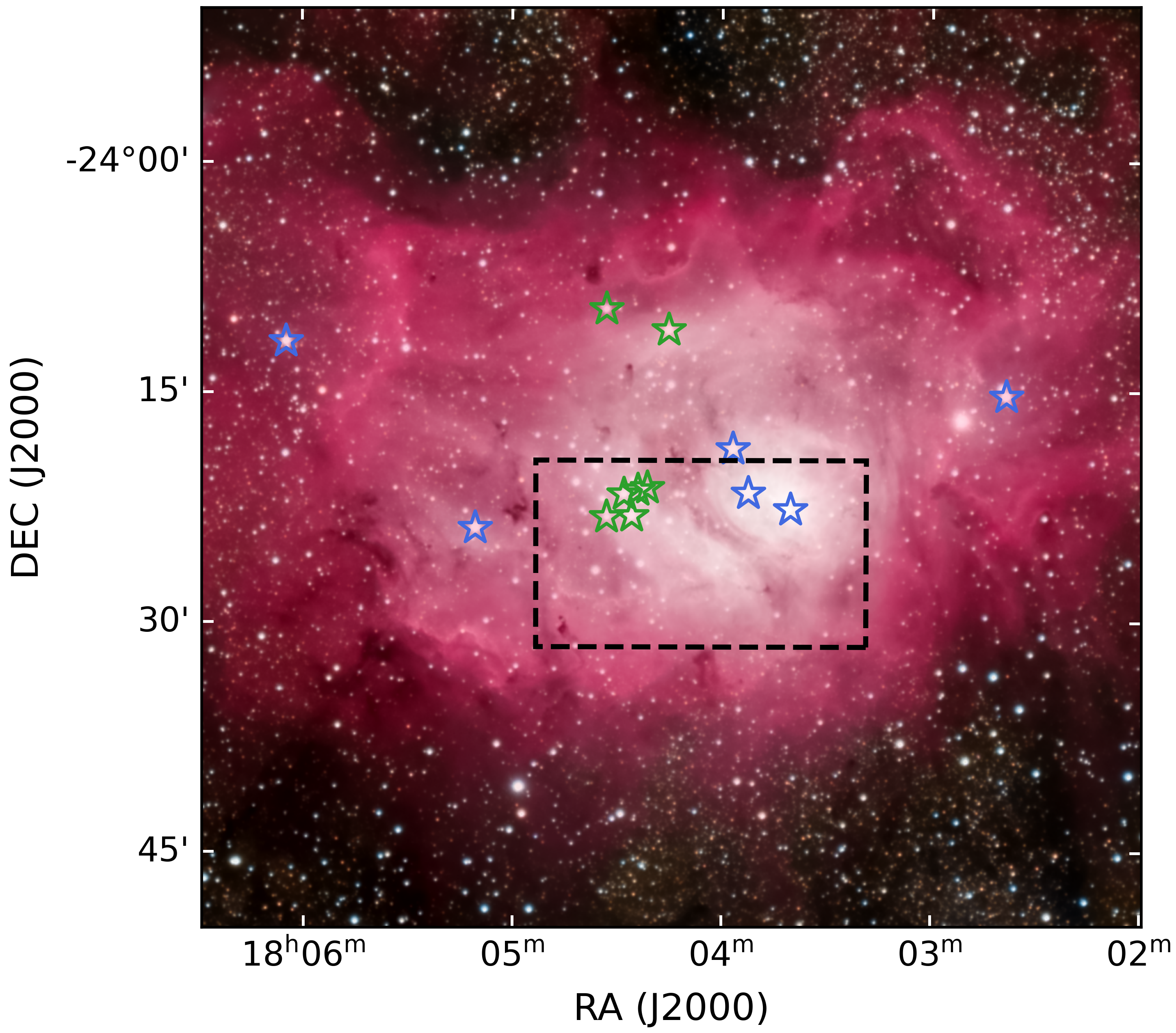}
                \caption[Optical image of the Lagoon Nebula]{Image of the Lagoon Nebula observed using B, V, and R broadband filters. Blue and green stars respectively mark the positions of O- and early B-type stars associated with the M8 region, according to Table 1 of~\citet{Wright2019OBstars}. The dashed black box marks the molecular gas examined by \citet{tothill2002structure}. Data credits: Felipe Mac Auliffe\footnotemark.}
                \label{fig:int:optical}
        \end{figure}
        \footnotetext{\url{https://clusteroneobservatory.com}}
        
        \begin{figure*}
                \begin{minipage}[c]{0.83\textwidth}
                        \includegraphics[width=0.999\textwidth]{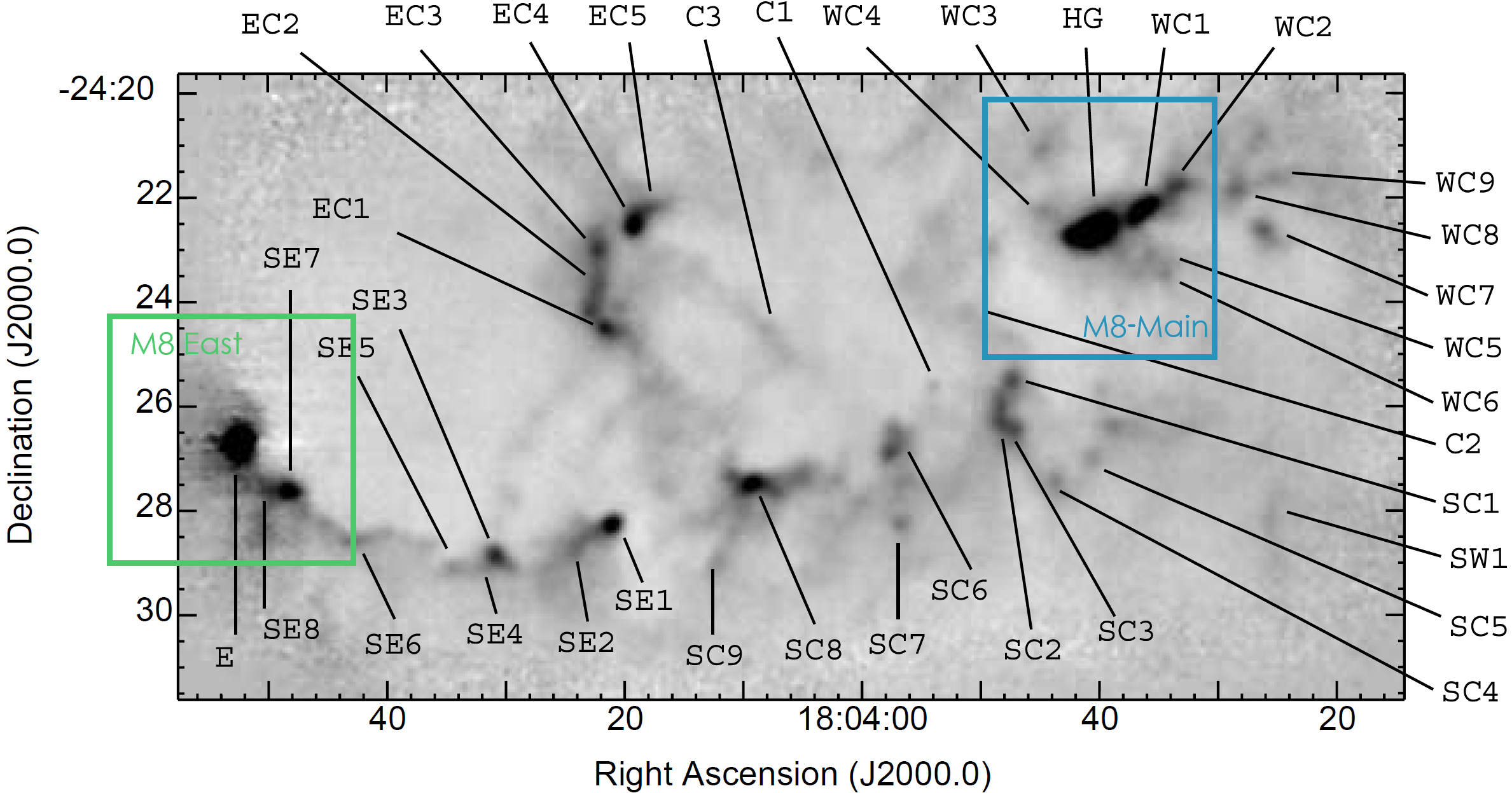}
                \end{minipage}\hfill
                \begin{minipage}[c]{0.16\textwidth}
                        \caption{SCUBA $\SI{850}{\micro\meter}$ image of the Lagoon Nebula adapted from~\citet{tothill2002structure}. Markers show the position of the molecular clumps in the Lagoon Nebula. The blue and green rectangles respectively indicate the regions examined by~\citet{tiwari2018M8HG} and~\citet{tiwari2020M8E}.} \label{fig:int:clumps}
                \end{minipage}
        \end{figure*}
        
        The Lagoon Nebula (Messier 8, M8) is an interesting target for investigating the effects of stellar feedback, as it is an \hii region associated with a star-forming~\citep{Kumar2010sf} molecular cloud complex.
        It is located at a distance of 1325\,$\si{pc}$~\citep{Damiani2019m8dist} in the Sagittarius-Carina arm. Its angular extent of about $\SI{40}{\arcminute}$ can be translated to a physical size of $\SI{15}{pc}$. The position of M8 corresponds to galactic coordinates of approximately $ l = \SI{6}{\degree}$, $b = \SI{-1.3}{\degree}$, which locates the region slightly below the inner Galactic plane, not far from the direction to the Galactic centre. Based on \textit{Gaia} proper motion data of associated cluster members,~\citet{Damiani2019m8dist} conclude that the cloud complex crossed the Galactic plane around $\SI{4}{Myr}$ ago, which may have been the initial trigger for its recent star formation activity.
        
        At optical wavelengths, M8 appears as a reddish emission nebula, with two particularly bright regions in the centre (see Fig.~\ref{fig:int:optical}). One of them is the prominent \hii region of the Lagoon Nebula, which is powered by the spectroscopic binary O star 9 Sagittarii~\citep[9\,Sgr;][]{Rauw20129sgr} and the multiple stellar system Herschel 36 \citep[Her\,36,][]{Arias2010her36} in the west. The other is the eastern illuminated region, associated with the open cluster NGC\,6530~\citep{prisinzano2005NGC6530}, which is part of M8 and contains several massive stars. An overview of high-mass stars (of O and early B types) identified in this region is available in~\citet[see their Table 1]{Wright2019OBstars}, and their positions are marked in Fig.~\ref{fig:int:optical} with coloured star symbols.

        Special attention has been paid to the interstellar medium (ISM) in the M8 region since~\citet{White1997brightCO} reported the Lagoon Nebula as the second brightest known CO-emitting source in our Galaxy. This study was followed up by~\citet{tothill2002structure}, who identified 37 individual dense clumps in the inner region of the Lagoon Nebula based on dust continuum maps at $\SI{450}{\micro\meter}$ and $\SI{850}{\micro\meter}$ observed with the \textit{James Clerk Maxwell} Telescope\footnote{\url{https://www.eaobservatory.org/jcmt}} (JCMT), towards which they observed the CO $J = 2-1$ line. As shown in Fig.~\ref{fig:int:clumps}, most of these clumps were loosely assigned to the west central region (WC), the east central ridge (EC), and two filaments in the southern part of the cloud (SE and SC). 
        
        Apart from the study conducted by \citet{tothill2002structure}, the molecular content of M8 and its environment has received little attention. However, this changed recently with \citet{tiwari2018M8HG} studying the massive star-forming WC (M8-Main); they report that the WC clumps are located behind the optically visible \hii region. Their spectroscopic observations revealed a photo-dissociation region (PDR) with a face-on geometry towards these sources, which is powered by Sgr\,9 and Her\,36 \citep[see e.g. Fig. 15 of][]{Tielens1985pdr}.
        
        A further study by~\citet{tiwari2020M8E} focused on the eastern part of the Lagoon Nebula (M8 East), which hosts clumps E, SE7, and SE8. This region is mainly illuminated by an embedded young stellar object (YSO), M8E-IR, which is likely to become a BO star \citep{Linz2008M8IR}. \citet{tiwari2020M8E} also observed an ionisation front moving in the south-east direction.
        These recent observations suggest that the remaining M8 clumps are also ideal targets for studying the effects of stellar feedback on the remnant gas in the region. Broad-bandwidth spectroscopic observations towards most molecular clumps in M8 are still missing. Through this work, we explore the complete sample of molecular clumps in M8 by combining the results of new spectroscopic observations at millimetre wavelengths with the information retrieved from archival infrared (IR) dust continuum images.
        
        Section~\ref{sec:obs} gives an overview of the observations and the data reduction strategy.
        In Sect.~\ref{sec:dust} the dust continuum images are analysed and the physical properties of the clumps are derived by fitting spectral energy distributions (SEDs) to the flux densities of all clumps.
        Section~\ref{sec:linesurvey} details the complete line survey of the $\SI{1.3}{\milli\meter}$ and $\SI{3}{\milli\meter}$ atmospheric windows towards the clumps in M8. The observed line emission is analysed in Sect.~\ref{sec:lineanalysis}. The results derived from the analysis of the dust continuum and the line survey are discussed in Sect.~\ref{sec:discussion}. Finally, Sect.~\ref{sec:summary} summarises the results of this study. 
        
        %
        %
        \section{Observations and data reduction}\label{sec:obs}
        We used archival IR to submillimetre wavelength data of the dust continuum emission from M8 in addition to new spectroscopic data taken with the Atacama Pathfinder EXperiment (APEX) 12 metre submillimetre telescope and the 30 metre telescope of the Institut de Radioastronomie Millimétrique (IRAM). The spectroscopic observations were conducted in on-off mode on the 37 molecular clumps identified by~\citet[see our Appendix~\ref{app:obs}]{tothill2002structure}. After inspecting the APEX spectra of all clumps, we decided to change the coordinates of WC3, SE8, and SC5 for the observations with the IRAM 30m telescope, in order to properly match the peak dust emission of the respective clumps. Therefore, the APEX observations of these three clumps were taken at the coordinates suggested by~\citet{tothill2002structure}, while the observations taken with the IRAM 30m telescope are offset by up to $1'$.
        
        We chose a fixed reference position M8REF at the coordinates RA=$18^\U{h}04^\U{m}40.0^\U{s}$, Dec.=$-24^\circ 23'00.0''$ (J2000), which is only slightly contaminated with line emission of $^{12}$CO and $^{13}$CO. In order to characterise the emission in M8REF, it was observed with a completely clean reference position at an offset of $(+2000'',-2000'')$ from M8REF.
        
        Inadvertently, first observations with APEX used the position switching mode with relative reference positions. These data were used to increase the sensitivity when the profiles of the thus observed lines agree with the data using the fixed reference position.
        
        \subsection{APEX observations with nFLASH230}
        The Atacama Pathfinder Experiment is a $\SI{12}{\meter}$ diameter submillimetre telescope located on the Llano de Chajnantor in the Chilean High Andes at an altitude of 5107\,$\si{\meter}$~\citep{gusten2006atacama}. The data were taken under project M-0107.F-9530C-2021 (P.I. Karl M. Menten) during several runs between 2021 July and October with the new FaciLity APEX Submillimetre Heterodyne instrument (nFLASH\footnote{\url{https://www.mpifr-bonn.mpg.de/5278273/nflash}}). The nFLASH receiver is a dual sideband (2SB) dual polarisation heterodyne receiver with two tunable frequency modules, of which we used the lower frequency nFLASH230 band for our observations. 
        The centres of the upper and lower sidebands are separated by $\SI{16}{\giga\hertz}$ and each cover a $\SI{7.9}{\giga\hertz}$ bandwidth in two polarisations. Four observed setups cover a total bandwidth of $\SI{58.3}{\giga\hertz}$ in a frequency range between $\SI{210}{\giga\hertz}$ and $\SI{280}{\giga\hertz}$ (see Appendix~\ref{app:obs}).
        
        The receiver was connected to modules of the APEX fast Fourier transform spectrometer (FFTS), which is an evolved version of the instrument described by \citep{Klein2012FFTS} and records each sideband and polarisation with partially overlapping $\SI{4}{\giga\hertz}$ wide FFTS processor units. These units provide each a total of $2^{16}$ channels per $\SI{4}{\giga\hertz}$ bandwidth, resulting in a channel spacing of $\SI{61.04}{\kilo\hertz}$. At our lowest and highest frequencies of $\SI{213.1}{\giga\hertz}$ and $\SI{279.7}{\giga\hertz}$, this results in velocity resolutions of $\SI{0.09}{\kilo\meter\per\second}$ and $\SI{0.07}{\kilo\meter\per\second}$, respectively. For analysing the data, each two adjacent channels are averaged. While this reduces the velocity resolution to values between $\SI{0.18}{\kilo\meter\per\second}$ and $\SI{0.14}{\kilo\meter\per\second}$, the resulting resolution is sufficient to resolve well all observed spectral lines. At this velocity resolution, the spectra show a typical average root mean square (RMS) noise of $\SI{28}{\milli\kelvin}$. 
        
        The system temperature during the observations typically ranged from $\SI{70}{\kelvin}$ to $\SI{250}{\kelvin}$, with a few scans at system temperatures of up to $\SI{460}{\kelvin}$. The conversion between antenna temperature $T_\U{A}^*$ and the main-beam brightness temperature $T_\U{MB}$ is given by \mbox{$T_\U{MB}=T_\U{A}^*\times \eta_\U{FW} / \eta_\U{MB}$}, where $\eta_\U{MB}$ is the main-beam efficiency and $\eta_\U{FW}$ the forward coupling efficiency. Based on Jupiter continuum pointings during the observation period of this project\footnote{The data of Jupiter are publicly available at \url{https://www.apex-telescope.org/telescope/efficiency}}, we estimate an average conversion factor of $\eta_\U{FW} / \eta_\U{MB} = 0.95/0.8$. The heterodyne line intensity monitoring\footnote{A regular line monitoring is performed with all heterodyne instruments of the APEX telescope. The results are made publicly available at \url{http://www.apex-telescope.org/grafana/d/-T6wuS_Mz/heterodyne-line-intensity-monitoring}} between 2021 July and October suggests a systematic calibration uncertainty of 5\% for the nFLASH230 observations. A more conservative estimate of 10\% is applied for the further analysis of the APEX data.
        
        The full width at half maximum (FWHM) of the APEX beam, $\theta_\U{B}$, at frequency $\nu$ (in GHz), in arcseconds is $\theta_\U{B} [\si{\arcsecond}] =7\rlap{.}''8 \times (800 / \nu[\U{GHz}])$ \citep{gusten2006atacama}. It thus varies at the observed frequencies between $\SI{22.3}{\arcsecond}$ and $\SI{29.3}{\arcsecond}$ (respectively corresponding to $0.14$\,pc and $0.18$\,pc at the distance of M8).
        
        \subsection{IRAM 30m telescope observations with EMIR 090}
        The IRAM 30m telescope is located in the Spanish Sierra Nevada on Pico Veleta at an altitude of 2850\,$\si{\meter}$\footnote{Institut de Radioastronomie Millimétrique, \url{https://www.iram-institute.org/EN/30-meter-telescope.php}}. The data were taken under project ID 141-21 (P.I. Friedrich Wyrowski) during several runs between 2022 March and June using the $\SI{3}{\milli\meter}$ band (`Band 1') of the heterodyne Eight MIxer Receiver \citep[EMIR\,090,][]{Carter2012EMIR}. Similar to nFLASH, EMIR is a 2SB two polarisation heterodyne receiver with a central sideband separation of $\SI{16}{\giga\hertz}$ and individual sideband bandwidths of $\SI{8}{\giga\hertz}$. Using three setups, the observations cover a total bandwidth of $\SI{40.3}{\giga\hertz}$ between $\SI{70}{\giga\hertz}$ and $\SI{117}{\giga\hertz}$. Additional on-off data of clump E taken by~\citet{tiwari2020M8E} were used to increase the sensitivity and frequency coverage  for this particular position. An overview of the frequency setups is shown in Appendix~\ref{app:obs}.
        
        EMIR was used in combination with the FFTS backend FTS200 that provides a total of 20737 frequency channels per $\SI{4}{\giga\hertz}$ wide sideband, resulting in a channel spacing of $\SI{192.89}{\kilo\hertz}$. This corresponds to $\SI{0.82}{\kilo\meter\per\second}$ and $\SI{0.50}{\kilo\meter\per\second}$ at our lowest and highest observed frequencies of $\SI{70.3}{\giga\hertz}$ and $\SI{116.5}{\giga\hertz}$, respectively. While this resolution is sufficient for the broader bright lines, weak and narrow spectral lines are only covered by a few channels. At this resolution, the data have an average RMS noise level of $\SI{0.17}{\kelvin}$.
        
        The EMIR system temperatures during the observations varied for frequencies below $\SI{80}{\giga\hertz}$ between $\SI{130}{\kelvin}$ and $\SI{290}{\kelvin}$, from $\SI{80}{\giga\hertz}$ to $\SI{105}{\giga\hertz}$ between $\SI{80}{\kelvin}$ and $\SI{200}{\kelvin}$, and above $\SI{105}{\giga\hertz}$ between $\SI{150}{\kelvin}$ and $\SI{480}{\kelvin}$. The conversion from antenna temperature $T_{\U{A}}^*$ to main-beam brightness temperature $T_\U{MB}$ is given by \mbox{$T_\U{MB}=T_\U{A}^*\times \eta_\U{FW} / \eta_\U{MB}$}, with the main-beam efficiency $\eta_\U{FW}$ and the forward coupling efficiency $\eta_\U{FW}$. Based on the average observed frequency of $\SI{93.4}{\giga\hertz}$, we assumed the default conversion factor\footnote{\label{note:iram}\url{https://publicwiki.iram.es/Iram30mEfficiencies}} described by $\eta_\U{FW} = 0.946$ and $\eta_\U{MB} = 0.797$ for the calibration.
        
        The FWHM, $\theta_\U{B}$, of the 30m telescope beam is\footref{note:iram} $\theta_\U{B}[\si{\arcsecond}] = 2460/\nu[\si{\giga\hertz}]$, with $\nu$ being the observed frequency in GHz. Therefore, the beam widths vary between $\SI{35.0}{\arcsecond}$ and $\SI{21.1}{\arcsecond}$ (corresponding to 0.22\,pc and 0.13\,pc at the distance of M8) for frequencies between $\SI{70.3}{\giga\hertz}$ and $\SI{116.5}{\giga\hertz}$, respectively.
        
        \subsection{Data reduction of spectroscopic data}
        The spectra taken with the APEX and the IRAM 30m telescope were reduced using the CLASS program of the GILDAS\footnote{\url{http://www.iram.fr/IRAMFR/GILDAS}} software package developed by IRAM. The spectra taken for each clump were combined and a first-order baseline was subtracted. This baseline was determined by averaged spectral channels located off, but in the vicinity, of the individual spectral lines. 
        
        In order to correct the CO and $^{13}$CO signal affected by a contaminated reference position, the spectrum observed at M8REF was re-added to the corresponding transitions. APEX observations obtained in frequency switching mode were compared to the on-off observations and combined if the residual between both spectra did not show emission with a significance above three times the baseline RMS.
        
        The observations taken with the IRAM 30m telescope in 2022 June were affected by a technical defect that caused a frequency and sideband-dependent shift of the observed frequencies by approximately $\SI{4}{\mega\hertz}$. This shift was corrected for all setups based on a comparison of the spectra at clump E taken before and in June. As this clump was observed at the start of each observing day, and also previously by~\citet{tiwari2020M8E}, it was possible to obtain a correction of the frequency scale for each band based on Gaussian fits to the strongest optically thin transitions. 
        
        \subsection{Archival continuum data}\label{subsec:obs:continuum}
        To derive the physical properties of the M8 clumps, we fitted their SEDs as in~\citet{urquhart2018atlasgal} to compare the M8 clumps to the sources identified through the APEX Telescope Large Area Survey of the Galaxy (ATLASGAL). For this, we used archival data from \textit{Spitzer}, \textit{Herschel}, APEX, the Midcourse Space Experiment~\citep[MSX;][]{price2001midcourse}, and the Wide-field Infrared Survey Explorer~\citep[WISE;][]{wright2010wise}. While WISE and MSX provide all-sky surveys, dedicated Galactic plane surveys have been performed with \textit{Spitzer}, \textit{Herschel}, and APEX. We used data from the GLIMPSE~\citep{Churchwell2009glimpse}, MIPSGAL~\citep{Carey2009mipsgal}, Hi-GAL~\citep{Molinari2010higal}, and ATLASGAL~\citep{Schuller2009atlasgal} surveys.
        
        Since M8 is located at a galactic latitude of $b=\SI{-1.3}{\degree}$, the Hi-GAL maps do not fully cover the nebula. Due to this, these surveys were supplemented with data from the AKARI~\citep{Doi2015akari} all-sky survey. In addition, we also used the $\SI{850}{\micro\meter}$ data of the Lagoon Nebula taken with the Submillimetre Common-User Bolometer Array~\citep[SCUBA;][]{Holland1999SCUBA} of the JCMT by~\citet{tothill2002structure}.
        
        The flux densities of each clump were extracted analogously to ~\citet{urquhart2018atlasgal} using several tools of the \texttt{astropy}~\citep{astropy:2022} and \texttt{Photutils}~\citep{bradley2022:photutils} packages for Python. For this, the flux density of each clump was extracted in an aperture of two or three times the clump size derived by \citet{tothill2002structure}, depending on the proximity of neighbouring clumps. This flux density was corrected for background flux based on the median flux density in an annulus around the respective clump. The RMS of the Gaussian noise was calculated for each band based on emission inside a defined mask of all annuli around the clumps, which excludes the clump emission.
        
        \begin{figure*}
                \begin{minipage}[c]{0.79\textwidth}
                        \includegraphics[width=0.999\textwidth]{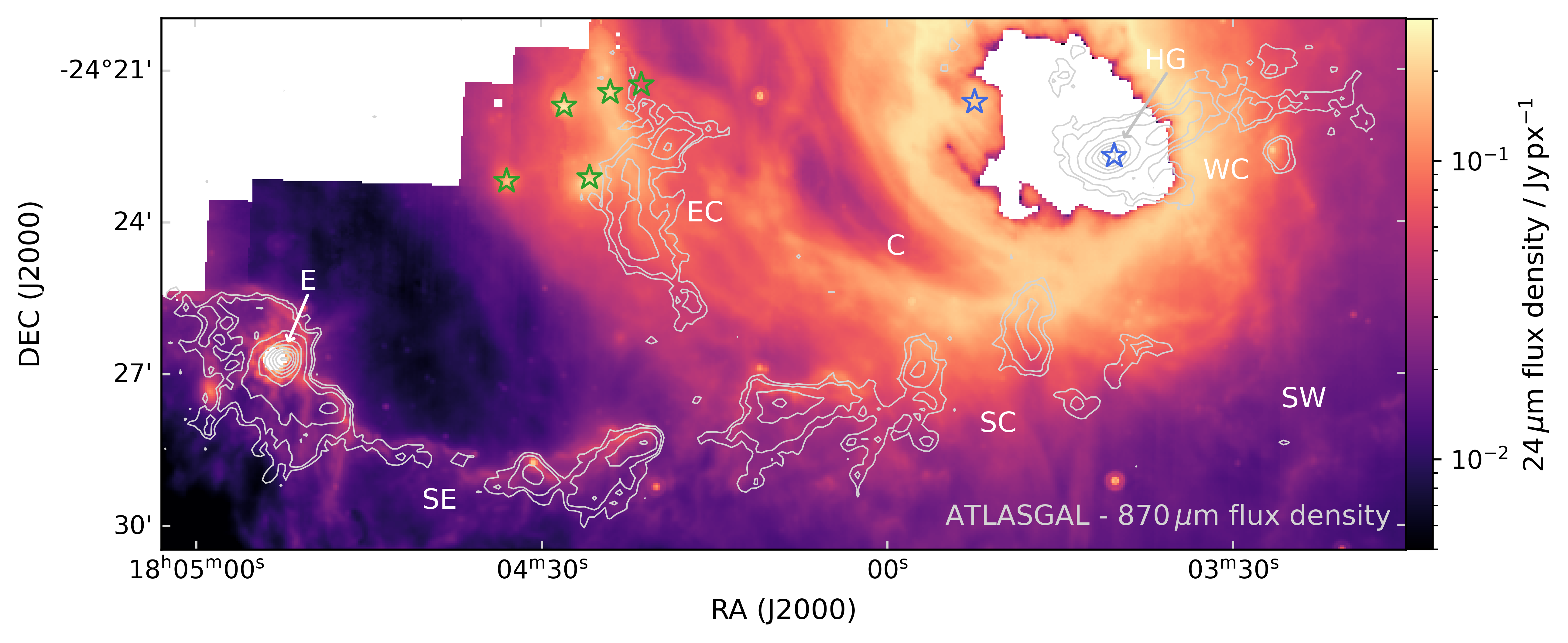}
                \end{minipage}\hfill
                \begin{minipage}[c]{0.20\textwidth}
                        \caption{Dust continuum image  of M8 at $\SI{24}{\micro\meter}$ from MIPSGAL. The ATLASGAL $\SI{870}{\micro\meter}$ emission is shown as contours. Blue and green stars show the positions of O- and early B-type stars in this region~\citep{Wright2019OBstars}, respectively. The image is saturated near the \hii region M8-Main.} \label{fig:dust:20micfull}
                \end{minipage}
        \end{figure*}
        The AKARI FWHM beam sizes of $\SI{63.4}{\arcsecond}$ at $\SI{65}{\micro\meter}$ (N60),  $\SI{77.8}{\arcsecond}$ at $\SI{90}{\micro\meter}$ (WIDE-S), and $\SI{88.3}{\arcsecond}$ for the $\SI{140}{\micro\meter}$ and $\SI{160}{\micro\meter}$ bands (WIDE-L and N160)~\citep{Takita2015akari} are not sufficient to resolve the M8 clumps that have FWHM sizes smaller than  $\SI{40}{\arcsecond}$. Due to this, the AKARI flux density was extracted at the exact position of the M8 clumps (see Table~\ref{tab:app:clumppositions}). This allowed the determination of the flux density within the corresponding AKARI beam, which covers most of the respective emission for the M8 clumps. In cases where multiple clumps are contained inside the extracted AKARI beam, the individual contribution of each clump was estimated based on the area fraction of each of the respective clumps inside the beam and the corresponding $\SI{350}{\micro\meter}$ Hi-GAL flux densities of the clumps. Due to the low resolution, we estimated the AKARI flux densities to have a measurement uncertainty of 50\% in addition to the uncertainty introduced by the RMS noise. In order to verify that the use of AKARI instead of Hi-GAL for wavelengths between $\SI{65}{\micro\meter}$ and $\SI{160}{\micro\meter}$ leads to results that are comparable to the ATLASGAL sample of clumps~\citep{urquhart2018atlasgal}, we tested the modified method on a sample of clumps in the NGC\,6334 cloud complex, which is similar in distance as M8. This comparison is presented in Appendix~\ref{app:akari}, where we find almost identical luminosities and only minor deviations in the derived masses, which can likely be attributed to the different methods in source size computation, instead of the usage of PACS rather than AKARI data.
        %
        %
        \section{Dust continuum emission at M8}\label{sec:dust}
        Figure~\ref{fig:dust:20micfull} shows the MIPSGAL 24\,$\mu$m image overlaid with the contours of the ATLASGAL $\SI{870}{\micro\meter}$ emission. In contrast to the optical image (see Fig.~\ref{fig:int:optical}), the dust continuum emission at $\SI{24}{\micro\meter}$ not only shows an emission peak at M8-Main (at clump HG, extending to WC1-6) but also a peak of similar brightness at clump E in the massive star-forming region M8 East. The position of HG is, within a few arcseconds, coincident with that of the O7.5V star Her\,36, one of main the ionisation sources. Additional fainter, point-like sources can be seen in the vicinity of the clumps WC7, SE2, SE3, SE7, SE8, and SC1.
        These IR-bright clumps may contain intermediate- to high-mass YSOs, of which the IR radiation penetrates the surrounding colder dust~\citep{konig2017atlasgal}. Further 24\,$\mu$m emission is located in the vicinity of the EC region, slightly offset from the M8 clumps and extending to the central region of the nebula. As this emission does not coincide with the $\SI{870}{\micro\meter}$ emission from the clumps, it likely originates from a diffuse foreground gas layer. Individual point-like 24\,$\mu$m sources in this region coincide with the positions of stars from the open cluster NGC\,6530 (see Fig.~\ref{fig:dust:20micfull}).
        
        \begin{figure}[tbp]
                \centering
                \includegraphics[width=0.499\textwidth]{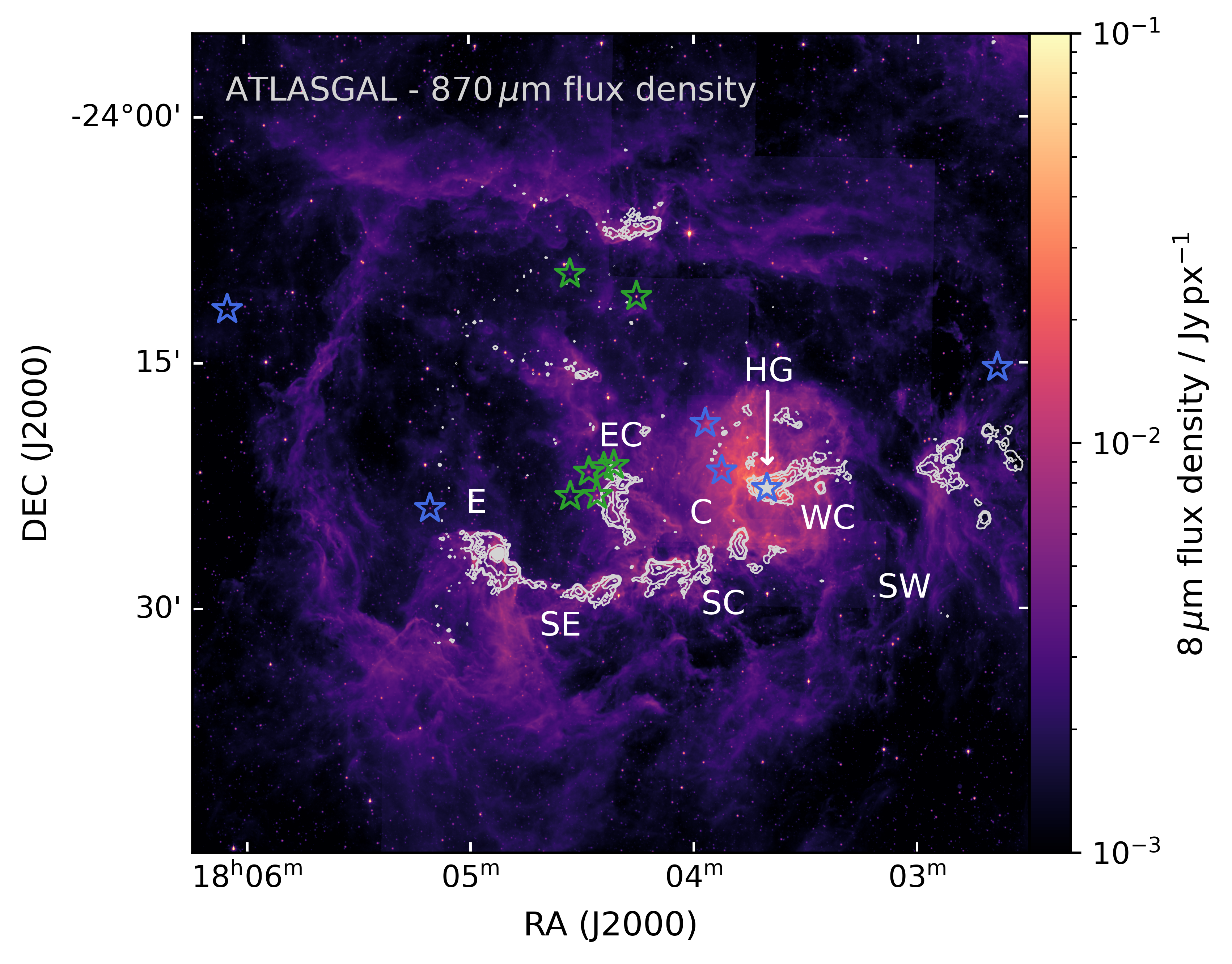}
                \caption[GLIMPSE $\SI{8}{\micro\meter}$ dust continuum image of the Lagoon Nebula]{$\SI{8}{\micro\meter}$ dust continuum image obtained from the GLIMPSE survey. Blue and green stars respectively mark the positions of present O- and B-type stars \citep{Wright2019OBstars}. Contours of the ATLASGAL $\SI{870}{\micro\meter}$ emission are shown in grey.}
                \label{fig:dust:8micfull}
        \end{figure}
        
        \begin{figure*}
                \begin{minipage}[c]{0.79\textwidth}
                        \includegraphics[width=0.999\textwidth]{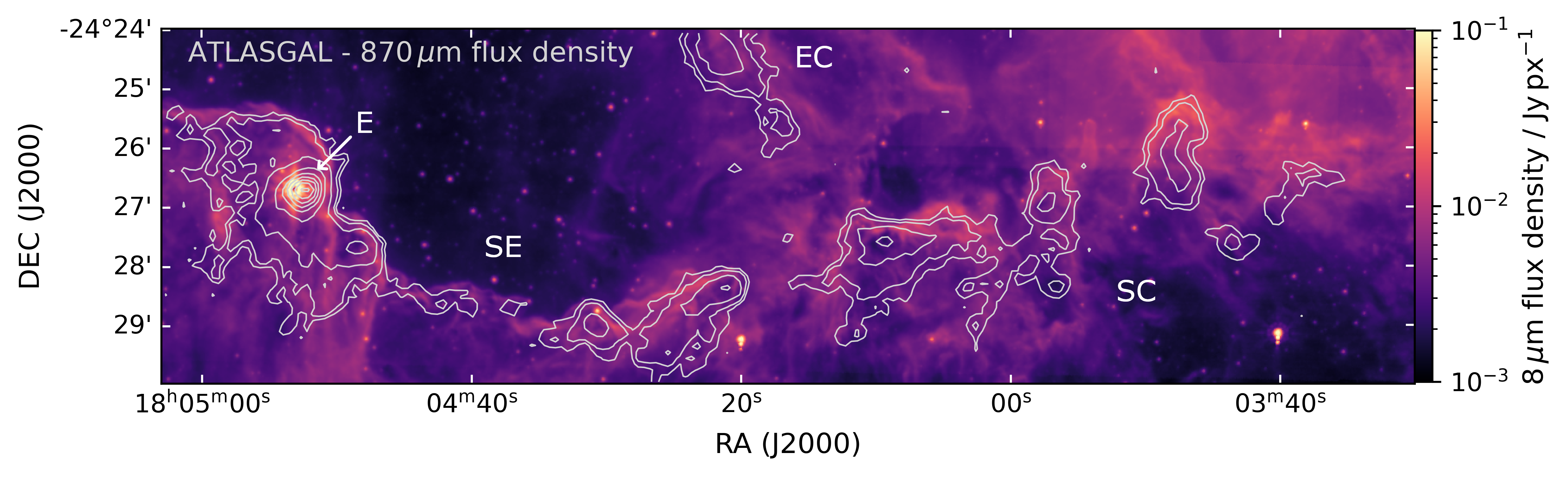}
                \end{minipage}\hfill
                \begin{minipage}[c]{0.20\textwidth}
                        \caption{$\SI{8}{\micro\meter}$ dust continuum image of the southern clumps in the Lagoon Nebula obtained from the GLIMPSE survey. Contours of the ATLASGAL $\SI{870}{\micro\meter}$ emission are shown in grey.}
                        \label{fig:dust:8micsouth}
                \end{minipage}
        \end{figure*}
        
        As shown in Fig.~\ref{fig:dust:8micfull}, the $\SI{8}{\micro\meter}$ emission extends in a bubble-like structure around the main condensations of the Lagoon Nebula (see \citealt{Deharveng2010bubbles} for other examples). Further inspections of the individual clumps additionally reveal that the emission is strongest on the edges of the clumps traced by the $\SI{870}{\micro\meter}$ emission (see Fig.~\ref{fig:dust:8micsouth}). The $\SI{8}{\micro\meter}$ band includes fluorescent emission from polycyclic aromatic hydrocarbons \citep[PAHs;][]{Draine2007PAHs}, which is pumped by far-UV photons radiated by the present O-type stars. This is a clear indicator of the feedback from the stars on the surrounding remnant gas. In particular, the emission peaking on the clump edges indicates the presence of PDRs on clump surfaces across the nebula.
        
        To examine the impact of stellar feedback on the physical properties of molecular clumps in the Lagoon Nebula, they were compared to the ATLASGAL sample of clumps examined by~\citet{urquhart2018atlasgal}. Analogously to these authors' analysis, we reconstructed the cold dust SED of the clumps using a modified blackbody model:
        \begin{equation}
                F_\lambda (T_\U{d}, \tau_{\lambda,\U{ref}})= \Omega_\U{d} \cdot B_\lambda(T_\U{d})\cdot\left[1-\exp\left(-\tau_{\lambda,\U{ref}}\left(\frac{\SI{870}{\micro\meter}}{\lambda}\right)^\beta \right) \right].
        \end{equation}
        In this model, the dust blackbody emission, $B_\lambda(T_\U{d})$,  at the dust temperature $T_\U{d}$ is modified by a factor that is composed of the opacity $\tau_{\lambda_\U{ref}}$ at the reference wavelength $\lambda_\U{ref} = \SI{870}{\micro\meter}$ and a wavelength-dependent power law.
        This reference wavelength was chosen such that it matches the value used by~\citet{urquhart2018atlasgal}. The spectral index $\beta$ was set to a fixed value of 1.75, which corresponds to the mean value across the dust models of~\citet{ossenkopf1994beta} for star-forming regions. The clump size $\Omega_\U{d}$ is given by $\pi R_\U{d}^2$ with the clump radius $R_\U{d}$, which we set to half of the FWHM source size of the respective clumps.
        
        We used a two-component model to describe the full SED of clumps, presuming internal heating based on the examination of the $\SI{24}{\micro\meter}$ emission. The second component resembles a black body spectrum of a hotter embedded object:
        \begin{equation}
                F_\lambda (T_\U{d}, \tau_{\lambda,\U{ref}}, T_\U{h},\Omega_\U{h})=         F_\lambda (T_\U{d}, \tau_{\lambda,\U{ref}}) + \Omega_\U{h} \cdot B_\lambda(T_\U{h})
        .\end{equation}
        In addition to the parameters used in the single-component model, the size of the compact blackbody component $\Omega_\U{h}$ and its temperature $T_\U{h}$ were determined.
        
        For the cold component fit, only data points with  longer than $\SI{65}{\micro\meter}$ were used. Due to the presence of diffuse warm gas in the vicinity of M8-Main and the EC region, mid-IR flux densities extracted at the associated clump positions are likely to be unrelated to the actual clump emission and were therefore also excluded for the fitting.
        
        As mentioned above, the $\SI{8}{\micro\meter}$ band is dominated by the emission of PAHs in the outer layers of the clumps. Flux detected at this wavelength was used as upper limit when fitting the hot SED component, to avoid overestimating the continuum flux originating from the embedded object. Since the $\SI{65}{\micro\meter}$ band might contain additional emission from very small grains~\citep{compi2010vsg}, the flux density at this wavelength was also used as an upper limit, in order to avoid an overestimation of the dust temperature. 
        
        Mass $M$ and H$_2$ column density $N(\mathrm{H}_2)$ of the clumps were calculated according to the equations (1) and (2) of~\citet{Schuller2009atlasgal}. The bolometric luminosity $L$ of the clumps was derived by integrating the flux density of the reconstructed models between $\SI{1}{\micro\meter}$ and $\SI{1000}{\micro\meter}$ and assuming isotropically radiating sources. The derived quantities and the corresponding SED plots can be found in Appendix~\ref{app:sed}. Fig.~\ref{fig:dust:L_M} illustrates the derived distribution of dust temperatures and clump masses. In particular, the central clumps C1--2 and clumps surrounding the central condensations in M8-Main show increased temperatures and comparably small clump masses. As discussed further in Sect.~\ref{sec:discussion}, these lower masses and increased dust temperatures are found for the whole sample of M8 clumps when comparing them to the ATLASGAL sample of clumps in the inner galaxy.
        
        \begin{figure}[t]
                {\centering
                        \includegraphics[width=0.499\textwidth]{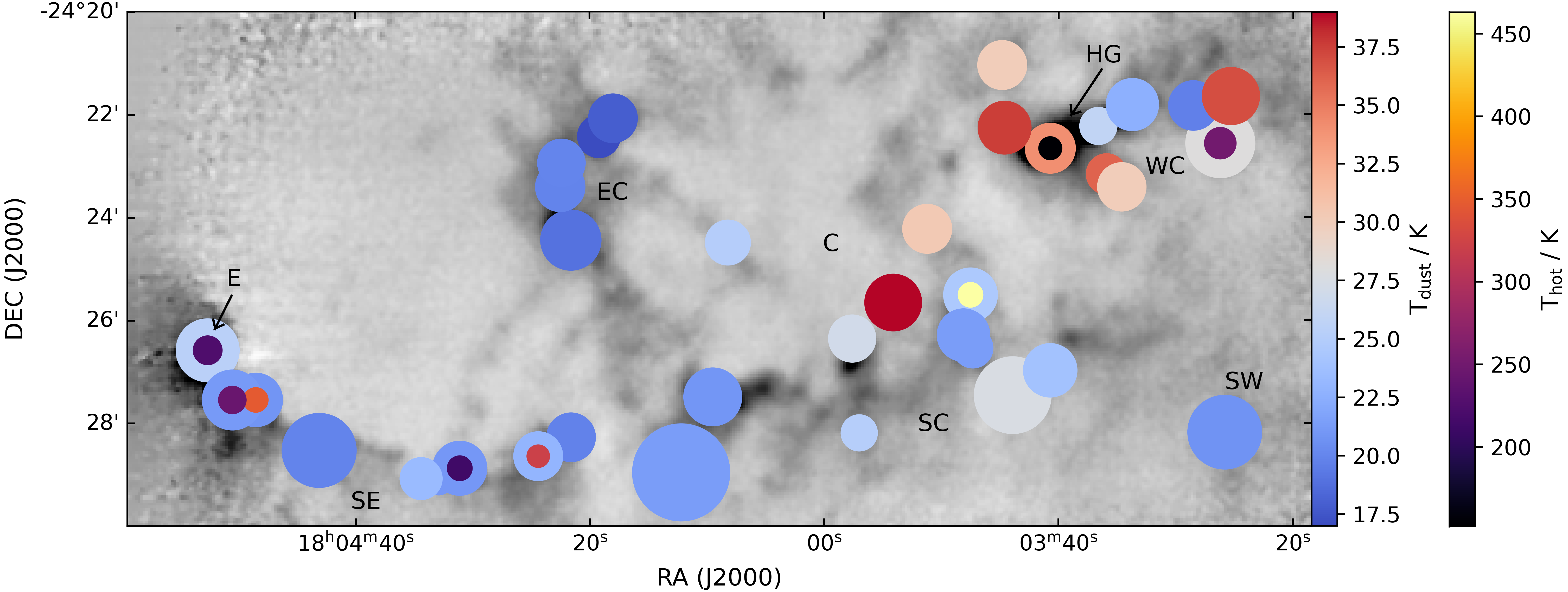}}
                \hspace*{0.01cm}\includegraphics[width=0.452\textwidth]{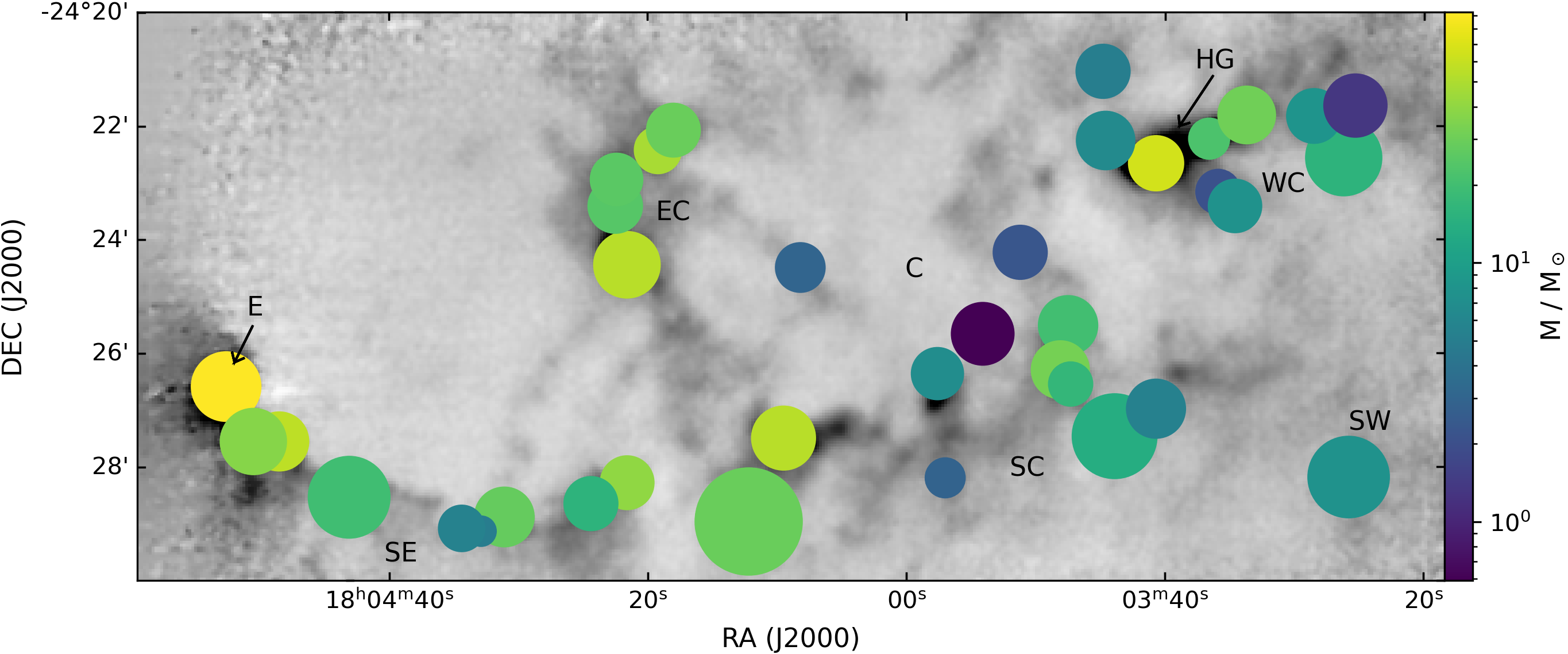}
                \caption[Dust temperatures and masses of clumps in the Lagoon Nebula.]{Dust temperatures (upper panel) and masses (lower panel) of clumps in the Lagoon Nebula. The circle size corresponds to the aperture size used for flux extraction in the analysis. For the temperatures, the smaller inner circles visualise the derived temperatures of the hot components. The grey-scale background image shows the JCMT SCUBA $\SI{870}{\micro\meter}$ dust continuum flux (see Fig.~\ref{fig:int:clumps}).}
                \label{fig:dust:L_M}
        \end{figure}
        
        %
        %
        \section{Line survey of M8 clumps}\label{sec:linesurvey}
        The rotational molecular line transitions in the M8 clumps were identified by matching rest frequencies from the Jet Propulsion Laboratory Line Catalog \citep[JPLC\footnote{\url{https://spec.jpl.nasa.gov}};][]{Pickett1998JPL} and the Cologne Database for Molecular Spectroscopy \citep[CDMS\footnote{\url{https://cdms.astro.uni-koeln.de}};][]{Mueller2001cdms,Mueller2005cdms,Endres2016cdms} to the observed spectra.
        In the first step, we visually inspected the spectrum of E, as we expected the lines to be brightest towards this massive star-forming region and because E has the longest integrated observing time of the clumps in the sample. Each significant line detected towards E was matched to a transition in the JPLC and CDMS with upper-level energy $E_\U{up} <\SI{200}{\kelvin}$. If transitions of multiple species have rest frequencies close to observed transitions, the identification favours species commonly present in the ISM and transitions with low $E_\U{up}$ and high Einstein A coefficients $A_\U{ij}$.
        
        For the remaining M8 clumps, we executed a \texttt{CLASS} script to examine the respective spectra for line emission. The script flagged emission inside a $\SI{10}{\kilo\meter\per\second}$ interval around the respective clump velocities (see Sect.~\ref{subsec:vlsr}) for all rest frequencies of transitions identified at E. All of the spectra were then visually inspected to identify any lines that are not present at E and to rule out false detections due to emission from unrelated nearby transitions. This visual inspection led to the detection of N$_2$D$^+$, which is not seen at E.
        
        A transition is considered to be detected if it has at least three adjacent channels with intensity higher than three times the baseline RMS noise for velocity resolutions of $\SI{0.4}{\kilo\meter\per\second}$ (APEX) or $\SI{0.7}{\kilo\meter\per\second}$ (IRAM 30m). Line candidates that show at least one channel above three times the RMS noise were confirmed based on their appearance in both polarisations and the presence of other transitions of the same species in the corresponding clump.
        
        Across all clumps in the nebula, it was possible to identify a total of 346 transitions of 70 molecular species, including isotopologues. Table~\ref{tab:identified_species} provides an overview of all the detected species in the M8 region.
        
        \begin{table*}[thbp]
                \caption{Molecular species detected in the M8 region.}
                \label{tab:identified_species}
                \centering
                \begin{tabular}{ccccccc}
                        \hline \hline
                        Carbon chains & S-bearing & O-bearing & N-bearing & Deuterated & COMs & Others \\\hline
                        c-C$_3$H$_2$    &SO                     &  H$_2$CO                  & CN            & HDCO          & CH$_3$OH          & SiO \\
                        C$_2$H          &$^{34}$SO              &  H$_2^{13}$CO             & $^{13}$CN     & C$_2$D        & $^{13}$CH$_3$OH   & CF$^+$\\
                        C$^{13}$CH      &H$_2$S                 &  HCO                      & N$_2$H$^+$    & DCO$^+$       & CH$_3$SH          &\\
                        C$_4$H          &HCS$^+$                &  HCO$^+$                  & HCN           & N$_2$D$^+$    & CH$_3$CHO & \\
                        HC$_3$N         &H$_2$CS                &  H$^{13}$CO$^+$           & H$^{13}$CN    & DNC           & CH$_3$C$_2$H & \\
                        H$^{13}$CCCN    &H$_2^{34}$CS           &  HC$^{18}$O$^+$           & HC$^{15}$N    & DCN           & CH$_3$CN   & \\
                        HC$^{13}$CCN    & SO$_2$                &  CO                       & HNC           & DC$_3$N       &           & \\
                        HCC$^{13}$CN    & SO$^+$                &  $^{13}$CO                & HN$^{13}$C    & NH$_2$D       &  &\\
                        HCCC$^{15}$N    &NS                     &  C$^{17}$O                & H$^{15}$NC    & HDCS          &  &\\
                        c-C$_3$H        & OCS                   &  C$^{18}$O                & HNCO          & & &\\
                        C$_3$H$^+$      &  CS                   &  $^{13}$C$^{18}$O         & HCNO          & & &\\
                        HC$_5$N         &  $^{13}$CS            &  H$_2$C$_2$O                       & NO            & & &\\
                        & C$^{33}$S             &  t-HCO$_2$H           && & &  \\
                        &  C$^{34}$S            &                           &               & & & \\
                        & $^{13}$C$^{34}$S      &                           &               & & &\\
                        & C$_2$S &                           &               & & &\\\hline

                \end{tabular}
                \tablefoot{With the exception of N$_2$D$^+$, all species are detected in clump E. Here, some of the listed N-bearing species also contain an O atom.}
        \end{table*}
        
        All species (except N$_2$D$^+$) and most of their emission lines were observed towards clump E, which hosts the embedded YSO M8E-IR. The large chemical richness of this object is explained by it being an early-stage massive star-forming region with an associated PDR. In addition, it was possible to detect also fainter transitions in the $\SI{3}{\milli\meter}$ band towards this clump, as the increased observing time at E reduces the RMS noise of the combined spectrum to $\SI{10}{\milli\kelvin}$, as compared to $\SI{17}{\milli\kelvin}$ on average for the other clumps. 
        
\begin{figure}[tbp]
                \centering
                \includegraphics[width=0.499\textwidth]{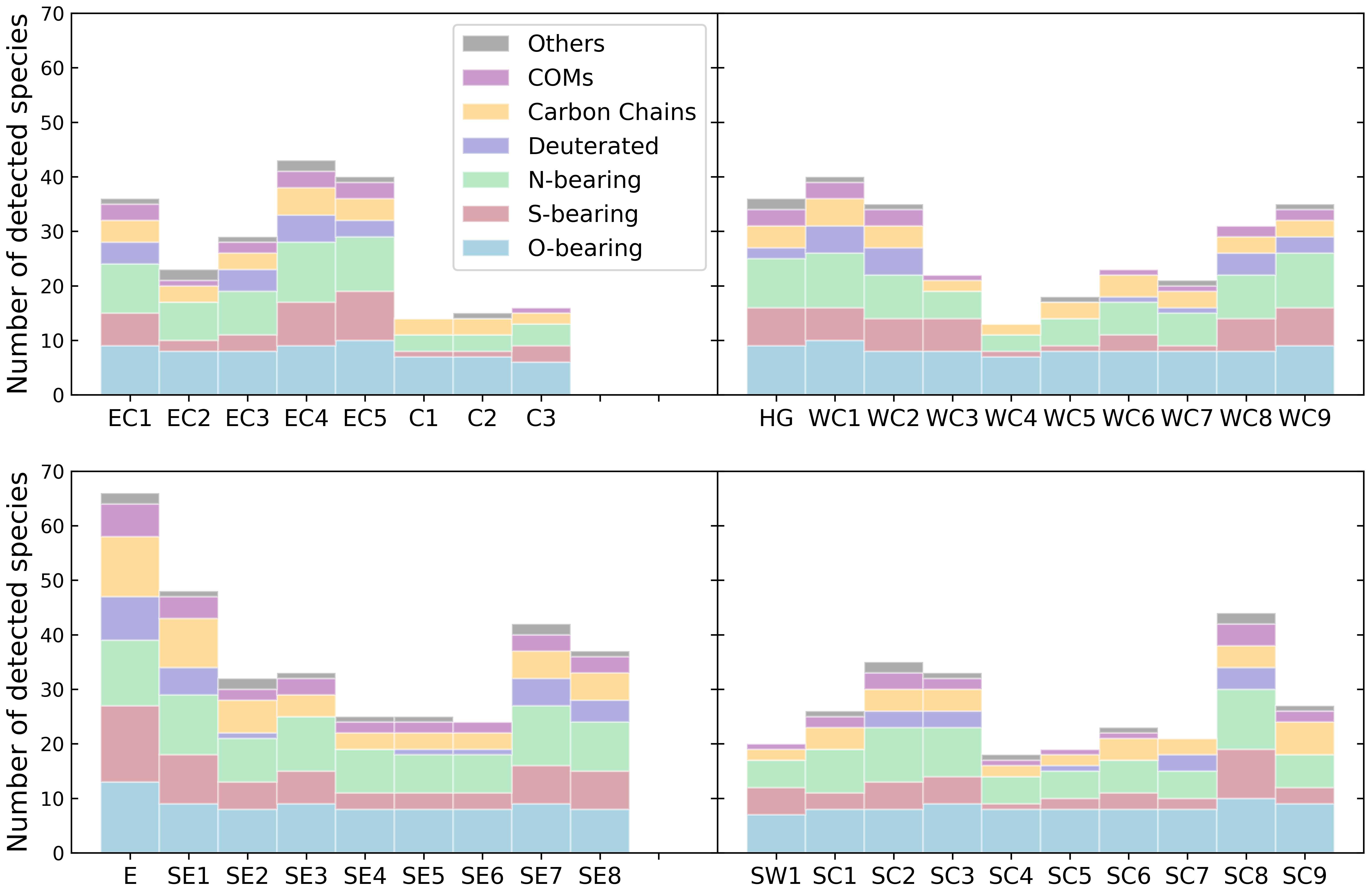}
                \caption[Overview of the number of detected species in each clump in M8]{Distribution of species across the M8 clumps. Different colours differentiate between S-, O-, and N-bearing species, deuterated species, carbon chains, complex organic molecules (COMs), and other species, as given in Table~\ref{tab:identified_species}.}
                \label{fig:lin:linenr}
        \end{figure}

        Fig.~\ref{fig:lin:linenr} shows that almost all M8 clumps have a complex chemistry that varies across the cloud with an average of 30 species detected at each position. While the category of O-bearing species contains mostly molecules with bright lines detected in the entire cloud, the clumps show interesting differences in the number of (and ratio between) detections of N-bearing, S-bearing, and deuterated species. For instance, the number of nitrogen- and sulphur-bearing species detected in the SE region is overall larger than in the SC clumps. In contrast, we observe each region to have a few individual clumps with a large number of deuterated species, without a pronounced trend of a whole filament with a higher deuteration fraction.
        
        The chemistry seen in the individual clumps is discussed further in Sects.~\ref{subsec:cd} and~\ref{subsec:disc:chemistry}. A detailed description of the individual detected transitions in each clump can be found in Appendix~\ref{app:lineidentification}, which lists the line parameters of each detected transition in Table~\ref{tab:app:idtransitions} as well as the derived line properties in Table~\ref{tab:app:line_properties}.
        
        \subsection{Line properties} \label{subsec:lin:properties}
        In order to further analyse the chemical properties of the clumps in M8, Gaussian profiles were fitted to all detected transitions. In general, the fits are performed using the \texttt{GAUSS} method of the \texttt{MINIMIZE} function of CLASS, which returns the integrated intensity, $I$, the FWHM line width, $\Delta v$, and the line of sight (LOS) velocity with respect to the local standard of rest (LSR) velocity, $\varv_\U{LSR}$, of a given line. Partially blended transitions and blended velocity components of the same transition were fitted simultaneously. If well resolved, a maximum of two Gaussian components of the same transition were fitted as possible additional components were almost exclusively detected for transitions of CO isotopes.
        
        The spectra partially resolve the hyperfine structure (HFS) due to the non-zero nuclear spin of $^{14}$N for DCN, HCN, H$^{13}$CN, N$_2$H$^+$, N$_2$D$^+$, and NH$_2$D. In addition, a splitting of the transitions of C$^{33}$S and C$^{17}$O is observed as a consequence of the non-zero nuclear spin of $^{33}$S and $^{17}$O. Corresponding transitions show a non-Gaussian line profile and were therefore fitted using the \texttt{HFS} method of \texttt{MINIMIZE}, which additionally returns the optical depths of the transitions. This function was also used to fit the well-resolved hyperfine components of lines from NS, NO, HCO, c-C$_3$H, CN, $^{13}$CN, C$_2$H, C$_2$D, and C$^{13}$CH. 
        
        Observed line intensities of the CN $\SI{226}{\giga\hertz}$ transitions do not match the relative intensities expected based on the HFS calculations provided by the CDMS. A similar behaviour is noted by~\citet{Kim2020pdrtracer} for the HFS transitions of CN at $\SI{113}{\giga\hertz}$ and HCO at $\SI{87}{\giga\hertz}$, which they attribute to optical depth effects and non-local thermodynamic equilibrium (LTE) excitation. As the corresponding transitions of CN and HCO are only weakly affected in the M8 clumps, we used the respective \texttt{HFS} fits for the column density estimation in Sect.~\ref{subsec:cd}. In contrast, we refrained from computing column densities for the CN $\SI{226}{\giga\hertz}$ transitions. 
        
        A complete overview of the fitted line profiles for detected transitions is provided in Table~\ref{tab:app:line_properties} of Appendix~\ref{app:lineidentification}. For emission lines with multiple velocity components that are blended due to small differences in their observed frequencies, we give the integrated intensities over the full line profiles. The results of this survey are analysed and discussed in Sects.~\ref{sec:lineanalysis} and~\ref{sec:discussion}.
        
        \subsection{Radio recombination lines} \label{subsec:lin:rrls}
        \begin{figure*}
                \begin{minipage}[c]{0.65\textwidth}
                        \includegraphics[width=0.999\textwidth]{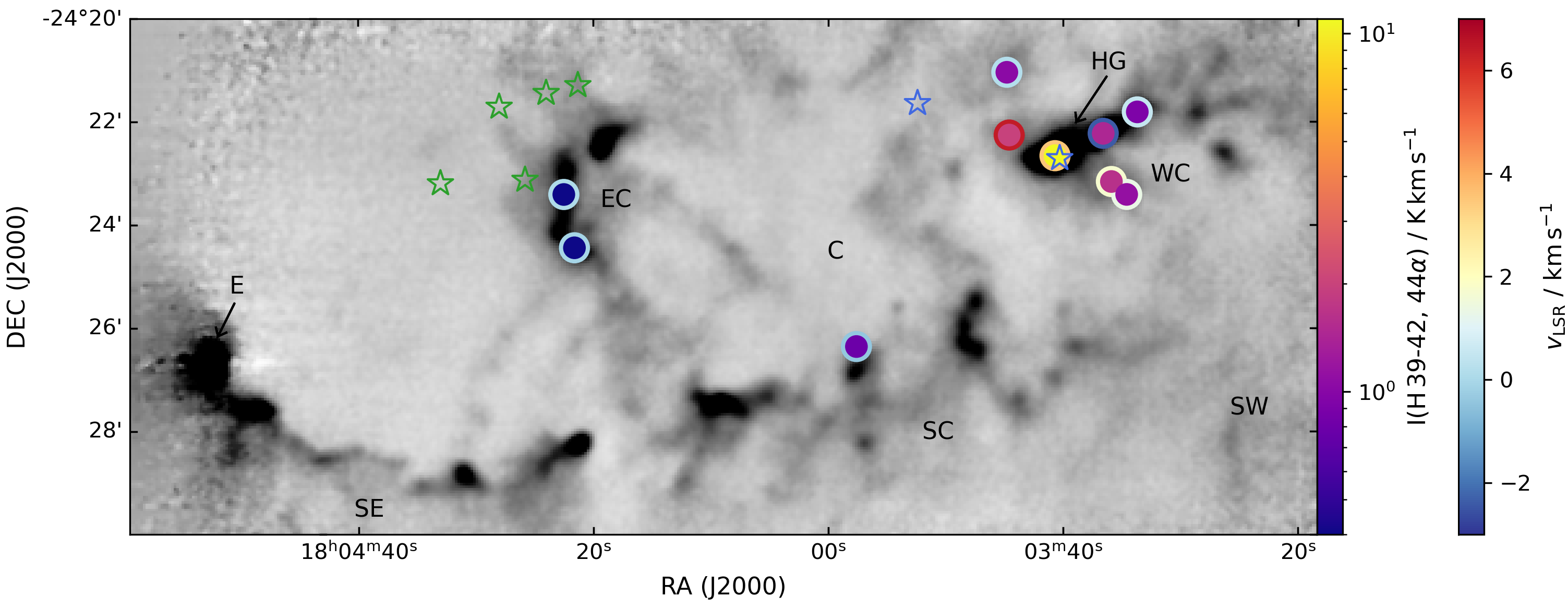}
                \end{minipage}\hfill
                \begin{minipage}[c]{0.33\textwidth}
                        \caption{Intensity and LOS velocities of the H\,39--44$\alpha$ transitions observed at the positions of the clumps in the Lagoon Nebula. The fillings of the coloured dots indicate the intensities and the borders describe the velocities. Blue and green stars respectively mark the positions of present O- and B-type stars. The grey-scale background image shows the JCMT SCUBA $\SI{870}{\micro\meter}$ dust continuum flux (see Fig.~\ref{fig:int:clumps}).}
                        \label{fig:lin:map_halpha}
                \end{minipage}
        \end{figure*}
        In addition to the molecular rotational transitions, radio recombination lines (RRLs) of hydrogen, helium, and carbon are detected in the vicinity of M8-Main. In particular, the frequency setups observed with the IRAM 30m telescope cover H\,N$\alpha$, He\,N$\alpha,$ and C\,N$\alpha$ recombination lines with N=39--42 and N=44. Towards HG, which coincides with the main ionisation source Her\,36, we additionally detect multiple H\,N$\beta$ and H\,N$\gamma$ transitions with N=48--56 and N=54--63, respectively. Using APEX, we also detect the H\,29--30$\alpha$ and H\,36--38$\beta$ lines at HG.
        
        For the analysis of the RRLs, we focused on the H\,N$\alpha$, He\,N$\alpha,$ and C\,N$\alpha$ transitions with N=39--42 and N=44. The covered transitions with different N were averaged along the velocity axis to create individual combined spectra of the respective RRL towards each clump. Gaussians were fitted to each combined spectrum, in order to derive the line properties for the detected RRLs. The results of these fits are given in Appendix~\ref{app:rrl}, where we also give the line parameters of the H\,N$\beta$ and H\,N$\gamma$ lines at HG. Figure~\ref{fig:lin:map_halpha} provides an overview of clumps with H\,39--44$\alpha$ emission and the respective line intensities.
        
        \begin{figure}[tbp]
                \centering
                \includegraphics[width=0.499\textwidth]{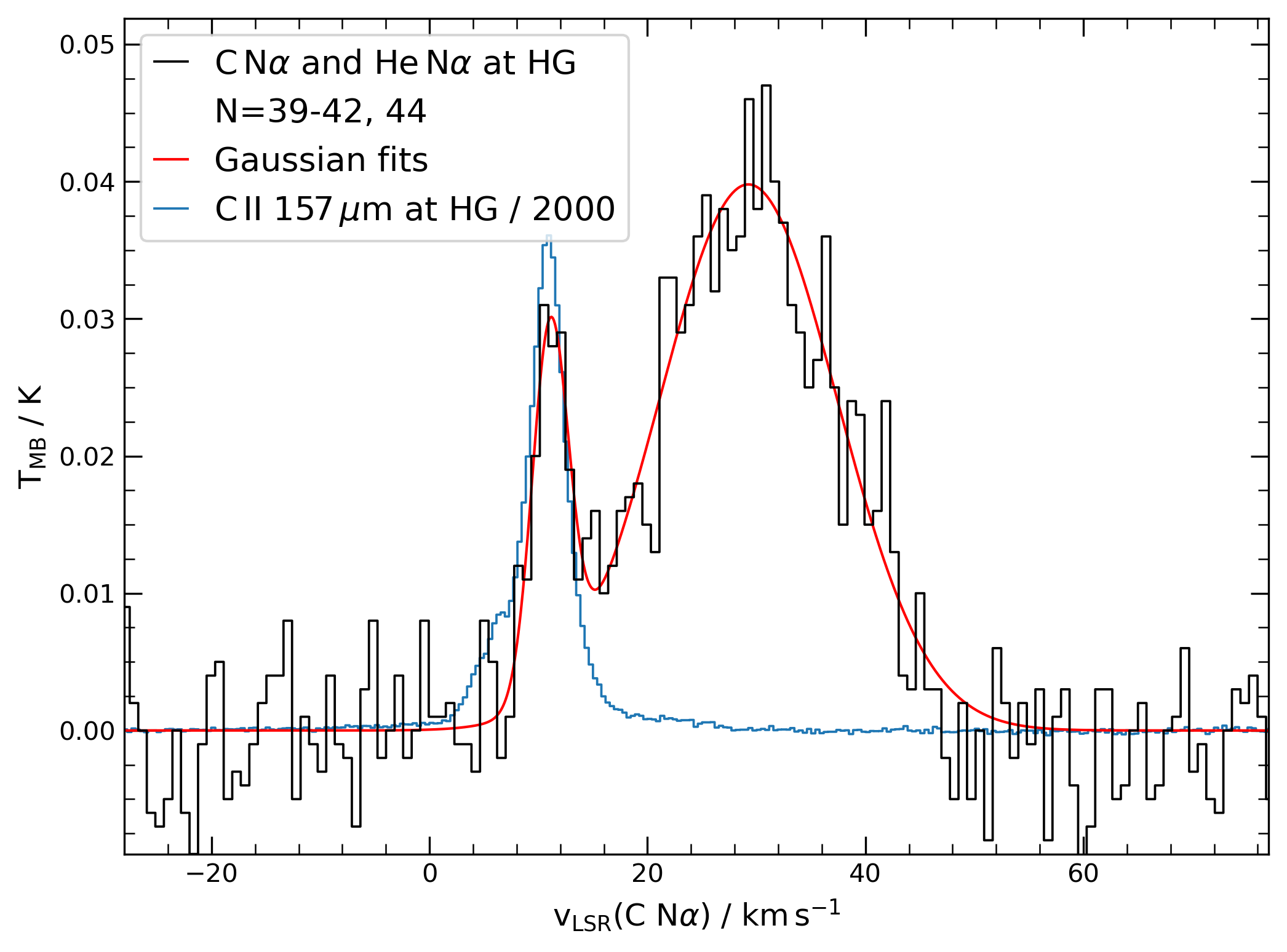}
                \caption[RRLs at HG]{C\,II $\SI{157}{\micro\meter}$ transition (blue) and stacked spectra of the He\,N$\alpha$ and C\,N$\alpha$ RRLs with N=39--42 and 44 (black) towards clump HG (Her\,36). The velocity scale of the RRLs is computed based on the rest frequency of the C\,N$\alpha$ lines. The Gaussians fitted to the RRLs are displayed in red. The temperature scale of the $158~\mu$m C\,II fine structure transition is divided by 2000 to compare it with the RRLs.}
                \label{fig:lin:Calpha}
        \end{figure}
        
        While the H\,N$\alpha$ and He\,N$\alpha$ RRLs observed towards HG have broad spectral line profiles and central velocities close to the systemic LSR velocity (between $\SI{2}{\kilo\meter\per\second}$ and $\SI{4}{\kilo\meter\per\second}$), the C\,N$\alpha$ lines are observed to be relatively narrow and shifted to velocities of $\SI{11.1}{\kilo\meter\per\second}$ (see Fig.~\ref{fig:lin:Calpha}). These findings are in agreement with the structure of M8-Main derived by~\citet{tiwari2018M8HG}, who place the dense molecular gas in the background of the \hii region at velocities between $\SI{10}{\kilo\meter\per\second}$ and $\SI{13}{\kilo\meter\per\second}$. The main fraction of ionised carbon hereby originates from the warm gas of the PDR between the dense molecular clump and the \hii region, as indicated by the narrow line widths and velocity shift of the C\,N$\alpha$ transitions. This is confirmed by the observations of the $\SI{157}{\micro\meter}$ C\,II transition at HG from~\citet{tiwari2018M8HG}, which is shown as the blue spectrum in Fig.~\ref{fig:lin:Calpha}. While the $\varv_\U{LSR}$ of the more strongly emitting C\,II component agrees with the velocities of the C\,N$\alpha$ RRLs, the weaker component at approximately $\SI{6}{\kilo\meter\per\second}$ is likely associated with the hotter foreground layer. This foreground gas contributes to most of the H\,N$\alpha$ and He\,N$\alpha$ emission and is expanding towards us with relative velocities between $\SI{2}{\kilo\meter\per\second}$ and $\SI{6}{\kilo\meter\per\second}$.
        
        In contrast to HG, the remaining clumps in the Lagoon Nebula either only show the H\,39--44$\alpha$ lines or no RRL emission at all (see Fig.~\ref{fig:lin:map_halpha}). Unsurprisingly, the brightest recombination lines are observed in the vicinity of HG, where the clumps, in projection, are closest to the O-type stars (Her\,36 and Sgr\,9). The velocities in this region hereby largely follow the trend observed for the CII emission in Fig. 6 of~\citet{tiwari2018M8HG}. Their channel maps indicate an enhanced ionisation of the molecular gas in WC4, which could possibly explain the low degree of chemical variety observed at this position (see Fig.~\ref{fig:lin:linenr}).
        Additional weak H\,39--44$\alpha$ emission is detected towards the EC clumps. The weak line strengths and the measured velocities of $\varv_\U{LSR}=\SI{0}{\kilo\meter\per\second}$ imply that the emission does not originate from the associated clumps, which we observe at velocities of order $\varv_\U{LSR}=\SI{16}{\kilo\meter\per\second}$. Instead, we might observe a less dense foreground gas layer, which is affected by the radiation of the nearby massive stars.
        
        \subsection{Methanol maser emission}
        Methanol masers are commonly associated with star formation. As described by~\citet{Menten1991maser}, these objects can be classified into two distinct classes. The Class II methanol masers are associated with the presence of high-mass protostars~\citep{Urquhart2015maserSF}, where they are presumably pumped by the radiation of the surrounding warm dust~\citep{Sobolv1997pumping}. In contrast, Class I methanol masers are collisionally pumped~\citep{Lees1973masercol}, due to which they are associated with the shocked material of protostellar outflows~\citep{Cyganowski2009classIoutlows}.
        
        \citet{Leurini2016classI} provide an overview of all known Class I maser transitions, some of which are also detected in the clumps of the Lagoon Nebula. Of these, the transitions with the highest detection rate for clumps in M8 are the lines at 84.5\,GHz, 95.2\,GHz, and 218.4\,GHz. The detection of these transitions alone does not automatically imply the presence of maser emission, as the corresponding transitions could also be thermally excited. In order to probe if this is the case, the line properties derived in Sect.~\ref{subsec:lin:properties} were used to examine the line widths of these transitions.
        
\begin{figure}[tbp]
                \centering
                \includegraphics[width=0.495\textwidth]{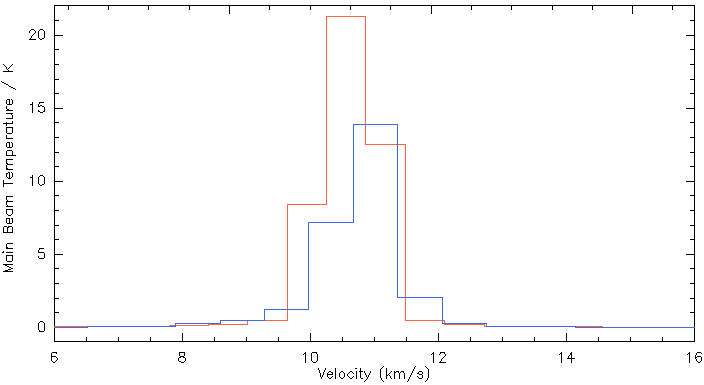}
                \includegraphics[width=0.495\textwidth]{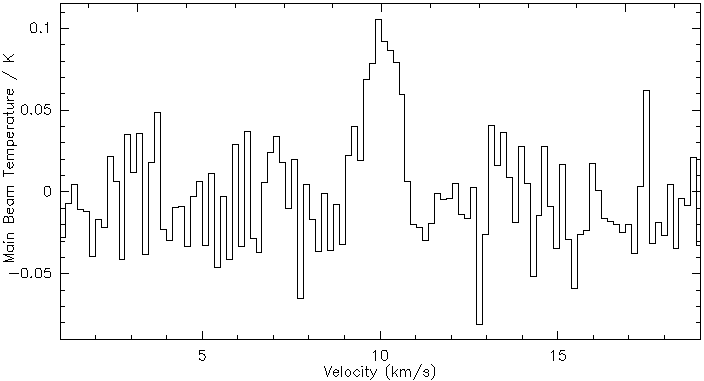}
                \caption[Typical line profiles of the methanol maser transitions]{Typical line profiles of the discussed maser transitions. Upper panel: Methanol maser emission of the $\SI{84.5}{\giga\hertz}$ and $\SI{95.2}{\giga\hertz}$ transitions at clump E, respectively shown in blue and red. Lower panel: Maser emission of the $\SI{218.4}{\giga\hertz}$ transition at the position of SE7.}
                \label{fig:ana:maser}
        \end{figure}

        As methanol masers amplify the emission from the respective transitions, typical line profiles are narrow and do not necessarily possess Gaussian shapes. Figure~\ref{fig:ana:maser} shows the line profiles of the maser transitions observed at E and SE7, which are characteristic of the line profiles observed at the remaining clumps. The methanol 84.5\,GHz and 95.2\,GHz transitions at E have FWHM line widths of $\SI{1.04}{\kilo\meter\per\second}$ and $\SI{1.17}{\kilo\meter\per\second}$, respectively, less than half the median line width of $\SI{2.45}{\kilo\meter\per\second}$ for non-masing methanol transitions at this source. The 218.4\,GHz transition at SE7 has a width of $\SI{1.06}{\kilo\meter\per\second}$, which is about $0.8$ times the median width of $\SI{1.37}{\kilo\meter\per\second}$ for non-masing methanol transitions at SE7.
        
        M8 East is a known host to Class I methanol masers~\citep[see, for example,][]{Kogan1998maser44,sarma2009}, which is in agreement with it showing the brightest 95.2\,GHz transition of the sample. The narrow line profiles indicate maser emission for the two transitions at 84.5\,GHz and 95.2\,GHz, originating from the vicinity of M8 East, the massive star-forming region containing clump E. Similar narrow line widths between $0.5$ and $0.8$ times the median line width of non-masing methanol lines are detected for the 95.2\,GHz transition at HG, WC1, and SE7, while only SE2 shows potential maser emission at 84.5\,GHz.
        
        In contrast to the bright transitions in the $\SI{3}{\milli\meter}$ atmospheric band, the potential maser transitions observed at $\SI{218.4}{\giga\hertz}$ are very faint (see Fig.~\ref{fig:ana:maser}). Given that only a few sources have been reported to detect $\SI{218.4}{\giga\hertz}$ maser emission~\citep{Hunter2014maser218,Leurini2016classI}, it is not surprising that the emission observed at M8 is also not very strong. Nevertheless, we potentially observe 218.4\,GHz masers at the positions of HG, EC4, EC5, SE1, SE7, and SC8, where the line widths of the $\SI{218.4}{\giga\hertz}$ transition are about $0.8$ times the respective median methanol line widths of non-masing transitions.
        
        The line widths provide a reliable indicator of the presence of methanol masers in M8. A confirmation of the maser activities in the clumps would require high angular resolution interferometric observations, supported by a detailed radiative transfer modelling of maser and thermally excited methanol transitions, which would go beyond the scope of this work. Furthermore, interferometry alone of the M8 clumps could confirm maser action.
        %
        %
        \section{Analysis}\label{sec:lineanalysis}
        \subsection{Systemic clump velocities}\label{subsec:vlsr}
        \noindent
        A first overview of the velocity structure of the Lagoon Nebula was obtained by~\citet{tothill2002structure} based on observations of $^{13}$CO and C$^{18}$O. While they noted the presence of double-peaked line profiles, they only give the central line velocity for the stronger component at each clump. To gain a more detailed view of the velocities in the Lagoon Nebula, the analysis of CO line profiles was repeated based on our new APEX data of the $J=2-1$ transitions at each clump.
        
        While the profiles of $^{12}$CO and $^{13}$CO lines show wings and optical depth effects, the transitions of the less abundant isotopes C$^{18}$O and C$^{17}$O are mostly optically thin. Due to this, we used the C$^{18}$O and C$^{17}$O $J=2-1$ transition data to derive the LOS velocities $\varv_\U{LSR}$ of the clumps (see Sect.~\ref{subsec:lin:properties}). For positions at which both transitions are detected, the weighted average of both shifts was computed. 
        For clumps with multiple velocity components observed in these optically thin transitions, we chose the peak velocity for the strongest two components. Table~\ref{tab:ana:vlsr} gives an overview of the derived $\varv_\U{LSR}$ at each clump position. As WC3, SE8, and SC5 have been observed at deviating coordinates with both telescopes, the velocities at these clump positions observed with the IRAM 30m telescope were estimated based on the $J=1-0$ transitions of the same species.
        
\begin{figure}[t]
                {\centering
                        \includegraphics[width=0.499\textwidth]{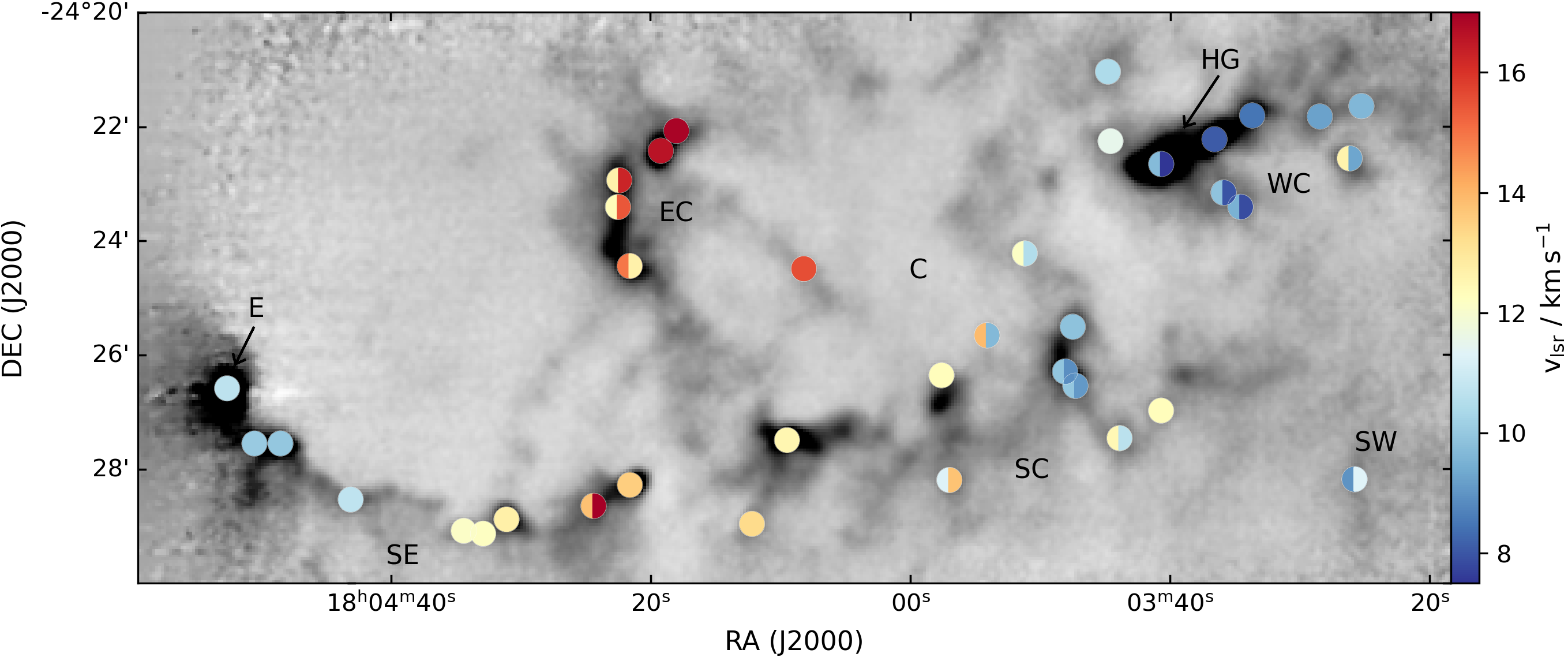}}
			
                \centering
                \hspace*{0.14cm}\includegraphics[width=0.496\textwidth]{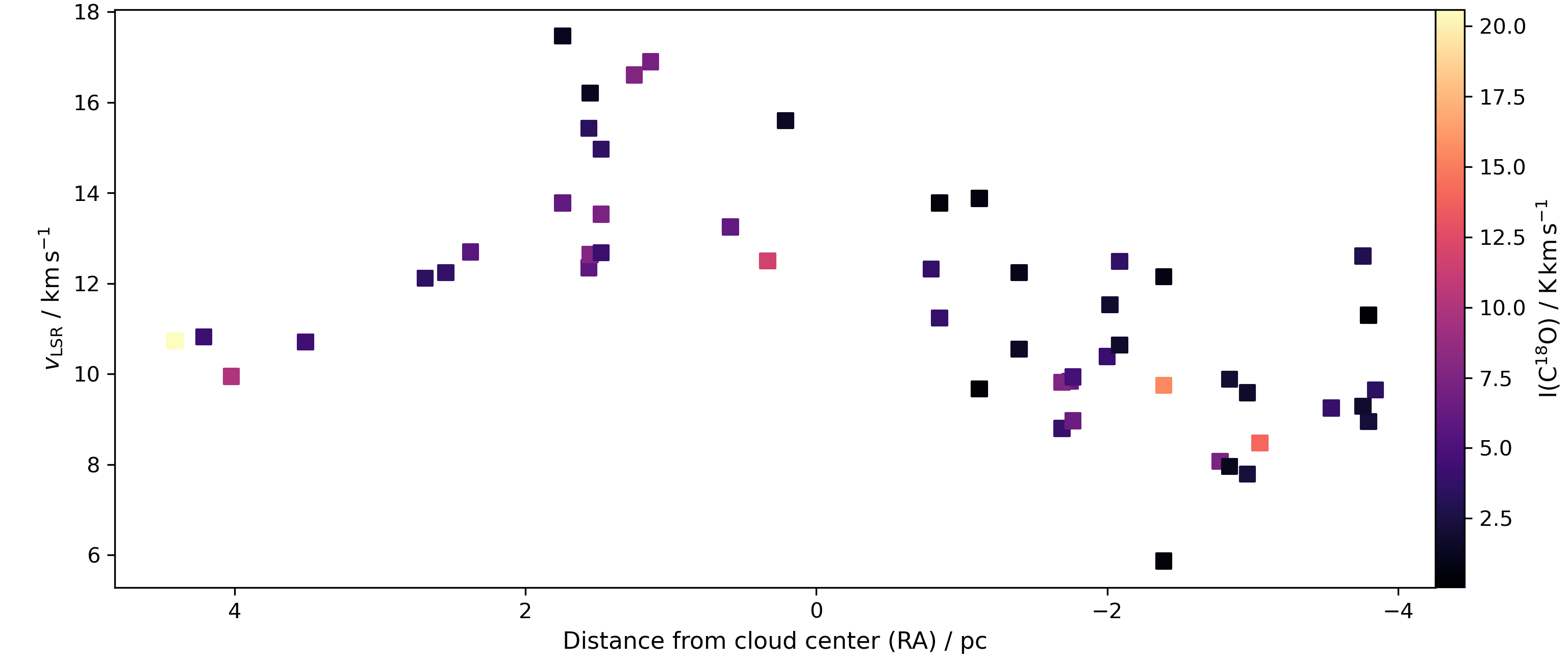}
                \caption[LOS velocities of clumps in the Lagoon Nebula]{LOS velocities of clumps in the Lagoon Nebula. Upper panel: Positions of the individual clumps (circles), with the grey-scale image showing the JCMT SCUBA $\SI{870}{\micro\meter}$ dust continuum flux (see Fig.~\ref{fig:int:clumps}). The circle size corresponds to the beam size of the on-off observations with APEX and the IRAM 30m telescope. Bi-coloured circles visualise distinct velocity components observed in the same beam. Lower panel: LOS velocity of the M8 clumps as a function of RA separation from the cloud centre at RA=18$^\U{h}$04$^\U{m}$06$^\U{s}$. Multiple velocity components inside one beam are shown as individual points.}
                \label{fig:ana:vlsr_map}
        \end{figure}

        As can be seen in the upper panel of Fig.~\ref{fig:ana:vlsr_map}, the clumps in M8 show velocity gradients along the filaments. This suggests that the clumps in the respective cloud parts are likely to be kinematically related. In contrast, the systemic velocities of the individual filaments differ across the nebula. With respect to the southern clumps, the WC clumps in M8-Main show blue-shifted emission, while the EC clumps of the central ridge show significantly high redshifted velocities. This relatively large-scale velocity gradient in the western half of M8 is also apparent in the position-velocity (PV) diagram shown in the lower panel of Fig.~\ref{fig:ana:vlsr_map} and might be caused by the radiation or mechanical feedback of the massive stars on the remnant gas. The SE clumps do not seem to follow this trend, as they branch out to lower velocities in the PV diagram. The cloud-scale kinematics in M8 is discussed further in Sect.~\ref{subsec:disc:kinematics}. 
        
        \begin{table}[htbp]
                \caption{LOS velocities of the M8 clumps.}
                \label{tab:ana:vlsr}
                \centering
                \begin{tabular}{ccc}
                        \hline
                        \hline
                        Clump & $\varv_\mathrm{1, LSR}$ & $\varv_\mathrm{2, LSR}$ \\ 
                        & (km\,s$^{-1}$) & (km\,s$^{-1}$) \\ \hline
                        HG  &        9.75 &       5.87 \\
                        WC1  &        8.09 &   - \\
                        WC2  &        8.48 &   - \\
                        WC3*  &       10.40 &   - \\
                        WC4  &       11.53 &   - \\
                        WC5  &        9.89 &       7.96 \\
                        WC6  &        9.59 &       7.79 \\
                        WC7  &       12.61 &       9.29 \\
                        WC8  &        9.25 &   - \\
                        WC9  &        9.65 &   - \\
                        SW1  &        8.95 &      11.30 \\
                        EC1  &       14.97 &      12.68 \\
                        EC2  &       12.35 &      15.43 \\
                        EC3  &       12.64 &      16.21 \\
                        EC4  &       16.61 &   - \\
                        EC5  &       16.90 &   - \\
                        E  &       10.73 &   - \\
                        SE1  &       13.53 &   - \\
                        SE2  &       13.78 &      17.47 \\
                        SE3  &       12.70 &   - \\
                        SE4  &       12.24 &   - \\
                        SE5  &       12.12 &   - \\
                        SE6  &       10.71 &   - \\
                        SE7  &        9.95 &   - \\
                        SE8*  &       10.83 &   - \\
                        SC1  &        9.84 &   - \\
                        SC2  &        9.82 &       8.80 \\
                        SC3  &        9.94 &       8.97 \\
                        SC4  &       12.49 &      10.64 \\
                        SC5*  &       12.55 &      11.16 \\
                        SC6  &       12.32 &   - \\
                        SC7  &       11.24 &      13.78 \\
                        SC8  &       12.50 &   - \\
                        SC9  &       13.25 &   - \\
                        C1  &       13.88 &       9.67 \\
                        C2  &       12.24 &      10.55 \\
                        C3  &       15.60 &   - \\
                        \hline
                \end{tabular}
                \tablefoot{These values are based on the line profiles of the $^{18}$CO (2-1) and $^{17}$CO (2-1) transitions.  The clumps marked with a * have been observed at deviating coordinates with APEX and the IRAM 30m telescope. The table lists the LOS velocities on the position observed with the IRAM 30m telescope based on the $^{18}$CO and $^{17}$CO transitions with J=1-0.}
        \end{table}

        \subsection{Kinetic temperatures and H\textsubscript{2} volume densities from para-formaldehyde}\label{subsec:temps}
        The dust temperatures of the clumps in M8 were derived in Sect.~\ref{sec:dust} using the SEDs obtained from the dust continuum images. In order to complement the derived temperatures with values from the line emission, we estimate the rotational temperatures of para-formaldehyde (p-H$_2$CO), acetonitrile (CH$_3$CN), and methyl acetylene (CH$_3$C$_2$H) in the following two sections.
        
        Formaldehyde in particular has been shown in the past to be a good thermometer for dense molecular clumps~\citep{Mangum1993h2co}. As this molecule is a slightly asymmetric rotor (described by $J_{K_a,K_c}$), each respective energy level is split further into multiple levels with different $K$ values as a consequence of different projections of the rotational axis on the symmetry axis of the molecule. While line ratios involving transitions with different angular momentum quantum numbers, $J$, are sensitive to density deviations, transitions with the same $J$ can be used to obtain reliable temperature estimates of the gas~\citep{Mangum1993h2co}.
        
        The abundances of ortho- and para-formaldehyde are not equal, which is why transitions of the same symmetry state have to be considered when deriving temperatures and densities. As a consequence, we limited this analysis to the p-H$_2$CO transitions $3_{0,3} - 2_{0,2}$, $3_{2,2} - 2_{2,1}$, $3_{2,1} - 2_{2,0}$, and $1_{0,1} - 0_{0,0}$. While the $J=3-2$ transitions enable the derivation of rotational temperature $T$ and p-H$_2$CO column density $N$, adding the $J=1-0$ transition allowed us to estimate the H$_2$ volume density $n_{\U{H}_2}$ in the M8 clumps. 
        
        In order to derive the physical properties of the M8 clumps, we followed the approach introduced by~Christensen et al. (in prep.) by utilising the Python wrapper \texttt{pyradex} for the non-LTE radiative transfer code RADEX~\citep{Tak2007radex} in combination with the \texttt{emcee}~\citep{Yang2017emcee} package for Python, which implements a Markov chain Monte Carlo (MCMC) algorithm. To obtain the line parameters of p-H$_2$CO with \texttt{pyradex}, we assumed a background temperature of $T_\U{bkg}= \SI{2.7315}{\kelvin}$ and used the collisional rate coefficients calculated by~\citet{Wiesenfeld2013collisionrates}. We fixed the line width in the computation to the weighted average line width of the p-H$_2$CO transitions at the individual clumps. This line width should have minimal variations from line to line as all transitions should be probing the same gas. The line properties in M8 derived in Sect.~\ref{subsec:lin:properties} were then fitted to obtain the physical parameters of volume density, kinetic gas temperature and p-H2CO column density. The starting position for the MCMC was obtained by \texttt{scipy.curve\_fit}, after which the MCMC algorithm explored the parameter space within $n_{\U{H}_2} = 10 - 10^7$\,cm$^{-3}$, $T = 10- 300$\,K and $N = 10^{10}-10^{17}$\,cm$^{-2}$. For each clump, 1100 steps were taken where the first 100 were discarded, and the last 1000 steps were converging on the best fit physical conditions. This process is detailed in Christensen et al. (in prep.). Derived temperatures and column densities as well as the posterior probability distributions visualising the explored parameter space are shown in Appendix~\ref{app:temp_cd}.
        
\begin{figure}[tbp]
                \centering
                \includegraphics[width=0.499\textwidth]{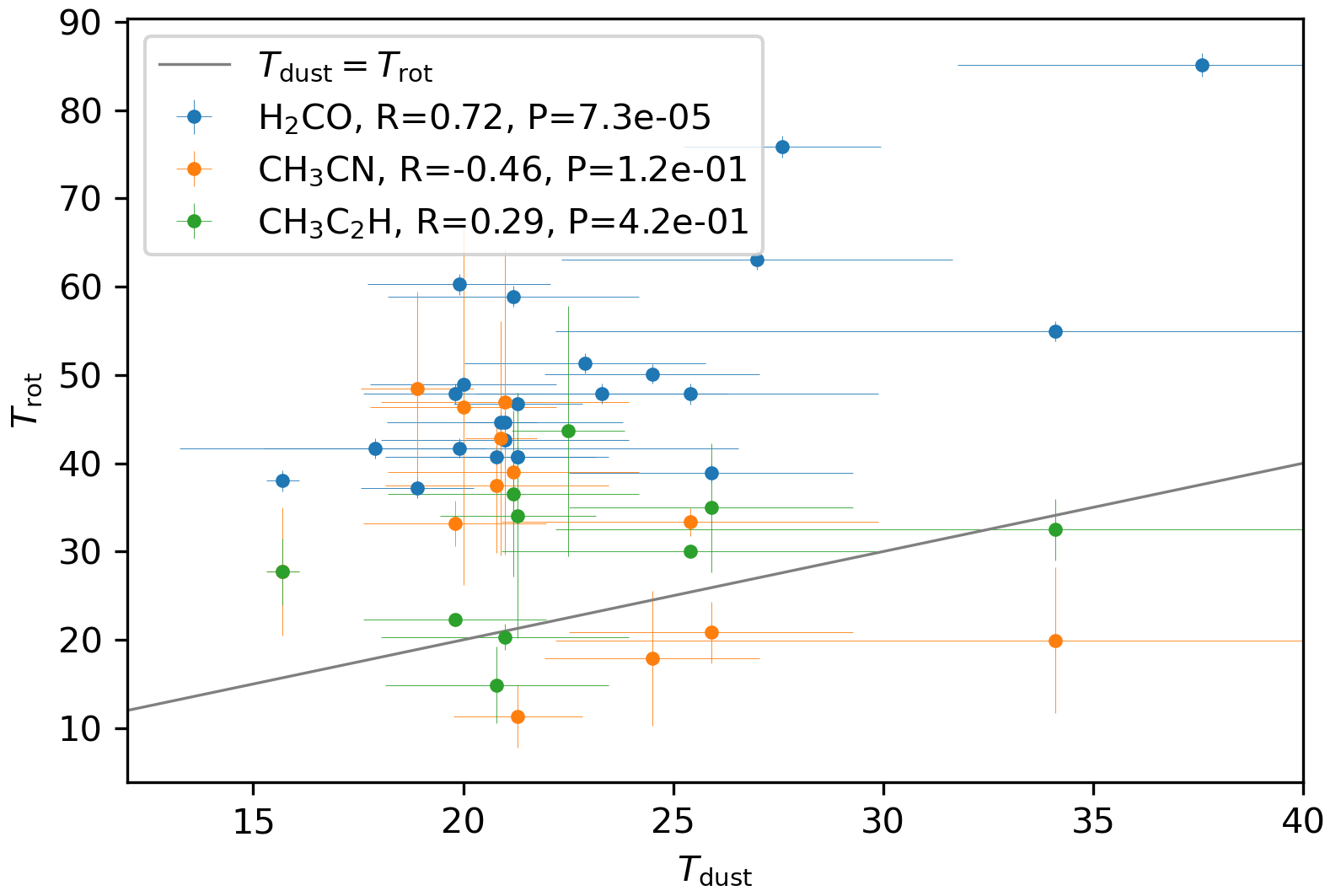}
                \caption{Temperatures derived from the rotational transitions of para-formaldehyde, acetonitrile, and methyl acetylene as a function of dust temperatures at the clumps in M8. The legend in the upper-left corner provides the Pearson correlation coefficient R and the P values for the respective samples.}
                \label{fig:ana:temps_corr}
        \end{figure}

        Formaldehyde is ubiquitous in the ISM and its transitions studied by us probe gas layers with temperatures $<\SI{100}{\kelvin}$~\citep{Mangum1993h2co}. Hence, it is not surprising that it has a high detection rate among the clumps in M8. 
        As can be seen in Fig.~\ref{fig:ana:temps_corr}, the temperatures derived from para-H$_2$CO data are on average $\SI{20}{\kelvin}$ higher than the dust temperatures. To probe the presence of a linear correlation between dust and rotational temperatures, we calculated the corresponding Pearson correlation coefficient R and P values with the python module \texttt{scipy.stats}~\citep{2020SciPy-NMeth}. Given the Pearson coefficient R of 0.72, the dust temperatures and p-H$_2$CO rotational temperatures are linearly correlated in the M8 sample. This indicates that the analysed formaldehyde transitions probe the clump envelope.

        As the upper-level energy of both the H$_2$CO $3_{2,2}-2_{2,1}$ and $3_{2,1}-2_{2,0}$ transitions is $\SI{68}{\kelvin}$ above the para ground state, only the warmer gas contributes to the H$_2$CO temperature measurement. In particular, the WC4 clump shows relatively high temperatures compared to the surrounding clumps. As discussed in Sect.~\ref{subsec:lin:rrls}, \citet{tiwari2018M8HG} detect an influence of the ionised gas on the position of WC4, which might also attribute to the heating of the clump.
        
        \subsection{Rotational temperatures of acetonitrile and methyl acetylene}\label{subsec:temps_ch3cn_ch3cch}
        \begin{figure*}
                \begin{minipage}[c]{0.65\textwidth}
                        \includegraphics[width=0.999\textwidth]{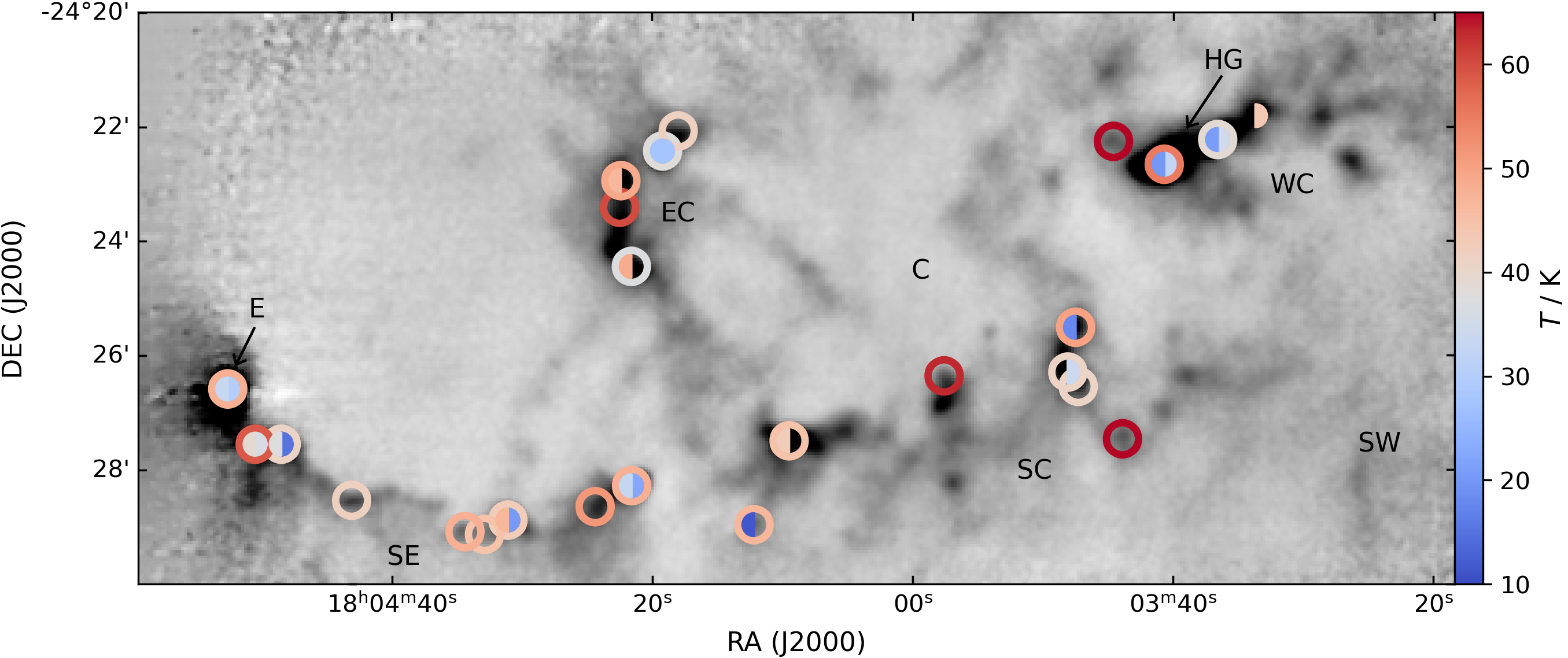}
                \end{minipage}\hfill
                \begin{minipage}[c]{0.33\textwidth}
                        \caption{Temperatures in clumps in the Lagoon Nebula derived from different molecular species. Coloured circles show the kinetic temperatures from para-formaldehyde (annuli) and the rotational temperatures of acetonitrile (left inner circle) and methyl acetylene (right inner circle). The size of the inner circles corresponds to the approximate beam size of the observations. The grey-scale background image shows the JCMT SCUBA $\SI{870}{\micro\meter}$ dust continuum flux (see Fig.~\ref{fig:int:clumps}).}
                        \label{fig:ana:temps}
                \end{minipage}
        \end{figure*}
        Both acetonitrile and methyl acetylene are symmetric top molecules (described by $J_{K}$) that show multiple transitions with the same angular momentum quantum number $J$ and with different values for the projected angular momentum $K$ at similar frequencies. Thus, both species have been found to provide good temperature estimates for astrophysical environments~\citep[e.g.][]{Askne1984acetylene,bisshop2007ch3cn}. A study of galactic molecular clumps by \citet{Giannetti2017temps} reveals that both these species trace warm gas in the clumps, while higher-energy CH$_3$CN transitions additionally trace even warmer embedded hot cores. Based on their findings, all the transitions we detected towards the M8 clumps are arising from the extended warm component.
        
        We used rotation diagrams with a simple least-squares fit to derive temperatures and column densities of CH$_3$CN and CH$_3$C$_2$H assuming LTE. This was done for the M8 clumps towards which we detect at least three spectral lines. A detailed explanation of the method, the derived physical parameters, and the respective rotation diagrams are shown in Appendix~\ref{app:temp_cd}. Figure~\ref{fig:ana:temps} visualises the rotational temperatures of acetonitrile and methyl acetylene alongside the kinetic temperatures from para-formaldehyde.
        
        The analysed transitions of CH$_3$C$_2$H and CH$_3$CN are excited either internally by star formation or externally by the feedback from the surrounding massive stars. In contrast to the p-H$_2$CO temperatures, the temperatures derived with CH$_3$C$_2$H and CH$_3$CN do not correlate with the dust temperatures (see Fig.~\ref{fig:ana:temps_corr}), which may suggest that we primarily observe internal heating. Overall, we find similar to slightly higher rotational temperatures for acetonitrile as compared to methyl acetylene. In particular, we measure comparably high acetonitrile temperatures for the clumps EC1, EC3, and SC8, which do not have a temperature estimate based on methyl acetylene. Finding higher acetonitrile temperatures is compatible with the results of~\citet{Giannetti2017temps}, who conclude that the gas layers traced by CH$_3$C$_2$H extend further to the outer parts of the clump core than the regions traced by CH$_3$CN. Interestingly, the opposite is seen for the HG and WC1 clumps in the vicinity of M8-Main, where methyl acetylene traces higher temperatures. A possible explanation may be an influence of the nearby \hii region, which externally heats the outer layers of these clumps, where methyl acetylene is more common than acetonitrile. Overall, the temperatures derived from CH$_3$C$_2$H and CH$_3$CN are within the expected values for the warm gas component surrounding clump cores derived by~\citet{Giannetti2017temps}.
        
        Despite being bright in the $\SI{24}{\micro\meter}$ dust continuum (see Fig.~\ref{fig:dust:20micfull}), no CH$_3$CN emission is detected towards the WC7 core. This missing line emission hints at the absence of a hot core at this position, which implies that the observed mid-IR emission at this position is unrelated to the clump. This is further confirmed by the methanol emission at WC7, which we only detect in the lowest energy transitions at 96.7\,GHz. These transitions are typically very strong and therefore also detected in most IR-dark clouds~\citep[e.g.][]{Leurini2007CH3OHIRDCs}. The non-detection of higher-energy methanol transitions towards this clump suggests a low excitation of this species and therefore the absence of a hot core at WC7.
        
        \subsection{Column densities}\label{subsec:cd}
        Using the line intensities, $I$, derived in Sect.~\ref{subsec:lin:properties}, column densities of the detected species can be computed.
        Assuming optically thin emission and LTE, the column density, $N$, can be described as a function of $I$, the clump temperature $T$, and the background temperature $T_\U{bg}$ according to
        
        \begin{align}\label{eq:cd_final}
                \begin{split}
                        N_\U{total}^\U{thin}(T, I, T_\U{bg}) &= \frac{Q(T)}{g_\U{u}}\cdot \frac{\exp \left(\frac{E_\U{u}}{k_\U{B} T }\right)}{\exp \left(\frac{h\nu}{k_\U{B} T }\right)-1} \\ &\cdot \frac{1}{J_\nu T-J_\nu(T_\U{bg})} \cdot \frac{8 \pi \nu^3 }{c^3 A_\U{ul}} \frac{I}{f},
                \end{split}
        \end{align}
        by introducing the Rayleigh-Jeans equivalent temperature $J_\nu(T)$=$h\nu/[k_\U{B}(\exp(h\nu/(k_\U{B}T))-1)]$ \citep{Mangum2015rotdiag}. Additional parameters are the Planck constant, $h$, the Boltzmann constant, $k_\U{B}$, the speed of light, $c$, and the transition-specific frequency, $\nu$, upper-level energy, $E_\U{u}$, upper-level degeneracy, $g_\U{u}$, and spontaneous Einstein A coefficient, $A_\U{ul}$. In order to account for the clump sizes $\theta_\U{S}$, the measured intensities were corrected by the beam filling factor $f=\theta_\U{S}^2/(\theta_\U{S}^2+\theta_\U{B}^2)$ with the FWHM beam width, $\theta_\U{B}$. The corresponding clump sizes for this were adopted from~\citet{tothill2002structure}. Analogous to \citet{Kim2020pdrtracer}, we approximated the kinetic temperature, $T,$ in Eq.~\ref{eq:cd_final} with the dust temperatures, $T_\U{dust}$, derived in Sect.~\ref{sec:dust}, as they are available for the full sample of M8 clumps. This approximation can lead to an underestimation of the derived column densities, as the dust temperature of a respective clump tends to be lower than its actual kinetic temperature, especially for species that trace the warmer gas component, such as H$_2$CO, CH$_3$CN, and CH$_3$C$_2$H. A more precise approximation for the kinetic temperatures of the clumps are the excitation temperatures derived from rotational transitions of certain molecular species (see Sect.~\ref{subsec:temps}). Nevertheless, we used the dust temperatures here since they are available for all M8 clumps, while deriving rotational temperatures was only possible for a subsample of them.
        
        \begin{figure}[tbp]
                \centering
                \includegraphics[width=0.499\textwidth]{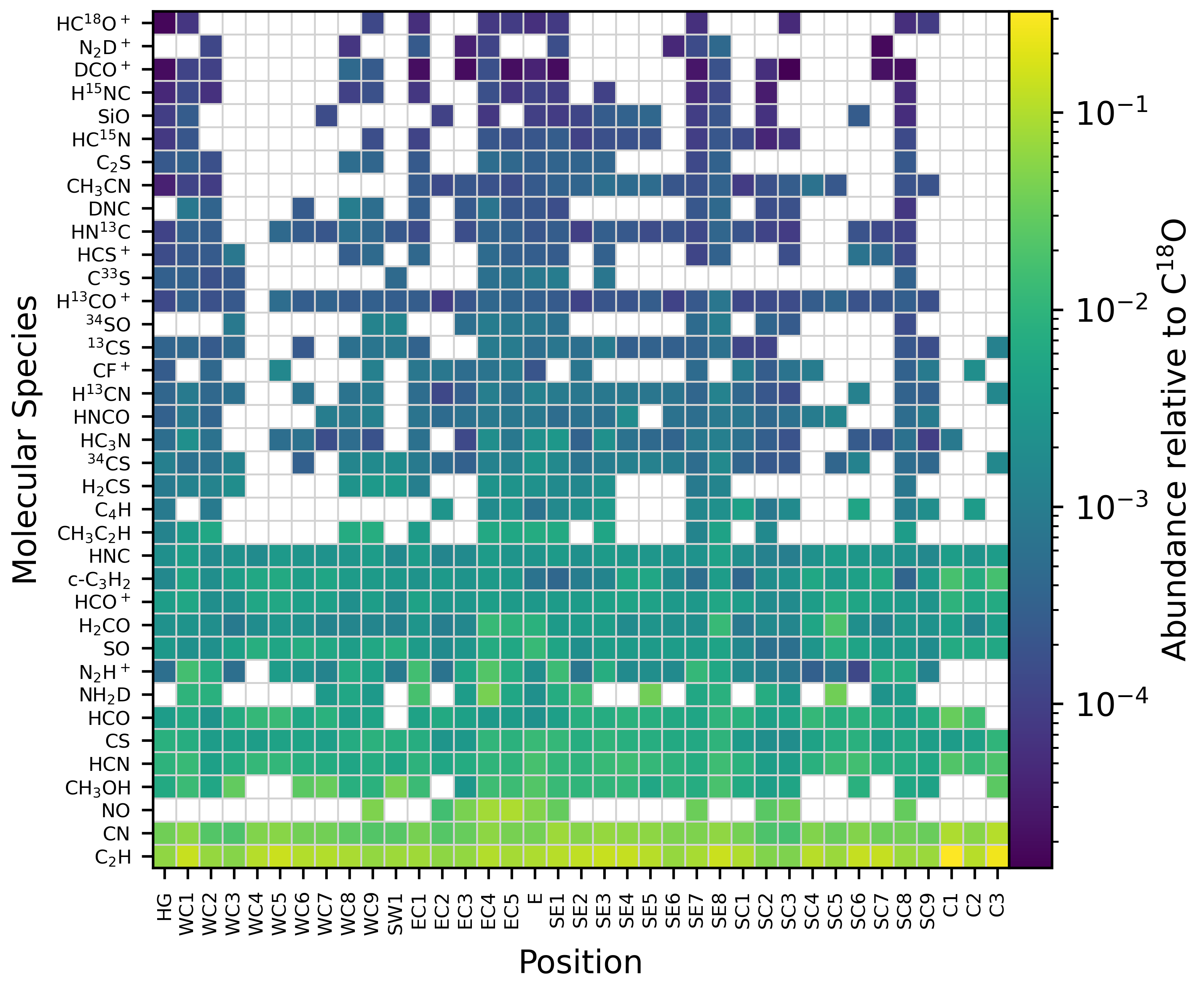}
                \caption[Column densities in M8]{Column densities of species detected in at least ten M8 clumps. The colour scale gives the relative abundance of a respective species to the column density of C$^{18}$O at the same position. White cells indicate the non-detection of a species in the associated clump.}
                \label{fig:ana:cd}
        \end{figure}
        
        The assumptions mentioned above do not hold in many cases, for example where emission is optically thick like in CO, HCN, or CS. As the optical depth $\tau$ of transitions with fitted HFS is known (see Table~\ref{tab:app:line_properties}), the derived column densities for the corresponding species were corrected according to
        \begin{equation}\label{eq:cd_tau}
                N_\U{total} = N_\U{total}^\U{thin}\cdot \frac{\tau}{1-e^{-\tau}}.
        \end{equation}
        For other optically thick transitions, the derived column densities, $N_\U{total}^\U{thin}$, act as lower limits to the actual column density of the species. 
        
        If multiple transitions of a species are detected in a molecular clump, the median of the column densities derived from each line was computed. For species that remain undetected in some clumps, we estimated upper limits to their column densities by using the RMS at lines of these species detected in other clumps of M8. For all possible transitions of the species, we independently calculated an upper limit of the column density based on the median line width at the respective clump and a peak intensity equal to three times the spectrum RMS close to the non-detected transition. The lowest limit obtained with this method was then considered to be the upper limit of the column density.
        
        All derived column densities and upper limits are presented in Table~\ref{tab:app:cd}. Figure~\ref{fig:ana:cd} shows an overview of all species that have been detected in at least ten clumps of M8. 
        
        In addition to the most common tracers of dense clumps in the ISM, we also detect PDR tracers such as HCO, c-C$_3$H$_2$, CN, and C$_2$H (see \citealt{Kim2020pdrtracer} and the references therein), in a large fraction of clumps in M8. This is consistent with the widespread $\SI{8}{\micro\meter}$ emission detected on the surfaces of the M8 clumps (see Sect.~\ref{sec:dust}). About half of the clumps also show the presence of shocks, as indicated by the detection of SiO~\citep[e.g.][]{Bergin2007sf,schilke1997sio,Bachiller1991SiOShock}. In contrast, some clumps also show the presence of cold and dense gas tracers such as NH$_2$D, N$_2$D$^+$, and DNC. A more detailed comparison of the chemical conditions in the M8 clumps is given in Sects.~\ref{subsec:disc:chemistry} and~\ref{subsec:disc:cd}.
        %
        %
        \section{Discussion}\label{sec:discussion}
        \subsection{Dust continuum clump properties}\label{subsec:disc:dust}
        The physical properties of the M8 clumps were derived in Sect.~\ref{sec:dust}. In this section, they are compared to the overall population of clumps in the inner Galactic plane examined by~\citet{urquhart2018atlasgal}. Their sample of clumps is based on the ATLASGAL survey of the Galactic plane and therefore contains a large variety of different sources. 
        
        ATLASGAL has the highest source densities at distances between 2 and $\SI{4}{kpc}$, further away than the M8 clumps, which have a distance of $\SI{1.3}{kpc}$. This imposes constraints on the spatial resolution and sensitivity, implying that the typical ATLASGAL source will have a larger physical size and is more massive compared to the clumps in M8. At these larger distances, it is even possible that the corresponding ATLASGAL sources contain sub-structure similar to multiple M8 clumps. In order to compare similar objects, we therefore also compared the clumps of M8 with a distance-limited sample of the ATLASGAL clumps, containing only sources that are at distances smaller than $\SI{1.5}{kpc}$.
        
        Furthermore, the ATLASGAL survey is likely to be incomplete for clumps below a few hundred\,M$_\odot$. For instance, out of the 37 molecular clumps studied in this work, only 14 are retrieved in the ATLASGAL survey. While the comparably bright clumps such as SC1-3 might have been rejected due to their extended elongated shapes, other adjacent clumps like EC4 and 5 are recognised as a single clump. In contrast, the $\SI{870}{\micro\meter}$ peak emission from the weaker clumps SW1 and C1-3 is below the detection threshold of about 0.4\,Jy\,beam$^{-1}$, which was applied for the source extraction of the ATLASGAL compact source catalogue~\citep{contreeras2013CSC}. To account for these differences in source selection, we compared the distance-limited ATLASGAL sample with both the full sample of M8 clumps and a subsample of M8 clumps that do have an ATLASGAL counterpart.

        \begin{figure}[tbp]
                \centering
                \includegraphics[width=0.499\textwidth]{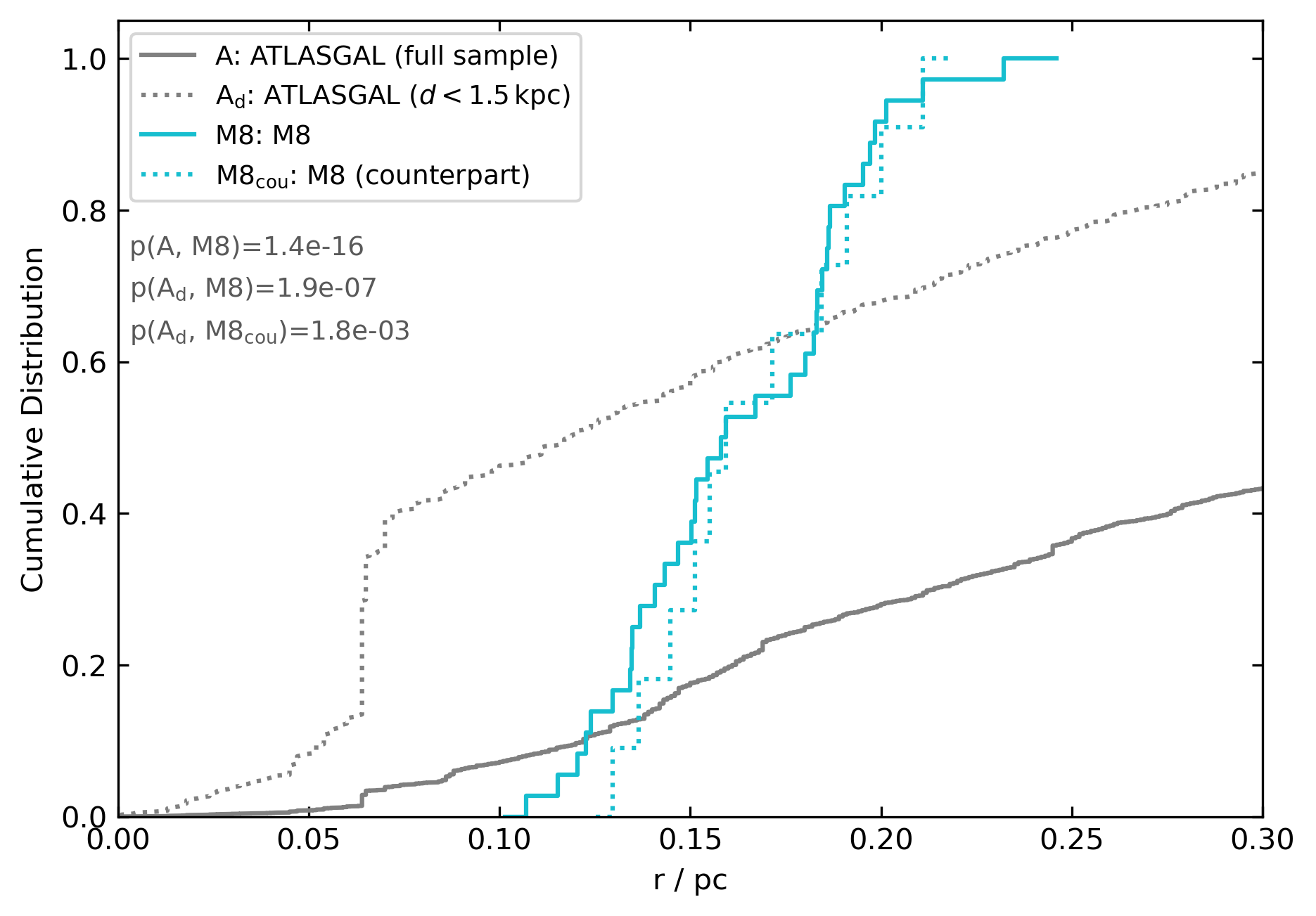}
		
                \includegraphics[width=0.499\textwidth]{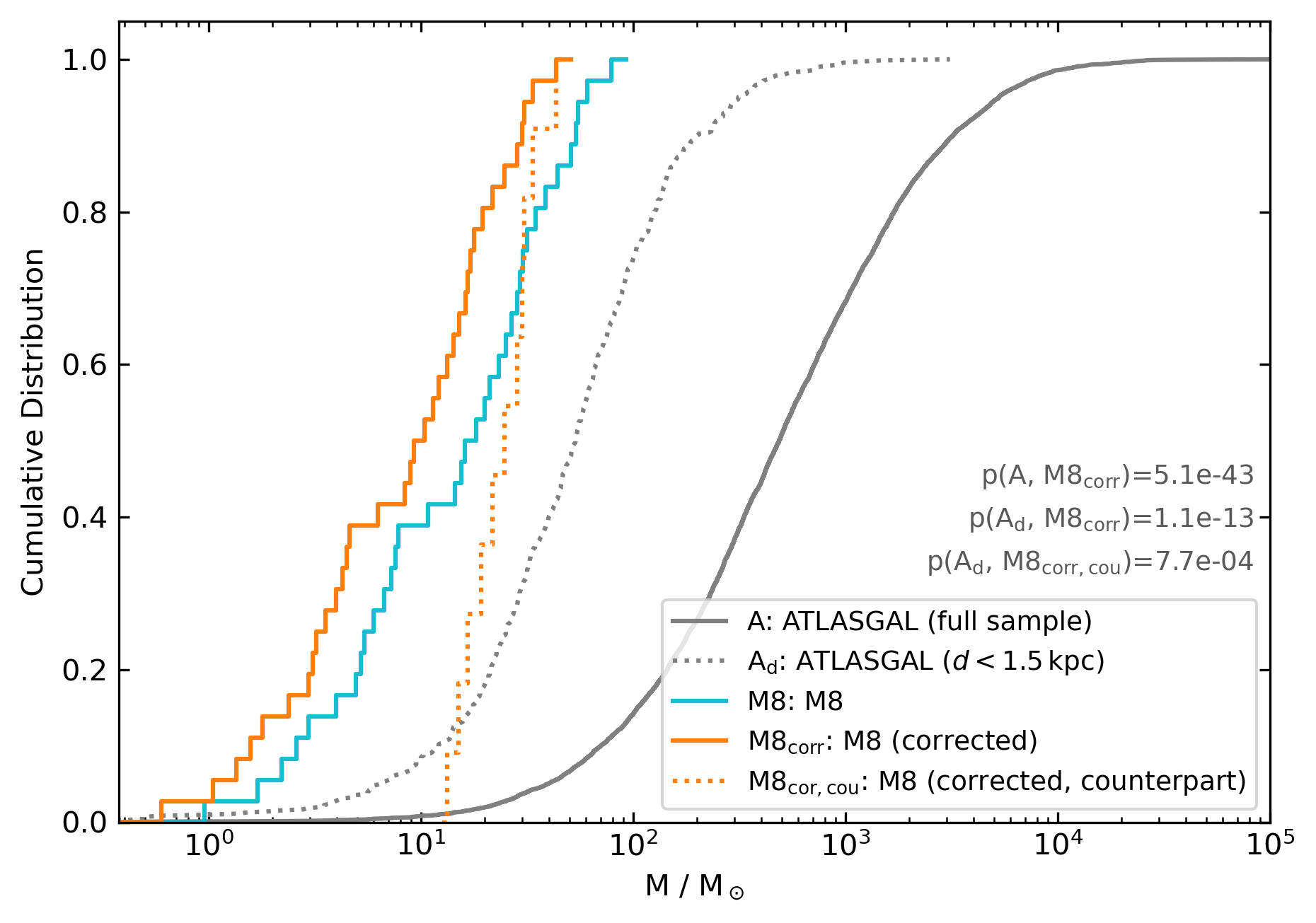}
                \caption[Cumulative distributions of the clump radii and masses in M8 and for the full sample of ATLASGAL clumps]{Cumulative distributions of the clump radii derived by~\citet[upper panel]{tothill2002structure} and masses (lower panel) in M8 and for the full and distance-limited sample of ATLASGAL clumps. The M8 clump masses shown in orange colour are corrected based on the results of Appendix~\ref{app:akari}. The dotted orange distribution only contains M8 clumps that have a nearby counterpart in the ATLASGAL sample. The p-values of KS tests between selected samples are provided as grey text.}
                \label{fig:disc:dust_mass}
        \end{figure}
        
\begin{figure}[tbp]
                \centering
                \includegraphics[width=0.49\textwidth]{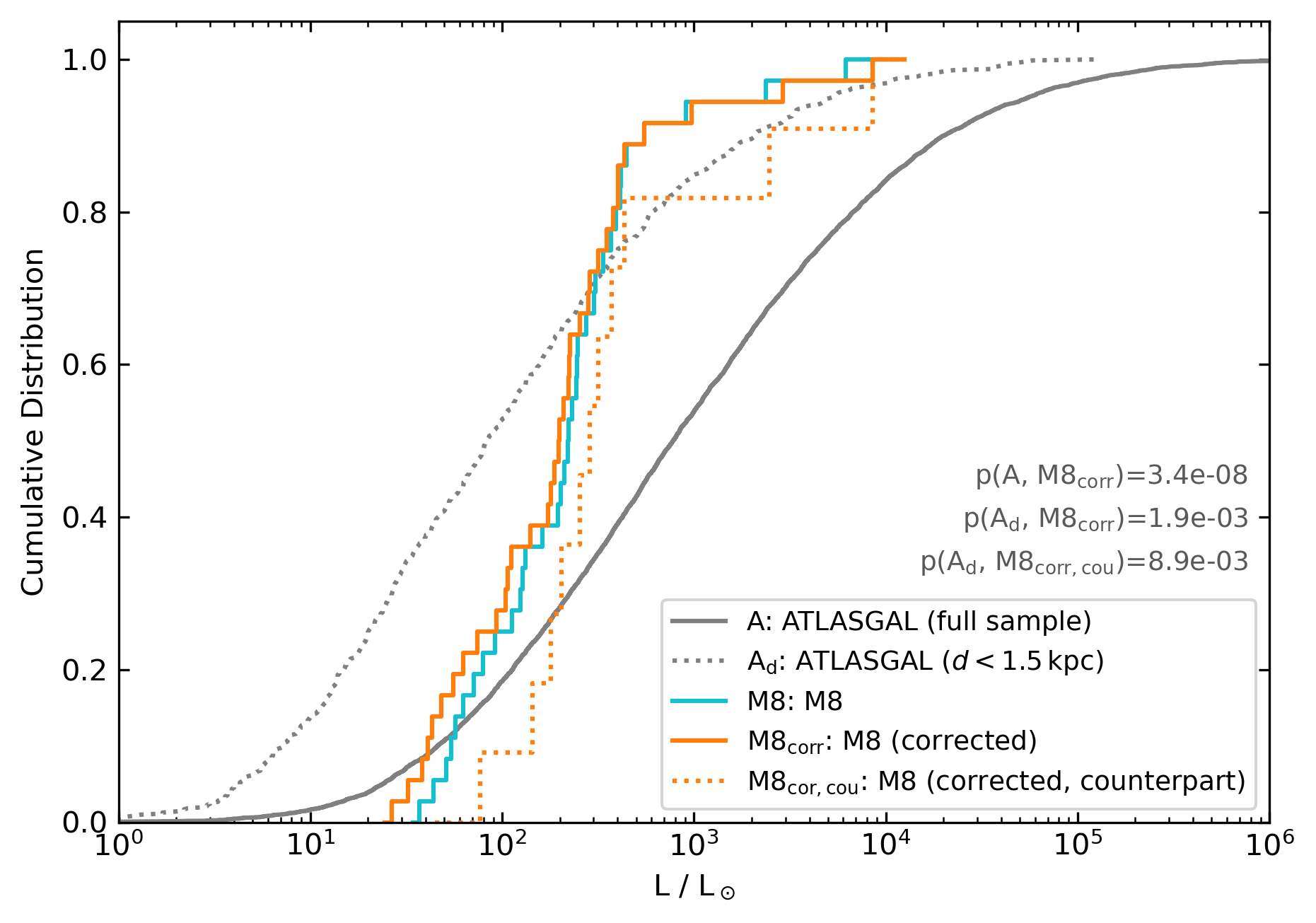}
		
                \includegraphics[width=0.486\textwidth]{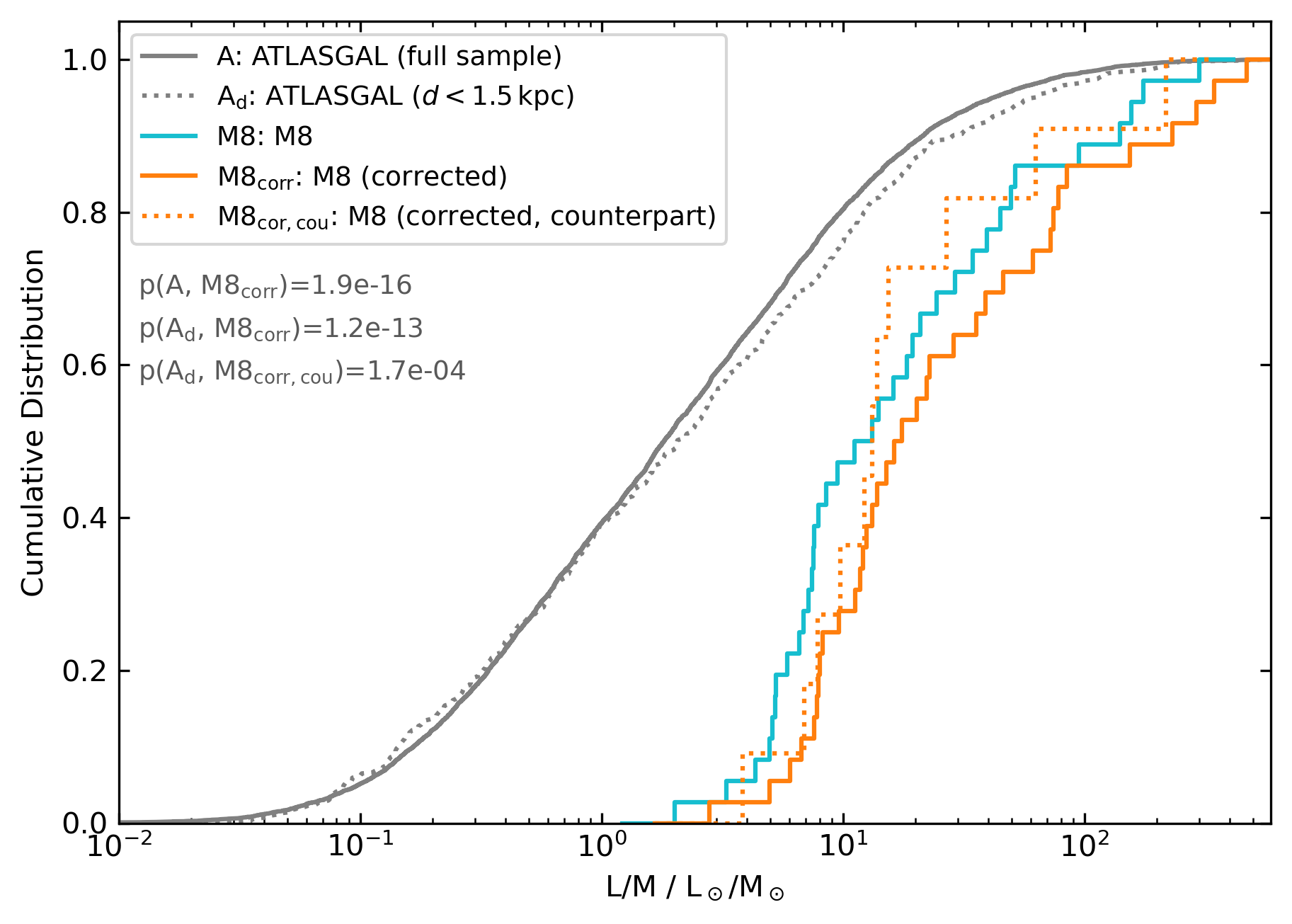}
		
                \includegraphics[width=0.49\textwidth]{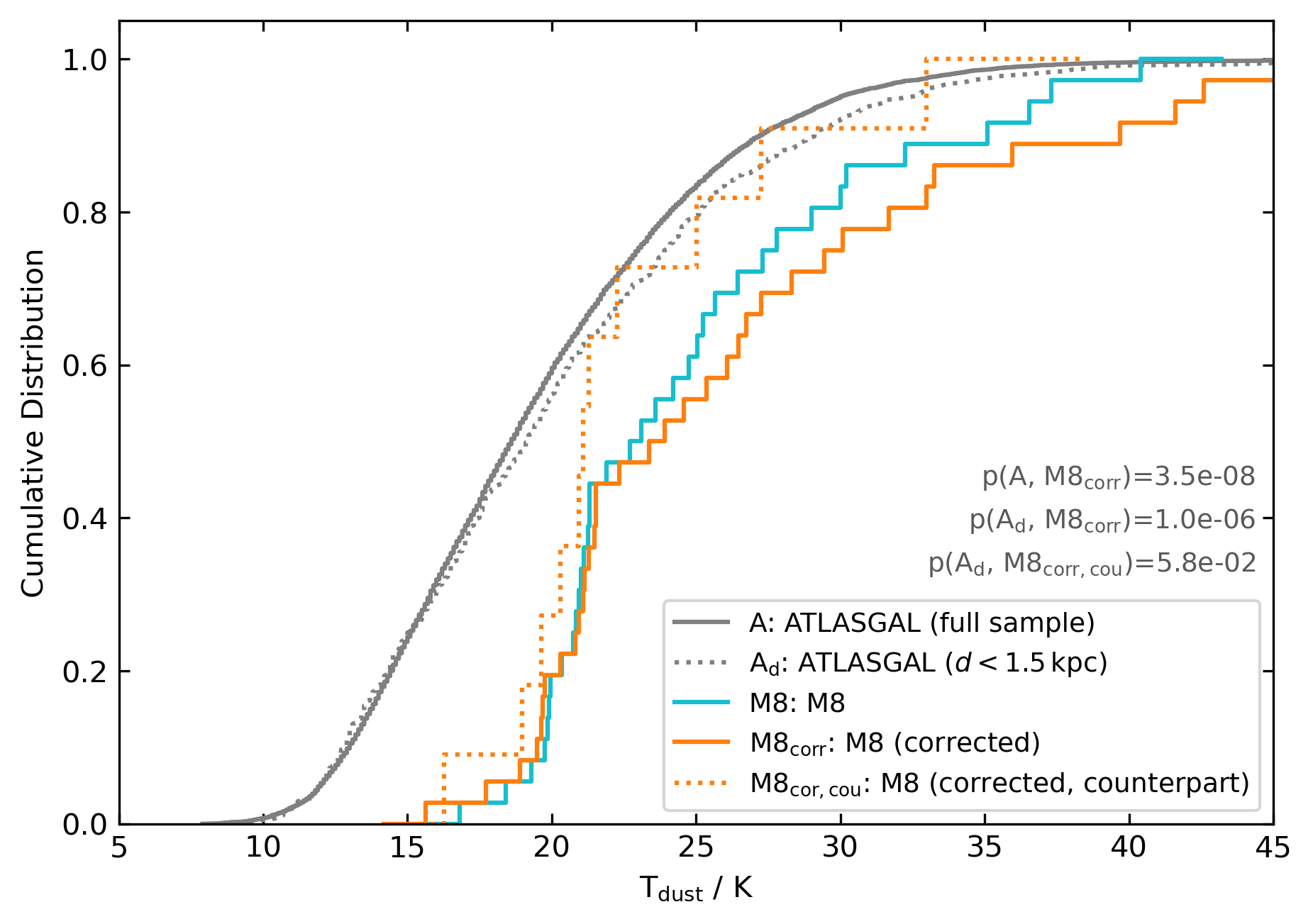}
                \caption[Cumulative distributions of the clump luminosities and dust temperatures in M8 and for the full sample of ATLASGAL clumps]{Cumulative distributions of the clump luminosities (upper panel), the $L/M$ ratios of the respective clumps (middle panel), and clump dust temperatures (lower panel) in M8 and for sources of the full and distance-limited ATLASGAL sample. The M8 distributions shown in the orange colour are corrected based on the results of Appendix~\ref{app:akari}. The dotted orange distribution only contains M8 clumps that have a nearby counterpart in the ATLASGAL sample. The p-values of KS tests between selected samples are provided as grey text.}
                \label{fig:disc:L_T}
        \end{figure}

        We performed two-sample Kolmogorov-Smirnov (KS) tests between physical properties of clumps in the ATLASGAL and the M8 samples using the Python module \texttt{scipy.stats}. These tests estimate the p-value, which indicates the likelihood that the properties of both samples are distributed equally. A low p-value indicates that both samples originate from different distributions, while values above $p>0.0013$ suggest that the samples have the same underlying distribution~\citep[][]{urquhart2018atlasgal}. Cumulative distribution plots of the clumps' radii, masses, luminosities, $L/M$ ratios, and dust temperatures are shown alongside the corresponding p-values in Figs.~\ref{fig:disc:dust_mass} and~\ref{fig:disc:L_T}.
        
        The clump masses and physical radii are compared in Fig.~\ref{fig:disc:dust_mass}. Effective radii for the ATLASGAL sources are provided by~\citet[see their Table 5]{urquhart2018atlasgal}, who derive the values by multiplying the geometric mean of the deconvolved semi-major and semi-minor axes by $2.4$~\citep{contreeras2013CSC}. The radii of the M8 clumps were calculated analogously based on the distance of $\SI{1.3}{kpc}$ and the clump sizes derived by~\citet[see their Table 1]{tothill2002structure}.
        
        Both masses and radii are significantly lower in the M8 clumps compared to the full ATLASGAL sample of clumps. While for the clumps in M8, the median radius and mass is about $\SI{0.16}{pc}$ and $\SI{10}{M_\odot}$, respectively, the ATLASGAL clumps tend to be larger and more massive with a median radius of $\SI{0.37}{pc}$ and a median mass of about $\SI{500}{M_\odot}$. As argued above, this discrepancy is less pronounced when comparing with the distance-limited sample that has median radii of $\SI{0.12}{pc}$ and masses of $\SI{50}{M_\odot}$.
        
        The sample of M8 clumps with ATLASGAL counterparts has a median mass of $\SI{25}{M_\odot}$, which is only a factor of two smaller than the median mass of the distance-limited ATLASGAL clumps. However, the p-value of $p=7.7\times 10^{-4}$, obtained when comparing these samples, suggests that the underlying distribution of masses still differs. A possible physical mechanism that could cause these smaller clump masses is the fragmentation of the filaments by the radiation pressure of the nearby O stars in M8. Advocation of this mechanism, however, is in contrast with a recent study by ~\citet{Mazumdar2021G305clumps}, who examined the clump properties in the star-forming complex G305. They find increased clump masses as a result of the collect and collapse feedback mechanism. We believe this contradiction is due to the differences in the ages and star formation histories of M8 and G305. While the ages of the stellar clusters in the G305 complex vary between 1\,Myr and 3\,Myr~\citep{Davies2012G305age}, the initial trigger of the star formation in M8 is estimated to have occurred around 4\,Myr~ago \citep{Damiani2019m8dist}. As a consequence, the M8 cloud might be allowing for a stronger dispersal of the clumps, while in the G305 complex the fragmentation has only started recently, maybe as a consequence of a previously triggered star formation phase. Furthermore, the distance of G305 of about $\SI{4}{kpc}$ suggests that it is not possible to resolve the sub-structures in that complex in the same detail as in M8.
        
        With a median value of $\SI{200}{L_\odot}$, luminosities in the M8 sample are only lower by a factor of four compared to the typical luminosities of about $\SI{800}{L_\odot}$ observed for the full sample of clumps in the inner Galactic plane. Additionally, the M8 luminosities are slightly higher when compared to nearby ATLASGAL clumps, which have a median luminosity of $\SI{83}{L_\odot}$ (see the upper panel of Fig.~\ref{fig:disc:L_T}). As shown in the middle panel of Fig.~\ref{fig:disc:L_T}, the combination of small measured masses and comparable luminosities leads to increased $L/M$ ratios with a median of $\SI{17}{L_\odot\,M_\odot^{-1}}$ for the M8 clumps as compared to about $\SI{2}{L_\odot\,M_\odot^{-1}}$ for the ATLASGAL samples.
        
        Moreover, the derived dust temperatures of the M8 clumps, shown in the bottom panel of Fig.~\ref{fig:disc:L_T}, are higher by approximately $\SI{5}{\kelvin}$ with respect to the ATLASGAL sources. The full and close-by samples of ATLASGAL clumps have median dust temperatures of $\SI{18.6}{\kelvin}$ and $\SI{19.2}{\kelvin}$. In contrast, the median dust temperature of M8 clumps is $\SI{23.6}{\kelvin}$, indicating that the temperatures are approximately 25\% higher. The evolutionary sequence for dense clumps introduced by~\citet{konig2017atlasgal} and refined by~\citet{urquhart2018atlasgal} predicts these conditions of increased temperatures and large $L/M$ ratios only for clumps associated with evolved, massive YSOs and \hii regions. While such sources exist in the major star-forming regions (M8-Main and M8 East) of the Lagoon Nebula, arguably not every observed clump is host to a massive stellar object. For instance, some of the warmest clumps are the central clumps C1-2, which due to their low masses below 3\,M$_\odot$ are not capable of harbouring a massive object.

        In addition to evolved star-forming regions, a likely cause for the higher temperatures and luminosities is the presence of an externally heating source. As shown in Sect.~\ref{sec:dust}, almost all clumps in the Lagoon Nebula are being exposed to the radiation from the nearby massive stars, which is revealed by the presence of numerous PDRs traced by $\SI{8}{\micro\meter}$ PAH emission, which is the fluorescent result of UV radiation from the nearby O stars that heats the outer regions of the clumps. This external heating increases the luminosities of low-mass M8 clumps, which intrinsically are likely to have low luminosities characteristic for clumps of the inner Galactic plane.
        
        When only considering the M8 sample of clumps that have an ATLASGAL counterpart, this temperature offset seemingly vanishes at higher temperatures. A possible reason for this is that the higher temperature clumps also correspond to the less massive clumps offset from the filament that were rejected in the ATLASGAL survey. We therefore conclude that especially these less massive clumps are affected by the external heating.

        \subsection{Kinematics in the Lagoon Nebula}\label{subsec:disc:kinematics}
        The velocities of the individual LOS components towards the M8 clumps were examined in Sect.~\ref{subsec:vlsr} based on the position of the peak intensities of the optically thin C$^{18}$O and C$^{17}$O spectral lines. The $\varv_\U{LSR}$ overview in Fig.~\ref{fig:ana:vlsr_map} shows velocity gradients of the clumps along the filaments. For example, the LOS velocity gradually changes from $\SI{13.5}{\kilo\meter\per\second}$ to $\SI{12.1}{\kilo\meter\per\second}$ for the SE1-SE5 clumps. The only exception is the other relatively faint redshifted component of SE2 that likely originates from a more diffuse gas component.
        
        The distinct velocity components observed in lines from many species towards M8-Main were examined in detail by~\citet{tiwari2018M8HG}, who observe the dense gas components to be seen with higher systemic velocities above $\SI{10}{\kilo\meter\per\second}$, while the blue-shifted gas is accelerated towards us and is powered by the radiation of the nearby O-type stars. The two velocity components found in our observations are consistent with that interpretation. We note that we only analysed one bright component of the WC1 clumps, while the CO line profiles suggest the presence of multiple weak components in this region.
        
\begin{figure}[tbp]
                \centering
                \subfigure[]{\includegraphics[width=0.495\textwidth]{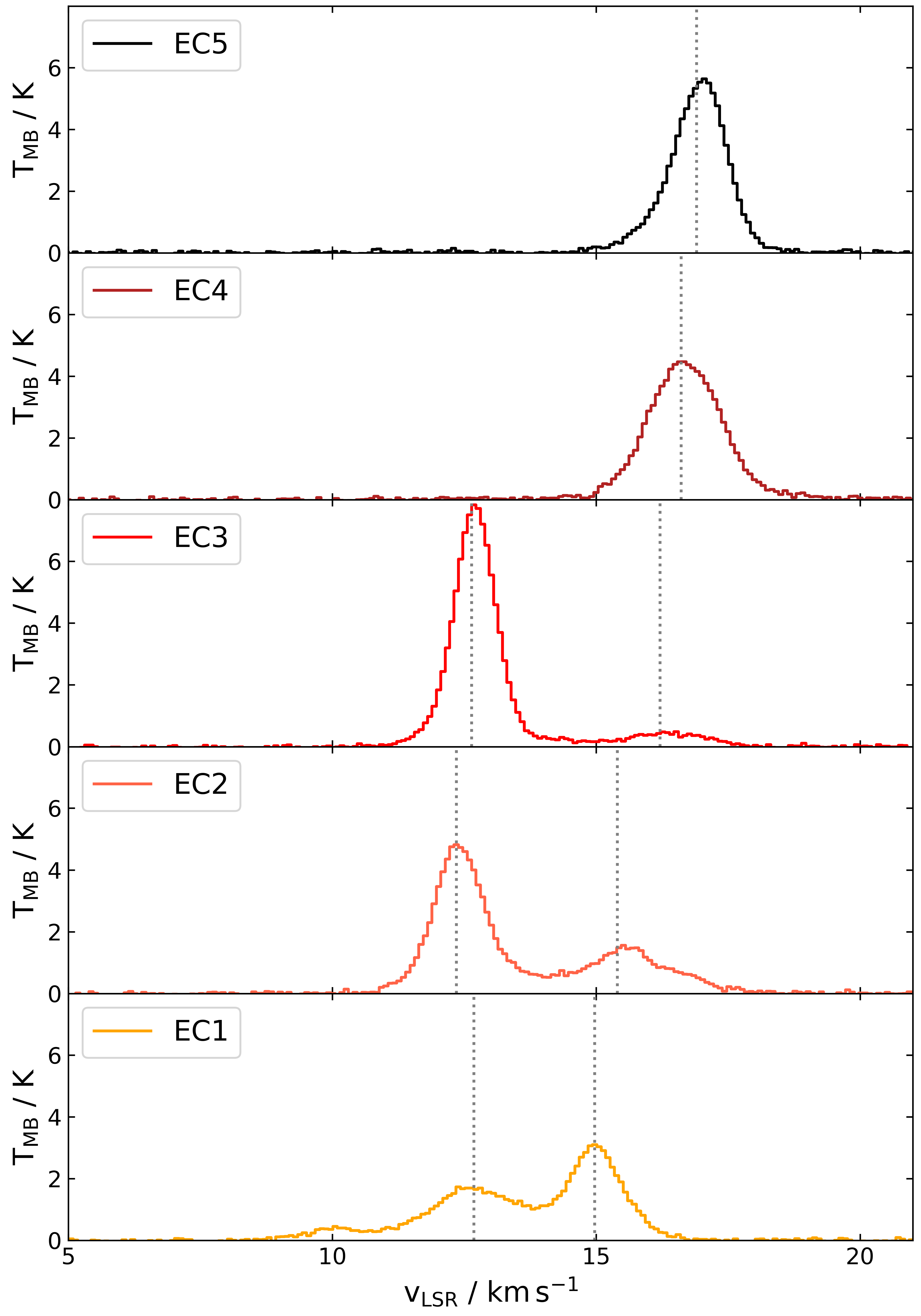}}
                \caption[Line profiles of the C$^{18}$O~(2-1) transition for clumps located in the EC ridge and the western part of the SC filament]{Line profiles of the C$^{18}$O~(2-1) transition for clumps located in the EC ridge.}
                \label{fig:disc:lineprofiles}
        \end{figure}

        Another striking example of velocity and intensity gradients is observed towards the EC ridge whose C$^{18}$O~(2-1) line profiles are shown in Fig.~\ref{fig:disc:lineprofiles}. While the component at $\SI{12}{\kilo\meter\per\second}$ shows approximately the same LOS velocity for all EC1-EC3 clumps, the peak of the redshifted component at about $\SI{15}{\kilo\meter\per\second}$ shifts by $\SI{1.3}{\kilo\meter\per\second}$ along the ridge. It is also interesting to examine the intensities of the respective components, as the brightness of the $\SI{12}{\kilo\meter\per\second}$ component increases from south to north, while the $\SI{15}{\kilo\meter\per\second}$ emission gets fainter along this direction. The EC4 and EC5 clumps follow the velocity shift in the northern direction, with bright emission centred at approximately $\SI{17}{\kilo\meter\per\second}$. In contrast to the three southern clumps, these clumps show no emission component at $\SI{12}{\kilo\meter\per\second}$.
        
        As both EC velocity components are seen in high-density tracers such as HNC, it is probable that both components correspond to dense gas layers. Nevertheless, it is difficult to judge if the two components actually interact, as these findings are purely based on selected on-off observations, while a complete mapping of the region is still missing. The overall redshifted emission of the EC region may be explained by a location of the clumps behind the massive stars of the open cluster NGC\,6530. It is possible that the radiation from the associated massive stars transfers momentum to the dense molecular gas, which would explain the observed high redshift velocity component with receding clump motions.
        
        In contrast to the EC region, the velocity components observed in the SC2 and SC3 clumps are less separated and approximately show the same intensities for both clumps. The similarity in the spectral line profiles for both clumps is caused by an overlap of the observed beams. In contrast, the close proximity of the two line components shows that the two gas layers have a similar LOS velocity. This suggests similar motions of both gas layers, indicating that they are exposed to the same external forces.
        
        On larger scales, the velocity distribution in the ISM around M8 shows two distinct northern regions, with the blueshifted components between $\SI{7}{\kilo\meter\per\second}$ and $\SI{12}{\kilo\meter\per\second}$ in the fore- and background of M8-Main, and the redshifted EC ridge in the background of the NGC\,6530 cluster with velocities between $\SI{12}{\kilo\meter\per\second}$ and $\SI{17}{\kilo\meter\per\second}$. In contrast, the LOS velocities of the southern clumps are more clustered around a systemic velocity of 10 to $\SI{12}{\kilo\meter\per\second}$. This small variation in velocity for the southern clumps only reflects their motion along the LOS. It is therefore possible that the southern filament instead moves perpendicular to the LOS, which may be due to an acceleration in southern direction caused by the massive stars located in the north.
        
        \subsection{Chemistry of the M8 clumps}\label{subsec:disc:chemistry}
        The line survey towards all the clumps in M8 was described in Sect.~\ref{sec:linesurvey}, based on which we estimated the column densities of all the detected species in Sects.~\ref{subsec:temps},~\ref{subsec:temps_ch3cn_ch3cch}, and~\ref{subsec:cd}. The detection and column density distributions of the observed species across all the clumps are summarised in Table~\ref{tab:identified_species} and Figs.~\ref{fig:lin:linenr} and~\ref{fig:ana:cd}, respectively.
        
        \begin{figure*}
                \begin{minipage}[c]{0.65\textwidth}
                        \includegraphics[width=0.999\textwidth]{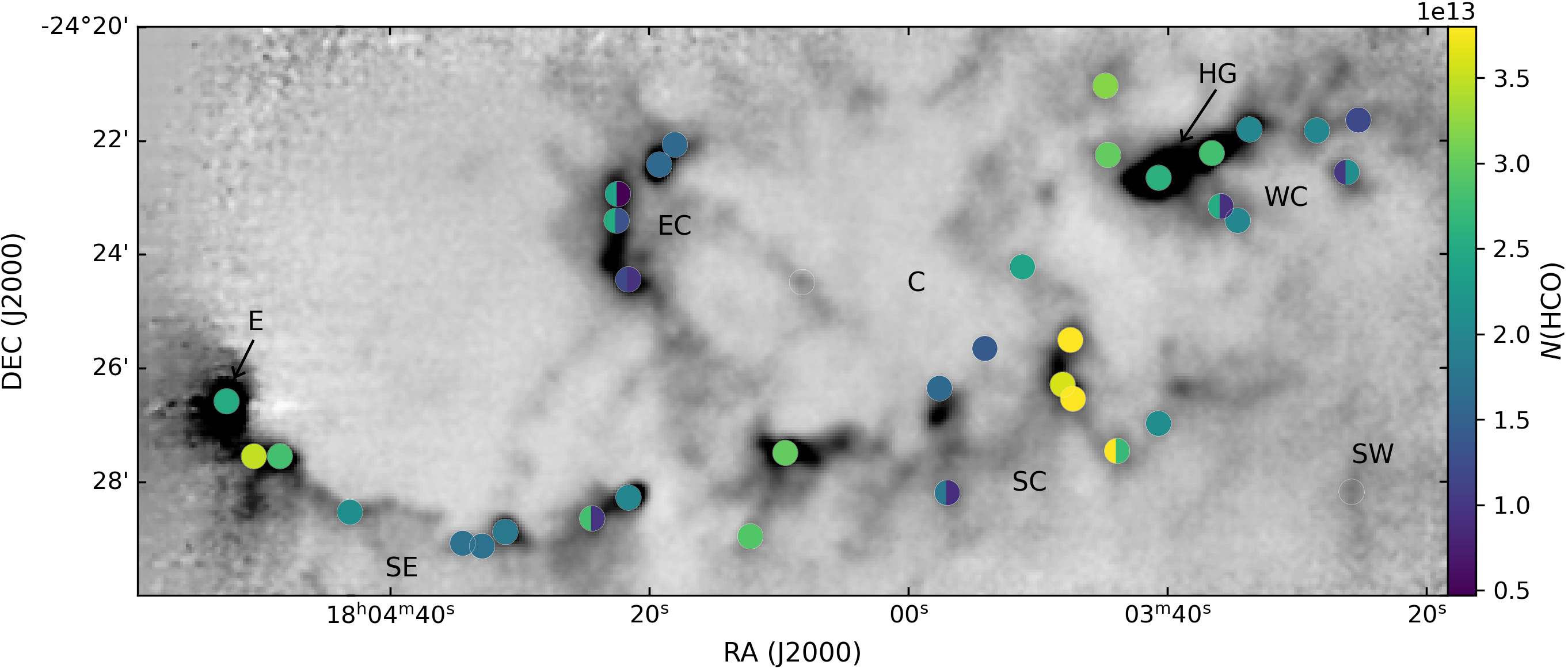}
                \end{minipage}\hfill
                \begin{minipage}[c]{0.33\textwidth}
                        \caption{HCO column densities at the M8 clumps. The circle markers correspond to the positions of the individual clumps, with the grey-scale image showing the JCMT SCUBA $\SI{870}{\micro\meter}$ dust continuum flux (see Fig.~\ref{fig:int:clumps}). The circle size corresponds to the beam size of the on-off observations with APEX and the IRAM 30m telescope. Two-coloured circles visualise distinct velocity components observed in the same beam.}
                        \label{fig:disc:hco}
                \end{minipage}
        \end{figure*}
        In general, a large variety of molecular species is detected towards all the clumps, including complex organic molecules (COMs) and deuterated species. M8 East in particular shows an overall higher number of detected species, which can attributed to its hosting of an embedded massive star-forming region~\citep{tothill2008lagoon} and a PDR on the associated clump surface~\citep{tiwari2020M8E}. The impact of these physical conditions extends to the nearby clumps SE7 and SE8, which also show a large number of observed species. In general, the south-eastern filament is chemically the richest, which is likely caused by the presence of PDRs on the northern clump surfaces (see Fig.~\ref{fig:dust:8micfull}) that alter their chemistry. The high number of detected species in the SE1 clump hereby may be due to its location at the western end of the filament. As large parts of its surface are exposed to the incoming radiation from the O- and B-type stars, a large fraction of the observed beam might be occupied with the PDR towards this clump. As PDR tracers are detected towards all of the clumps in M8, they are discussed in detail in Sect.~\ref{subsec:disc:cd}.
        
        In other regions of M8, the individual clumps EC4, EC5, and SC8 also show an enhanced number of species, possibly hinting at a rich chemistry. As discussed further in Sect.~\ref{subsec:disc:SF}, the EC4 and SC8 clumps are found to be forming low-mass stars. This can also be seen in their chemical composition, which contains cold and dense gas tracers such as N$_2$D$^{+}$ and DCO$^{+}$ in addition to the shock tracer SiO, possibly tracing an outflow from a low-mass protostar. In contrast, we detect neither SiO nor N$_2$D$^{+}$ in EC5, indicating that the condensation may not yet have cooled enough to initiate star formation.
        
        Towards the C1-3 and WC4 clumps, we only detect a few species that are mostly limited to those with the highest detection rates in the M8 cloud. As mentioned previously in Sect.~\ref{subsec:lin:rrls}, the position of WC4 is strongly affected by the ionising radiation of the nearby massive stars, possibly halting more complex chemical mechanisms in this clump. Comparing them to the remaining M8 clumps, all WC3-7 clumps show a decrease in chemical complexity, which is likely also caused by the radiation of the M8-Main \hii region.
        While the C1-3 clumps are located at a larger distance from the massive stars, their small masses (see Table~\ref{tab:app:clumpproperties}) enable the ionising radiation to penetrate the clump surfaces, reducing the chemical complexity of the clumps. Moreover, it is possible that the incoming radiation could lead to a complete disintegration of the clumps, due to their small masses.
        
        A cold and dense gas tracer, the N$_2$D$^{+}$ emission line, is detected in a total of ten clumps, which are distributed over various regions across the nebula. Bright emission of N$_2$D$^{+}$ is detected in the SE6-8 clumps, which are neighbouring the M8 East star-forming region. This further strengthens the argument of~\citet{tiwari2020M8E}, who suggest that the compression of gas introduced by the ionisation front north of M8 East may trigger star formation in this region. Interestingly, we do not detect N$_2$D$^{+}$ directly at the E clump. As E is associated with an IR-bright source, the absence of N$_2$D$^{+}$ could indicate a more evolved protostellar object in the clump core~\citep{Emprechtinger2009N2D+}.
        
        The shocked regions traced by SiO generally coincide well with the positions towards which we detect star formation (see Sect.~\ref{subsec:disc:SF}). In addition, this species is also observed in almost all clumps of the SE filament. This indicates that the ionisation front that may be triggering star formation in M8 East also extends to the SE clumps located in the west.
        
        \begin{figure*}[tbhp]
                \centering
                \includegraphics[width=0.495\textwidth]{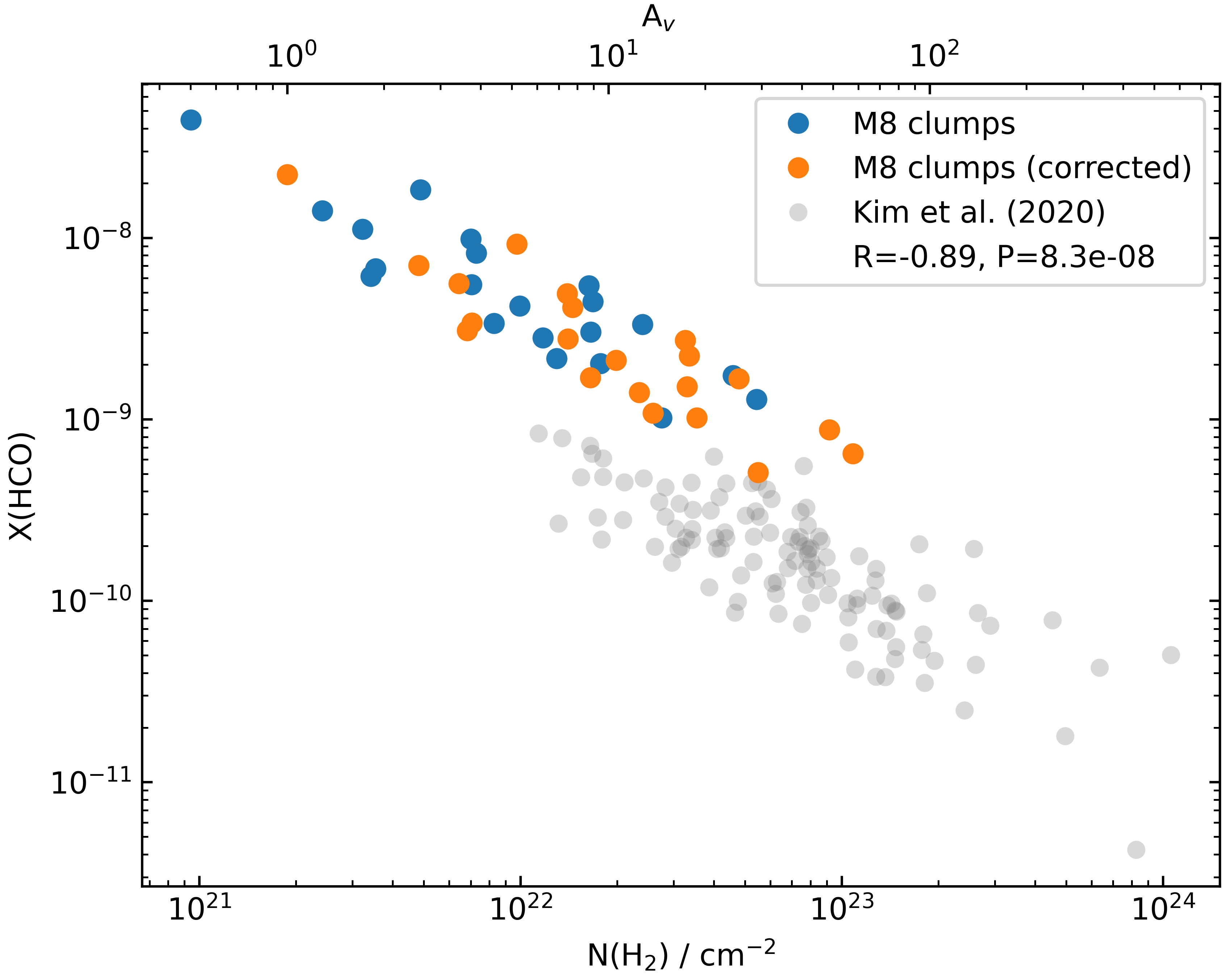}
                \includegraphics[width=0.495\textwidth]{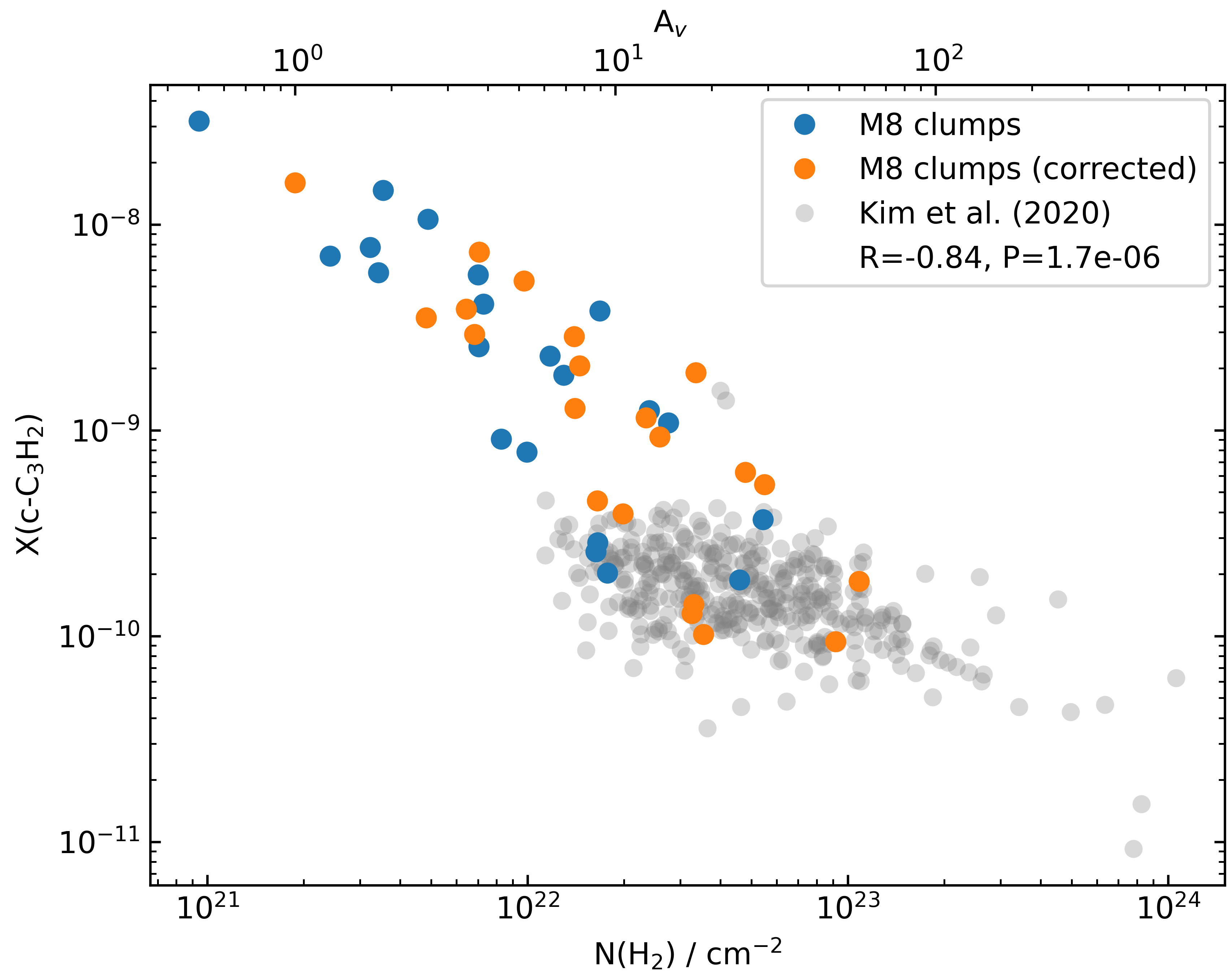}
                \caption[Abundance of HCO and c-C$_3$H$_2$ as a function of H$_2$ column density]{Abundance of HCO (left panel) and c-C$_3$H$_2$ (right panel) as a function of H$_2$ column density. The Pearson correlation coefficient R and the P value for the M8 clumps are given in the upper-right corner.}
                \label{fig:disc:pdr}
        \end{figure*}

        \subsection{PDR tracers in the Lagoon Nebula}\label{subsec:disc:cd}
        PDR tracers such as HCO and c-C$_3$H$_2$, CN, and C$_2$H are detected towards all the clumps in M8. As shown in Fig.~\ref{fig:disc:hco}, clumps associated with the known PDRs at M8-Main and M8 East show higher HCO column densities as compared to the surrounding clumps. In addition, the southern clumps SC8 and SC9 have similar column densities when compared to the known PDR regions M8-Main and M8 East, while the column densities observed towards the SC1--SC4 clumps are significantly higher. Comparing the location of these clumps with the $\SI{8}{\micro\meter}$ emission in Fig.~\ref{fig:dust:8micfull}, it can be seen that these SC clumps are associated with bright PAH emission coming from a structure that looks like an ionisation front, which is receding away from the HG region. The C2 clump can also be associated with this structure, which may explain the high HCO column densities in this overall lower density clump.
        
        \citet{Kim2020pdrtracer} examined PDR tracer abundances for a sample of massive clumps in the inner Galactic plane and find a relation of decreasing abundances with increasing H$_2$ column densities. This relation can be interpreted to mean that these species are more abundant in the less shielded outer parts of the PDRs, which are associated with stronger UV emission from external sources.
        
        To investigate whether this anti-correlation is also seen in the clumps of M8, Fig.~\ref{fig:disc:pdr} shows the PDR tracer abundances (X(HCO) =$N$(HCO)/$N$(H$_2$) and X(c-C$_3$H$_2$)=$N$(c-C$_3$H$_2$)/$N$(H$_2$)) as a function of the Hydrogen column densities ($N$(H$_2$)) determined in Sect.~\ref{sec:dust}. As explained in Appendix~\ref{app:akari}, the method used to calculate H$_2$ column densities overestimates the corresponding values by approximately a factor of 10$^{0.3}$ for source sizes around $\SI{30}{\arcsecond}$. Due to this, Fig.~\ref{fig:disc:pdr} also shows data points computed with H$_2$ column densities that were corrected by this factor. The top axis of this figure displays the visual extinction $A_v$, as derived according to the conversion $A_v = N(\U{H}_2)/1.88\times10^{21}\,\si{\per\centi\meter\squared}$~\citep{Bohlin1978h2, Frerking1982h2}.
        
        It can be seen that the observed c-C$_3$H$_2$ and HCO abundances in the M8 clumps are in agreement with the trend observed by~\citet{Kim2020pdrtracer}. Additionally, this trend can also be seen in the clumps with relatively low observed column densities in M8. This is an extension of the study by \citet{Kim2020pdrtracer}, who purely rely on massive clumps. The HCO column densities observed towards the clumps in M8 are systematically higher than the values derived by~\citet{Kim2020pdrtracer}. While we used all transitions of the HCO hyperfine structure line for deriving the column density, \citet{Kim2020pdrtracer} only consider the weakest component.  
        
        In addition to HCO and c-C$_3$H$_2$, the estimated abundances for CN and C$_2$H also follow the same trend as the clumps observed by~\citet{Kim2020pdrtracer}. This further confirms the suggestion that PDR species are located at the outer edges of the dense molecular clumps.
        
        \subsection{Star formation in M8}\label{subsec:disc:SF}
        \begin{figure}[tbp]
                \centering
                \includegraphics[width=0.49\textwidth]{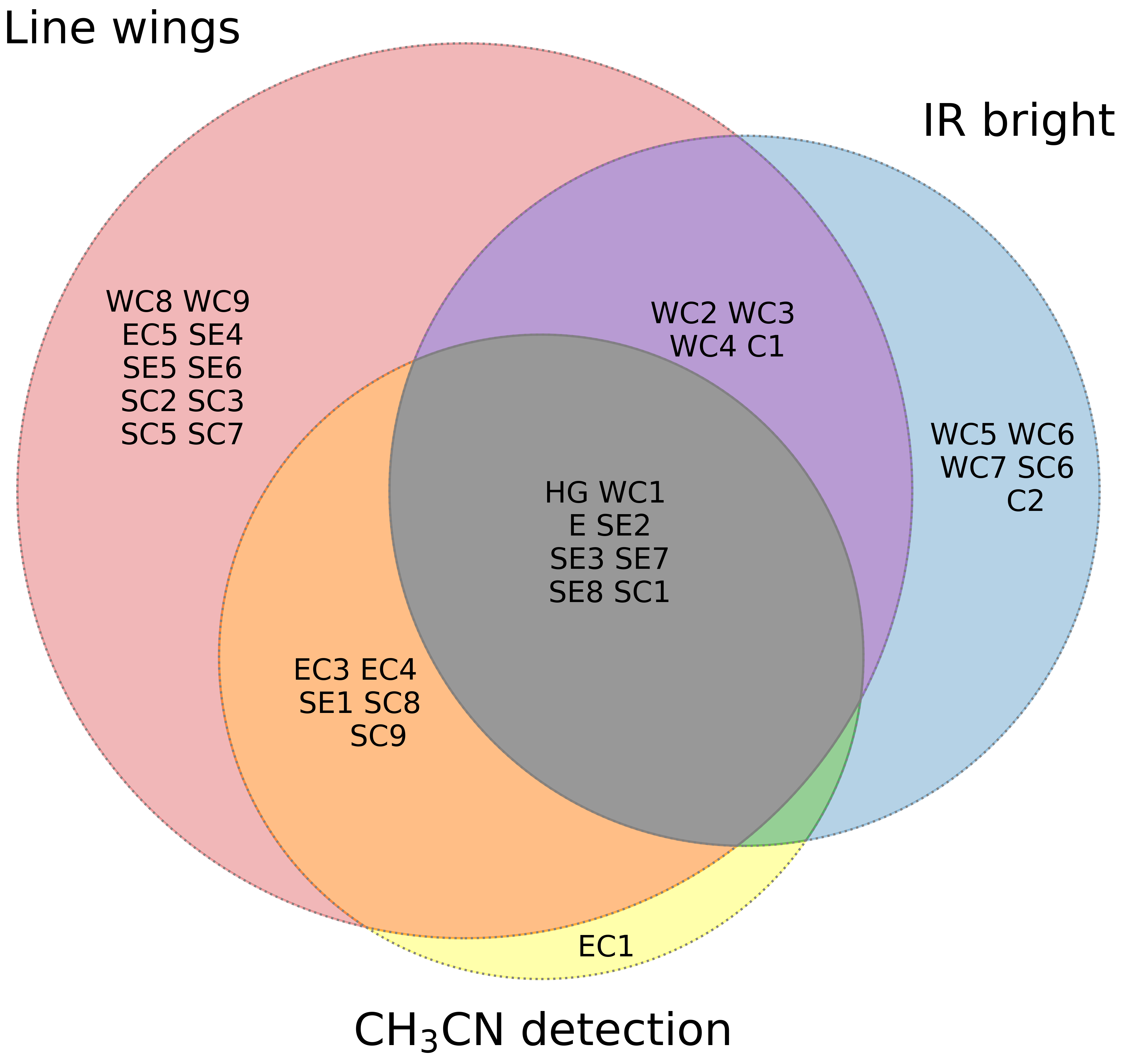}
                \caption[Venn diagram visualising the detected signs of star formation in the M8 clumps]{Venn diagram visualising the detected signs of star formation in the M8 clumps. Clumps shown in the overlapping regions show multiple tracers of star formation.}
                \label{fig:disc:venn}
        \end{figure}
        \begin{figure}[tbp]
                \centering
                \includegraphics[width=0.495\textwidth]{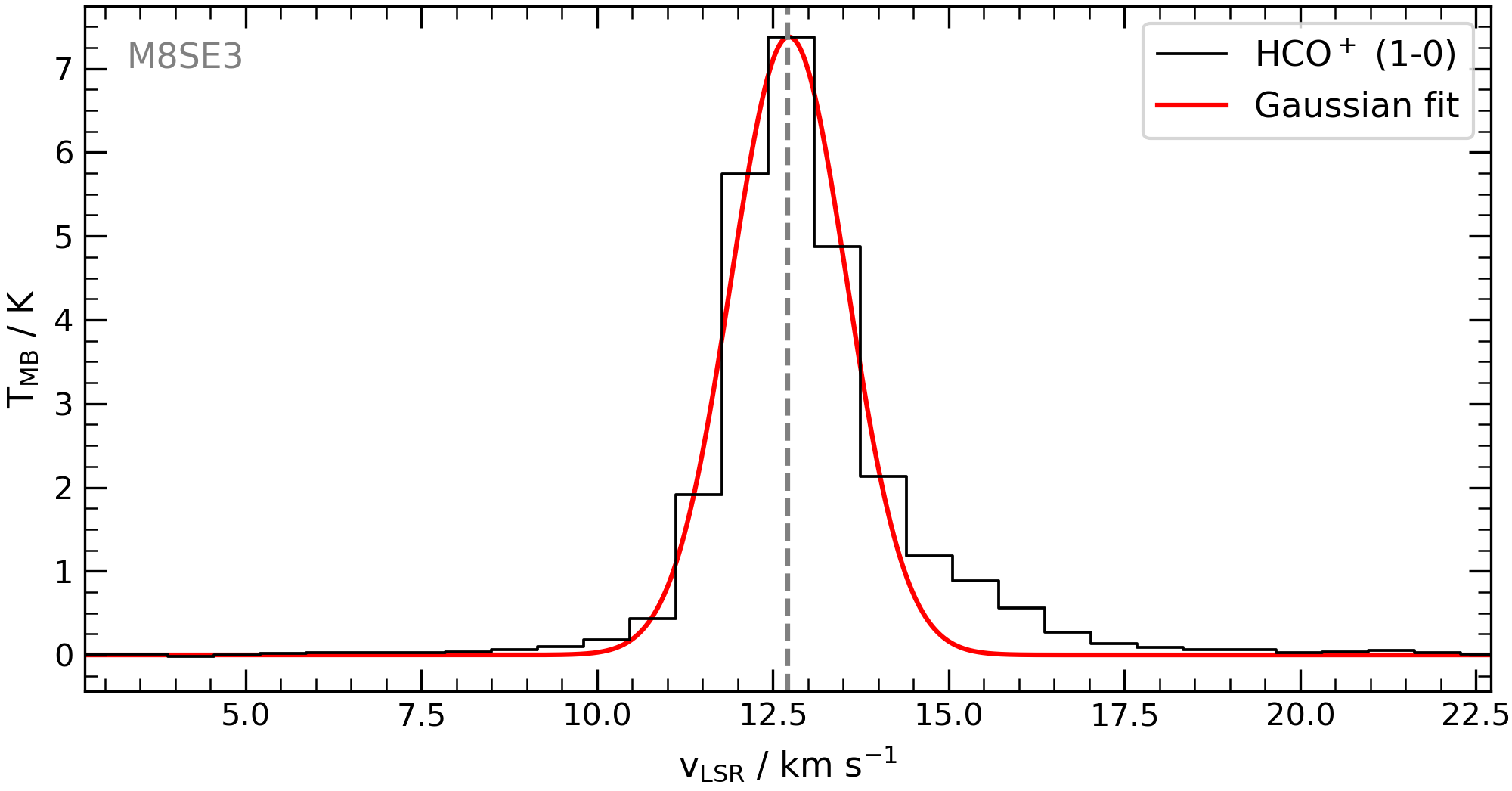}
		
                \includegraphics[width=0.495\textwidth]{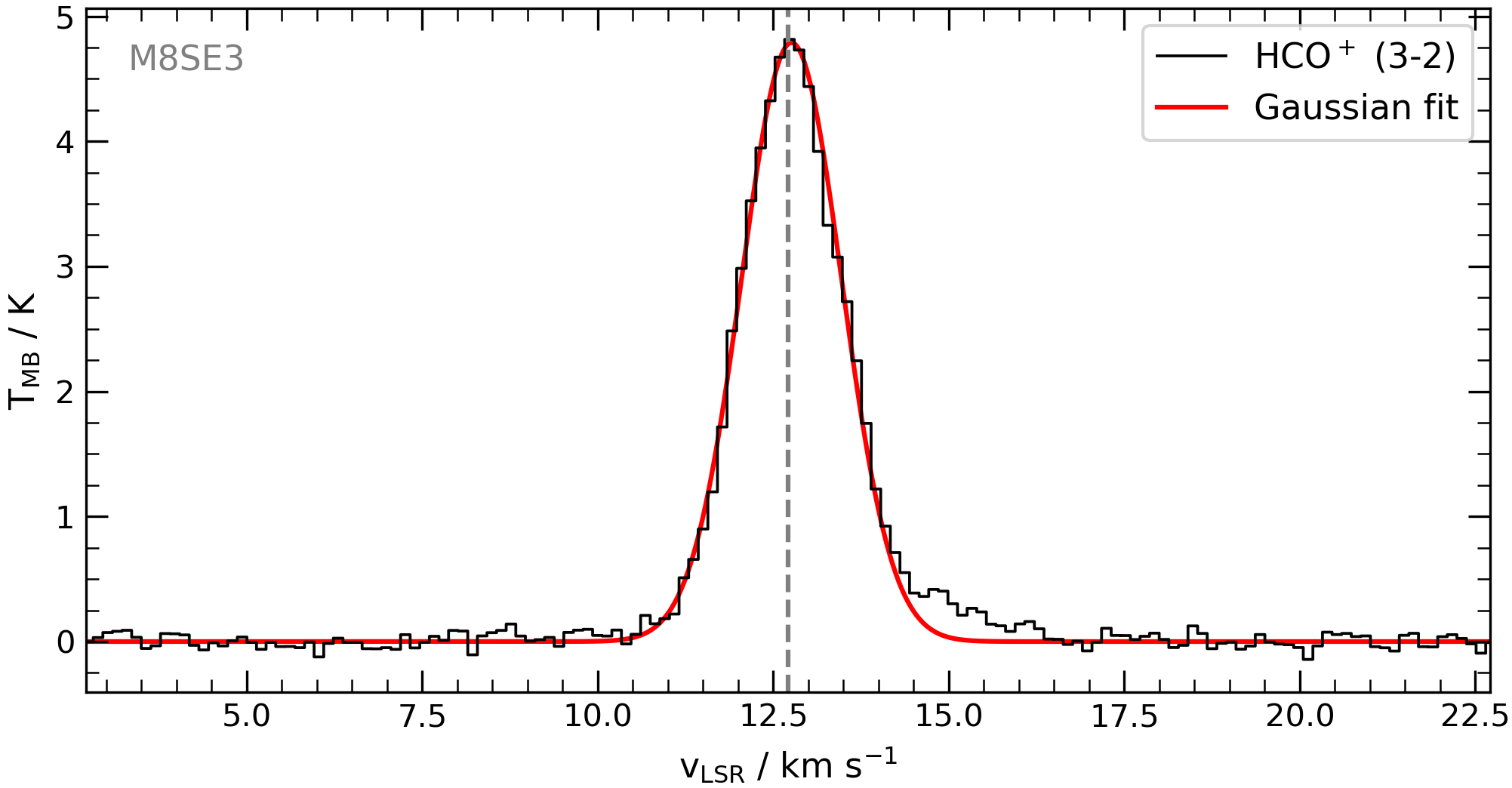}
                \caption{Line profiles of the HCO$^+$(1-0) and (3-2) transitions at SE3 with the corresponding Gaussian fits displayed in red. Excess emission can be seen around $\SI{15}{\kilo\meter\per\second}$.}
                \label{fig:disc:HCO+}
        \end{figure}
        This study of the Lagoon Nebula illustrates several signatures of star formation in the clumps based on the analysis of the dust continuum emission and the observed chemical species. In order to provide an overview of the star-forming clumps in M8, the most reliable probes are introduced in this section and the results are compared in Fig.~\ref{fig:disc:venn}.
        
        The large abundance of $^{12}$CO in the ISM makes the line wings of its transition commonly used tracers to identify protostellar outflows~\citep[e.g.][]{Duarte2013outflow_cygnusX,Kahle2022iras16293}. In M8, the spectral line profiles of $^{12}$CO are too complex for such an analysis due to the presence of many different velocity components along the LOS. As a consequence, we instead used formylium (HCO$^+$) as a probe for outflow activities, since the line profiles of this species are also very sensitive to motions of the associated gas layers~\citep[e.g.][]{Wyrowski2016outflows}. Excess emission in the line wings of the HCO$^+$ therefore indicates the presence of a protostellar outflow in the clump and that of a protostellar object driving it.
        
        For examining the M8 clumps, spectral line profiles of the HCO$^+$ (1-0) and (3-2) transitions from each clump were inspected for this excess emission by comparing the spectra of the line wings with fitted Gaussian profiles for the line emission. As an example, excess emission in the line profile at SE3 is shown in Fig.~\ref{fig:disc:HCO+}. Among the clumps where we detect excess emission, about half show a one-sided excess at redshifted velocities like SE3, while the other half show excess emission in both line wings. The kinematic complexity of the region makes it difficult to pinpoint the origin of this excess. While it is possible that excess emission is hidden by the presence of a second velocity component, weak emission from additional gas layers might be mistaken for excess emission originating from molecular outflows.
        
        Examining the mid-IR dust continuum towards M8 in Sect.~\ref{sec:dust} resulted in the identification of IR-bright sources that comprise all the clumps in the M8-Main region and several fainter point-like sources. The associated clumps might contain a bright internal heating source whose IR emission is strong enough to escape the molecular clump. As a consequence, the IR-bright clumps might be sites of intermediate to high-mass star formation. All IR-bright clumps are included in the diagram shown in Fig.~\ref{fig:disc:venn}, although it is not clear if the emission observed towards the clumps in the M8-Main region originates from the clumps themselves or from the surrounding \hii region. 
        
        CH$_3$CN has often been used to probe hot molecular cores~\citep[e.g.][]{bisshop2007ch3cn}. While we find temperatures in the clumps of M8 that are lower than in typical hot cores (which are about $\SI{100}{\kelvin}$), CH$_3$CN does seem to probe warm parts of the clumps that do not correlate with their general dust envelope (see Sect.~\ref{subsec:temps}). Therefore, we used the CH$_3$CN emission here as a probe for additional heating within the clumps.
        
        In contrast to the mid-IR emission, the millimetre line emission of CH$_3$CN escapes the molecular clumps even if the contained protostar is not very bright. The detection of this species therefore allows us to additionally trace ongoing low-mass star formation in the M8 clumps.
        
        Based on these criteria, at least eight of the M8 clumps show signs of intermediate to high-mass star formation. In addition to the known sites of high-mass star formation in HG and E, the clumps WC1, SE2, SE3, SE7, SE8, and SC1 are likely to contain a protostellar object. WC1 and SC1 correspond to some of the closest clumps to the O star Her\,36, which drives the \hii region at M8-Main. Due to this, the star formation in these objects might have been triggered by the compression introduced by the radiation of this star. The SE7 and SE8 clumps are located in the region observed by~\citet{tiwari2020M8E}, who find signs of triggered star formation across M8 East. The independent observation of star formation in this region presented in this study validates their findings.
        
        In addition to sites of intermediate to high-mass star formation, the clumps EC3, EC4, SE1, SC8, and SC9 are likely to contain low-mass protostellar objects. This follows from the detection of CH$_3$CN towards these clumps together with signs of outflows, while the absence of IR emission hints at very weakly emitting embedded objects. While the presence of excess emission could not be completely justified towards EC1 due to its complex velocity structure (see Fig.~\ref{fig:disc:lineprofiles}), the CH$_3$CN emission towards this clump indicates the presence of a hot core.
        
        \begin{figure}[tbp]
                \centering
                \includegraphics[width=0.499\textwidth]{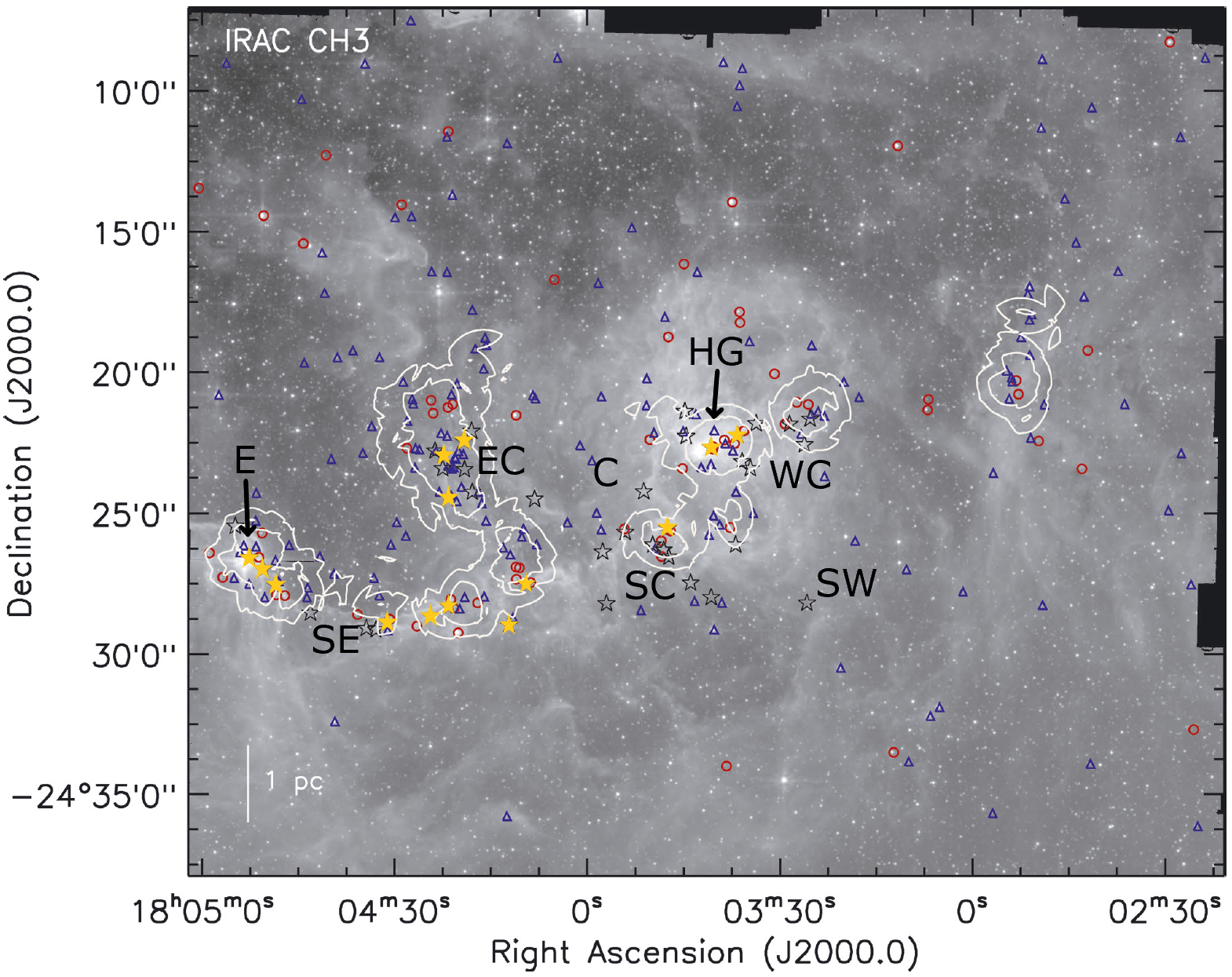}
                \caption[Star-forming clumps and IRAC class 0/I and class II sources in M8]{\textit{Spitzer} IRAC3 $\SI{5.8}{\micro\meter}$ image of the Lagoon Nebula adapted from~\citet{Kumar2010sf}. Red circles and blue triangles mark the positions of IRAC class 0/I and class II sources, respectively. Stars show the position of M8 clumps as indicated in Table 1 of~\citet{tothill2002structure}. Clumps associated with star formation based on the analysis presented here are shown in yellow. White contours show YSO densities of 5\,YSO\,pc$^{-2}$ and 10\,YSO\,pc$^{-2}$.}
                \label{fig:disc:sf}
        \end{figure}
        
        The clumps WC2, WC3, WC4, and C1 show signs of outflows and mid-IR emission. Nevertheless, the lack of CH$_3$CN emission implies that the corresponding clumps do not harbour a hot core. Consequently, the observations do not confirm star formation in these clumps. It is likely that the associated IR emission corresponds to the foreground \hii region, while lines that suggest an origin in outflows rather trace emission from unrelated gas layers.
        
        As a tracer species for shocked gas, SiO is also commonly used to identify outflows of protostars~\citep[e.g.][]{Bachiller1991SiO,Hirano2001SiO}. We detect it in most of the possibly star-forming clumps discussed above, except for EC1, EC3, SC1, and SC9. The absence of SiO may indicate that the observed excess emission in the line wings of HCO$^+$ is not related to outflow activity at these clumps. However, this does not rule out weaker embedded outflows, as the detection of CH$_3$CN implies that these clumps host warm cores. In addition to the star-forming clumps, SiO is also detected at WC7, EC2, SC2, SC6, SE4, and SE5. As we do not expect the presence of protostellar outflows at these clumps, their gas is likely shocked by external factors.

        Recent star formation in the Lagoon Nebula was previously examined by~\citet{Arias2007sf} and~\citet{Kumar2010sf} based on optical and near-IR observations. Both studies found a variety of pre-main sequence objects in the Lagoon Nebula and the associated cluster NGC\,6530. Fig.~\ref{fig:disc:sf} shows an overview of the YSOs identified by~\citet{Kumar2010sf} in the Lagoon Nebula. We have highlighted the clumps (in yellow) where star formation is observed in our study. It can be seen that most of the clumps that show signs of active star formation are located inside regions with the highest number of YSOs. While the EC clumps are associated with nearby IRAC class 0/I sources, the remaining clumps are primarily found in the vicinity of class II sources. As the IRAC class 0/I and II sources are associated with class 0/I and II protostars~\citep{billot2010classes}, this implies that the protostars contained in the EC clumps may be less evolved than objects in the remaining M8 clumps.
        %
        %
        \section{Summary}\label{sec:summary}
        We have presented the first spectroscopic observations towards 37 dense molecular clumps in the Lagoon Nebula since their identification by~\citet{tothill2002structure}. 
        Using the heterodyne receivers nFLASH230 and EMIR at APEX and the IRAM 30m telescope, we conducted pointed on-off observations in the complete frequency ranges from 210\,GHz to 280\,GHz and from 70\,GHz to 117\,GHz. 
        
        We identified a total of 346 transitions from 70 different molecular species towards the dense clumps, confirming the chemical complexity of the nebula. For every spectral line, we determined its parameters, which were then used to estimate temperatures and column densities towards every clump, applying the optically thin approximation and assuming LTE. 
        
        Combining the insights from the spectral line analysis with archival dust continuum maps, we investigated the morphology and kinematics of the region.
        We observe velocity gradients along the filaments, with many clumps showing multiple emission components, which may originate from kinematically related gas layers. On larger scales, we find a velocity gradient between the central EC clumps and the western WC clumps, which may be attributed to the momentum transfer from the massive stars into the surrounding gas. The intermediate LOS velocities of the clumps in the southern filaments indicate they are moving to the south.
        
        Heating from the O- and B-type stars in M8 is responsible for radiative feedback, due to which extended PDRs are observed throughout the nebula. Apart from the widespread \textit{Spitzer} 8\,$\si{\micro\meter}$ emission, we detected PDR tracers such as HCO, c-C$_3$H$_2$, CN, and C$_2$H towards all the clumps.
        Furthermore, the ionising radiation in the vicinity of M8-Main is traced by the detection of hydrogen, helium, and carbon RRLs.
        
        We examined the clumps in M8 for signs of low- and high-mass star formation, based on the presence of possible outflow signatures in the line profiles of HCO$^{+}$, the detection of mid-IR emission, and the detection of a warm core by observing CH$_3$CN emission. We find that 38\% of the M8 clumps are likely to host a protostellar object. 
        
        Using archival dust continuum maps from \textit{Spitzer}, MSX, WISE, \textit{Herschel}, AKARI, APEX, and the JCMT, we modelled SEDs for the clumps in M8 at wavelengths between $\SI{8}{\micro\meter}$ and $\SI{870}{\micro\meter}$. Based on these SEDs, we estimated the 
        masses, luminosities, dust temperatures, H$_2$ column densities, and upper limits on the virial parameter for all the clumps. We compared these clump properties to those in the ATLASGAL survey of clumps in the inner Galactic plane, finding that the M8 clumps show higher dust temperatures and $L/M$ ratios, which are likely caused by external heating of the clumps by the surrounding O- and B-type stars. 
        
        The clumps in M8 are found to be slightly less massive than the clumps of the ATLASGAL sample at comparable distances. This suggests a possible fragmentation of the remnant gas in M8 by the radiation pressure on the massive stars. But since the ATLASGAL sample of clumps is probably incomplete for lower masses, this scenario is not clearly proven. 
        %
        %
        \begin{acknowledgements}
                The authors would like to thank Nick Tothill for kindly providing the dust continuum maps of the Lagoon Nebula at $\SI{450}{\micro\meter}$ and  $\SI{850}{\micro\meter}$ wavelengths. We also thank James Urquhart for an early reading of the manuscript and for valuable suggestions. Furthermore, we thank the anonymous referee, whose many useful comments have considerably improved this work. This publication is based on data acquired with the Atacama Pathfinder Experiment (APEX) under programme ID [M-0107.F-9530C-2021]. APEX is a collaboration between the Max-Planck-Institut f\"{u}r Radioastronomie, the European Southern Observatory, and the Onsala Space Observatory. In addition, this work is based on observations carried out under project number 141-21 with the IRAM 30m telescope. IRAM is supported by INSU/CNRS (France), MPG (Germany) and IGN (Spain). This work was partially funded by the Collaborative Research Council 956 "Conditions and impact of star formation" funded by the Deutsche Forschungsgemeinschaft (DFG). 
        \end{acknowledgements}

        \bibliographystyle{aa} 
        \bibliography{M8} 
        
        \begin{appendix}
                \section{Observational parameters: Coordinates and setups}\label{app:obs}
\begin{figure}[htbp]
                        \centering
                        \includegraphics[width=0.499\textwidth]{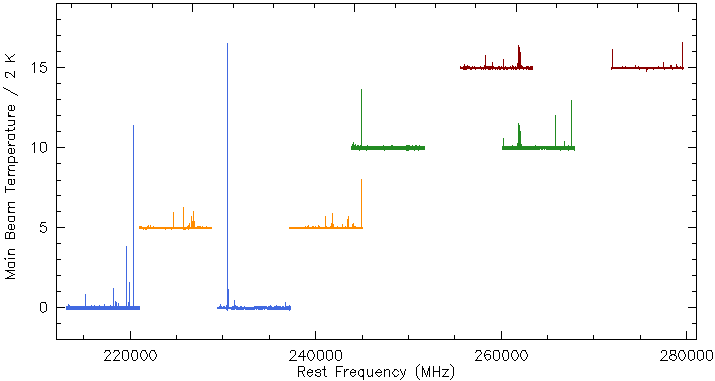}
                        \includegraphics[width=0.499\textwidth]{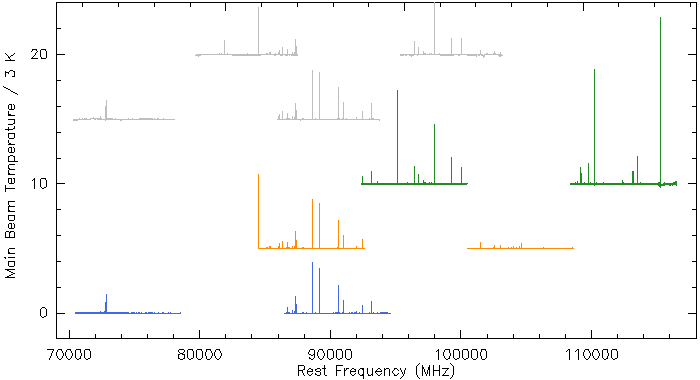}
                        \caption[Visualisation of the frequency coverage]{Visualisation of the frequency coverage. The spectra originate from the observations of E, which are divided by two (upper panel) or three (lower panel) and shifted on the temperature scale for illustrative purposes. Upper panel: The four setups used in the APEX observations. Lower panel: Setups (in colour) of the new observations conducted with the IRAM 30m telescope. Grey spectra show the additional frequency bands covered by the on-off observations of~\citet{tiwari2020M8E}, which are used to supplement our data at the position of clump E.}
                        \label{app:fig:setup}
                \end{figure}

                The on-off observations of the Lagoon Nebula were conducted towards 37 clumps identified by~\citet{tothill2002structure}. Coordinates of the observed positions are given in Table~\ref{tab:app:clumppositions}.
                Further inspection of the dust continuum emission and the APEX spectra revealed that the clumps WC3, SE8, and SC5 are at an offset from the observed positions given in  \citet{tothill2002structure}.
                Thus, the coordinates for these clumps were adjusted for the new observations taken with the IRAM 30m telescope. The frequency setups used with each telescope are shown in Fig.~\ref{app:fig:setup}.
                
                \begin{table}[htbp]
                        \caption{Coordinates of all the clumps observed with the APEX and the IRAM 30m telescopes.}
                        \label{tab:app:clumppositions}
                        \centering
                        \begin{tabular}{ccc}
                                \hline
                                \hline
                                Clump & RA (J2000) & Dec. (J2000) \\ \hline
                                HG & 18$^\U{h}$03$^\U{m}$40.7$^\U{s}$  & -24$^{\circ}$22$'$40$''$ \\
                                WC1 & 18$^\U{h}$03$^\U{m}$36.6$^\U{s}$  & -24$^{\circ}$22$'$14$''$  \\
                                WC2 & 18$^\U{h}$03$^\U{m}$33.7$^\U{s}$  & -24$^{\circ}$21$'$49$''$  \\
                                WC3* & 18$^\U{h}$03$^\U{m}$44.8$^\U{s}$  & -24$^{\circ}$21$'$23$''$  \\
                                WC4 & 18$^\U{h}$03$^\U{m}$44.6$^\U{s}$  & -24$^{\circ}$22$'$16$''$  \\
                                WC5 & 18$^\U{h}$03$^\U{m}$35.9$^\U{s}$  & -24$^{\circ}$23$'$10$''$  \\
                                WC6 & 18$^\U{h}$03$^\U{m}$34.6$^\U{s}$  & -24$^{\circ}$23$'$25$''$  \\
                                WC7 & 18$^\U{h}$03$^\U{m}$26.2$^\U{s}$  & -24$^{\circ}$22$'$34$''$  \\
                                WC8 & 18$^\U{h}$03$^\U{m}$28.5$^\U{s}$  & -24$^{\circ}$21$'$50$''$  \\
                                WC9 & 18$^\U{h}$03$^\U{m}$25.3$^\U{s}$  & -24$^{\circ}$21$'$39$''$  \\
                                SW1 & 18$^\U{h}$03$^\U{m}$25.8$^\U{s}$  & -24$^{\circ}$28$'$11$''$  \\
                                EC1 & 18$^\U{h}$04$^\U{m}$21.6$^\U{s}$  & -24$^{\circ}$24$'$27$''$  \\
                                EC2 & 18$^\U{h}$04$^\U{m}$22.5$^\U{s}$  & -24$^{\circ}$23$'$25$''$  \\
                                EC3 & 18$^\U{h}$04$^\U{m}$22.4$^\U{s}$  & -24$^{\circ}$22$'$57$''$  \\
                                EC4 & 18$^\U{h}$04$^\U{m}$19.2$^\U{s}$  & -24$^{\circ}$22$'$26$''$  \\
                                EC5 & 18$^\U{h}$04$^\U{m}$18.0$^\U{s}$  & -24$^{\circ}$22$'$05$''$  \\
                                E & 18$^\U{h}$04$^\U{m}$52.6$^\U{s}$  & -24$^{\circ}$26$'$35$''$  \\
                                SE1 & 18$^\U{h}$04$^\U{m}$21.6$^\U{s}$  & -24$^{\circ}$28$'$17$''$  \\
                                SE2 & 18$^\U{h}$04$^\U{m}$24.4$^\U{s}$  & -24$^{\circ}$28$'$39$''$  \\
                                SE3 & 18$^\U{h}$04$^\U{m}$31.1$^\U{s}$  & -24$^{\circ}$28$'$53$''$  \\
                                SE4 & 18$^\U{h}$04$^\U{m}$32.9$^\U{s}$  & -24$^{\circ}$29$'$08$''$  \\
                                SE5 & 18$^\U{h}$04$^\U{m}$34.4$^\U{s}$  & -24$^{\circ}$29$'$05$''$  \\
                                SE6 & 18$^\U{h}$04$^\U{m}$43.1$^\U{s}$  & -24$^{\circ}$28$'$32$''$  \\
                                SE7 & 18$^\U{h}$04$^\U{m}$48.5$^\U{s}$  & -24$^{\circ}$27$'$33$''$  \\
                                SE8* & 18$^\U{h}$04$^\U{m}$50.5$^\U{s}$  & -24$^{\circ}$26$'$59$''$  \\
                                SC1 & 18$^\U{h}$03$^\U{m}$47.5$^\U{s}$  & -24$^{\circ}$25$'$31$''$  \\
                                SC2 & 18$^\U{h}$03$^\U{m}$48.1$^\U{s}$  & -24$^{\circ}$26$'$18$''$  \\
                                SC3 & 18$^\U{h}$03$^\U{m}$47.3$^\U{s}$  & -24$^{\circ}$26$'$33$''$  \\
                                SC4 & 18$^\U{h}$03$^\U{m}$43.9$^\U{s}$  & -24$^{\circ}$27$'$28$''$  \\
                                SC5* & 18$^\U{h}$03$^\U{m}$40.7$^\U{s}$  & -24$^{\circ}$27$'$59$''$  \\
                                SC6 & 18$^\U{h}$03$^\U{m}$57.6$^\U{s}$  & -24$^{\circ}$26$'$22$''$  \\
                                SC7 & 18$^\U{h}$03$^\U{m}$57.0$^\U{s}$  & -24$^{\circ}$28$'$12$''$  \\
                                SC8 & 18$^\U{h}$04$^\U{m}$09.5$^\U{s}$  & -24$^{\circ}$27$'$30$''$  \\
                                SC9 & 18$^\U{h}$04$^\U{m}$12.2$^\U{s}$  & -24$^{\circ}$28$'$58$''$  \\
                                C1 & 18$^\U{h}$03$^\U{m}$54.1$^\U{s}$  & -24$^{\circ}$25$'$40$''$  \\
                                C2 & 18$^\U{h}$03$^\U{m}$51.2$^\U{s}$  & -24$^{\circ}$24$'$14$''$  \\
                                C3 & 18$^\U{h}$04$^\U{m}$08.2$^\U{s}$  & -24$^{\circ}$24$'$30$''$  \\
                                \hline
                        \end{tabular}
                        \tablefoot{For clumps marked with an asterisk, we used corrected coordinates when observing with the IRAM 30m telescope and when extracting flux density from the dust continuum maps (WC3: 18$^\U{h}$03$^\U{m}$44.8$^\U{s}$, -24$^{\circ}$21$'$03$''$; SE8: 18$^\U{h}$04$^\U{m}$50.5$^\U{s}$, -24$^{\circ}$27$'$33$''$; SC5: 18$^\U{h}$03$^\U{m}$40.7$^\U{s}$, $-24^{\circ}$26$'$59$''$).}
                \end{table}

                \section{Testing AKARI data as a substitute for Hi-GAL-PACS data in SEDs}\label{app:akari}
                In order to derive SEDs for the clumps in M8, data from the AKARI satellite was used to estimate flux densities between $\SI{65}{\micro\meter}$ and $\SI{160}{\micro\meter}$. In comparison to data from the Hi-GAL survey, the AKARI images have a lower resolution, with FWHM beam widths of $\SI{63.4}{\arcsecond}$ at $\SI{65}{\micro\meter}$ (N60),  $\SI{77.8}{\arcsecond}$ at $\SI{90}{\micro\meter}$ (WIDE-S), and $\SI{88.3}{\arcsecond}$ for the $\SI{140}{\micro\meter}$ and $\SI{160}{\micro\meter}$ bands (WIDE-L and N160)~\citep{Takita2015akari}. As a consequence, the M8 clumps with FWHM sizes lower than $\SI{38.3}{\arcsecond}$ are not spatially resolved. Due to this, the flux extraction method used here for the AKARI images differs from the method applied by \citet{urquhart2018atlasgal}. Instead of extracting the AKARI flux in an aperture with a radius based on the source size, the flux density within an AKARI beam was extracted at the exact position of the M8 clump (see Table~\ref{tab:app:clumppositions}), which accounts for most of a clump's emission. 
                
                In order to compare the M8 clumps to the sample of ATLASGAL clumps located in the inner Galactic plane examined by~\citet{urquhart2018atlasgal}, it is necessary to identify possible differences in the results of the two flux
                retrieval methods. For this, the SEDs of a sample of clumps in the NGC\,6334 region were computed with the AKARI images as a replacement for the Hi-GAL PACS data. The clumps in this region provide ideal targets for testing the usability of
                AKARI data for our cause, as NGC\,6334 is located at a distance of $\SI{1.34}{pc}$~\citep{urquhart2018atlasgal}, which is similar to the distance of the Lagoon Nebula. This ensures that the method is tested on objects with similar properties as the M8 clumps.
                
                From the clumps covered by both AKARI and Hi-GAL, a selection was made considering only bright sources that are well separated from other objects in the region. While AKARI conducted an all-sky survey that covers 99\% of the sky, the data have a large image defect covering the brightest sources of NGC\,6334, limiting the sample size of sources suitable for this test to 16.
                
                The flux densities of the NGC\,6334 clumps were extracted analogously to the M8 clumps, as explained in Sect.~\ref{subsec:obs:continuum}. Since there is no available SCUBA data from the JCMT that covers the whole NGC\,6334 region, the $\SI{870}{\micro\meter}$ flux density was derived purely based on the ATLASGAL data. Because the $\SI{870}{\micro\meter}$ data of the JCMT and ATLASGAL are in good agreement for the M8 clumps, the SEDs should be unaffected by this change. 
                
                \begin{figure}[thbp]
                        \centering
                        \subfigure{\includegraphics[width=0.502\textwidth]{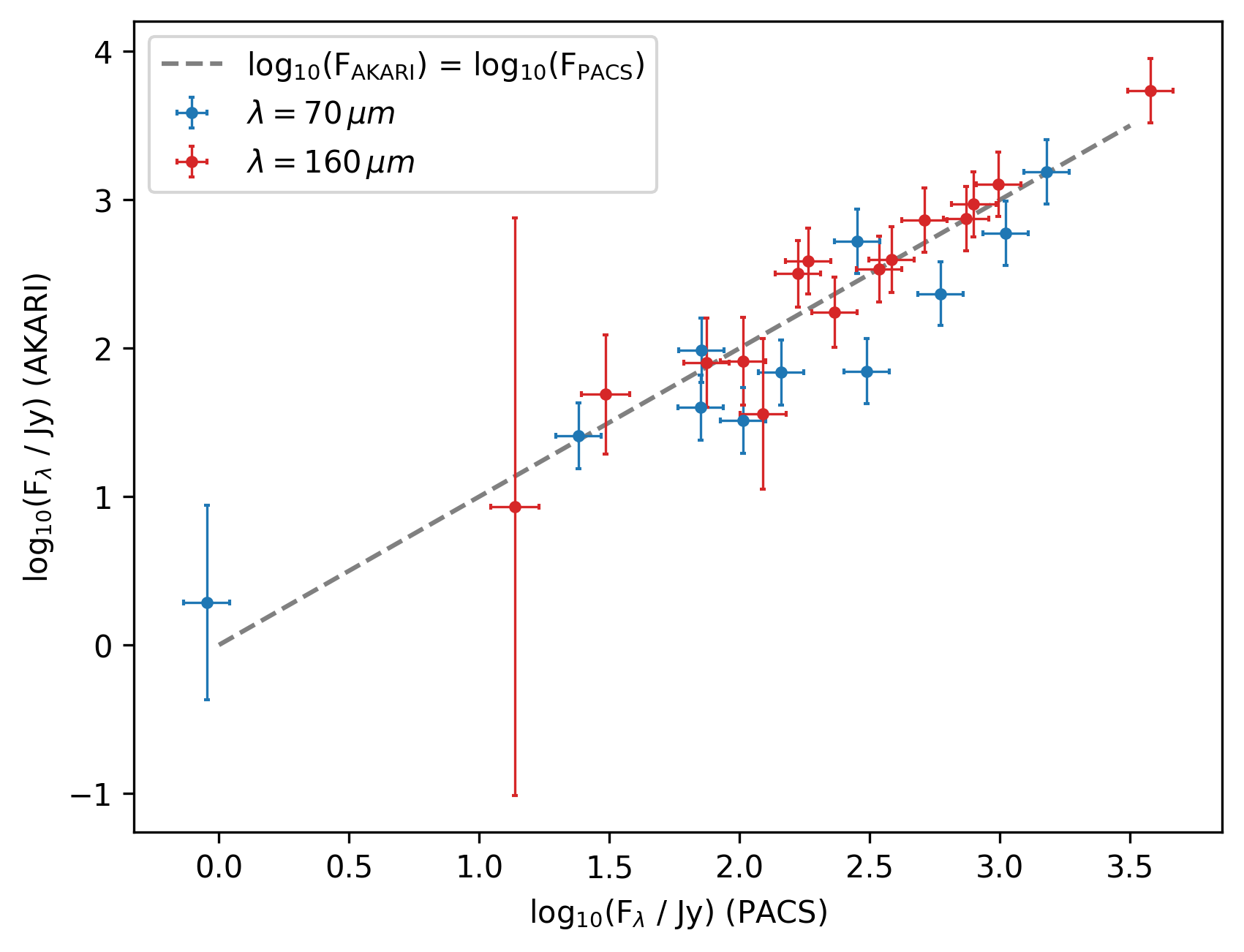}}
                        \subfigure{\includegraphics[width=0.488\textwidth]{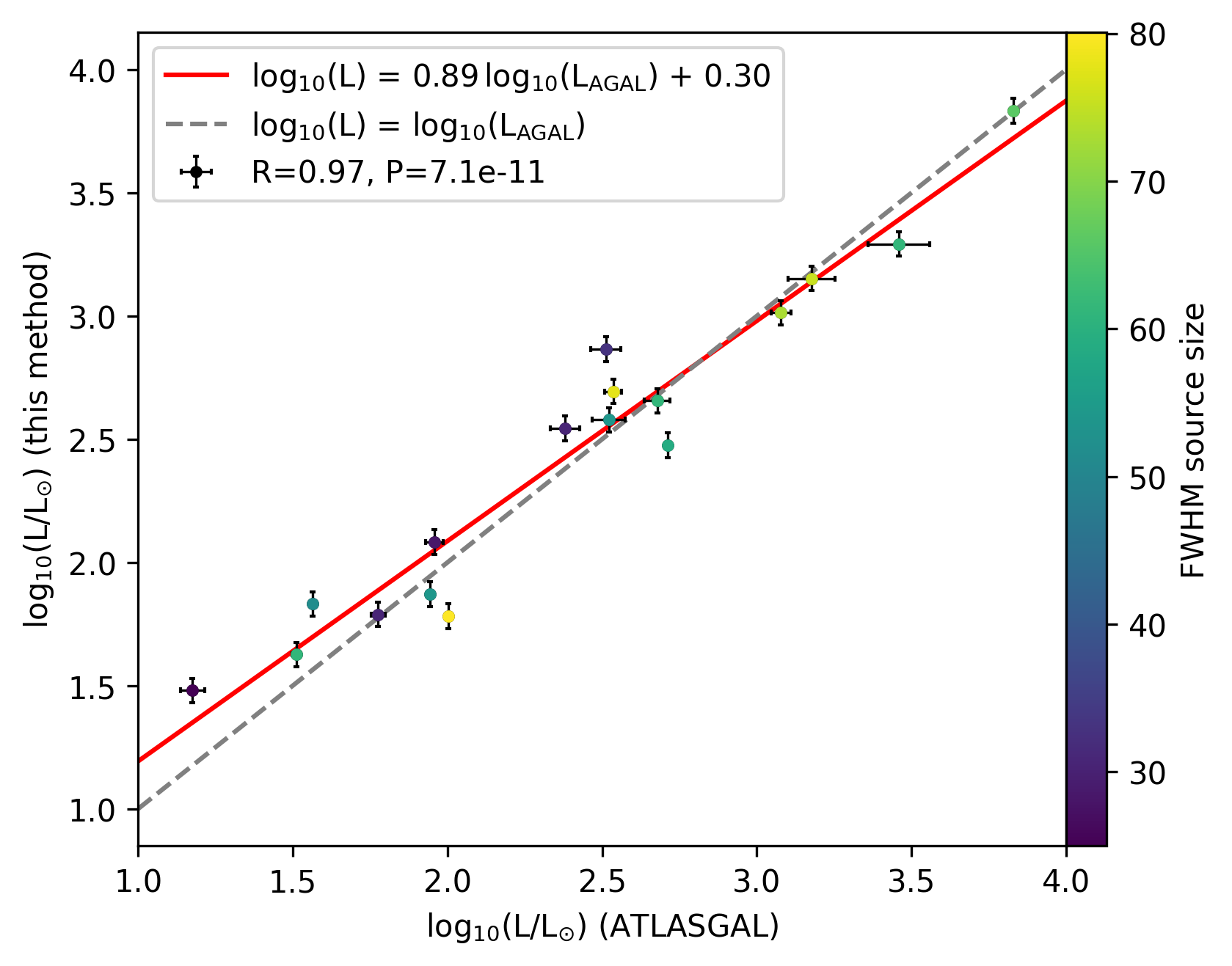}}
                        \caption[NGC 6334: Comparison between AKARI and PACS -- flux density and luminosity]{Upper panel: Comparison of the flux density extracted inside a $\SI{40.9}{\arcsecond}$ beam for the AKARI and PACS images of selected clumps in NGC\,6334. Lower panel: Luminosities of these clumps as derived from fitting SEDs with the AKARI data as compared to the luminosities derived by~\citet{urquhart2018atlasgal} for the same clumps. The fitted linear relation and the Pearson correlation coefficient R and the P value are given in the upper-left corner.}
                        \label{fig:dust:akari_cal}
                \end{figure}
                
                Firstly, we verified the calibration of the AKARI data in order to exclude the possibility of a systematic under- or overestimation of the derived flux densities. This is done by extracting the flux densities measured by AKARI-FIS and Hi-GAL-PACS instruments according to the method explained in Sect.~\ref{subsec:obs:continuum}. As shown in the upper panel of Fig.~\ref{fig:dust:akari_cal}, the flux densities extracted from the Hi-GAL datasets at $\SI{70}{\micro\meter}$ and $\SI{160}{\micro\meter}$ are in good agreement with the flux densities derived from AKARI images. The slightly lower flux density values of AKARI in the $\SI{70}{\micro\meter}$ band are likely due to the different wavelengths covered by the respective receivers. While the AKARI band is centred on $\SI{65}{\micro\meter}$, the PACS band is centred on $\SI{70}{\micro\meter}$. This difference is also seen when comparing the spectral response functions of both instruments shown in Fig. 1 of~\citet{shirahata2009akaribeam} and Fig. 6 of~\citet{poglitsch2010pacs}.
                
                To test the influence of the different extraction methods applied for the AKARI bands, the SED fitting procedure was applied to all sample sources, and the physical parameters were derived analogously to~\citet{urquhart2018atlasgal} using the equations given by~\citet{Schuller2009atlasgal}. The lower panel of Fig.~\ref{fig:dust:akari_cal} shows a comparison between the computed luminosities and the values derived by~\citet{urquhart2018atlasgal} for the same sources. The luminosities derived using both methods are in good agreement for the examined range between 10 and $10^4\,\si{M_\odot}$. A linear fit leads to a slope slightly below 1, indicating that lower luminosities might be overestimated, while higher luminosities could be underestimated. This can be explained by the constant extraction size of one AKARI beam size, which for larger and brighter sources may not cover the full source emission. Thus, when dealing with much larger sources, it might be beneficial to extract the AKARI emission with a defined aperture analogous to the other bands. In contrast, accurate luminosity values for weak sources cannot be assured if multiple sources are located within the AKARI beam.
                
                \begin{figure}[tbhp]
                        \centering
                        \subfigure{\includegraphics[width=0.49\textwidth]{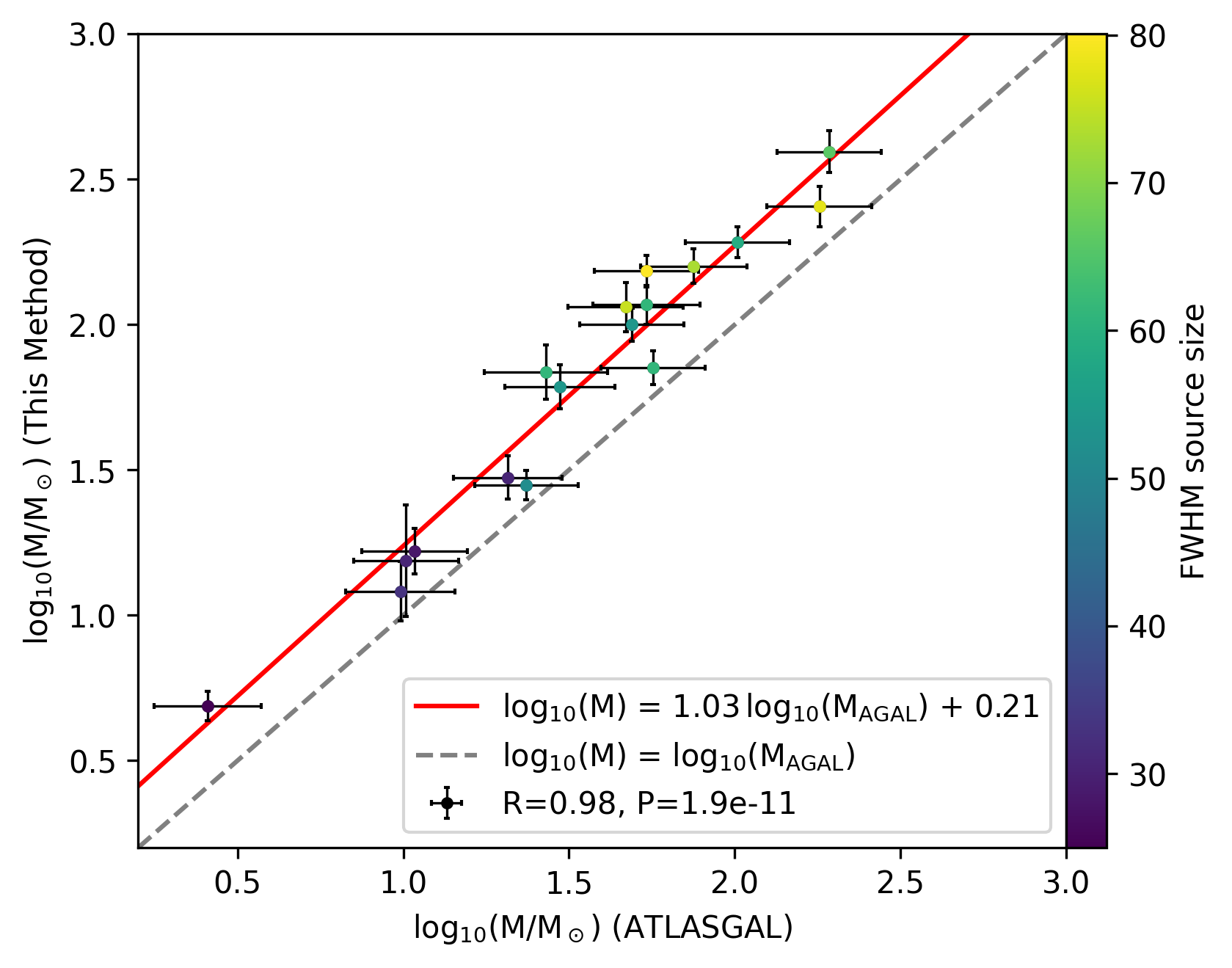}}
                        \subfigure{\includegraphics[width=0.50\textwidth]{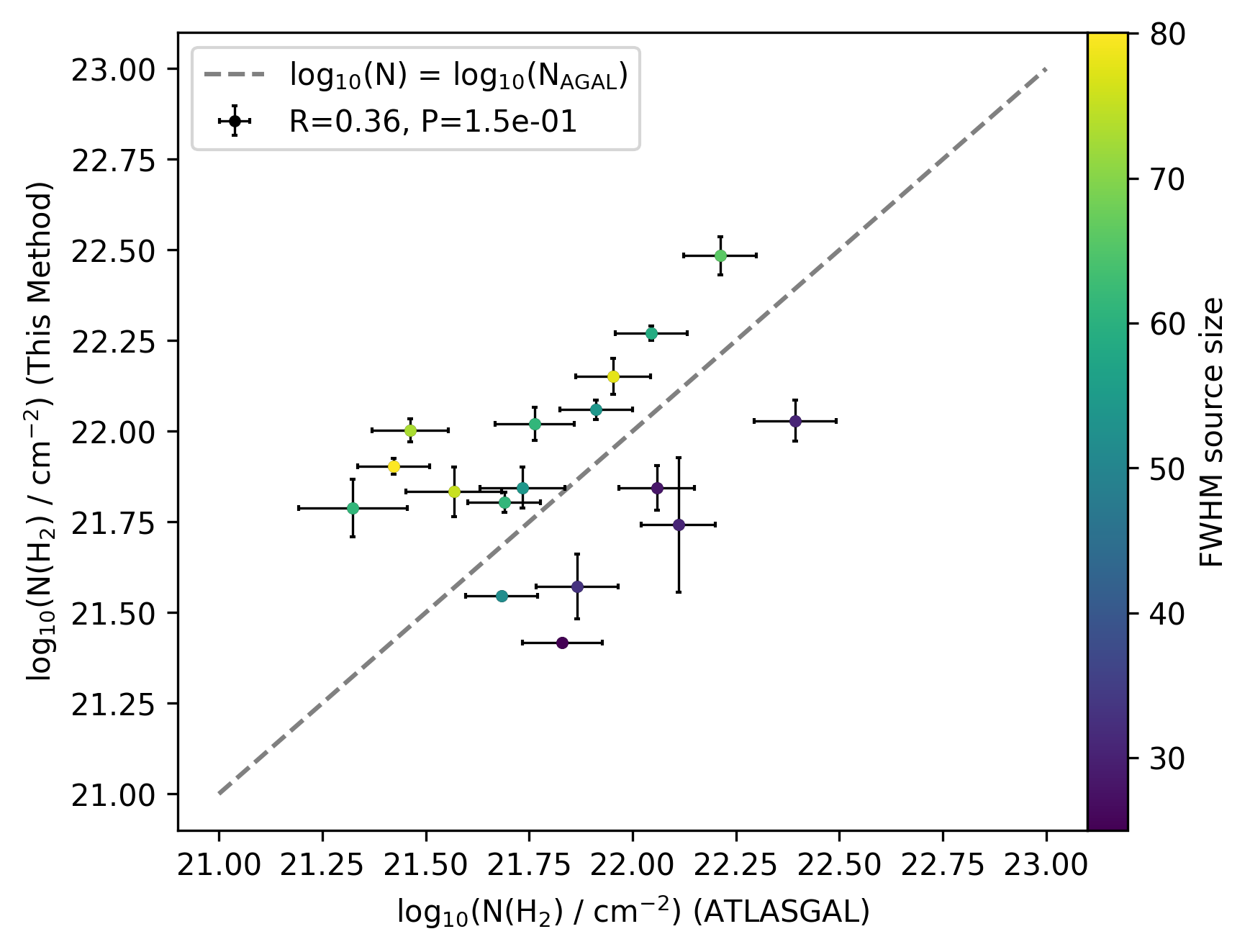}}
                        \caption[NGC\,6334: Comparison of masses and column densities]{Comparison of the computed masses (upper panel) and column densities (lower panel) for selected clumps in NGC\,6334 to the values derived by~\citet{urquhart2018atlasgal} for the same clumps. The Pearson correlation coefficient, R, and the P value are given in the lower-right and upper-left corner, respectively, where also the fitted linear relations are indicated.}
                        \label{fig:dust:comp_mn}
                \end{figure}
                
                As shown in the upper panel of Fig.~\ref{fig:dust:comp_mn}, the applied method introduces a slight overestimation of the clumps masses, which is similar across the considered clump masses between $3$ and $300\,\si{M_\odot}$. Nevertheless, there is a good correlation between the derived masses, and the values computed by~\citet{urquhart2018atlasgal}. On the contrary, The derived H$_2$ column densities do not show a clear correlation to the column densities derived by~\citet[see the lower panel of our Fig.~\ref{fig:dust:comp_mn}]{urquhart2018atlasgal}. While the column density is overestimated for intermediate to larger clumps, the column density is underestimated for smaller clumps. Both quantities are beam-specific quantities, and might differ due to the different methods used to derive the source sizes. While we used the source FWHM based on the ATLASGAL measurements as size estimate, \citet{urquhart2018atlasgal} based their size estimate on the extent of the $\SI{350}{\micro\meter}$ emission inside the extraction aperture.
                
                \begin{figure}[thbp]
                        \centering
                        \includegraphics[width=0.499\textwidth]{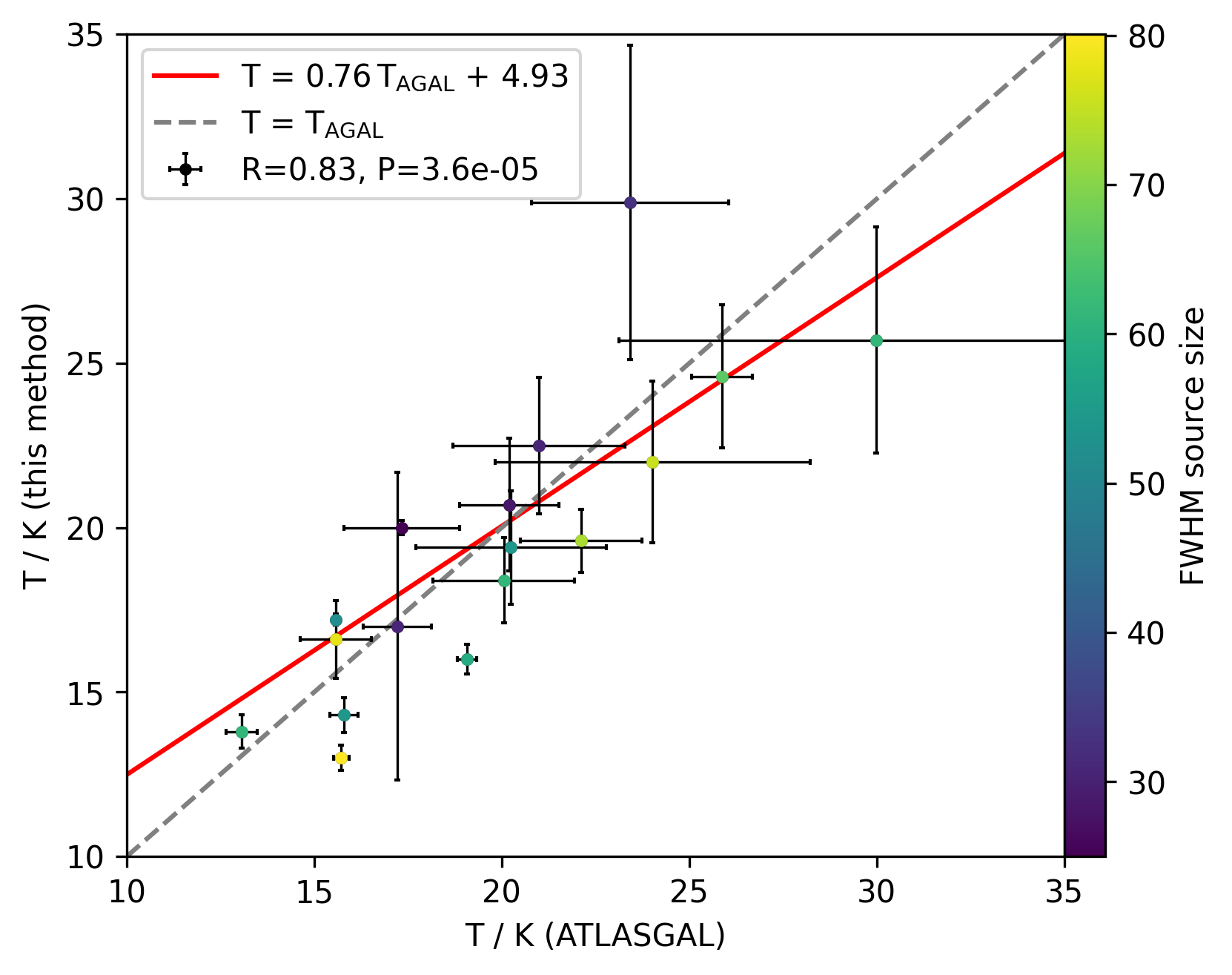}
                        \caption[NGC\,6334: Comparison of dust temperatures]{Comparison of the computed dust temperatures for selected clumps in NGC\,6334 to the values derived by~\citet{urquhart2018atlasgal} for the same clumps. The fitted linear relation and Pearson correlation coefficient R and the P value are given in the upper-left corner.}
                        \label{fig:dust:comp_t}
                \end{figure}
                
                Finally, the dust temperatures between $\SI{15}{\kelvin}$ and $\SI{30}{\kelvin}$ derived by both methods are consistent with one another, independent of the respective source sizes (see Fig.~\ref{fig:dust:comp_t}). This is expected, as the cold gas component is usually broadly extended, such that it fills the beam in all considered cases. 
                
                This analysis demonstrates that AKARI has the potential to be used for deriving SEDs of clumps away from the Galactic plane, which were not covered by the PACS instrument of \textit{Herschel}. We note that the lower resolutions of the AKARI images, as compared to Hi-GAL, limits the use of the data in crowded regions, as the large beam may capture emission from neighbouring sources. While it is possible to correct for the influence of additional sources inside the AKARI beam as described in Sect.~\ref{subsec:obs:continuum}, this method may become inaccurate for regions with a higher number of sources than in M8. We note that for large sources it will be beneficial to extract the AKARI flux density inside an aperture that is based on the source size, as applied for the data of other surveys. This would prevent emission in the outer areas of these extended sources from being excluded.
                
                \onecolumn
                
                \section{Physical properties of the M8 clumps}\label{app:sed}
\begin{figure}[htbp]
                        \centering
                        \begin{minipage}[b]{0.4\linewidth}
                                \centering
                                \includegraphics[width=\linewidth]{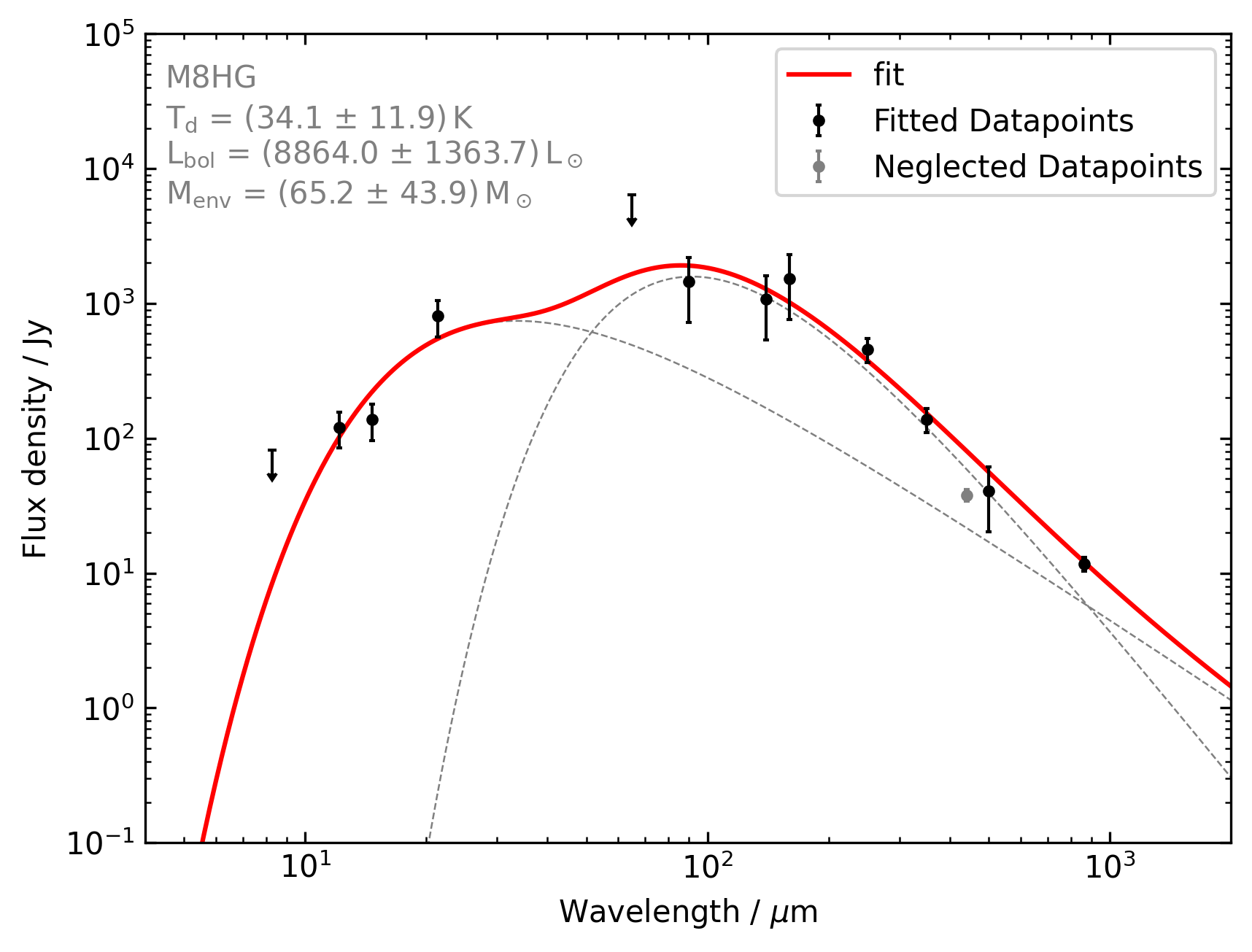}
                        \end{minipage}
                        \begin{minipage}[b]{0.4\linewidth}
                                \centering
                                \includegraphics[width=\linewidth]{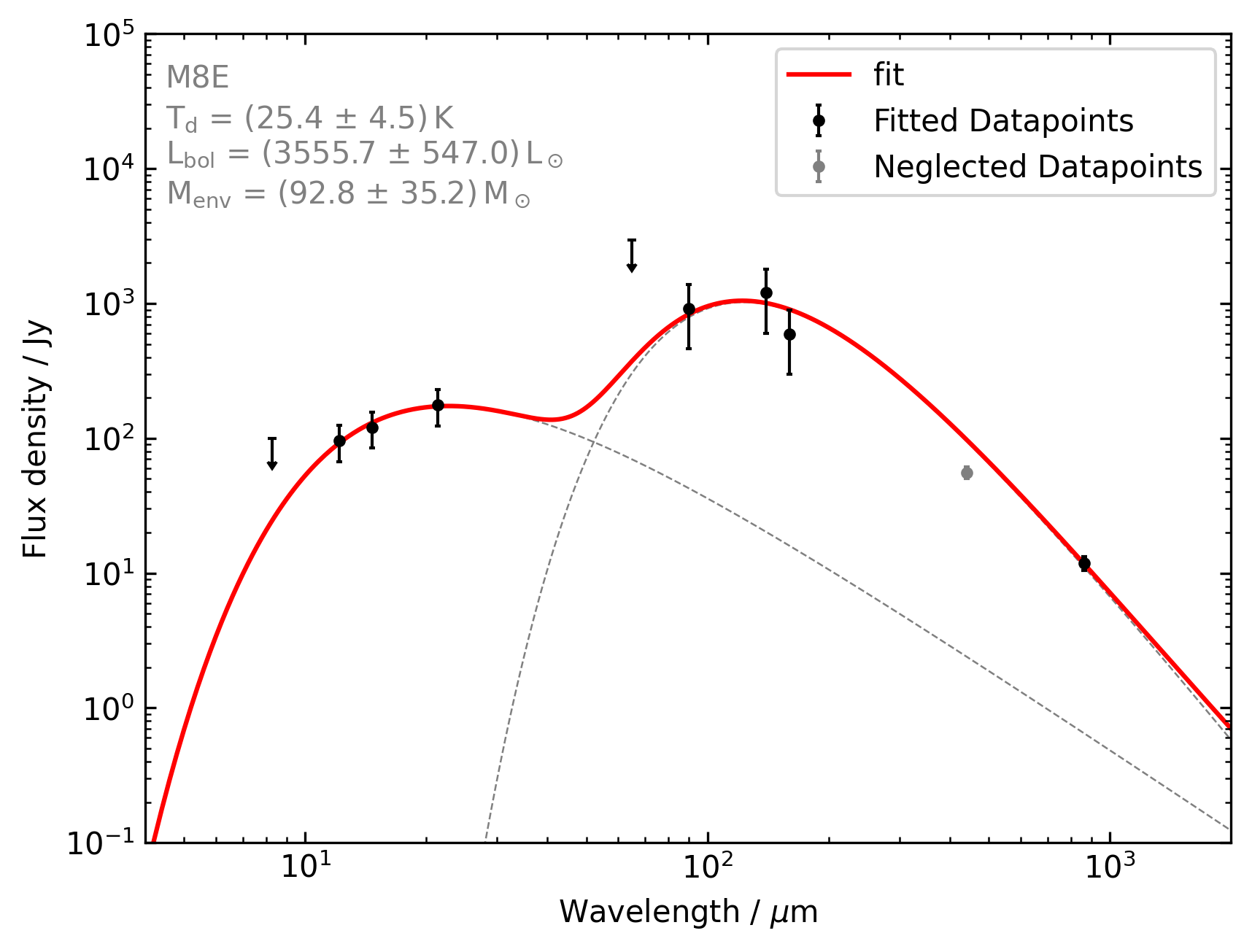}
                        \end{minipage}
                        \\
                        \begin{minipage}[b]{0.4\linewidth}
                                \centering
                                \includegraphics[width=\linewidth]{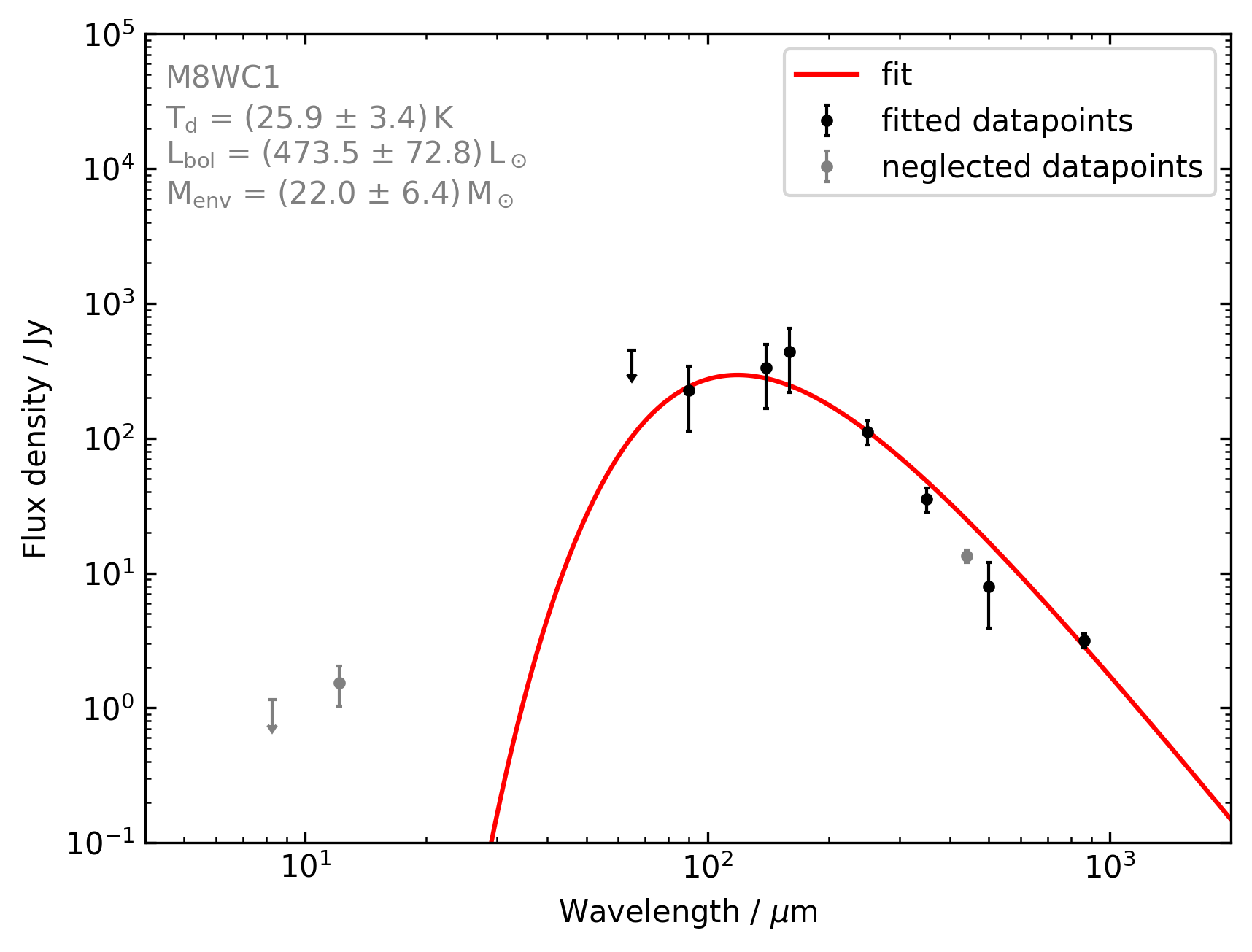}
                        \end{minipage}
                        \begin{minipage}[b]{0.4\linewidth}
                                \centering
                                \includegraphics[width=\linewidth]{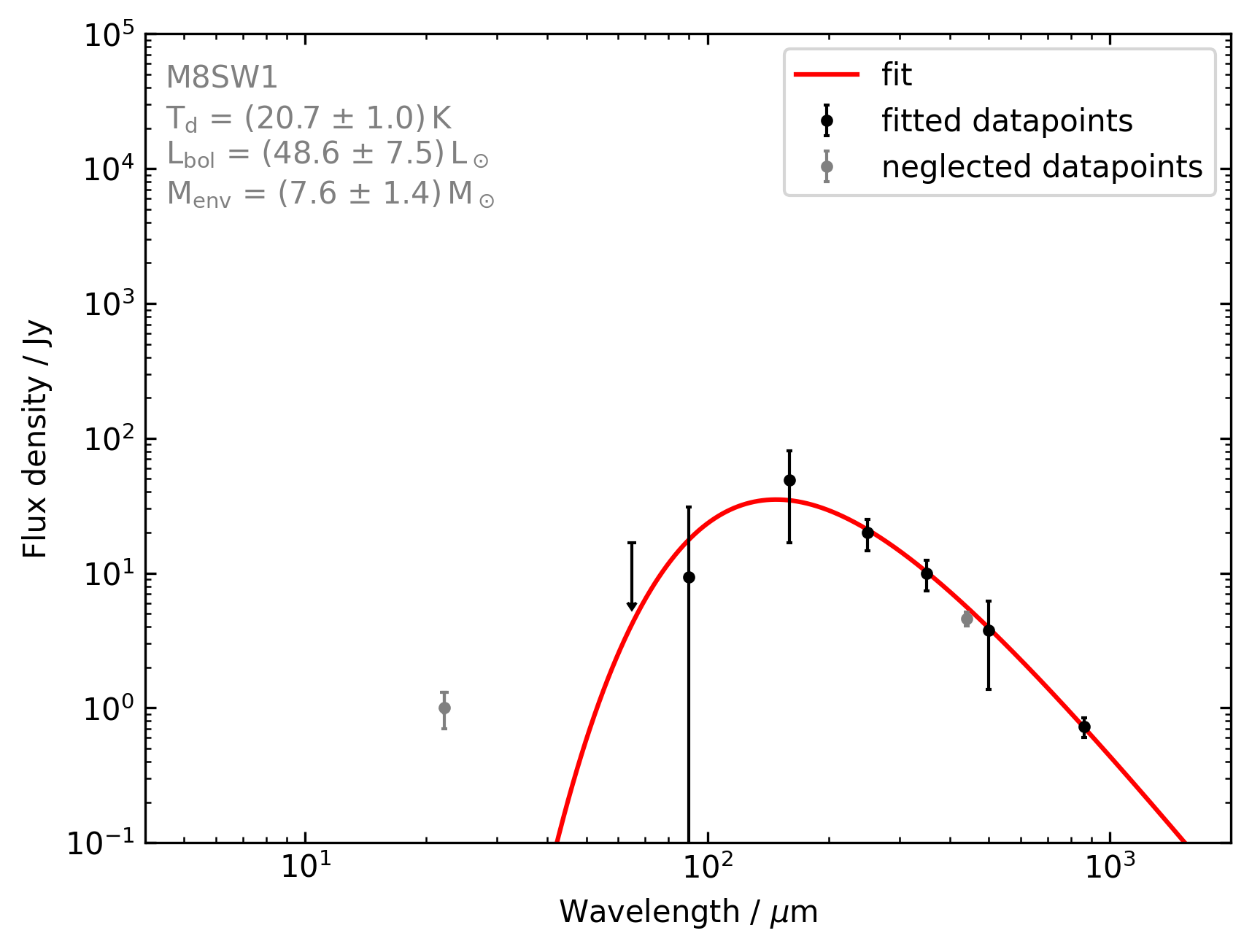}
                        \end{minipage}
                        \\
                        \begin{minipage}[b]{0.4\linewidth}
                                \centering
                                \includegraphics[width=\linewidth]{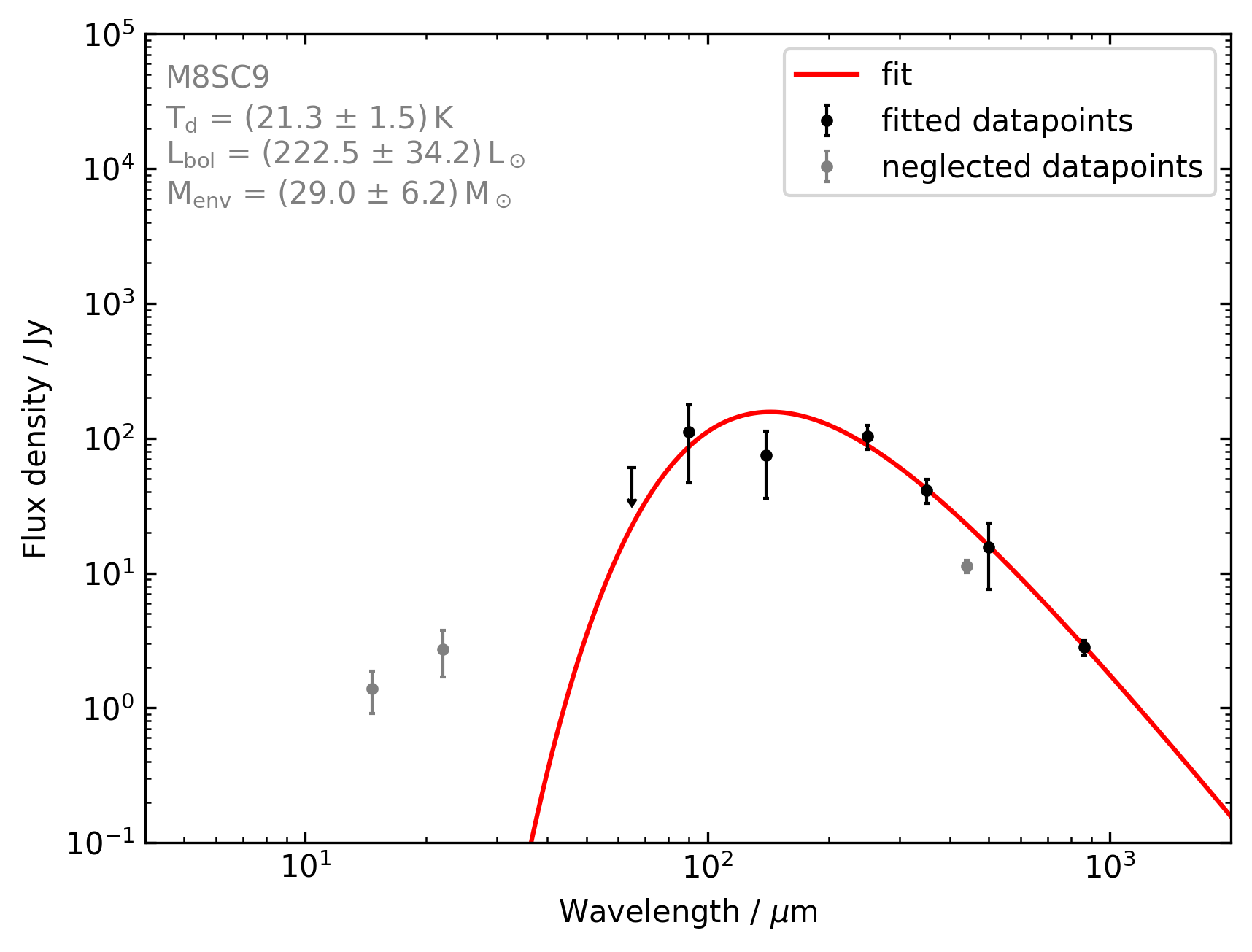}
                        \end{minipage}
                        \caption{Fit to the SEDs of HG, E, WC1, SC9, and SW1. Single-component fits only consider flux at wavelengths longer than $\SI{65}{\micro\meter}$ to avoid a contribution from unrelated diffuse warm gas to the clumps' SEDs. SCUBA $\SI{450}{\micro\meter}$ emission was not considered, as we found it to systematically underestimate the flux of all clumps. Flux densities at $\SI{8}{\micro\meter}$ and $\SI{65}{\micro\meter}$ are considered as upper limits due to the possible contributions of PAHs and very small grains.}
                        \label{fig:app:SED_HG}
                \end{figure}

\begin{figure}[htbp]
                        \centering
                        \begin{minipage}[b]{0.4\linewidth}
                                \centering
                                \includegraphics[width=\linewidth]{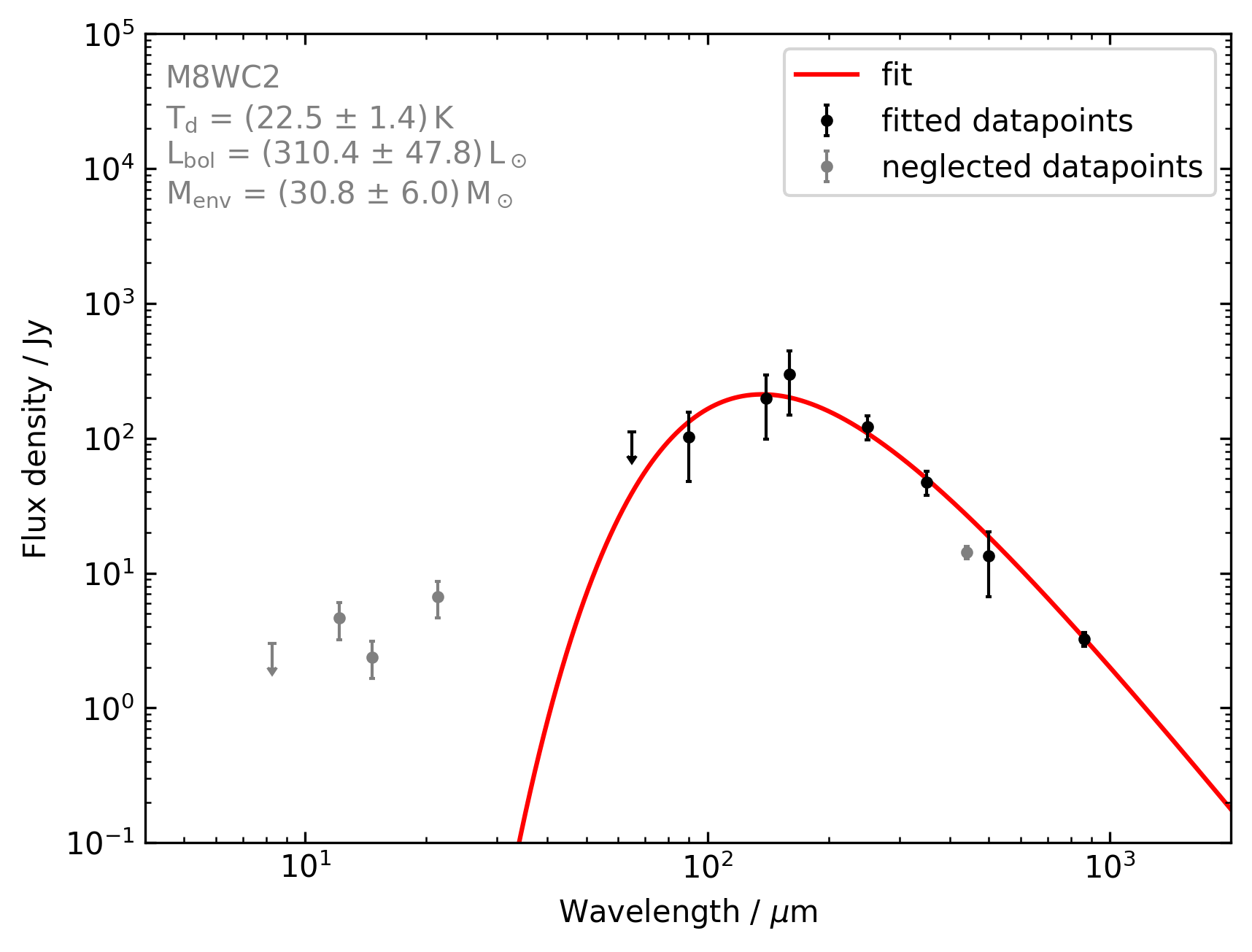}
                        \end{minipage}
                        \begin{minipage}[b]{0.4\linewidth}
                                \centering
                                \includegraphics[width=\linewidth]{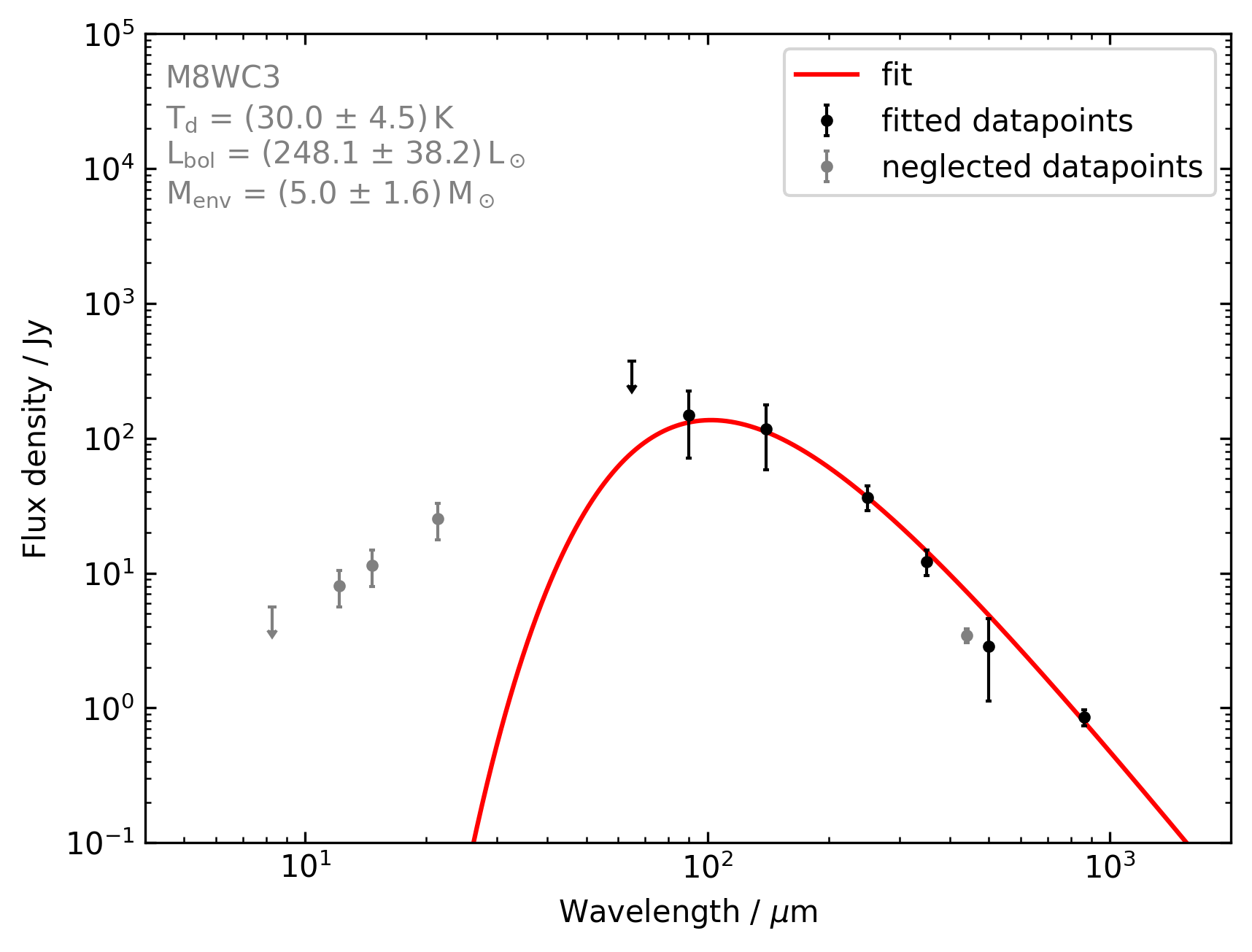}
                        \end{minipage}
                        \\
                        \begin{minipage}[b]{0.4\linewidth}
                                \centering
                                \includegraphics[width=\linewidth]{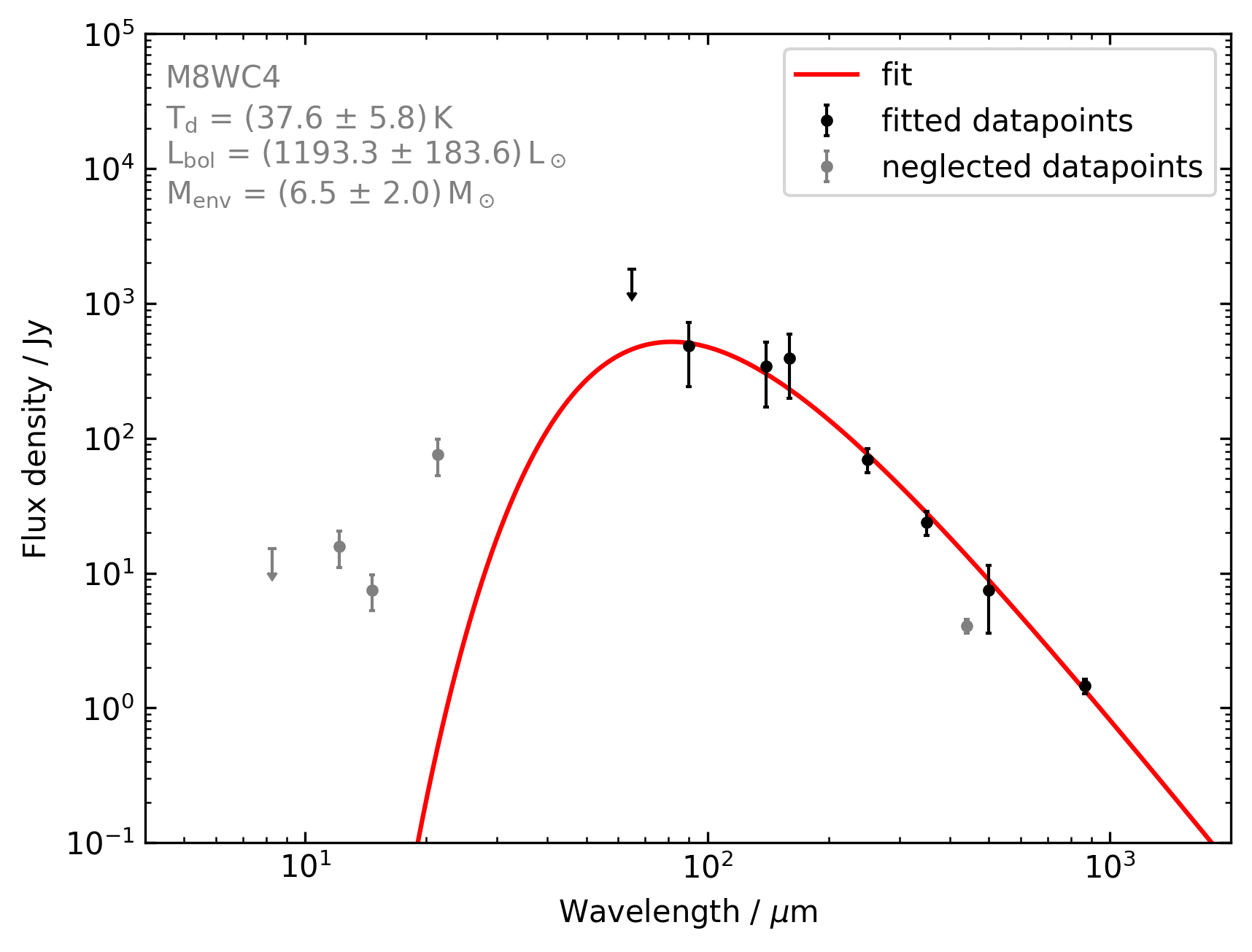}
                        \end{minipage}
                        \begin{minipage}[b]{0.4\linewidth}
                                \centering
                                \includegraphics[width=\linewidth]{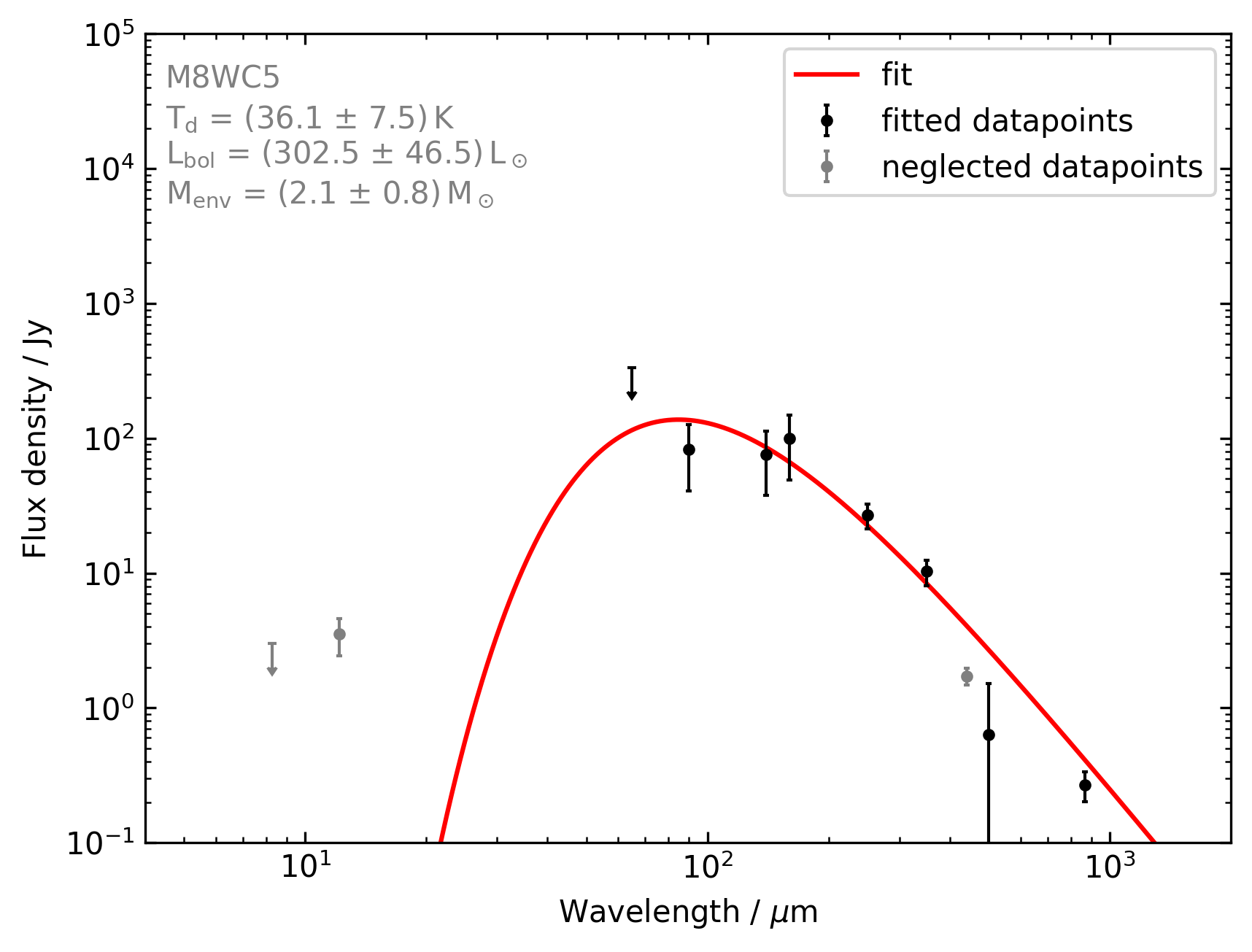}
                        \end{minipage}
                        \\
                        \begin{minipage}[b]{0.4\linewidth}
                                \centering
                                \includegraphics[width=\linewidth]{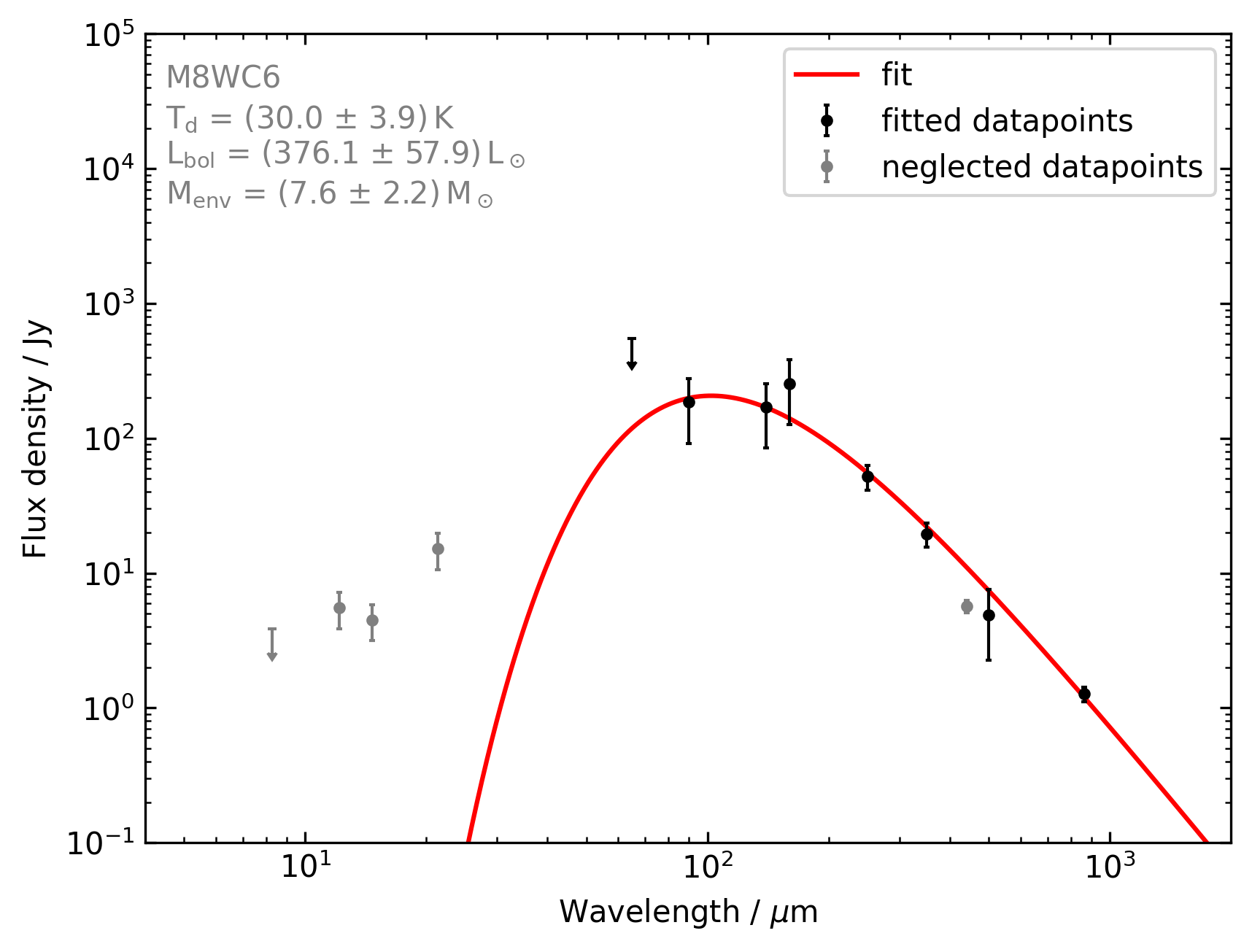}
                        \end{minipage}
                        \begin{minipage}[b]{0.4\linewidth}
                                \centering
                                \includegraphics[width=\linewidth]{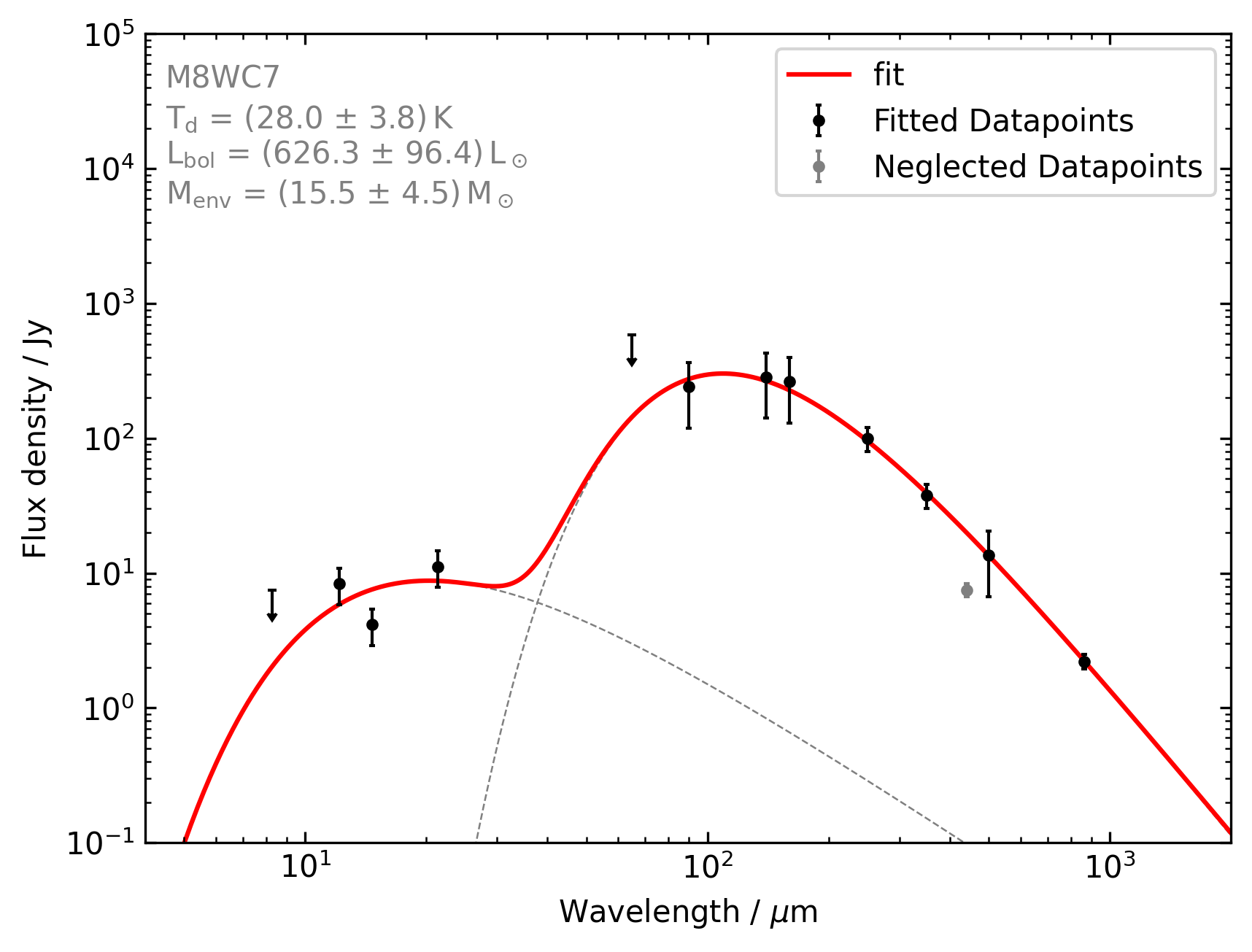}
                        \end{minipage}
                        \\
                        \begin{minipage}[b]{0.4\linewidth}
                                \centering
                                \includegraphics[width=\linewidth]{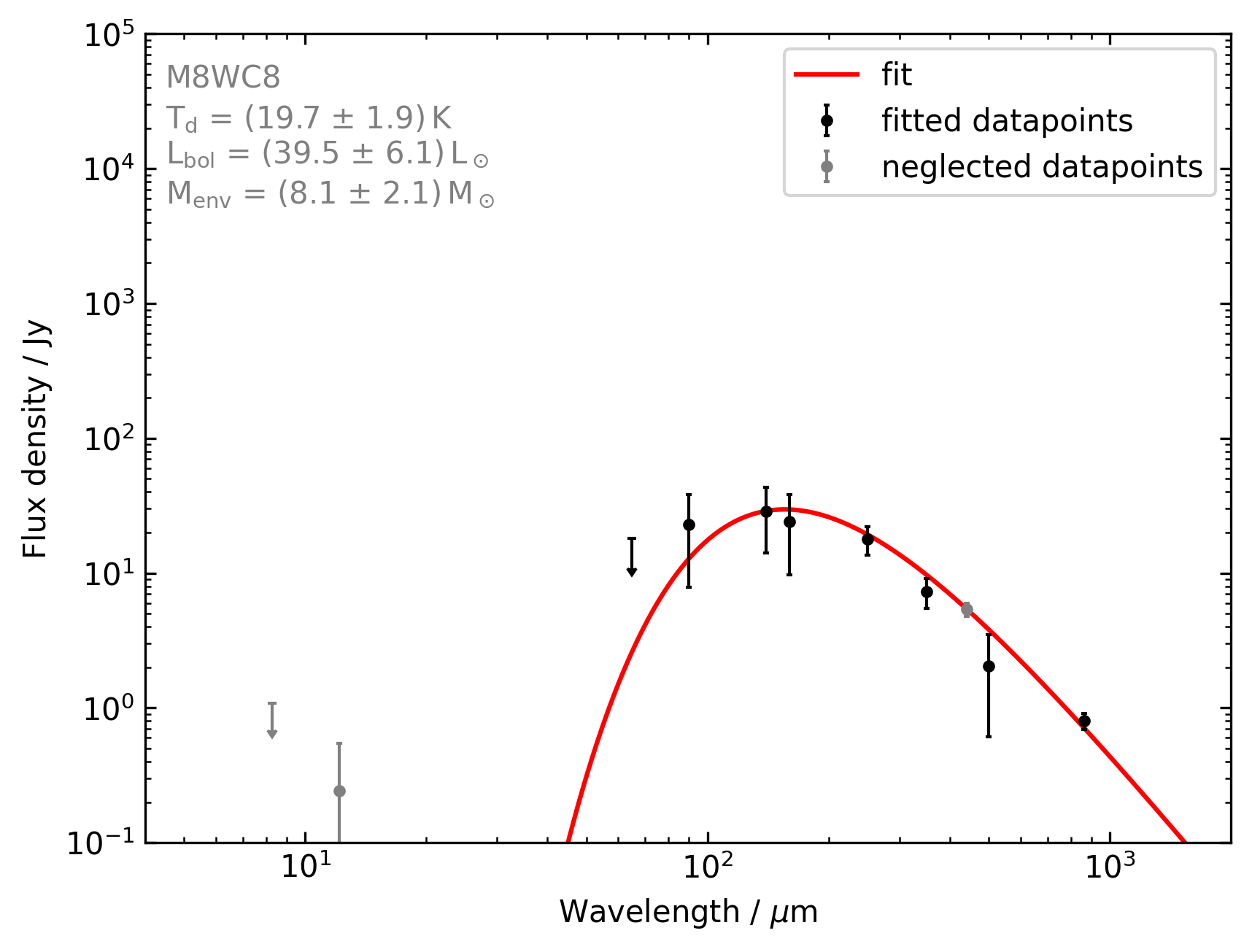}
                        \end{minipage}
                        \begin{minipage}[b]{0.4\linewidth}
                                \centering
                                \includegraphics[width=\linewidth]{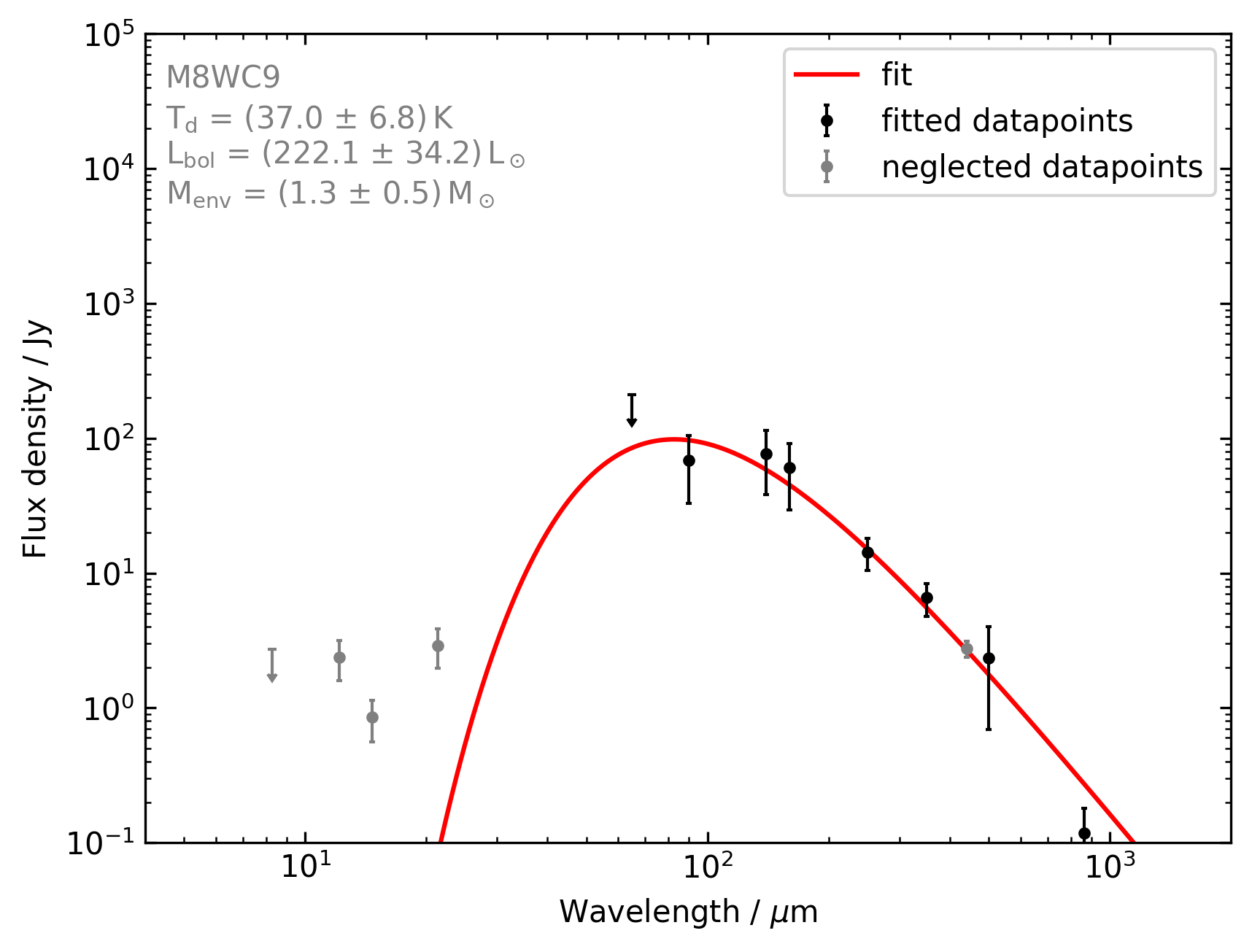}
                        \end{minipage}
                        \caption{Fits to the SEDs of the WC2--9 clumps. Single-component fits only consider flux at wavelengths longer than $\SI{65}{\micro\meter}$ to avoid a contribution from unrelated diffuse warm gas to the clumps' SEDs. SCUBA $\SI{450}{\micro\meter}$ emission was not considered as we found it to systematically underestimate the flux of all clumps. Flux densities at $\SI{8}{\micro\meter}$ and $\SI{65}{\micro\meter}$ are considered as upper limits due to the possible contributions of PAHs and very small grains.}
                        \label{fig:app:SED_WC}
                \end{figure}

\begin{figure}[htbp]
                        \centering
                        \begin{minipage}[b]{0.4\linewidth}
                                \centering
                                \includegraphics[width=\linewidth]{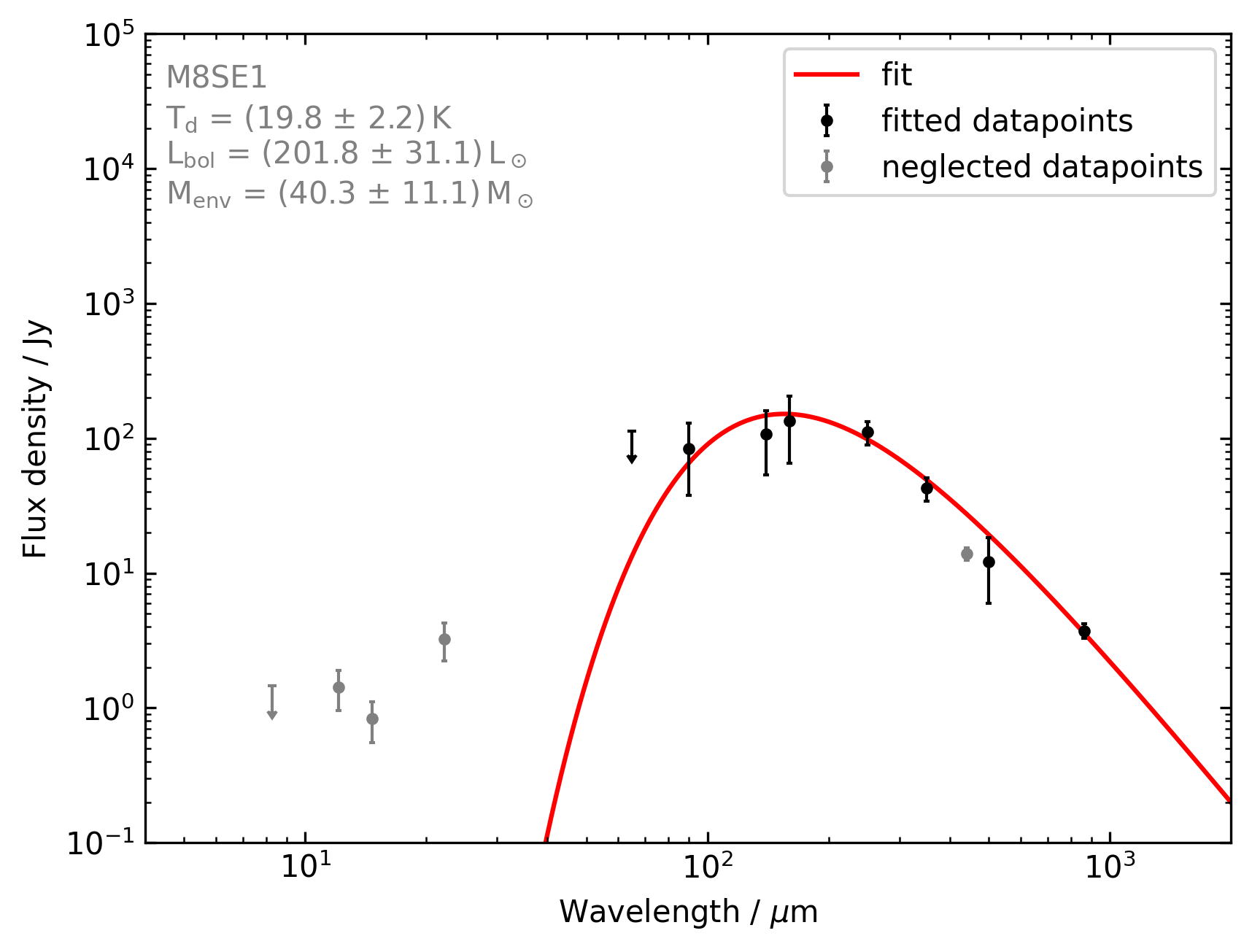}
                        \end{minipage}
                        \begin{minipage}[b]{0.4\linewidth}
                                \centering
                                \includegraphics[width=\linewidth]{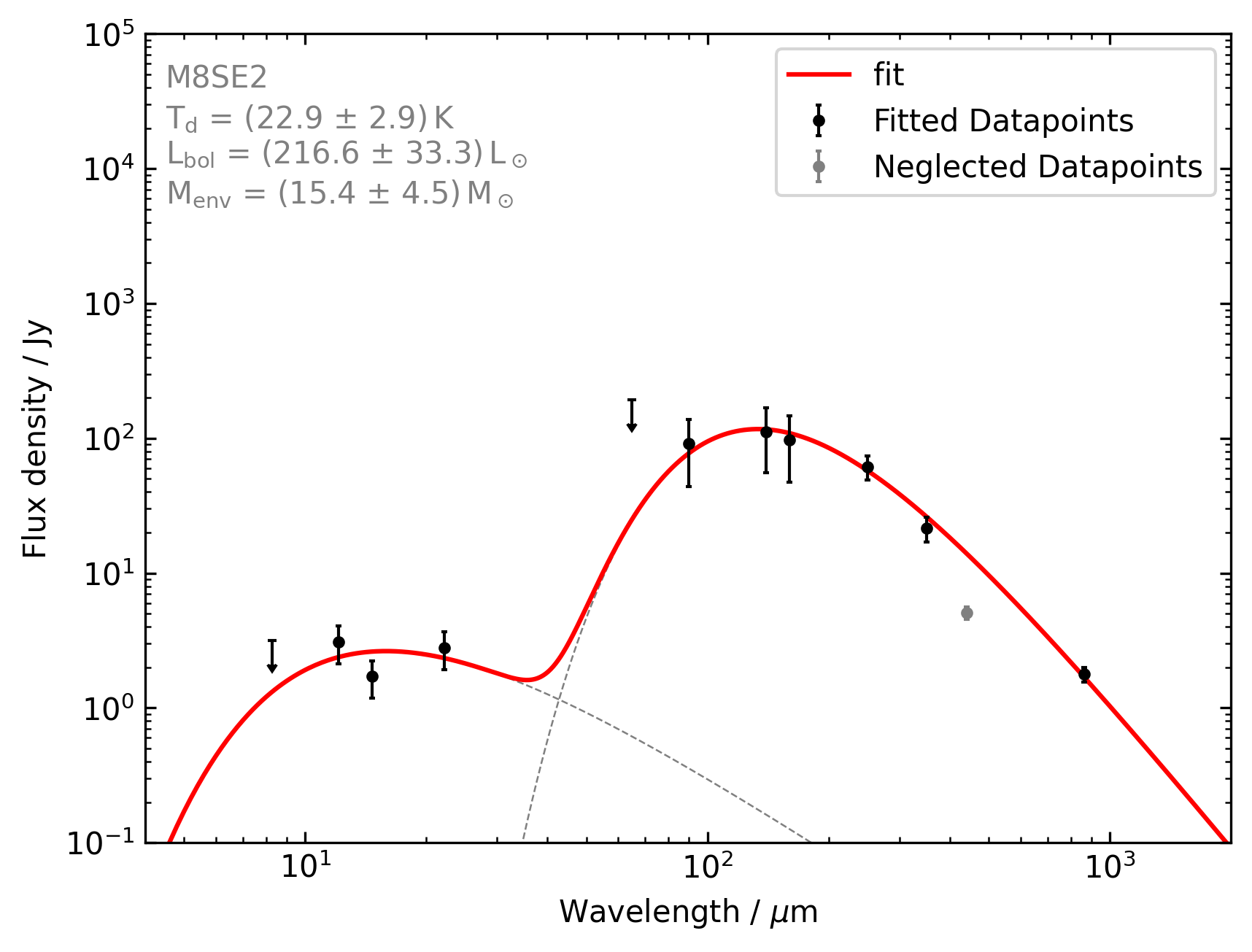}
                        \end{minipage}
                        \\
                        \begin{minipage}[b]{0.4\linewidth}
                                \centering
                                \includegraphics[width=\linewidth]{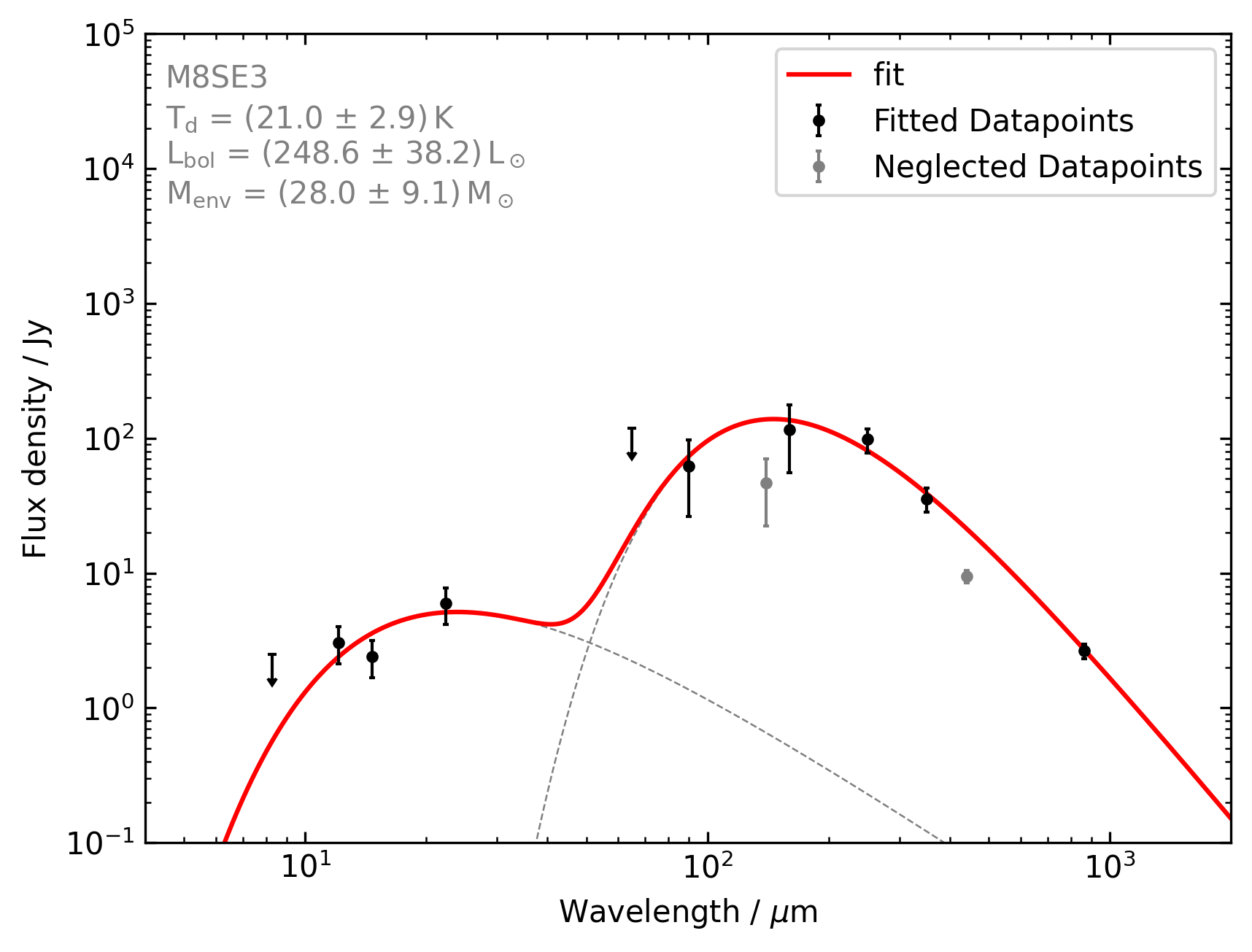}
                        \end{minipage}
                        \begin{minipage}[b]{0.4\linewidth}
                                \centering
                                \includegraphics[width=\linewidth]{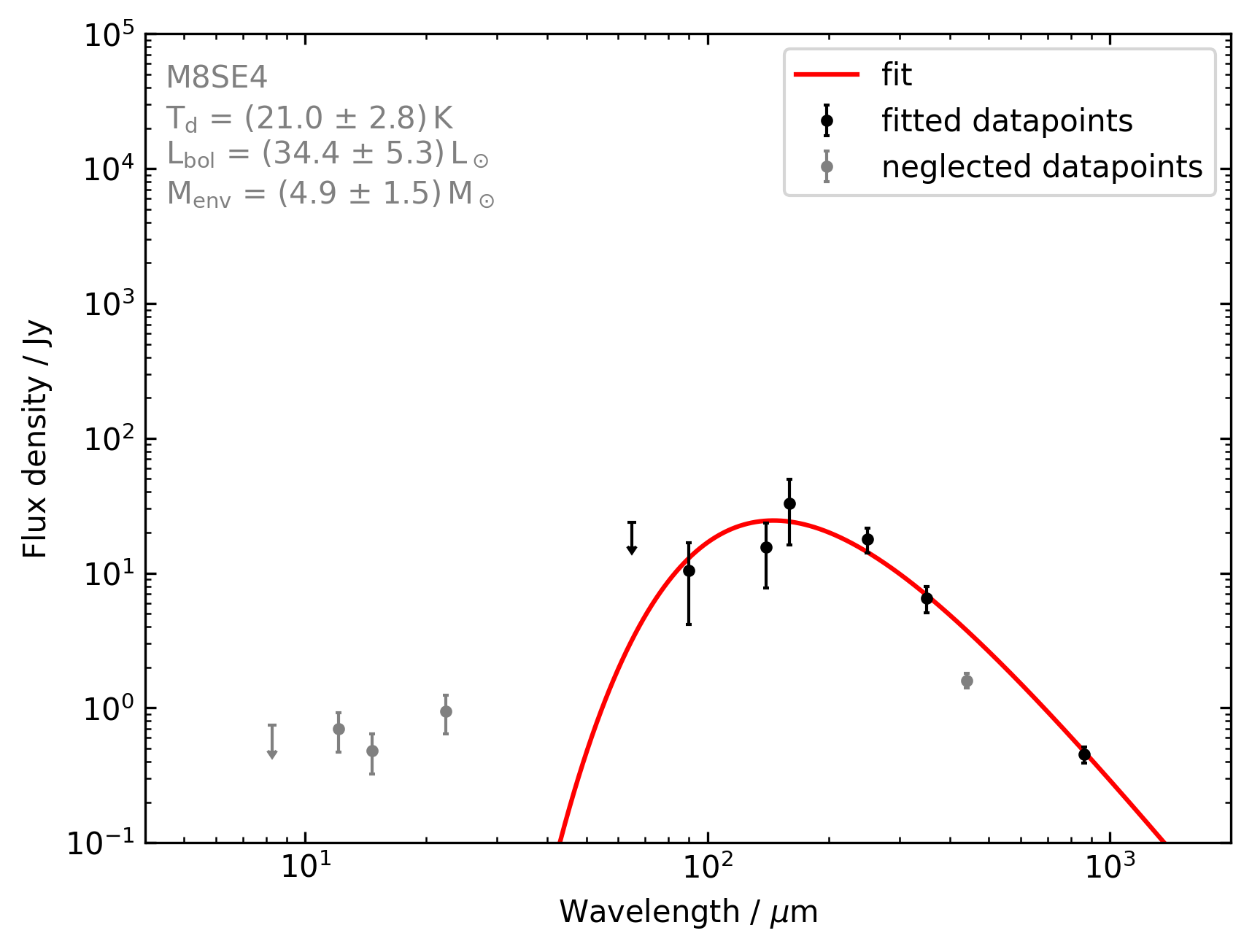}
                        \end{minipage}
                        \\
                        \begin{minipage}[b]{0.4\linewidth}
                                \centering
                                \includegraphics[width=\linewidth]{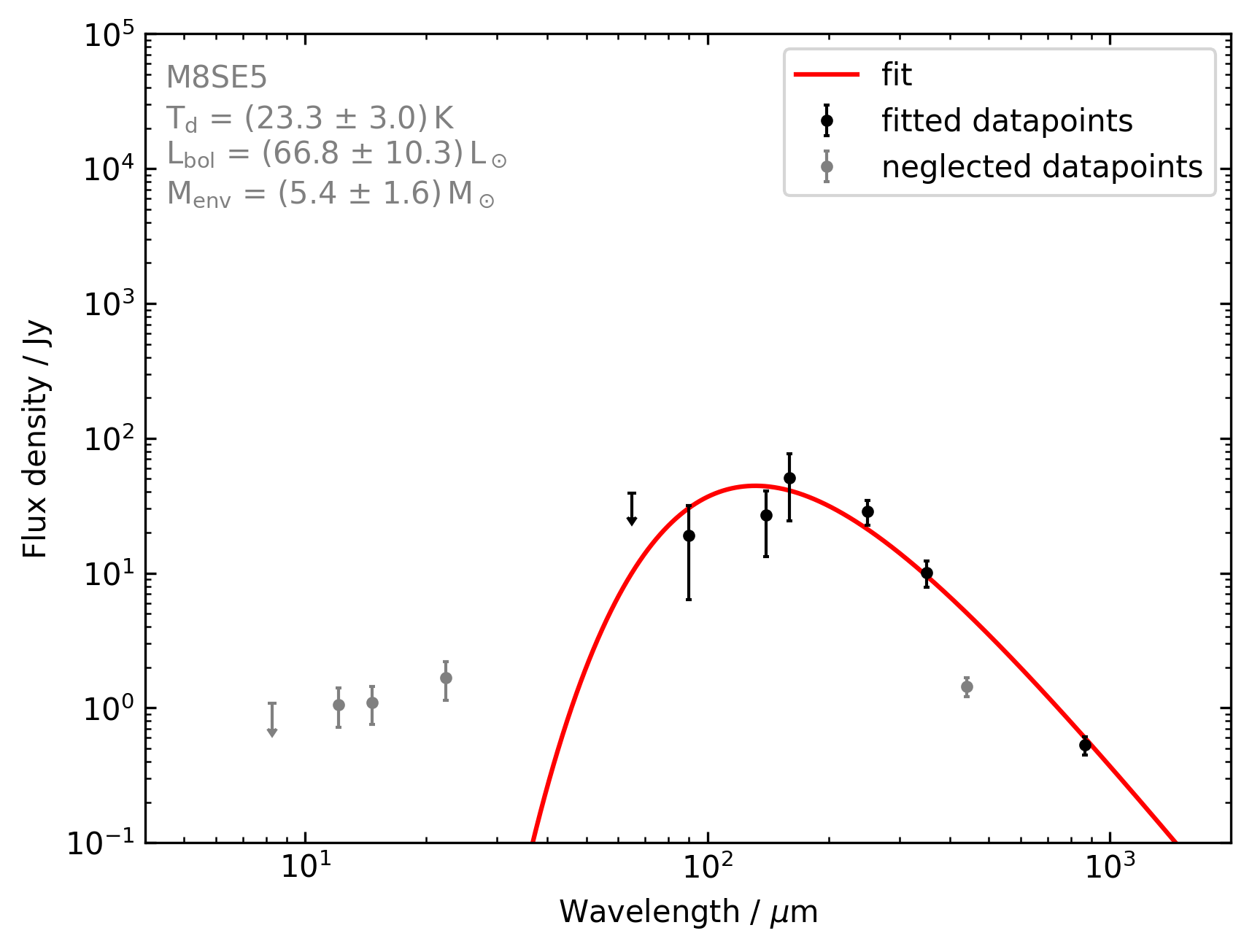}
                        \end{minipage}
                        \begin{minipage}[b]{0.4\linewidth}
                                \centering
                                \includegraphics[width=\linewidth]{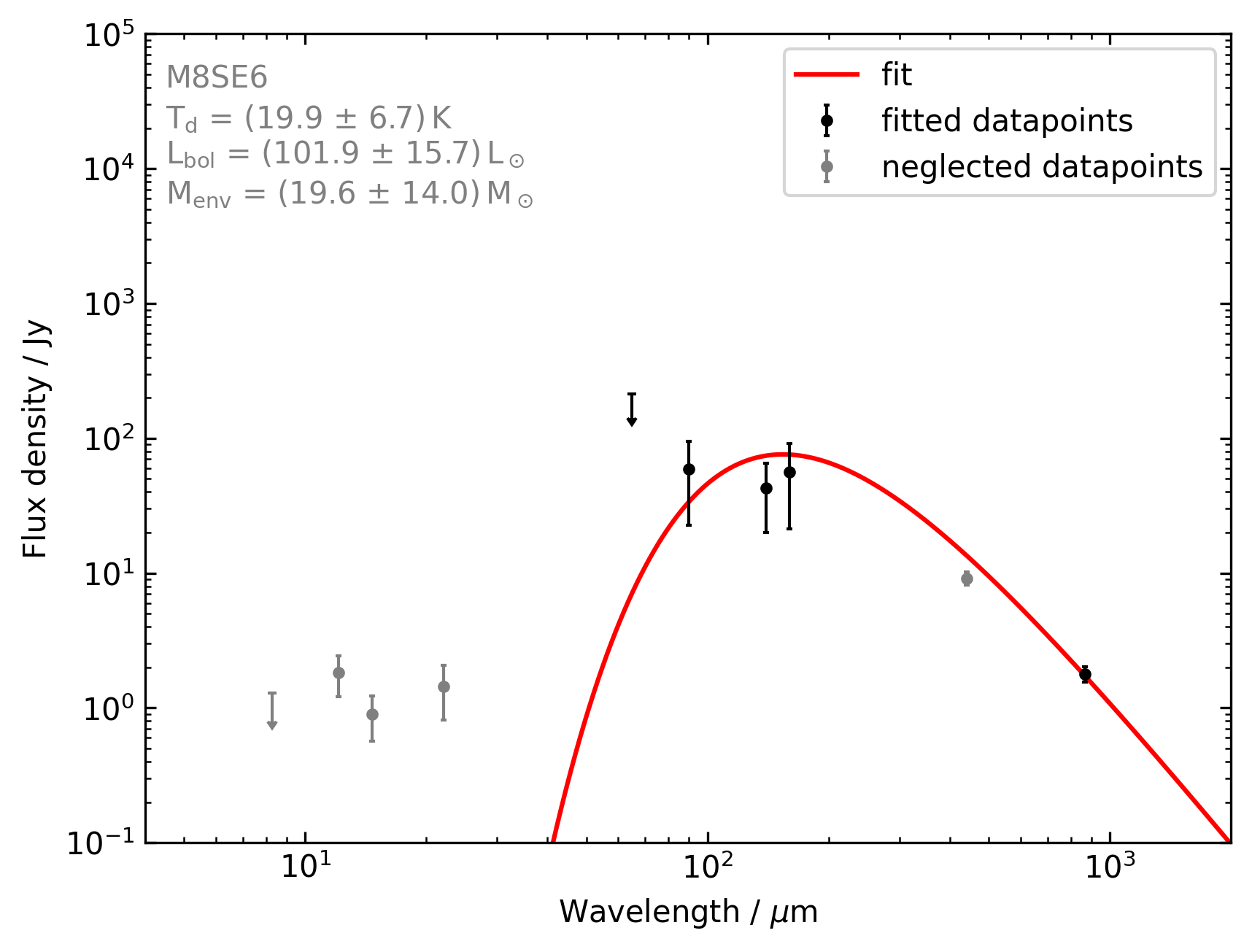}
                        \end{minipage}
                        \\
                        \begin{minipage}[b]{0.4\linewidth}
                                \centering
                                \includegraphics[width=\linewidth]{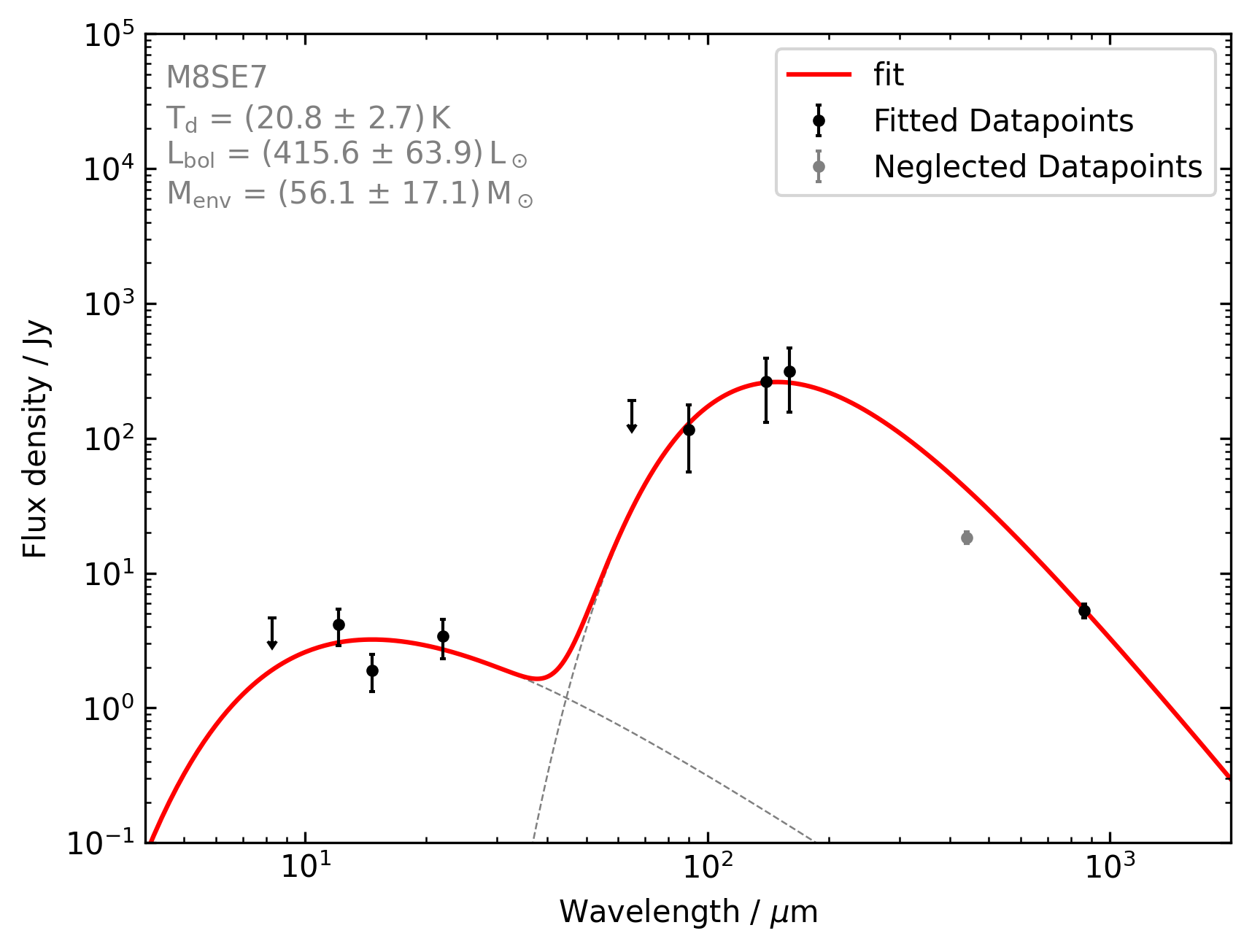}
                        \end{minipage}
                        \begin{minipage}[b]{0.4\linewidth}
                                \centering
                                \includegraphics[width=\linewidth]{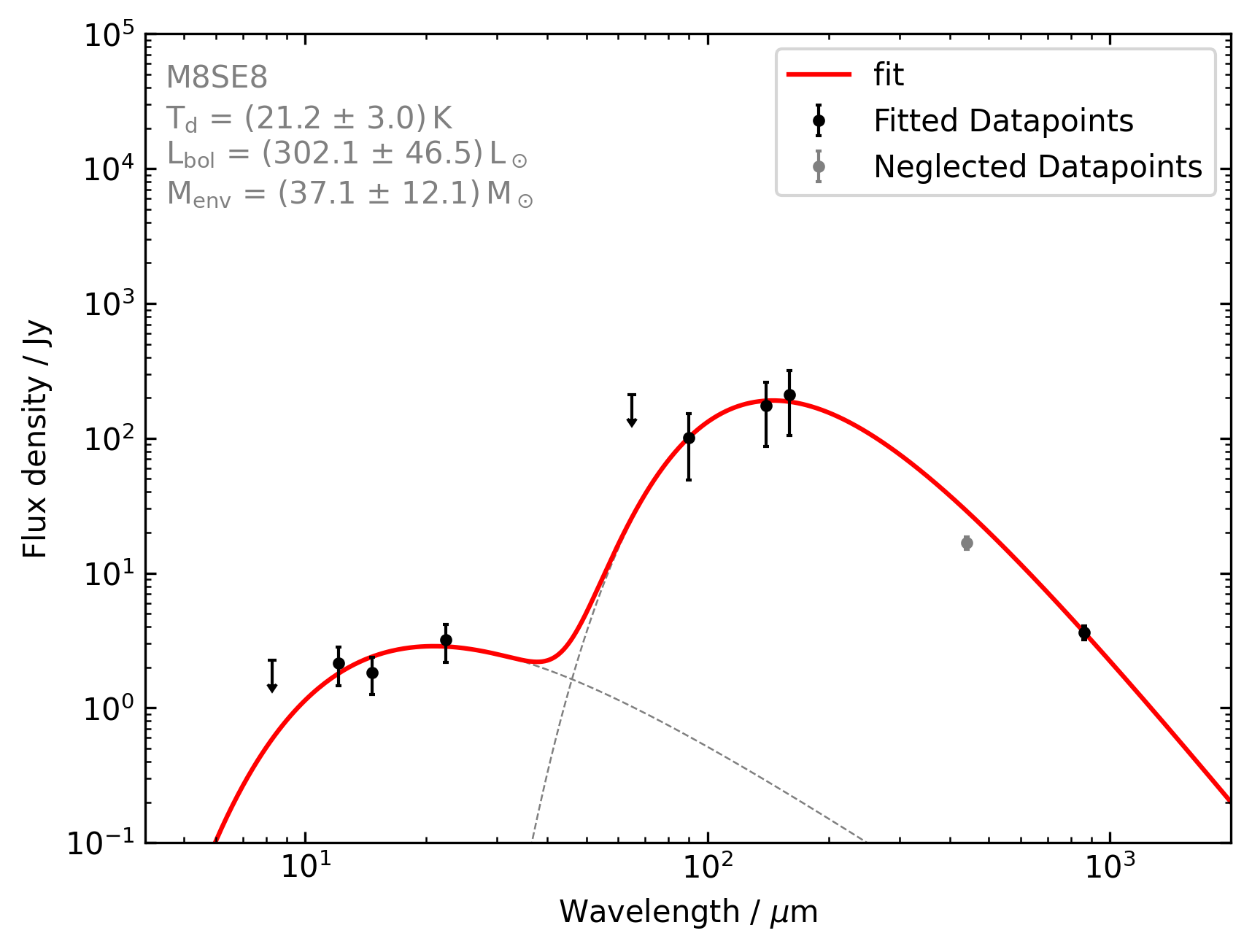}
                        \end{minipage}
                        \caption{Fits to the SED of the SE1--8 clumps. Single-component fits only consider flux at wavelengths longer than $\SI{65}{\micro\meter}$ to avoid a contribution from unrelated diffuse warm gas to the clumps' SEDs. SCUBA $\SI{450}{\micro\meter}$ emission was not considered as we found it to systematically underestimate the flux of all clumps. Flux densities at $\SI{8}{\micro\meter}$ and $\SI{65}{\micro\meter}$ are considered as upper limits due to the possible contributions of PAHs and very small grains.}
                        \label{fig:app:SED_SE}
                \end{figure}

\begin{figure}[htbp]
                        \centering
                        \begin{minipage}[b]{0.4\linewidth}
                                \centering
                                \includegraphics[width=\linewidth]{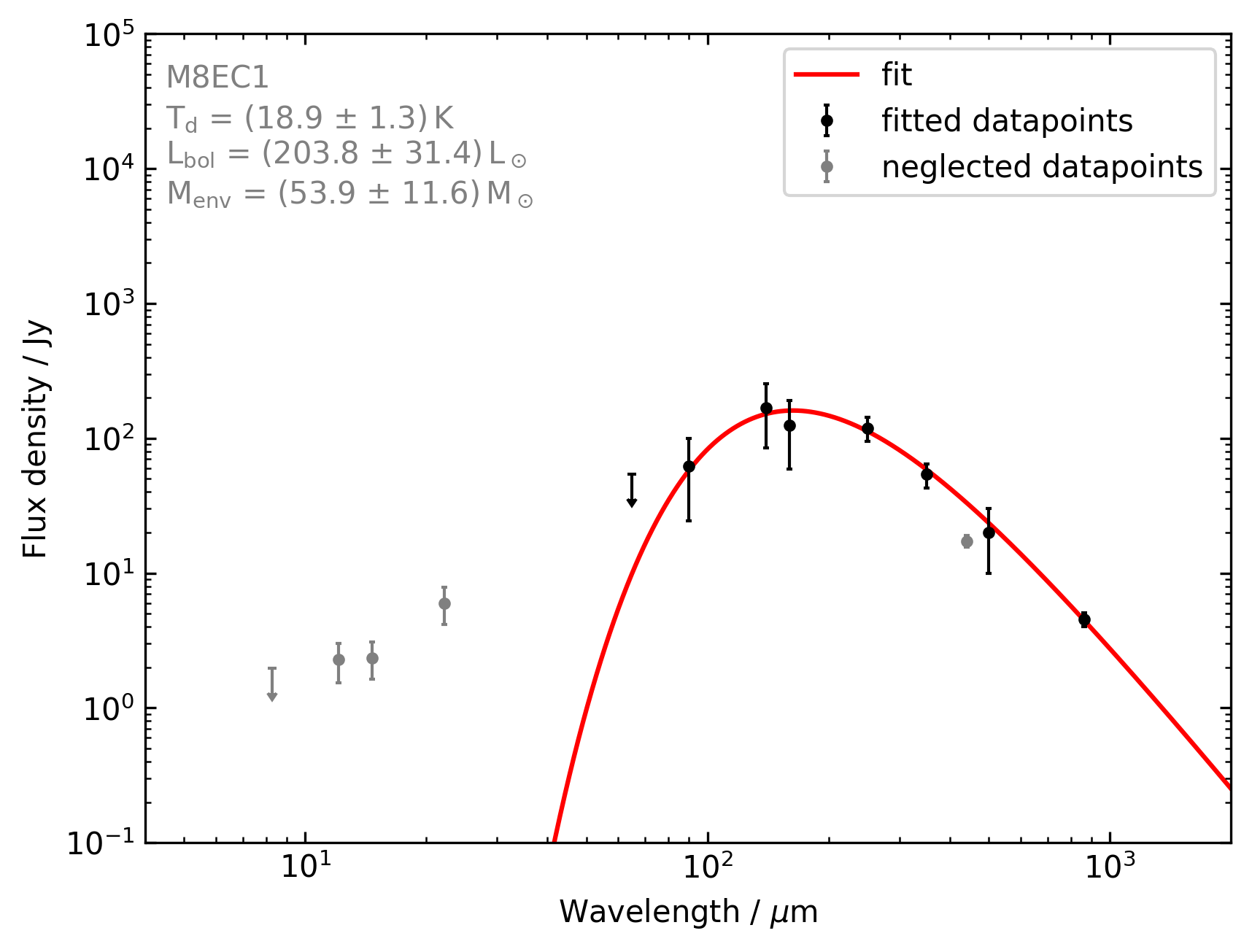}
                        \end{minipage}
                        \begin{minipage}[b]{0.4\linewidth}
                                \centering
                                \includegraphics[width=\linewidth]{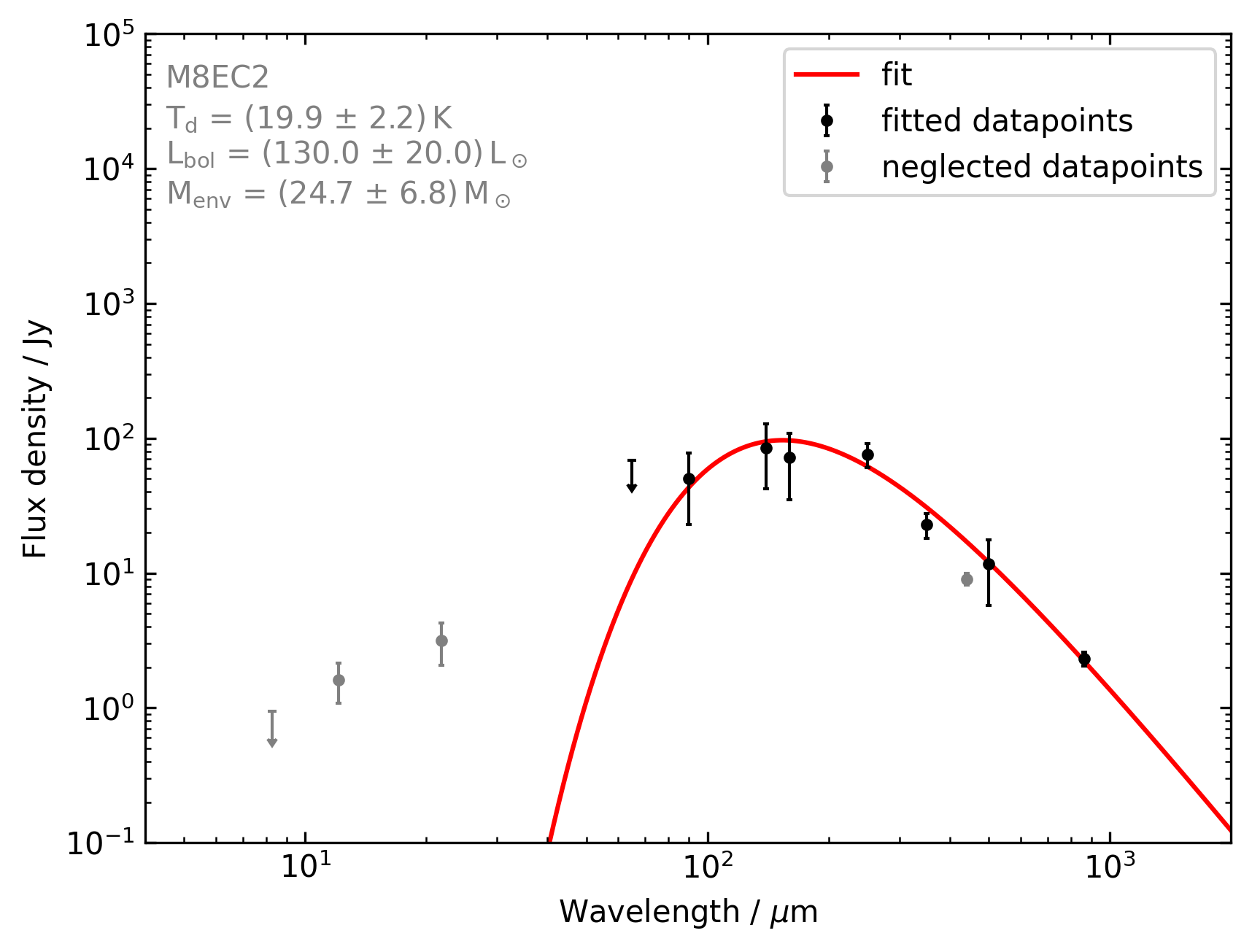}
                        \end{minipage}
                        \\
                        \begin{minipage}[b]{0.4\linewidth}
                                \centering
                                \includegraphics[width=\linewidth]{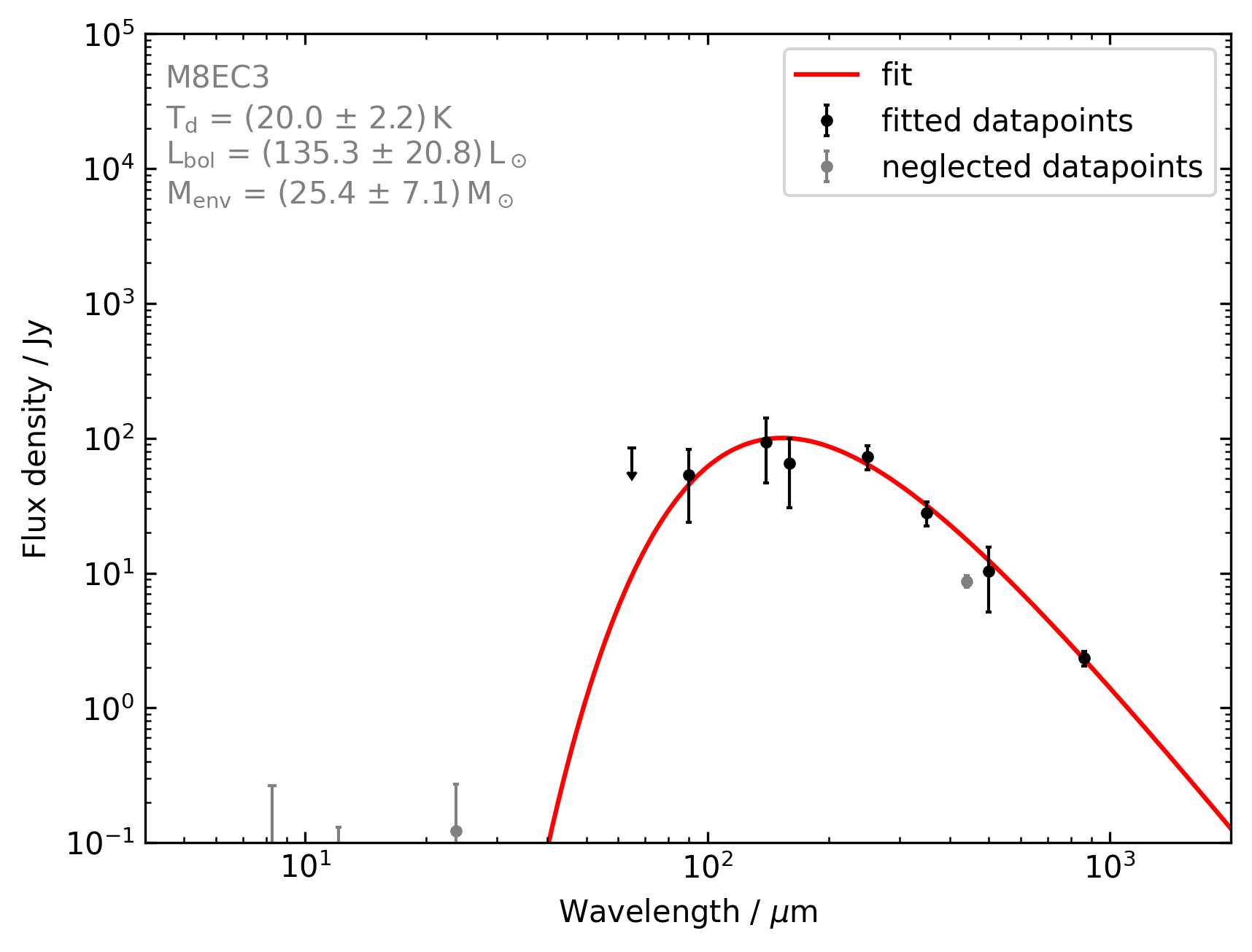}
                        \end{minipage}
                        \begin{minipage}[b]{0.4\linewidth}
                                \centering
                                \includegraphics[width=\linewidth]{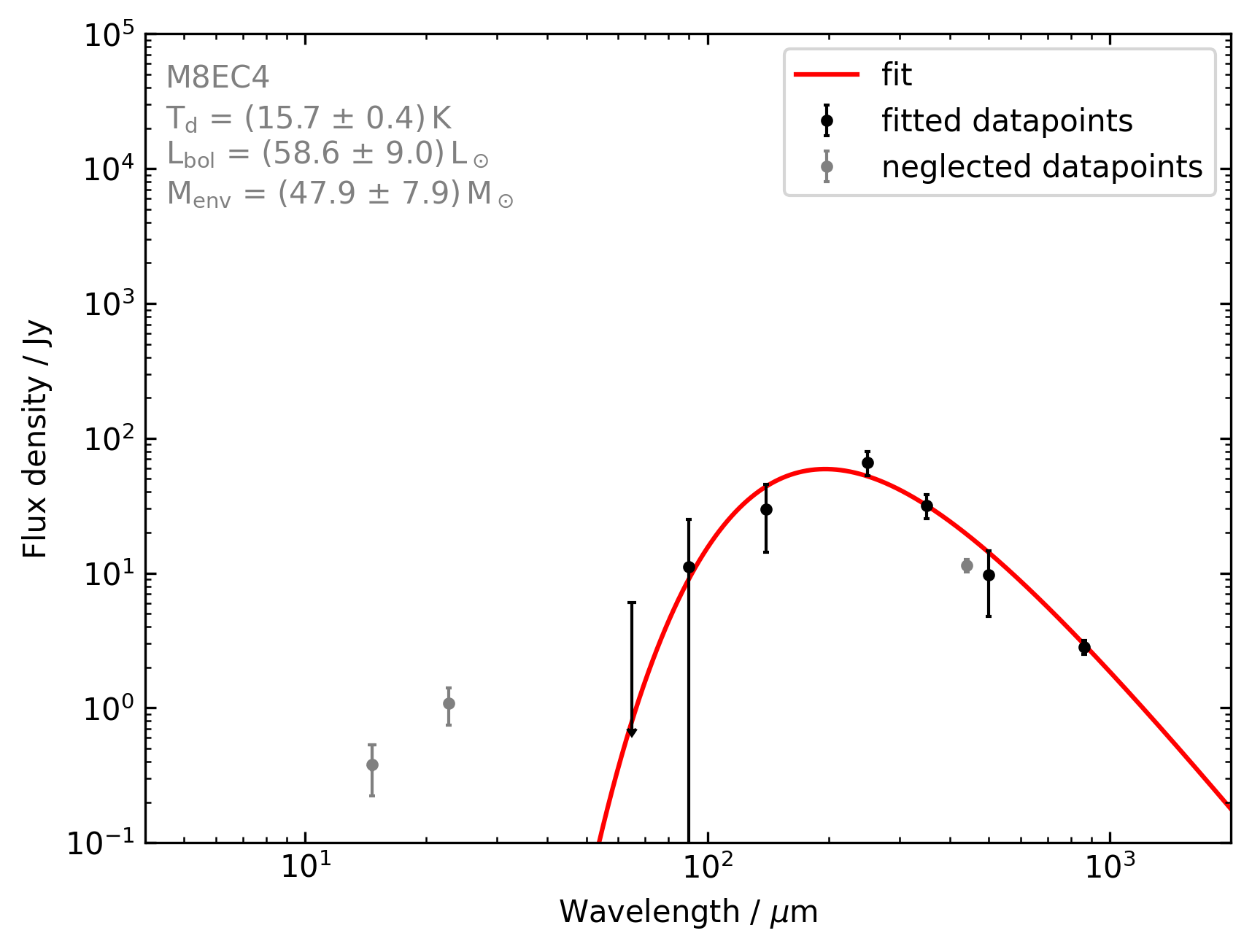}
                        \end{minipage}
                        \\
                        \begin{minipage}[b]{0.4\linewidth}
                                \centering
                                \includegraphics[width=\linewidth]{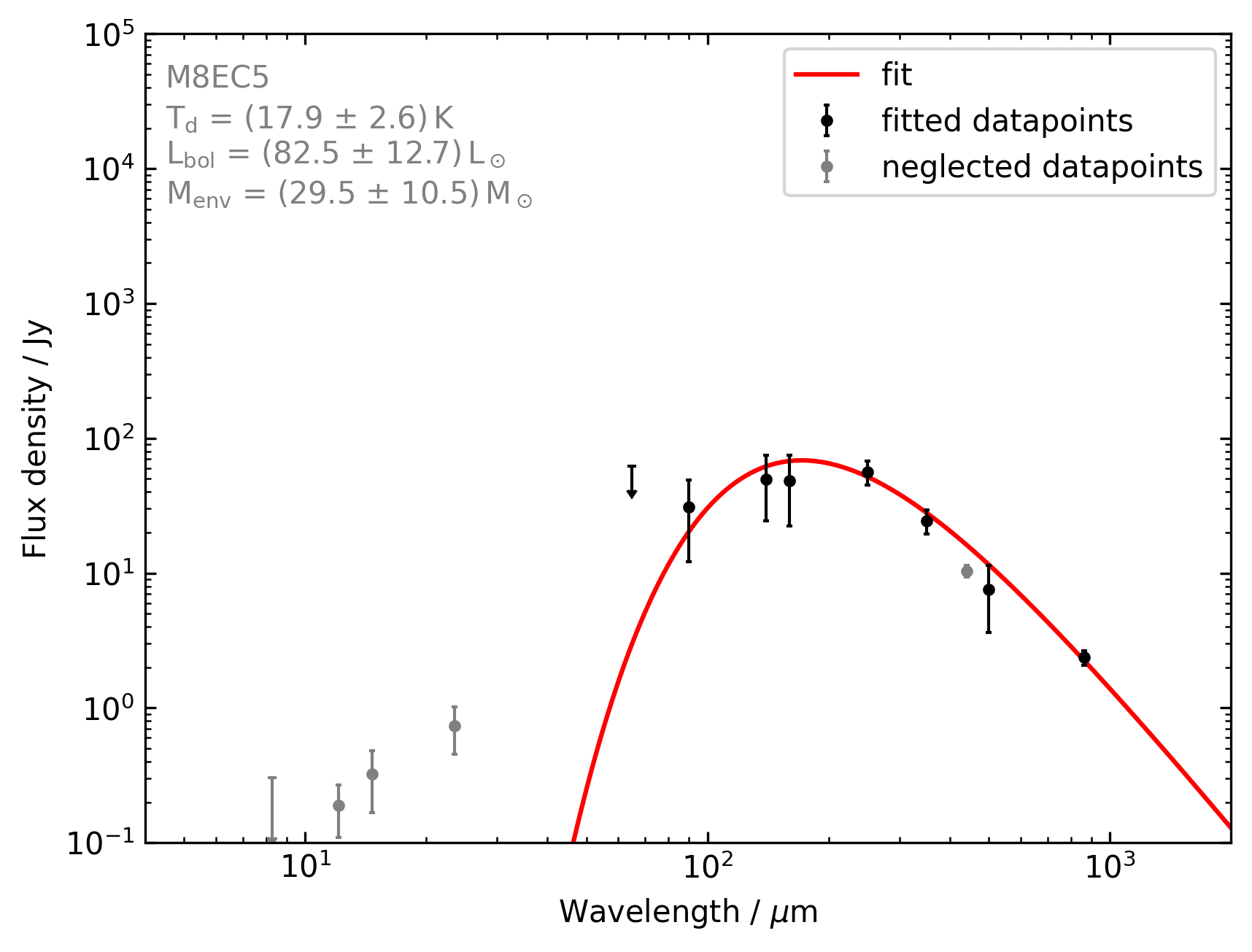}
                        \end{minipage}
                        \begin{minipage}[b]{0.4\linewidth}
                                \centering
                                \includegraphics[width=\linewidth]{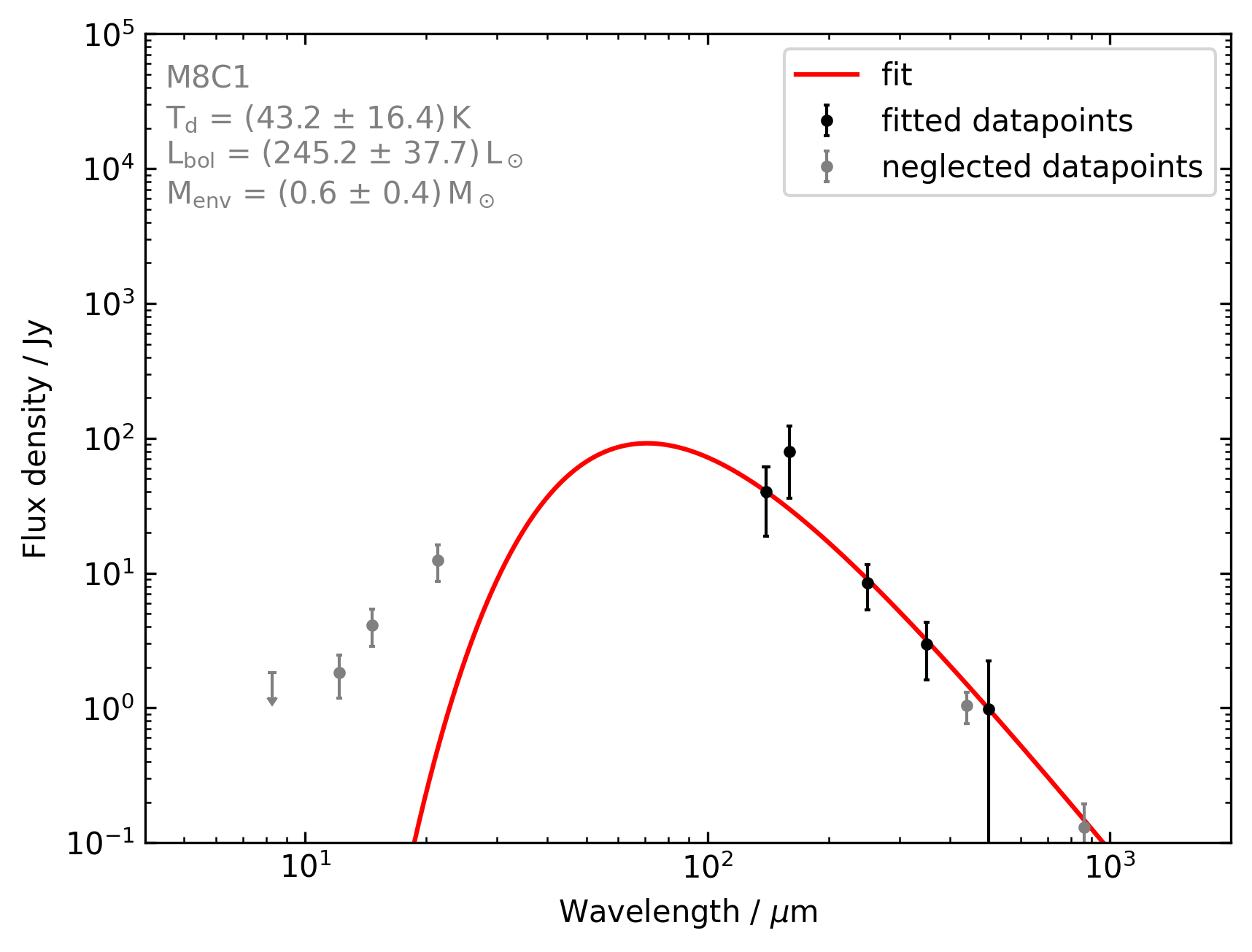}
                        \end{minipage}
                        \\
                        \begin{minipage}[b]{0.4\linewidth}
                                \centering
                                \includegraphics[width=\linewidth]{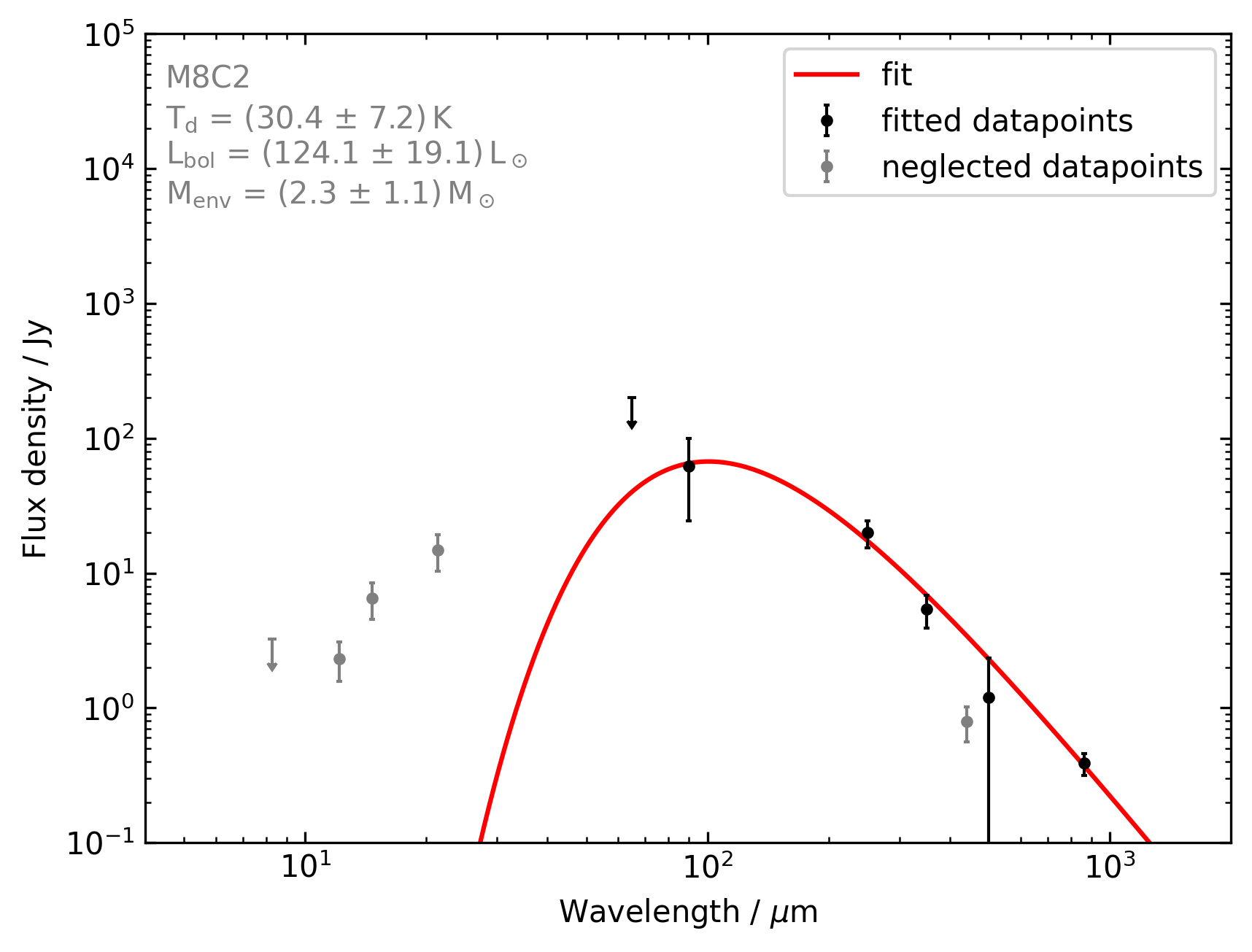}
                        \end{minipage}
                        \begin{minipage}[b]{0.4\linewidth}
                                \centering
                                \includegraphics[width=\linewidth]{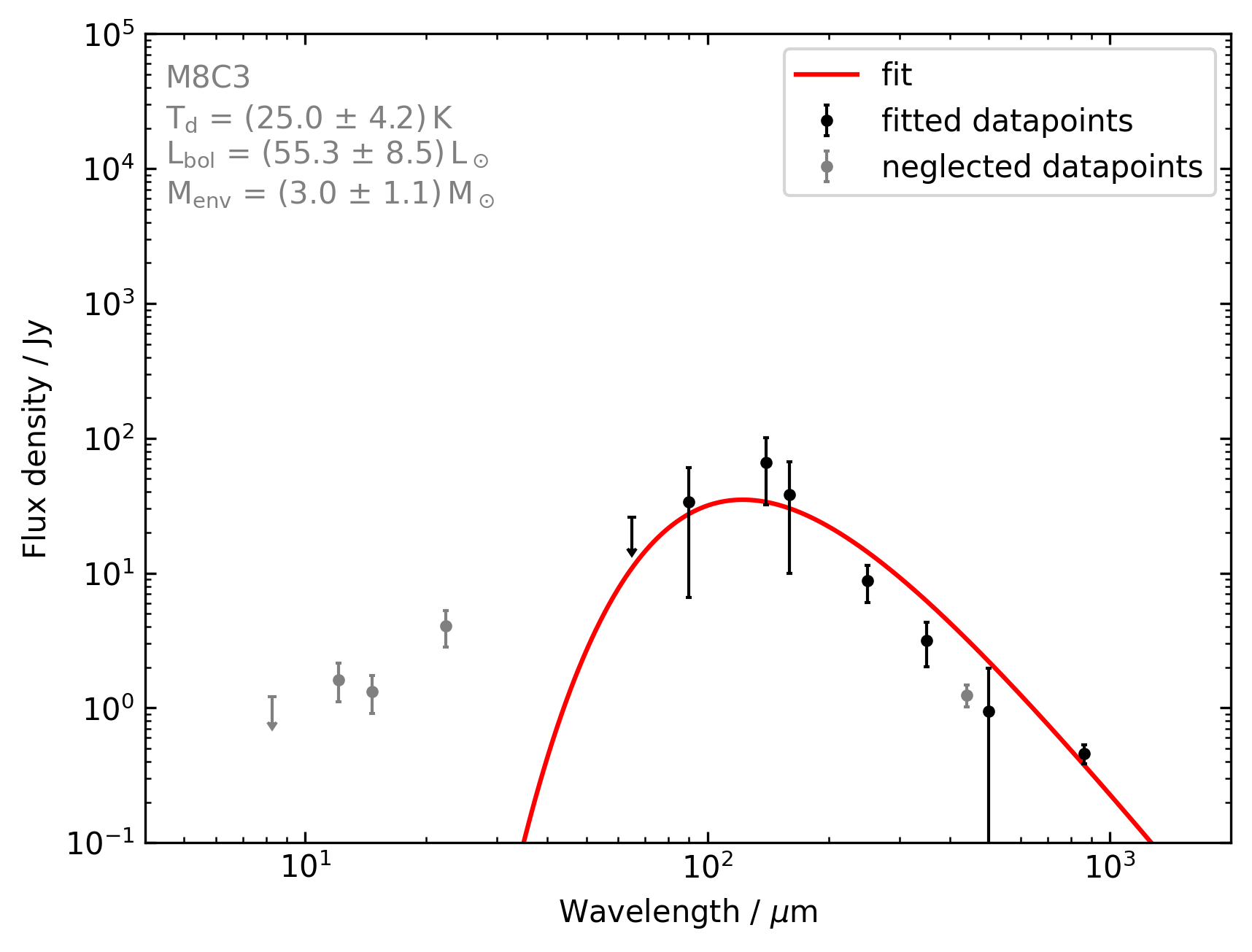}
                        \end{minipage}
                        \caption{Fits to the SED of the EC1--5 clumps and the clumps C1-3. Single-component fits only consider flux at wavelengths longer than $\SI{65}{\micro\meter}$ to avoid a contribution from unrelated diffuse warm gas to the clumps' SEDs. SCUBA $\SI{450}{\micro\meter}$ emission was not considered as we found it to systematically underestimate the flux of all clumps. Flux densities at $\SI{8}{\micro\meter}$ and $\SI{65}{\micro\meter}$ are considered as upper limits due to the possible contributions of PAHs and very small grains.}
                        \label{fig:app:SED_EC}
                \end{figure}

\begin{figure}[htbp]
                        \centering
                        \begin{minipage}[b]{0.4\linewidth}
                                \centering
                                \includegraphics[width=\linewidth]{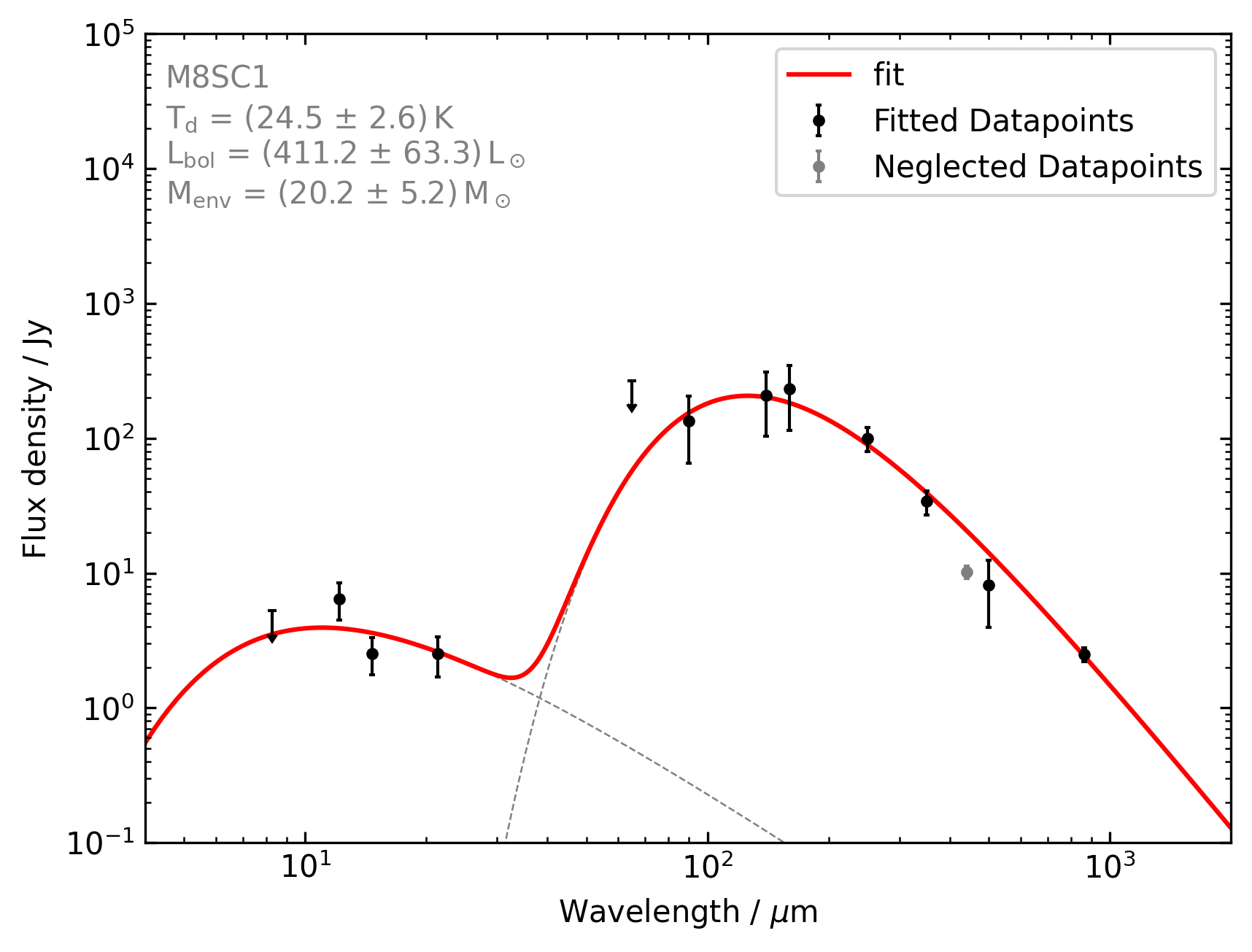}
                        \end{minipage}
                        \begin{minipage}[b]{0.4\linewidth}
                                \centering
                                \includegraphics[width=\linewidth]{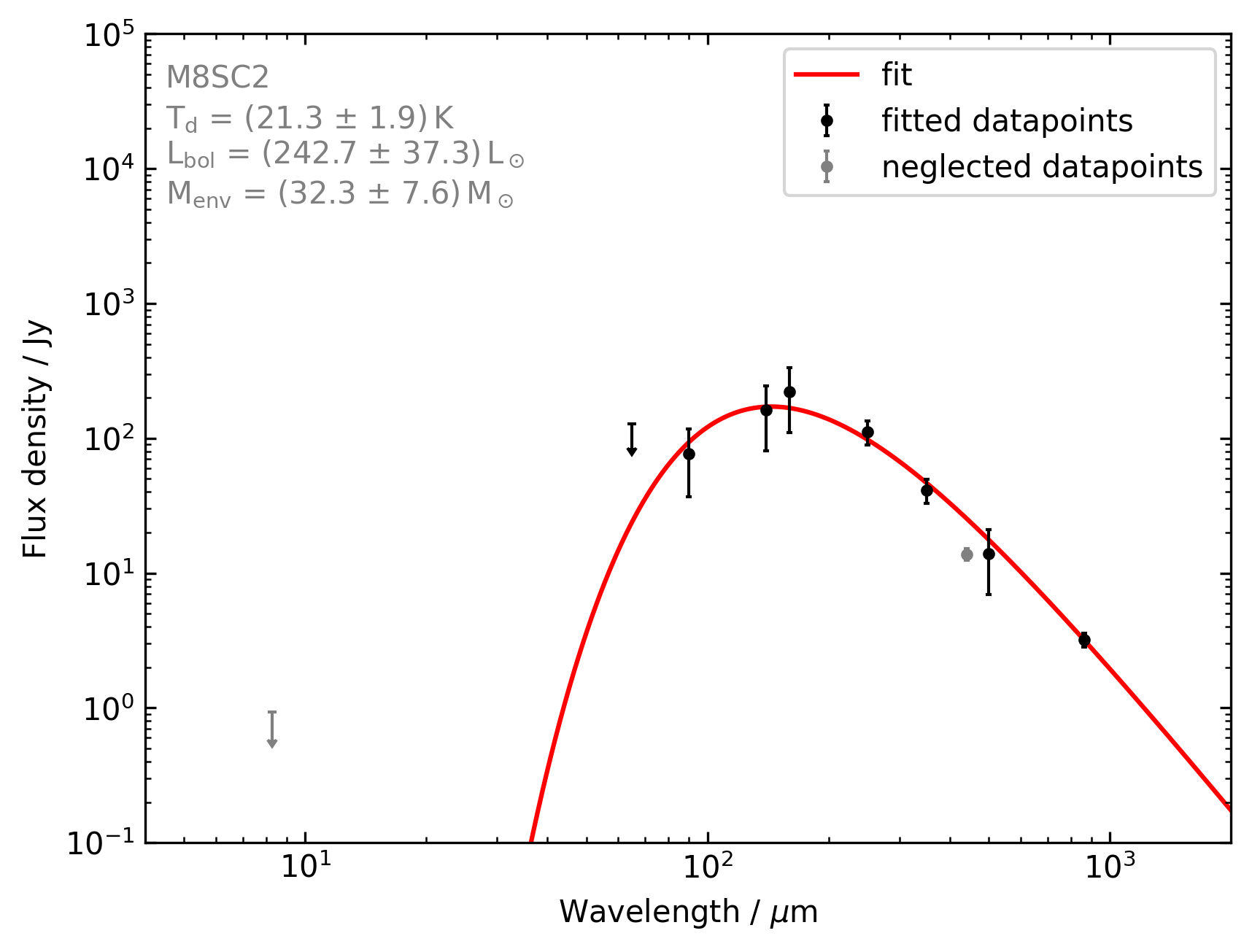}
                        \end{minipage}
                        \\
                        \begin{minipage}[b]{0.4\linewidth}
                                \centering
                                \includegraphics[width=\linewidth]{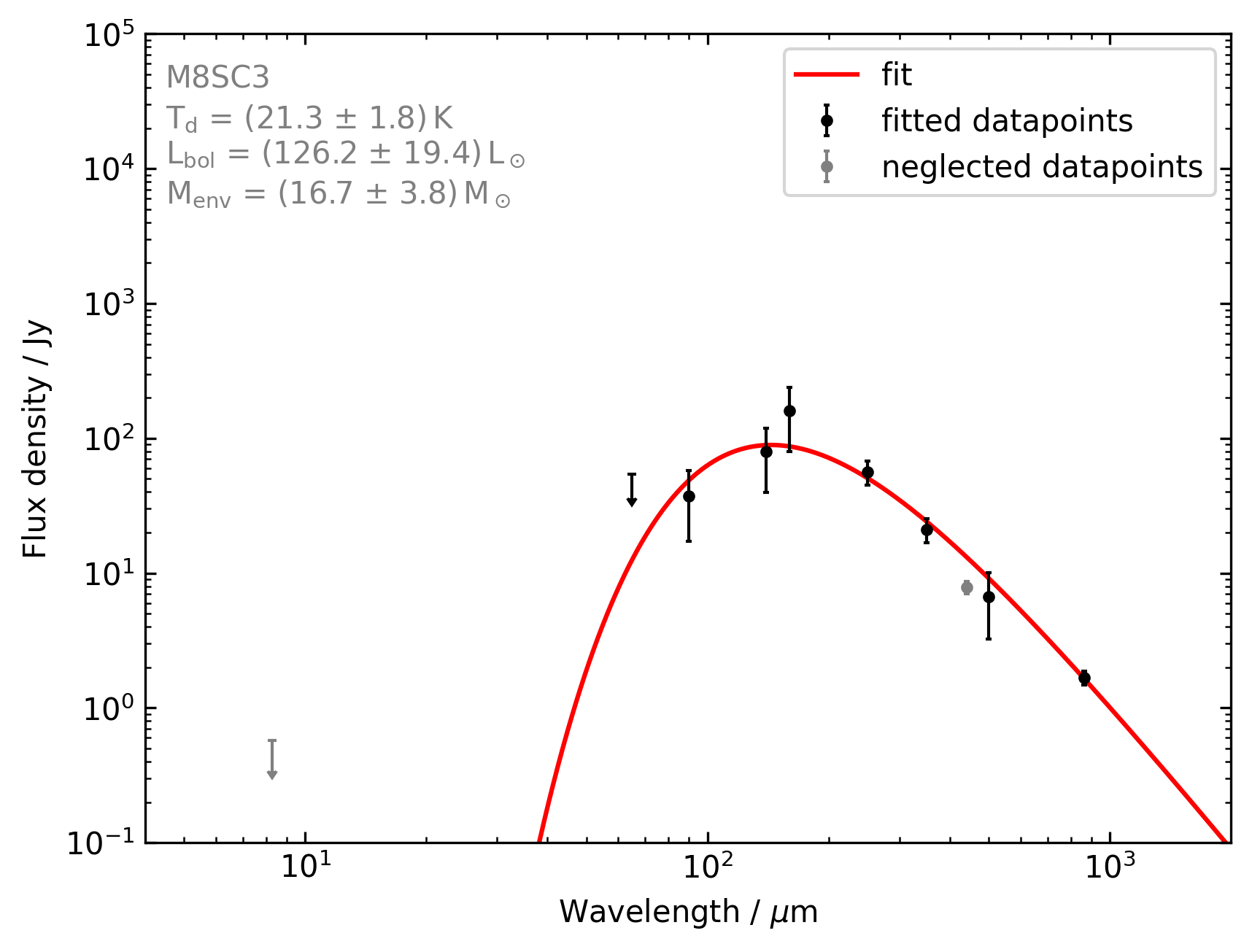}
                        \end{minipage}
                        \begin{minipage}[b]{0.4\linewidth}
                                \centering
                                \includegraphics[width=\linewidth]{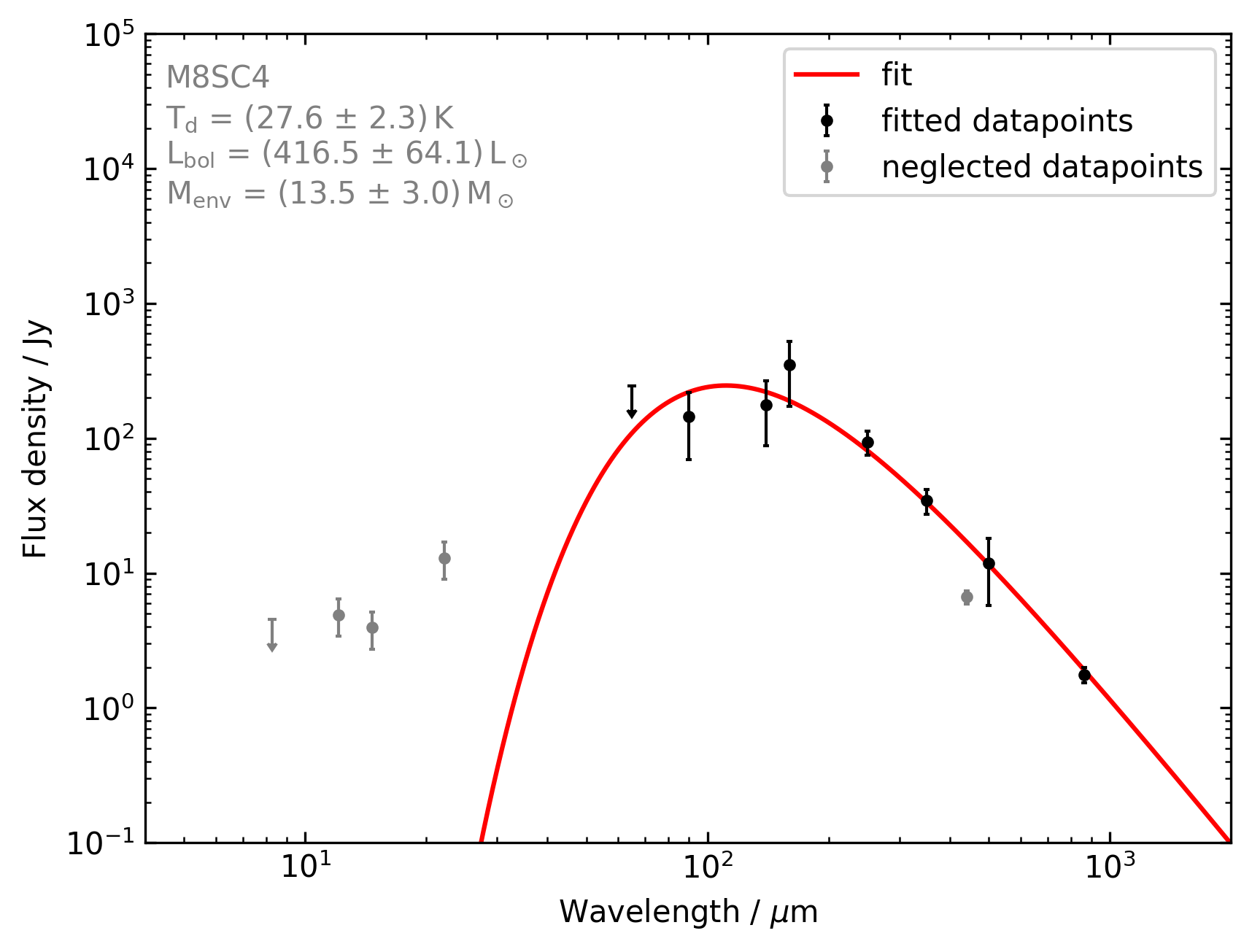}
                        \end{minipage}
                        \\
                        \begin{minipage}[b]{0.4\linewidth}
                                \centering
                                \includegraphics[width=\linewidth]{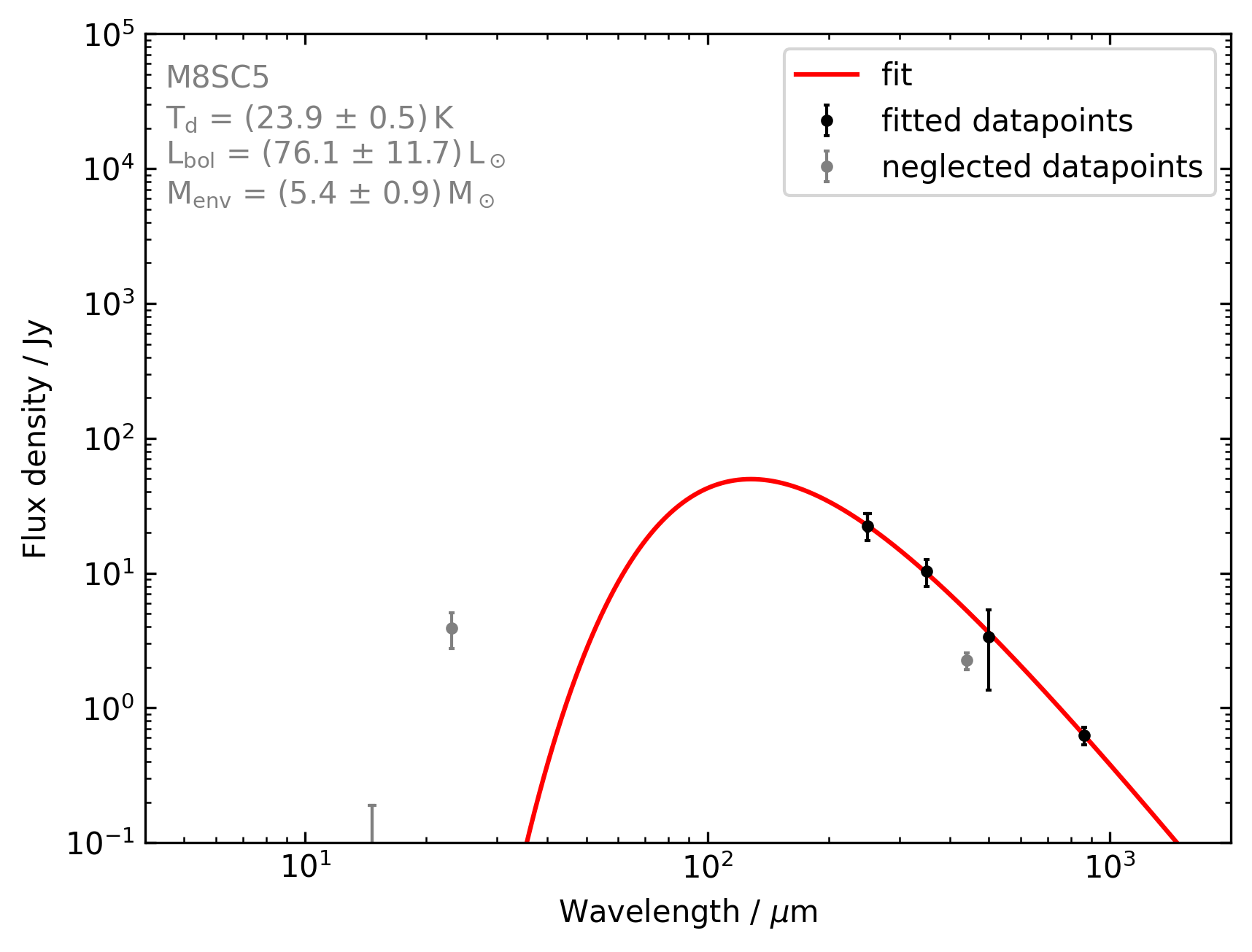}
                        \end{minipage}
                        \begin{minipage}[b]{0.4\linewidth}
                                \centering
                                \includegraphics[width=\linewidth]{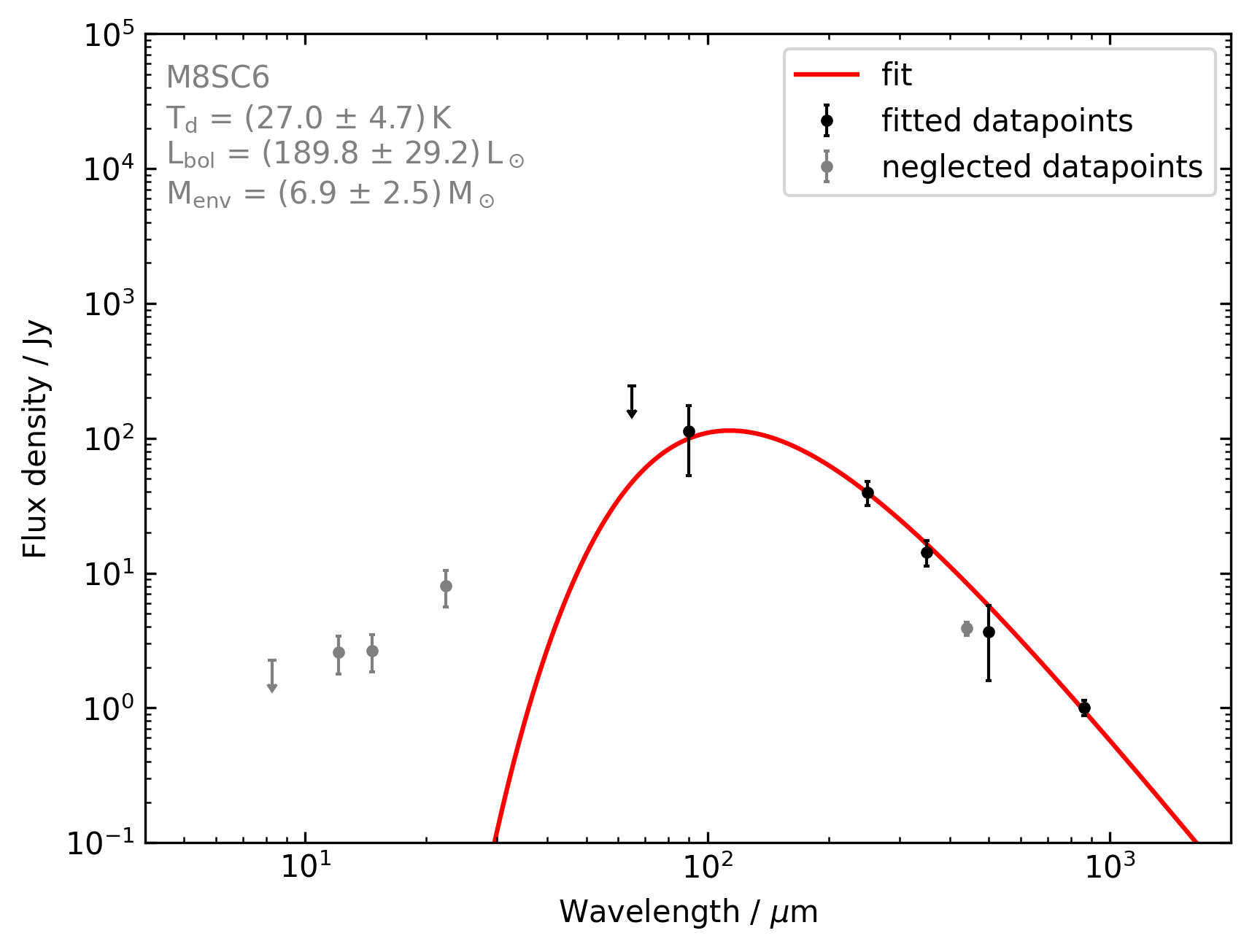}
                        \end{minipage}
                        \\
                        \begin{minipage}[b]{0.4\linewidth}
                                \centering
                                \includegraphics[width=\linewidth]{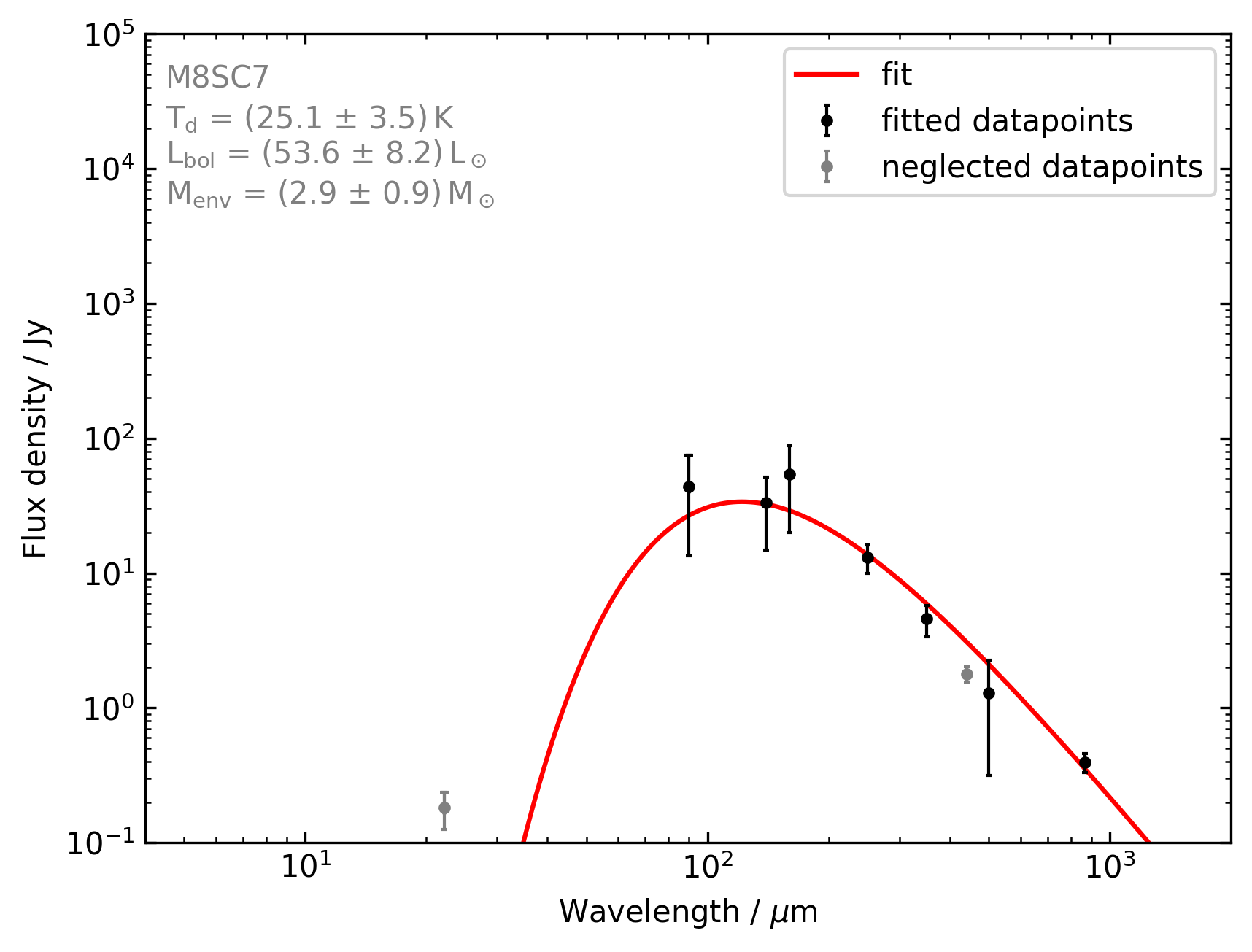}
                        \end{minipage}
                        \begin{minipage}[b]{0.4\linewidth}
                                \centering
                                \includegraphics[width=\linewidth]{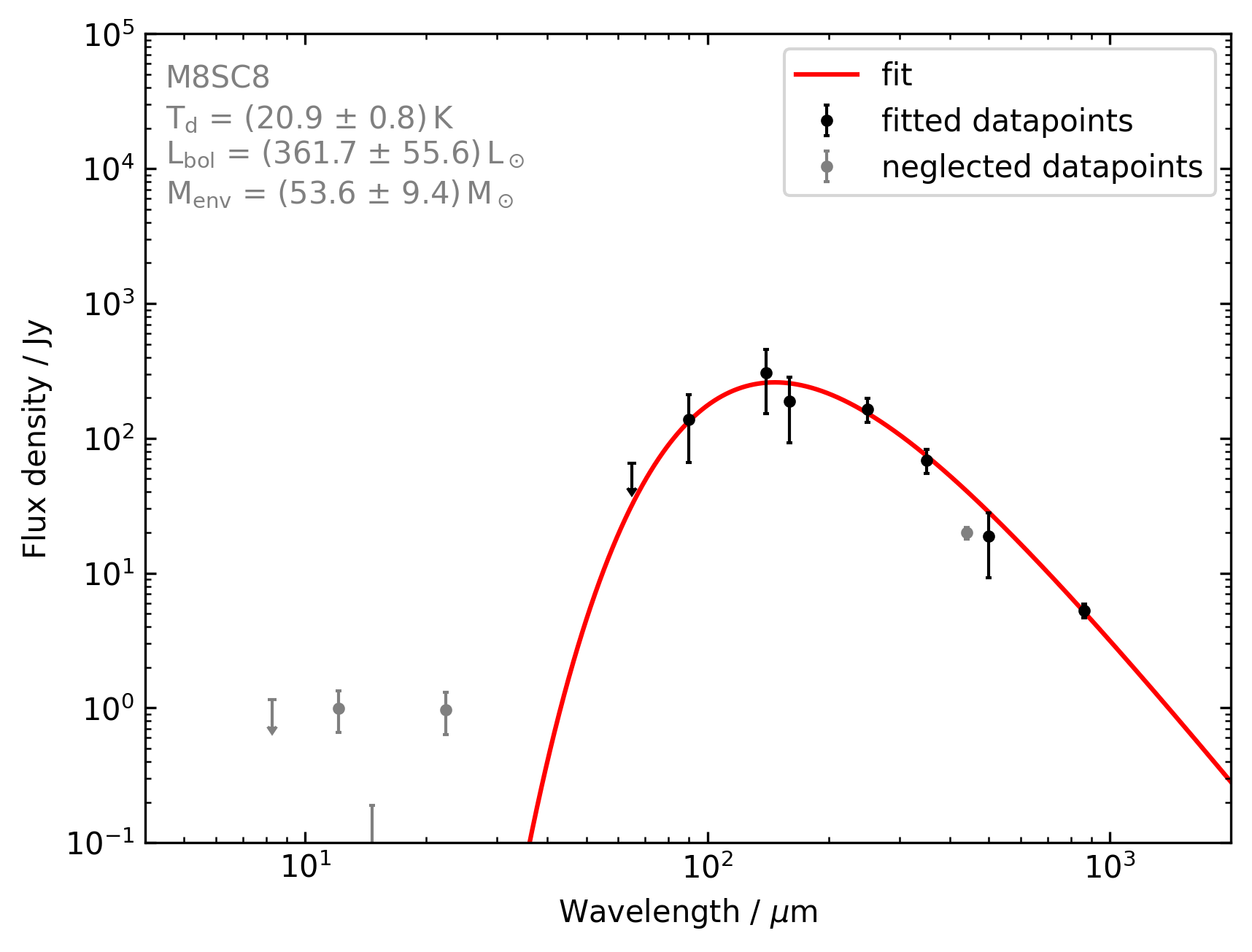}
                        \end{minipage}
                        \caption{Fits to the SED of the SC1--8 clumps. Single-component fits only consider flux at wavelengths longer than $\SI{65}{\micro\meter}$ to avoid a contribution from unrelated diffuse warm gas to the clumps' SEDs. SCUBA $\SI{450}{\micro\meter}$ emission was not considered as we found it to systematically underestimate the flux of all clumps. Flux densities at $\SI{8}{\micro\meter}$ and $\SI{65}{\micro\meter}$ are considered as upper limits due to the possible contributions of PAHs and very small grains.}
                        \label{fig:app:SED_SC}
                \end{figure}

                Sect.~\ref{sec:dust} describes the procedure we used for modelling the SEDs of the M8 clumps. This appendix provides an overview of the physical properties obtained from the modelling in Table~\ref{tab:app:clumpproperties}. The fitted SEDs for all clumps in M8 are shown in Fig.~\ref{fig:app:SED_HG}-\ref{fig:app:SED_SC}.

                \begin{table*}[htbp]
                        \caption{Clump properties derived from SED fits to the dust continuum emission.}
                        \label{tab:app:clumpproperties}
                        \centering
                        \begin{tabular}{cccccccc}
                                \hline
                                \hline
                                Clump & $T_\U{dust}$ & $\tau_{870\,\si{\micro\meter}}$ & $T_\U{hot}$ & $r_\U{hot}$ & $M$ & $L$  &  $n_{\U{H}_2}$  \\
                                & (K)          &    (10$^{-4}$)          &  (K)       &  (10$^{-2}\,\si{\arcsecond}$)    &  (M$_\odot) $ & (L$_\odot$)  & ($10^{20}$\,cm$^{-2}$)    \\ \hline
                                HG & 34.1 $\pm$ 11.9 & 22.1 $\pm$ 20.5 & 152.0 $\pm$ 21.8 & 38.9 $\pm$ 16.8 & 65.2 $\pm$ 43.9 & 8860.0 $\pm$ 1360.0 & 612.2 $\pm$ 401.7 \\
                                WC1 & 25.9 $\pm$ 3.4 & 12.0 $\pm$ 3.5 & - & - & 22.0 $\pm$ 6.5 & 473.0 $\pm$ 72.8 & 168.2 $\pm$ 42.0   \\
                                WC2 & 22.5 $\pm$ 1.4 & 8.4 $\pm$ 1.2 & - & - & 30.8 $\pm$ 6.0 & 310.0 $\pm$ 47.8 & 117.6 $\pm$ 14.2   \\
                                WC3 & 30.0 $\pm$ 4.5 & 3.5 $\pm$ 1.1 & - & - & 5.0 $\pm$ 1.6 & 248.0 $\pm$ 38.2 & 48.9 $\pm$ 13.6   \\
                                WC4 & 37.6 $\pm$ 5.8 & 1.7 $\pm$ 0.5 & - & - & 6.5 $\pm$ 2.0 & 1190.0 $\pm$ 184.0 & 24.2 $\pm$ 6.5   \\
                                WC5 & 36.1 $\pm$ 7.5 & 1.0 $\pm$ 0.4 & - & - & 2.1 $\pm$ 0.8 & 303.0 $\pm$ 46.5 & 13.4 $\pm$ 4.9   \\
                                WC6 & 30.0 $\pm$ 3.9 & 2.4 $\pm$ 0.6 & - & - & 7.6 $\pm$ 2.2 & 376.0 $\pm$ 57.9 & 33.9 $\pm$ 8.2  \\
                                WC7 & 28.0 $\pm$ 3.8 & 5.3 $\pm$ 1.5 & 251.0 $\pm$ 78.5 & 2.0 $\pm$ 1.2 & 15.5 $\pm$ 4.5 & 626.0 $\pm$ 96.4 & 75.2 $\pm$ 18.8  \\
                                WC8 & 19.7 $\pm$ 1.9 & 2.4 $\pm$ 0.6 & - & - & 8.1 $\pm$ 2.1 & 39.5 $\pm$ 6.1 & 34.2 $\pm$ 7.1  \\
                                WC9 & 37.0 $\pm$ 6.8 & 0.7 $\pm$ 0.3 & - & - & 1.3 $\pm$ 0.5 & 222.0 $\pm$ 34.2 & 9.4 $\pm$ 3.1   \\
                                SW1 & 20.7 $\pm$ 1.0 & 2.3 $\pm$ 0.3 & - & - & 7.6 $\pm$ 1.4 & 48.6 $\pm$ 7.5 & 32.1 $\pm$ 3.3   \\
                                EC1 & 18.9 $\pm$ 1.3 & 24.7 $\pm$ 4.8 & - & - & 53.9 $\pm$ 11.6 & 204.0 $\pm$ 31.4 & 345.2 $\pm$ 52.1  \\
                                EC2 & 19.9 $\pm$ 2.2 & 7.5 $\pm$ 2.2 & - & - & 24.7 $\pm$ 6.8 & 130.0 $\pm$ 20.0 & 105.7 $\pm$ 24.1  \\
                                EC3 & 20.0 $\pm$ 2.2 & 8.4 $\pm$ 2.5 & - & - & 25.4 $\pm$ 7.1 & 135.0 $\pm$ 20.8 & 117.5 $\pm$ 27.2  \\
                                EC4 & 15.7 $\pm$ 0.4 & 19.7 $\pm$ 2.0 & - & - & 47.9 $\pm$ 7.9 & 58.6 $\pm$ 9.0 & 275.6 $\pm$ 15.8   \\
                                EC5 & 17.9 $\pm$ 2.6 & 9.2 $\pm$ 3.9 & - & - & 29.5 $\pm$ 10.5 & 82.5 $\pm$ 12.7 & 129.7 $\pm$ 41.6  \\
                                E & 25.4 $\pm$ 4.5 & 36.6 $\pm$ 15.0 & 224.0 $\pm$ 87.8 & 10.5 $\pm$ 9.0 & 92.8 $\pm$ 35.2 & 3560.0 $\pm$ 547.0 & 543.1 $\pm$ 188.0  \\
                                SE1 & 19.8 $\pm$ 2.2 & 12.7 $\pm$ 3.7 & - & - & 40.3 $\pm$ 11.1 & 202.0 $\pm$ 31.1 & 177.3 $\pm$ 40.5  \\
                                SE2 & 22.9 $\pm$ 2.9 & 4.8 $\pm$ 1.4 & 321.0 $\pm$ 155.0 & 0.8 $\pm$ 0.6 & 15.4 $\pm$ 4.5 & 217.0 $\pm$ 33.3 & 67.3 $\pm$ 16.7  \\
                                SE3 & 21.0 $\pm$ 3.0 & 7.0 $\pm$ 2.4 & 213.0 $\pm$ 66.4 & 1.9 $\pm$ 1.4 & 28.0 $\pm$ 9.1 & 249.0 $\pm$ 38.2 & 99.6 $\pm$ 28.5  \\
                                SE4 & 21.0 $\pm$ 2.8 & 5.0 $\pm$ 1.8 & - & - & 4.9 $\pm$ 1.5 & 34.4 $\pm$ 5.3 & 70.2 $\pm$ 19.2  \\
                                SE5 & 23.3 $\pm$ 3.0 & 2.3 $\pm$ 0.7 & - & - & 5.4 $\pm$ 1.6 & 66.8 $\pm$ 10.3 & 32.3 $\pm$ 8.2   \\
                                SE6 & 19.9 $\pm$ 6.7 & 5.9 $\pm$ 5.1 & - & - & 19.6 $\pm$ 14.0 & 102.0 $\pm$ 15.7 & 82.8 $\pm$ 57.7 \\
                                SE7 & 20.8 $\pm$ 2.7 & 32.8 $\pm$ 10.9 & 347.0 $\pm$ 215.0 & 0.7 $\pm$ 0.7 & 56.1 $\pm$ 17.1 & 416.0 $\pm$ 63.9 & 458.8 $\pm$ 120.2  \\
                                SE8 & 21.2 $\pm$ 3.0 & 17.1 $\pm$ 6.2 & 244.0 $\pm$ 104.0 & 1.2 $\pm$ 1.0 & 37.1 $\pm$ 12.1 & 302.0 $\pm$ 46.5 & 240.1 $\pm$ 68.9  \\
                                SC1 & 24.5 $\pm$ 2.6 & 11.7 $\pm$ 2.8 & 463.0 $\pm$ 73.1 & 0.5 $\pm$ 0.2 & 20.2 $\pm$ 5.2 & 411.0 $\pm$ 63.3 & 163.6 $\pm$ 33.3  \\
                                SC2 & 21.3 $\pm$ 1.9 & 8.7 $\pm$ 1.9 & - & - & 32.3 $\pm$ 7.6 & 243.0 $\pm$ 37.3 & 121.5 $\pm$ 21.6   \\
                                SC3 & 21.3 $\pm$ 1.8 & 7.8 $\pm$ 1.7 & - & - & 16.7 $\pm$ 3.9 & 126.0 $\pm$ 19.4 & 109.6 $\pm$ 18.9  \\
                                SC4 & 27.6 $\pm$ 2.4 & 3.8 $\pm$ 0.7 & - & - & 13.5 $\pm$ 3.0 & 416.0 $\pm$ 64.1 & 52.9 $\pm$ 8.5  \\
                                SC5 & 23.9 $\pm$ 0.5 & 3.1 $\pm$ 0.2 & - & - & 5.4 $\pm$ 0.9 & 76.1 $\pm$ 11.7 & 44.1 $\pm$ 1.8   \\
                                SC6 & 27.0 $\pm$ 4.7 & 5.2 $\pm$ 1.8 & - & - & 6.9 $\pm$ 2.5 & 190.0 $\pm$ 29.2 & 72.9 $\pm$ 23.8  \\
                                SC7 & 25.1 $\pm$ 3.5 & 1.7 $\pm$ 0.5 & - & - & 2.9 $\pm$ 0.9 & 53.6 $\pm$ 8.2 & 23.3 $\pm$ 6.3   \\
                                SC8 & 20.9 $\pm$ 0.8 & 11.8 $\pm$ 1.2 & - & - & 53.6 $\pm$ 9.4 & 362.0 $\pm$ 55.6 & 165.5 $\pm$ 13.8  \\
                                SC9 & 21.3 $\pm$ 1.5 & 5.0 $\pm$ 0.9 & - & - & 29.0 $\pm$ 6.2 & 222.0 $\pm$ 34.2 & 70.5 $\pm$ 10.3  \\
                                C1 & 43.2 $\pm$ 16.4 & 0.3 $\pm$ 0.3 & - & - & 0.6 $\pm$ 0.4 & 245.0 $\pm$ 37.7 & 4.3 $\pm$ 2.8  \\
                                C2 & 30.4 $\pm$ 7.2 & 1.6 $\pm$ 0.8 & - & - & 2.3 $\pm$ 1.1 & 124.0 $\pm$ 19.1 & 22.6 $\pm$ 9.9   \\
                                C3 & 25.0 $\pm$ 4.2 & 2.5 $\pm$ 1.0 & - & - & 3.0 $\pm$ 1.1 & 55.3 $\pm$ 8.5 & 35.4 $\pm$ 11.6  \\
                                \hline
                        \end{tabular}
                        \tablefoot{Values that could not be derived are marked with a minus sign.}
                \end{table*}
                
                \clearpage
                
                \section{Identified transitions and line parameters}
                \label{app:lineidentification}
\begin{figure}[htbp]
                        \centering
                        \includegraphics[width=0.7\linewidth]{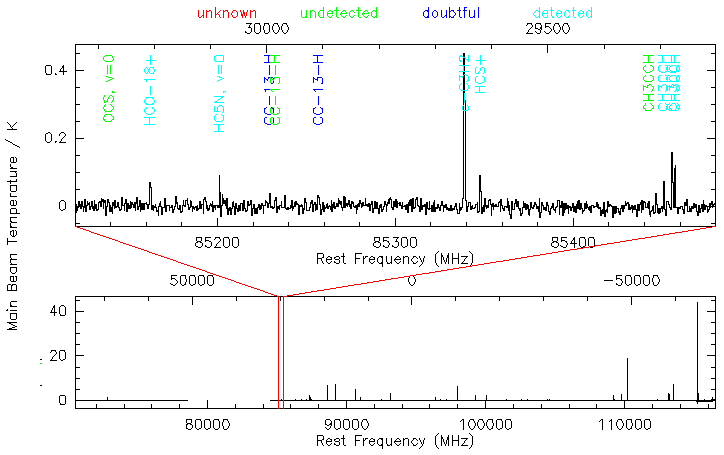}
                        \includegraphics[width=0.7\linewidth]{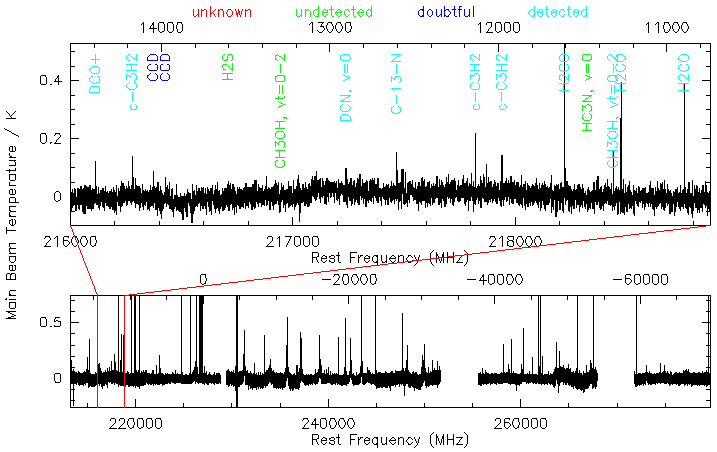}
                        \caption{Spectrum extracts of the data obtained with the IRAM 30m telescope (upper panel) and APEX (lower panel) at SE1.}
                        \label{fig:app:spectrum}
                \end{figure}

                This appendix provides a full overview of molecular rotational transitions observed in the Lagoon Nebula. As an example of the analysed data, spectrum extracts of SE1 are shown in Fig.~\ref{fig:app:spectrum}.
                
                Table~\ref{tab:app:idtransitions} lists the name of each species, the quantum numbers or identifiers of the transition, the associated frequency in MHz, the upper-level energy, $E_\U{up}$, in K, the Einstein A coefficient, $A_\U{ij}$, in s$^{-1}$ and the degeneracy $g_\mathrm{up}$ according to the information provided by the databases (last column). 
                
                The second column describes the transition according to the values stated by the respective database (usually $J$ or $J_F$ for linear molecules, $J_K$ for symmetric tops and $J_{K_a, K_c}$ for asymmetric tops). The third subscript number for NH$_2$D describes the symmetry state, where the anti-symmetric state is labelled 1 and the symmetric state is labelled 0. An exception is methanol, for which the symmetry state and, as subscript, $k$ for the E-type and $K$ for the A-type species are given. Lines from NO, NS, CN, C$_2$H, and C$_2$D are described as $N_{J,F}$, where the additional subscript of NO and NS states the parity. HCO and c-C$_3$H lines are described as $N_{K_a,K_c,J,F}$ in order to fully identify the transitions. The descriptions of $^{13}$CN and C$^{13}$CH lines additionally contain the intermediate quantum numbers as $N_{F_1,F_2,F}$ and $N_{J,F_1,F}$, respectively.
                
                Table \ref{tab:app:line_properties} gives a detailed overview of the derived line properties for detected emission from rotational transitions towards all M8 clumps. The full versions of Tables~\ref{tab:app:idtransitions} and~\ref{tab:app:line_properties} are available at the CDS.
                
                \begin{table}[ht]
                        \caption{Overview of the identified molecular transitions in the Lagoon Nebula (extract).}
                        \label{tab:app:idtransitions}
                        \centering
                        \begin{tabular}{ccccccc}
                                \hline
                                \hline
                                Molecule & Transition & Frequency  & $E_\U{up}$ & $A_\U{ij}$ & $g_\mathrm{up}$ & origin \\
                                &            & (MHz)      & (K)        & (s$^{-1}$)     &               &  \\ \hline
                                H$_2^{13}$CO & $1_{0,1}-0_{0,0}$ & 71024.788 (10) & 3.4 & 7.6e-06 & 3 & CDMS \\
                                HC$_5$N & $27-26$ & 71889.595 (0) & 48.3 & 4.0e-05 & 165 & CDMS \\
                                DCO$^+$ & $1-0$ & 72039.312 (1) & 3.5 & 2.2e-05 & 3 & CDMS \\
                                C$_2$S & $6_5-5_4$ & 72323.789 (20) & 19.2 & 1.6e-05 & 11 & CDMS \\
                                DCN & $1-0$ & 72414.694 (0) & 3.5 & 1.3e-05 & 9 & CDMS \\
                                \hline
                        \end{tabular}
                        \tablefoot{Overview of the identified molecular transitions in the Lagoon Nebula. Frequency, upper-level energy $E_\U{up}$, spontaneous emission Einstein (A) coefficient $A_\U{ij}$ and upper-level degeneracy $g_\mathrm{up}$ are listed for each transition according to the values stated by the database given in the last column. The frequency uncertainty for each transition is given in parenthesis in units of kHz and was rounded to integers. Partially blended transitions are indicated with superscript markers a) to u). The full table with all transitions is available at the CDS.}
                \end{table}
                
                \begin{table}[htbp]
                        \caption{Line parameters of transitions identified at the M8 clumps (extract).}
                        \label{tab:app:line_properties}
                        
                        \begin{tabular}{cccccccccc}
                                \hline
                                \hline
                                Clump & Species & Frequency & RMS & $I_\mathrm{integration}$ & $I_\mathrm{fit,1}$ & $\varv_1$ & $\Delta \varv_1$ & $T_\mathrm{peak,1}$ & $\tau_\U{1}$  \\
                                &  & (GHz) & (mK)     & ($\si{\kelvin\kilo\meter\per\second}$) & ($\si{\kelvin\kilo\meter\per\second}$) & ($\si{\kilo\meter\per\second}$) & ($\si{\kilo\meter\per\second}$) & (K)& \\ \hline
                                HG & H$_2^{13}$CO & 71.025 & 32.0 & - & - & - & - & - & - \\
                                HG & HC$_5$N & 71.890 & 27.0 & - & - & - & - & - & - \\
                                HG & DCO$^+$ & 72.039 & 24.0 & - & - & - & - & - & - \\
                                HG & C$_2$S & 72.324 & 26.0 & - & - & - & - & - & - \\
                                HG & DCN & 72.415 & 28.4 & 0.73 $\pm$ 0.12 & 0.42 $\pm$ 0.05 & 9.33 $\pm$ 0.25 & 2.50 $\pm$ 0.47 & - & 0.10 $\pm$ 3.39 \\
                                HG & HC$^{13}$CCN & 72.475 & 24.0 & - & - & - & - & - & - \\
                                HG & HC$_2^{13}$CN & 72.482 & 25.0 & - & - & - & - & - & -\\
                                HG & SO$_2$ & 72.758 & 31.0 & - & - & - & - & - & -\\
                                HG & HC$_3$N & 72.784 & 25.7 & 1.04 $\pm$ 0.11 & 0.78 $\pm$ 0.05 & 8.69 $\pm$ 0.07 & 2.24 $\pm$ 0.17 & 0.33 $\pm$ 0.03\\
                                HG & H$_2$CO & 72.838 & 21.1 & 3.08 $\pm$ 0.09 & 2.68 $\pm$ 0.11 & 9.90 $\pm$ 0.06 & 2.80 $\pm$ 0.14 & 0.90 $\pm$ 0.06 & -\\
                                \hline
                        \end{tabular}
                        \newline
                        \vspace*{1 cm}
                        \newline
                        \begin{tabular}{cccccccc}
                                \hline
                                \hline
                                Clump & Species & Frequency & $I_\mathrm{fit,2}$ & $\varv_2$ & $\Delta \varv_2$ & $T_\mathrm{peak,2}$ & $\tau_\U{2}$ \\
                                &  & (GHz)  & ($\si{\kelvin\kilo\meter\per\second}$) & ($\si{\kilo\meter\per\second}$) & ($\si{\kilo\meter\per\second}$) & (K) & \\ \hline
                                HG & H$_2^{13}$CO & 71.025 & - & - & - & - & - \\
                                HG & HC$_5$N & 71.890 & - & - & - & - & - \\
                                HG & DCO$^+$ & 72.039 & - & - & - & - & - \\
                                HG & C$_2$S & 72.324 & - & - & - & - & - \\
                                HG & DCN & 72.415 & - & - & - & - & - \\
                                HG & HC$^{13}$CCN & 72.475 & - & - & - & - & - \\
                                HG & HC$_2^{13}$CN & 72.482 & - & - & - & - & - \\
                                HG & SO$_2$ & 72.758 & - & - & - & - & - \\
                                HG & HC$_3$N & 72.784 & - & - & - & - & - \\
                                HG & H$_2$CO & 72.838 & 0.29 $\pm$ 0.12 & 6.06 $\pm$ 0.51 & 2.65 $\pm$ 1.55 & 0.10 $\pm$ 0.07 & - \\
                                \hline
                        \end{tabular}
                        \tablefoot{This table shows the line parameters of the first ten transitions at HG. Clump, species, and frequency of the corresponding transition are given in the first three columns. For lines with HFS, we give the central frequency of all considered components. The fourth column provides the RMS noise level in the vicinity of the respective transition in mK. The fifth column gives the integrated intensity of the whole line profile, including all possibly occurring components. The remaining columns state the fitted line parameters for the Gaussian and \texttt{HFS} fits: The line intensity $I_{\U{fit},i}$ (6, 11), the transition LOS velocity with respect to the LSR $\varv_i$ (7, 12), and the FWHM line width $\Delta \varv_i$ (8, 13). For Gaussian fits we additionally give the peak temperature of the respective transition $T_{\U{peak},i}$ (9, 14), while for the \texttt{HFS} fits we give the optical depth $\tau_i$ obtained by the fit (10, 15). If present, we fitted up to two line-components $i$ per clump. The columns 11 to 15 are presented in the lower part of the table, after a repetition of clump name, species, and frequency. Values that were not derived or where the transition has not been detected are marked with a minus (-). The full table with all transitions at all clumps is available at the CDS.}
                \end{table}
                \twocolumn
                \clearpage
                
                \section{Radio recombination lines detected in M8}
                \label{app:rrl}
                Table~\ref{tab:app:rrl} provides an overview of the computed Gaussian profiles for the combined spectra of the detected RRLs.
                \begin{table}[htbp]
                        \caption{Line parameters of identified RRLs in the M8 region.}
                        \label{tab:app:rrl}
                        \centering
                        \begin{tabular}{L{0.6cm}L{1.5cm}C{1.8cm}C{1.5cm}C{1.5cm}}
                                \hline \hline
                                Clump & Transition & $I$ & $\varv_\U{LSR}$ & $\Delta v$  \\ 
                                &            & K\,km\,s$^{-1}$ & km\,s$^{-1}$ & km\,s$^{-1}$  \\ \hline
                                HG  &  H\,39-44$\alpha$  &  12.69 $\pm$  0.04  &  3.5 $\pm$ 0.1  &  26.7 $\pm$  0.1 \\
                                &  He\,39-44$\alpha$ &  0.84 $\pm$  0.03  &  1.9 $\pm$ 0.3  &  19.7 $\pm$  0.8 \\
                                &  C\,39-44$\alpha$  &  0.11 $\pm$  0.02  &  11.1 $\pm$ 0.2  &  4.1 $\pm$  0.6 \\
                                &  H\,48-56$\beta$   &  3.78 $\pm$  0.09  &  3.3 $\pm$ 0.3  &  26.8 $\pm$  0.8 \\
                                &  H\,54-63$\gamma$  &  1.75 $\pm$  0.03  &  3.7 $\pm$ 0.3  &  26.6 $\pm$  0.7 \\
                                WC1  &  H\,39-44$\alpha$  &  1.46 $\pm$  0.04  &  -2.4 $\pm$ 0.3  &  22.0 $\pm$  0.7 \\
                                WC2  &  H\,39-44$\alpha$  &  0.92 $\pm$  0.05  &  0.5 $\pm$ 0.7  &  25.8 $\pm$  1.7 \\
                                WC3  &  H\,39-44$\alpha$  &  1.04 $\pm$  0.06  &  0.2 $\pm$ 0.6  &  22.8 $\pm$  1.4 \\
                                WC4  &  H\,39-44$\alpha$  &  1.97 $\pm$  0.06  &  6.4 $\pm$ 0.5  &  29.2 $\pm$  1.0 \\
                                WC5  &  H\,39-44$\alpha$  &  1.64 $\pm$  0.05  &  1.7 $\pm$ 0.4  &  23.8 $\pm$  0.8 \\
                                WC6  &  H\,39-44$\alpha$  &  1.12 $\pm$  0.05  &  1.3 $\pm$ 0.5  &  23.9 $\pm$  1.4 \\
                                EC1  &  H\,39-44$\alpha$  &  0.39 $\pm$  0.04  &  -0.1 $\pm$ 0.8  &  18.3 $\pm$  2.1 \\
                                EC2  &  H\,39-44$\alpha$  &  0.39 $\pm$  0.05  &  0.1 $\pm$ 1.0  &  16.0 $\pm$  2.4 \\
                                SC6  &  H\,39-44$\alpha$  &  0.79 $\pm$  0.05  &  -0.4 $\pm$ 0.8  &  21.9 $\pm$  1.4 \\ \hline

                        \end{tabular}
                        \tablefoot{Fits for the $\alpha$ RRLs with N=39-44 include lines with N=39-42 and N=44. Similarly, our observations do not cover RRLs with N=53 and 54 for $\beta$ and N=61 for $\gamma$ transitions, due to which these are not used for the combined spectra.}
                \end{table}

                \section{Temperatures and column densities of acetonitrile and methyl acetylene and formaldehyde}\label{app:temp_cd}
                
\begin{figure}[htbp]
                        \centering
                        \includegraphics[width=0.999\linewidth]{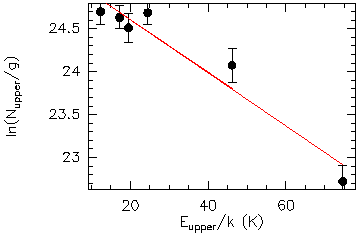}
                        \includegraphics[width=0.999\linewidth]{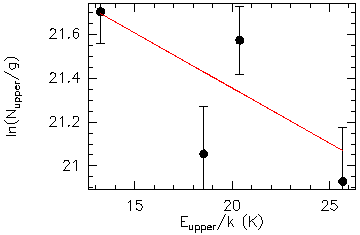}
                        \caption{Rotation diagrams of CH$_3$C$_2$H (upper panel) and CH$_3$CN (lower panel) at HG. Black data points are inferred from the line intensities, the red line shows the least squares linear fit.}
                        \label{fig:app:rd_M8HG}
                \end{figure}

\begin{figure}[htbp]
                        \centering
                        \includegraphics[width=0.999\linewidth]{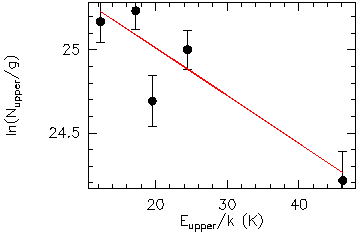}
                        \includegraphics[width=0.999\linewidth]{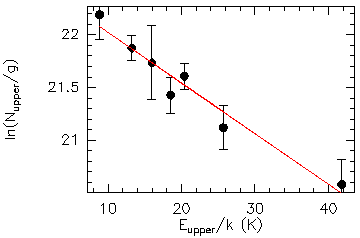}
                        \caption{Rotation diagrams of CH$_3$C$_2$H (upper panel) and CH$_3$CN (lower panel) at WC1. Black data points are inferred from the line intensities, the red line shows the least squares linear fit.}
                        \label{fig:app:rd_M8WC1}
                \end{figure}

\begin{figure}[htbp]
                        \centering
                        \includegraphics[width=0.999\linewidth]{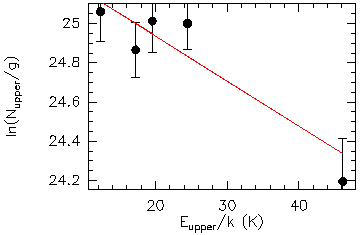}
                        \caption{Rotation diagram of CH$_3$C$_2$H at WC2. Black data points are inferred from the line intensities, the red line shows the least squares linear fit.}
                        \label{fig:app:rd_M8WC2}
                \end{figure}

\begin{figure}[htbp]
                        \centering
                        \includegraphics[width=0.999\linewidth]{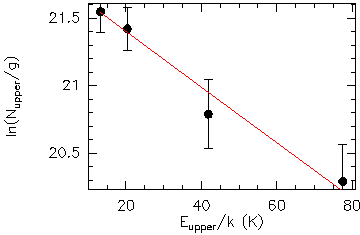}
                        \caption{Rotation diagram of CH$_3$CN at EC1. Black data points are inferred from the line intensities, the red line shows the least squares linear fit.}
                        \label{fig:app:rd_M8EC1}
                \end{figure}

\begin{figure}[htbp]
                        \centering
                        \includegraphics[width=0.999\linewidth]{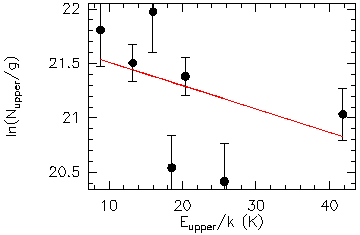}
                        \caption{Rotation diagram of CH$_3$CN at EC3. Black data points are inferred from the line intensities, the red line shows the least squares linear fit.}
                        \label{fig:app:rd_M8EC3}
                \end{figure}

\begin{figure}[htbp]
                        \centering
                        \includegraphics[width=0.999\linewidth]{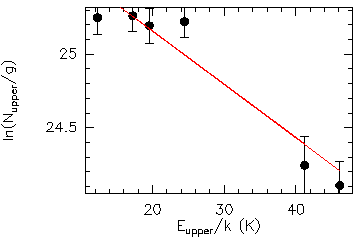}
                        \includegraphics[width=0.999\linewidth]{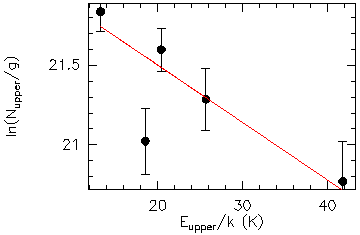}
                        \caption{Rotation diagrams of CH$_3$C$_2$H (upper panel) and CH$_3$CN (lower panel) at EC4. Black data points are inferred from the line intensities, the red line shows the least squares linear fit.}
                        \label{fig:app:rd_M8EC4}
                \end{figure}

\begin{figure}[htbp]
                        \centering
                        \includegraphics[width=0.999\linewidth]{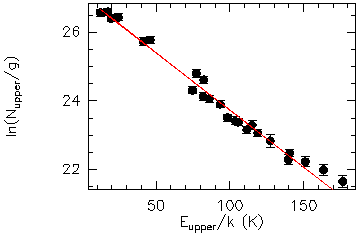}
                        \includegraphics[width=0.999\linewidth]{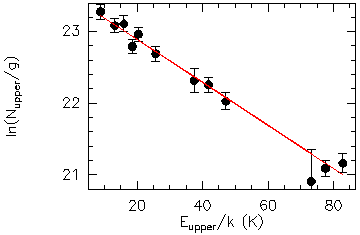}
                        \caption{Rotation diagrams of CH$_3$C$_2$H (upper panel) and CH$_3$CN (lower panel) at E. Black data points are inferred from the line intensities, the red line shows the least squares linear fit.}
                        \label{fig:app:rd_M8E}
                \end{figure}

\begin{figure}[htbp]
                        \centering
                        \includegraphics[width=0.999\linewidth]{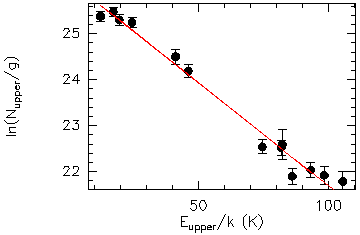}
                        \includegraphics[width=0.999\linewidth]{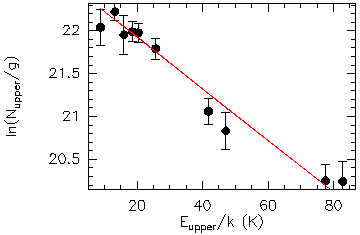}
                        \caption{Rotation diagrams of CH$_3$C$_2$H (upper panel) and CH$_3$CN (lower panel) at SE1. Black data points are inferred from the line intensities, the red line shows the least squares linear fit.}
                        \label{fig:app:rd_M8SE1}
                \end{figure}

\begin{figure}[htbp]
                        \centering
                        \includegraphics[width=0.999\linewidth]{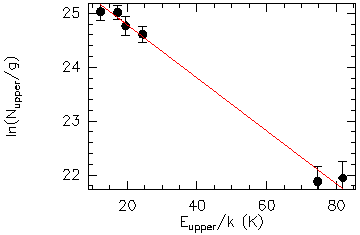}
                        \includegraphics[width=0.999\linewidth]{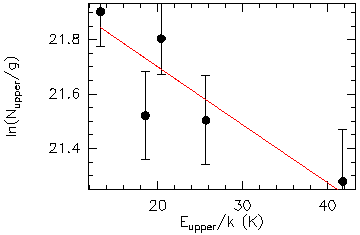}
                        \caption{Rotation diagrams of CH$_3$C$_2$H (upper panel) and CH$_3$CN (lower panel) at SE3. Black data points are inferred from the line intensities, the red line shows the least squares linear fit.}
                        \label{fig:app:rd_M8SE3}
                \end{figure}

\begin{figure}[htbp]
                        \centering
                        \includegraphics[width=0.999\linewidth]{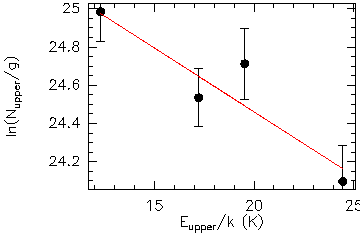}
                        \includegraphics[width=0.999\linewidth]{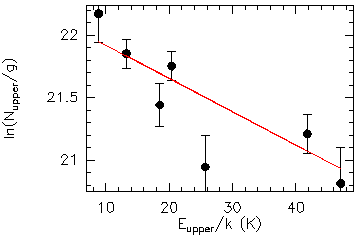}
                        \caption{Rotation diagrams of CH$_3$C$_2$H (upper panel) and CH$_3$CN (lower panel) at SE7. Black data points are inferred from the line intensities, the red line shows the least squares linear fit.}
                        \label{fig:app:rd_M8SE7}
                \end{figure}

\begin{figure}[htbp]
                        \centering
                        \includegraphics[width=0.999\linewidth]{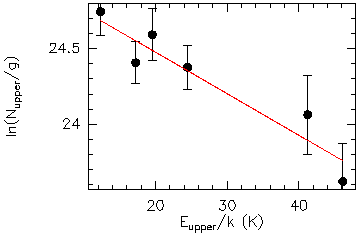}
                        \includegraphics[width=0.999\linewidth]{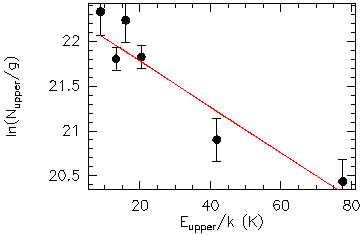}
                        \caption{Rotation diagrams of CH$_3$C$_2$H (upper panel) and CH$_3$CN (lower panel) at SE8. Black data points are inferred from the line intensities, the red line shows the least squares linear fit.}
                        \label{fig:app:rd_M8SE8}
                \end{figure}

\begin{figure}[htbp]
                        \centering
                        \includegraphics[width=0.999\linewidth]{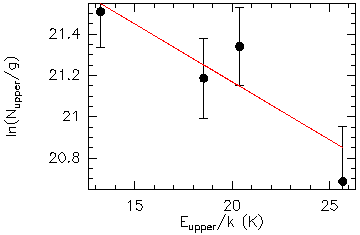}
                        \caption{Rotation diagram of CH$_3$CN at SC1. Black data points are inferred from the line intensities, the red line shows the least squares linear fit.}
                        \label{fig:app:rd_M8SC1}
                \end{figure}

\begin{figure}[htbp]
                        \centering
                        \includegraphics[width=0.999\linewidth]{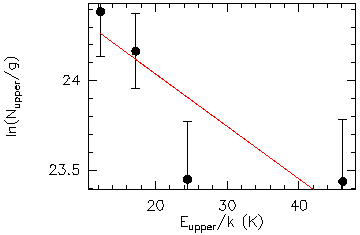}
                        \caption{Rotation diagrams of CH$_3$C$_2$H at SC2. Black data points are inferred from the line intensities, the red line shows the least squares linear fit.}
                        \label{fig:app:rd_M8SC2}
                \end{figure}

\begin{figure}[htbp]
                        \centering
                        \includegraphics[width=0.999\linewidth]{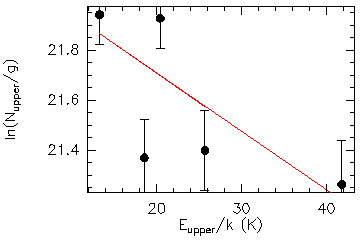}
                        \caption{Rotation diagram of CH$_3$CN at SC8. Black data points are inferred from the line intensities, the red line shows the least squares linear fit.}
                        \label{fig:app:rd_M8SC8}
                \end{figure}

\begin{figure}[htbp]
                        \centering
                        \includegraphics[width=0.999\linewidth]{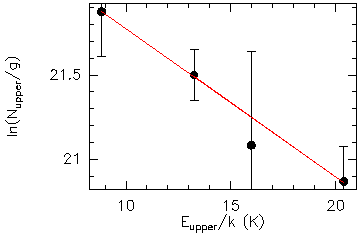}
                        \caption{Rotation diagram of CH$_3$CN at SC9. Black data points are inferred from the line intensities, the red line shows the least squares linear fit.}
                        \label{fig:app:rd_M8SC9}
                \end{figure}

\begin{figure}[htbp]
                        \centering
                        \includegraphics[width=0.999\linewidth]{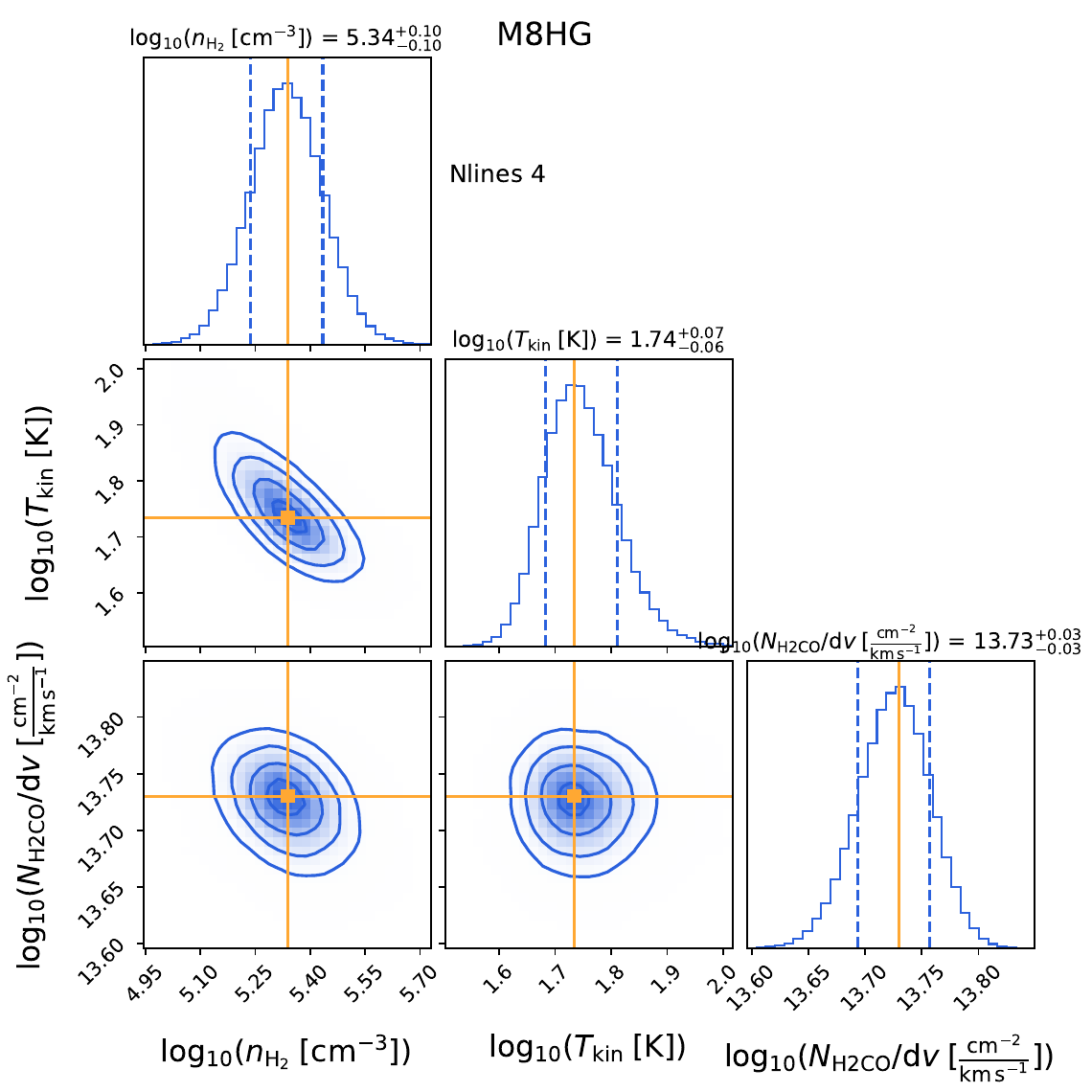}
                        \caption{Posterior probability distributions of the H$_2$ volume density ($n_{\U{H}_2}$), kinetic temperature ($T_\U{kin}$), and H$_2$CO column density per velocity bin ($N_\U{H2CO}/\U{dv}$) at HG.}
                        \label{fig:app:corner_M8HG}
                \end{figure}

\begin{figure}[htbp]
                        \centering
                        \includegraphics[width=0.999\linewidth]{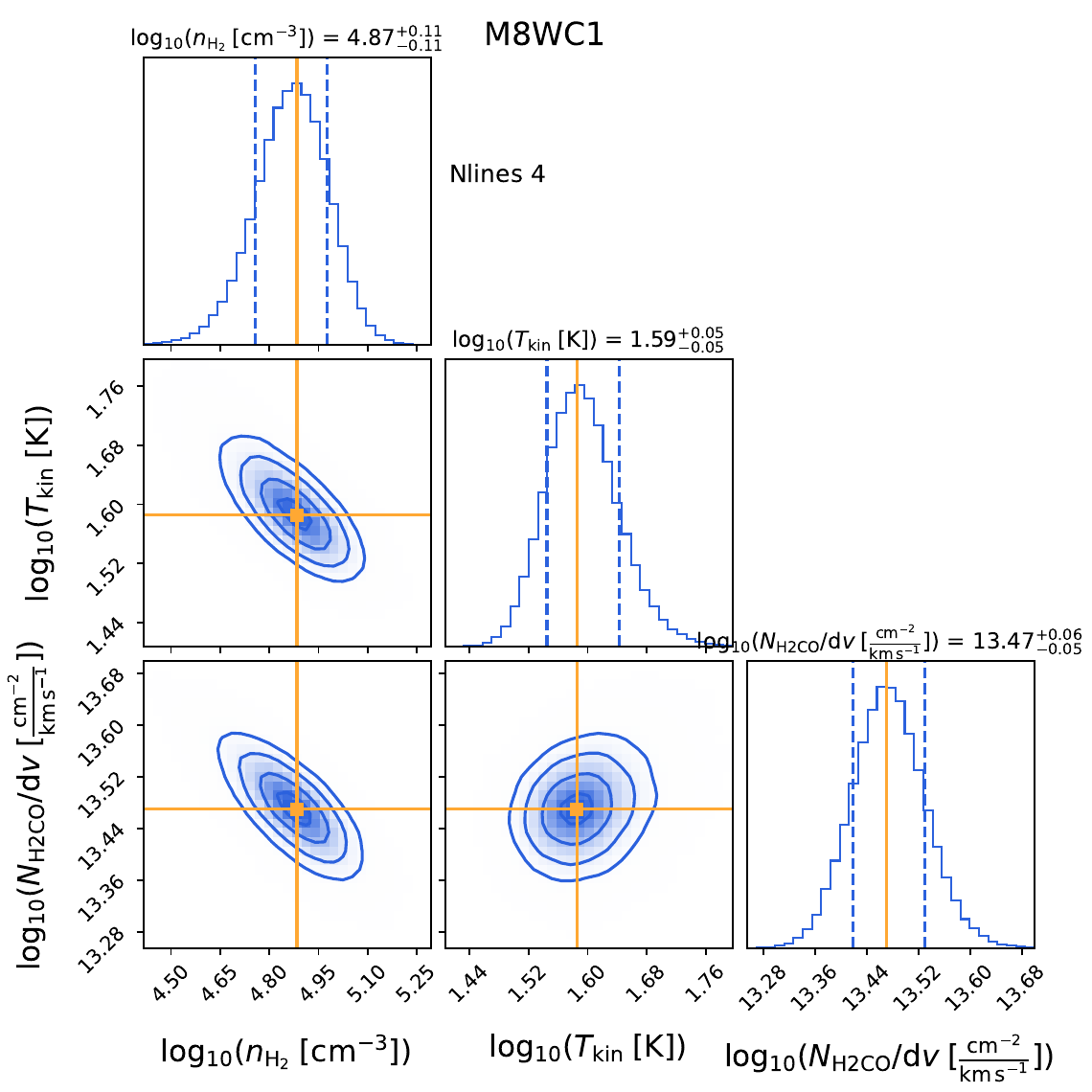}
                        \caption{Posterior probability distributions of the H$_2$ volume density ($n_{\U{H}_2}$), kinetic temperature ($T_\U{kin}$), and H$_2$CO column density per velocity bin ($N_\U{H2CO}/\U{dv}$) at WC1.}
                        \label{fig:app:corner_M8WC1}
                \end{figure}

\begin{figure}[htbp]
                        \centering
                        \includegraphics[width=0.999\linewidth]{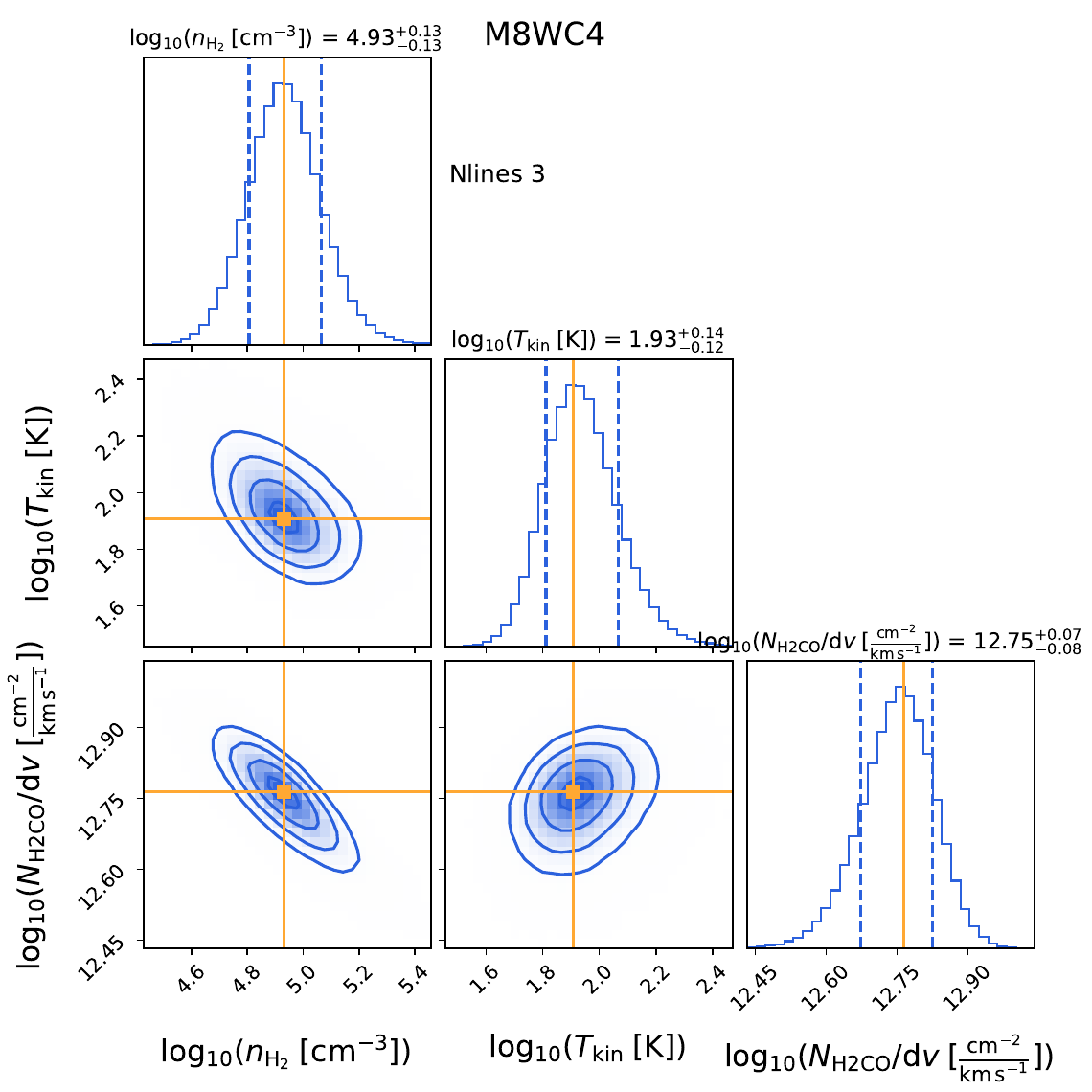}
                        \caption{Posterior probability distributions of the H$_2$ volume density ($n_{\U{H}_2}$), kinetic temperature ($T_\U{kin}$), and H$_2$CO column density per velocity bin ($N_\U{H2CO}/\U{dv}$) at WC4.}
                        \label{fig:app:corner_M8WC4}
                \end{figure}

\begin{figure}[htbp]
                        \centering
                        \includegraphics[width=0.999\linewidth]{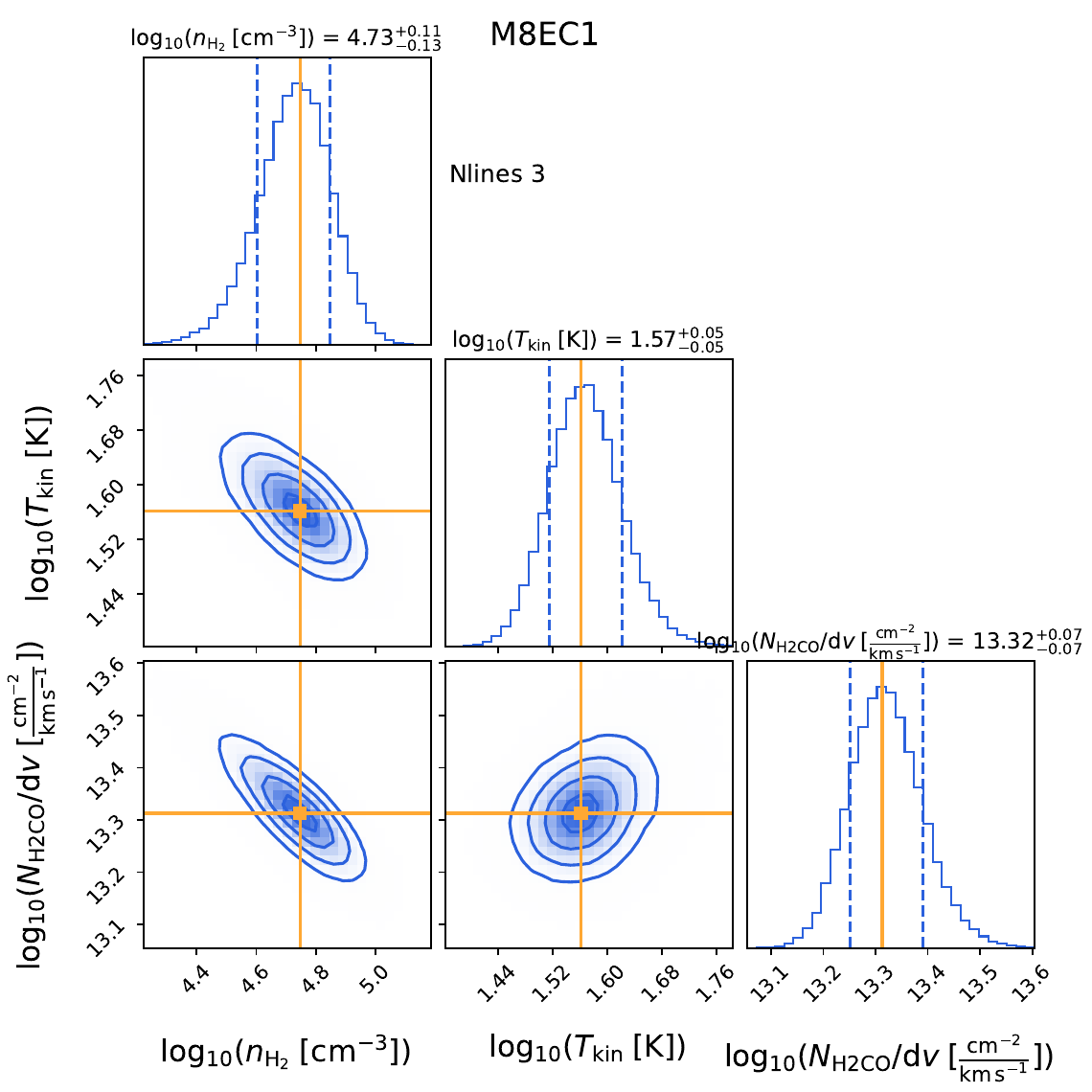}
                        \caption{Posterior probability distributions of the H$_2$ volume density ($n_{\U{H}_2}$), kinetic temperature ($T_\U{kin}$), and H$_2$CO column density per velocity bin ($N_\U{H2CO}/\U{dv}$) at EC1.}
                        \label{fig:app:corner_M8EC1}
                \end{figure}

\begin{figure}[htbp]
                        \centering
                        \includegraphics[width=0.999\linewidth]{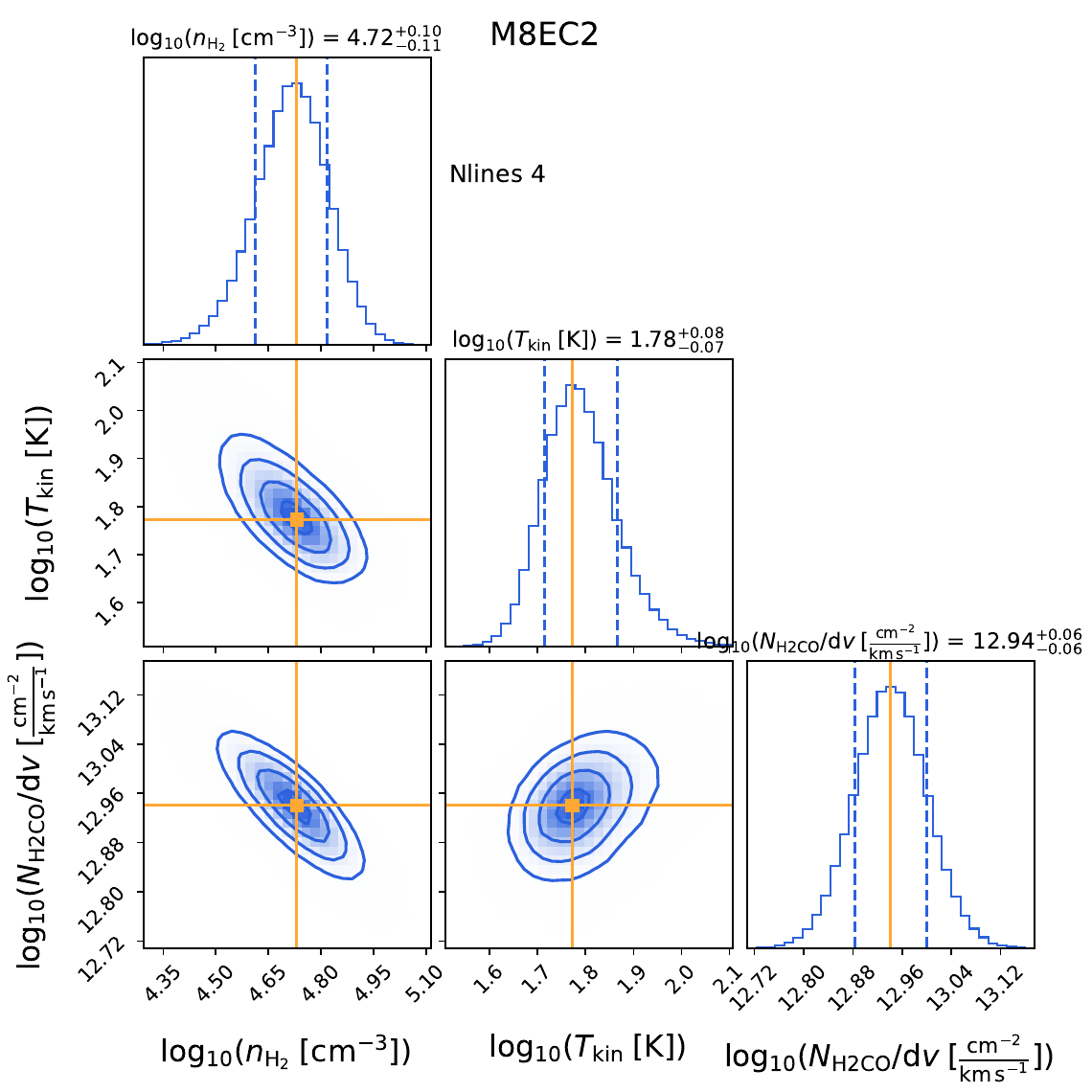}
                        \caption{Posterior probability distributions of the H$_2$ volume density ($n_{\U{H}_2}$), kinetic temperature ($T_\U{kin}$), and H$_2$CO column density per velocity bin ($N_\U{H2CO}/\U{dv}$) at EC2.}
                        \label{fig:app:corner_M8EC2}
                \end{figure}

\begin{figure}[htbp]
                        \centering
                        \includegraphics[width=0.999\linewidth]{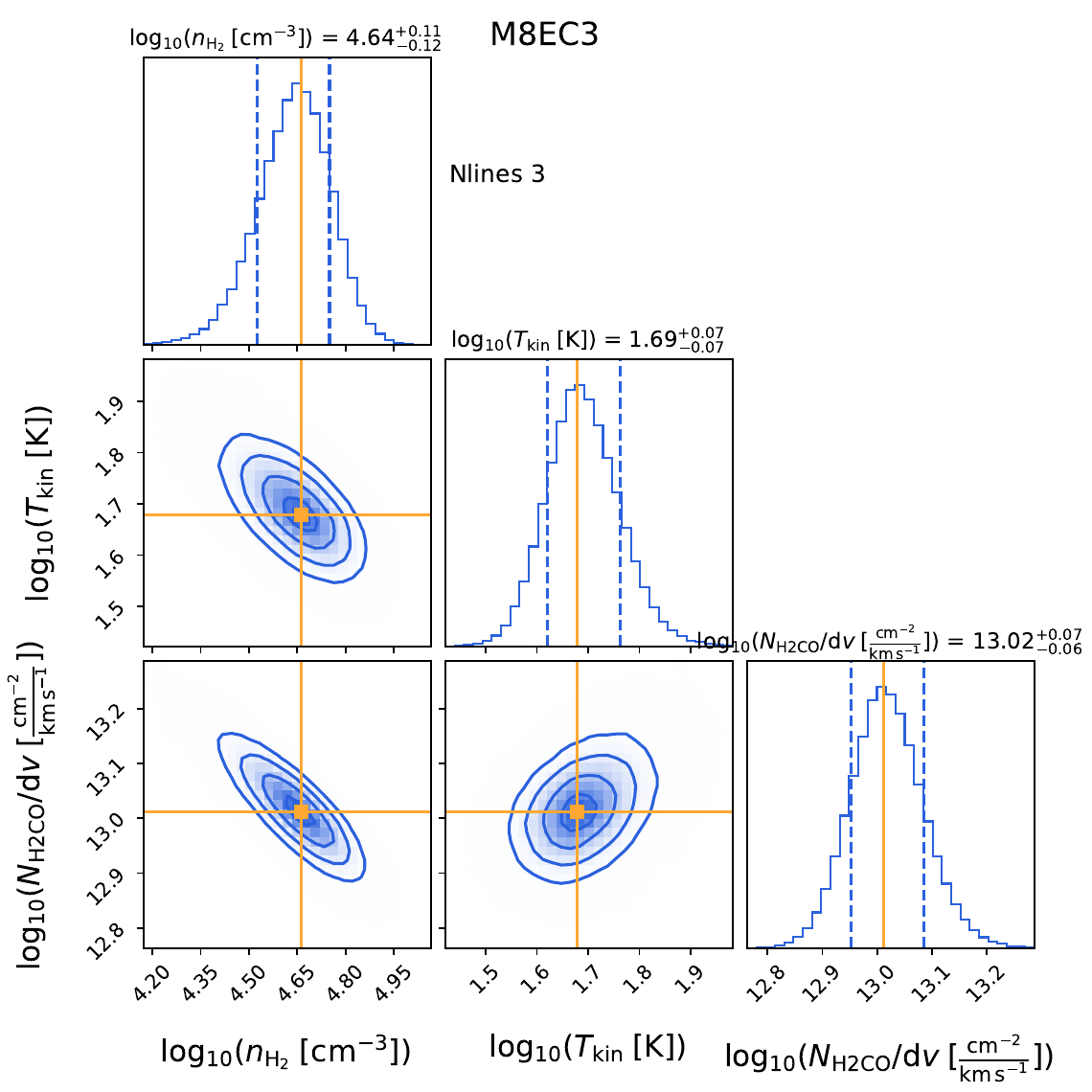}
                        \caption{Posterior probability distributions of the H$_2$ volume density ($n_{\U{H}_2}$), kinetic temperature ($T_\U{kin}$), and H$_2$CO column density per velocity bin ($N_\U{H2CO}/\U{dv}$) at EC3.}
                        \label{fig:app:corner_M8EC3}
                \end{figure}

\begin{figure}[htbp]
                        \centering
                        \includegraphics[width=0.999\linewidth]{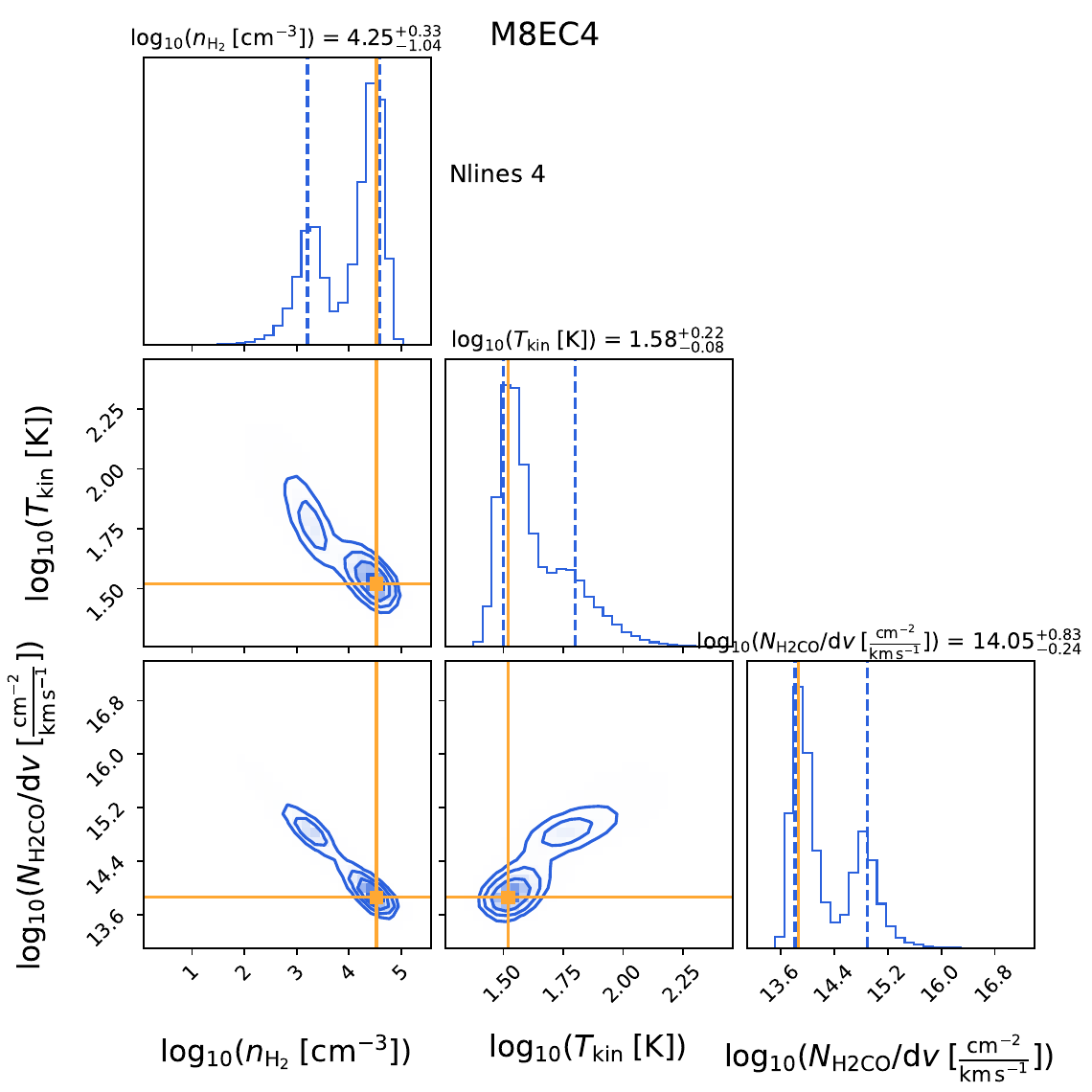}
                        \caption{Posterior probability distributions of the H$_2$ volume density ($n_{\U{H}_2}$), kinetic temperature ($T_\U{kin}$), and H$_2$CO column density per velocity bin ($N_\U{H2CO}/\U{dv}$) at EC4.}
                        \label{fig:app:corner_M8EC4}
                \end{figure}

\begin{figure}[htbp]
                        \centering
                        \includegraphics[width=0.999\linewidth]{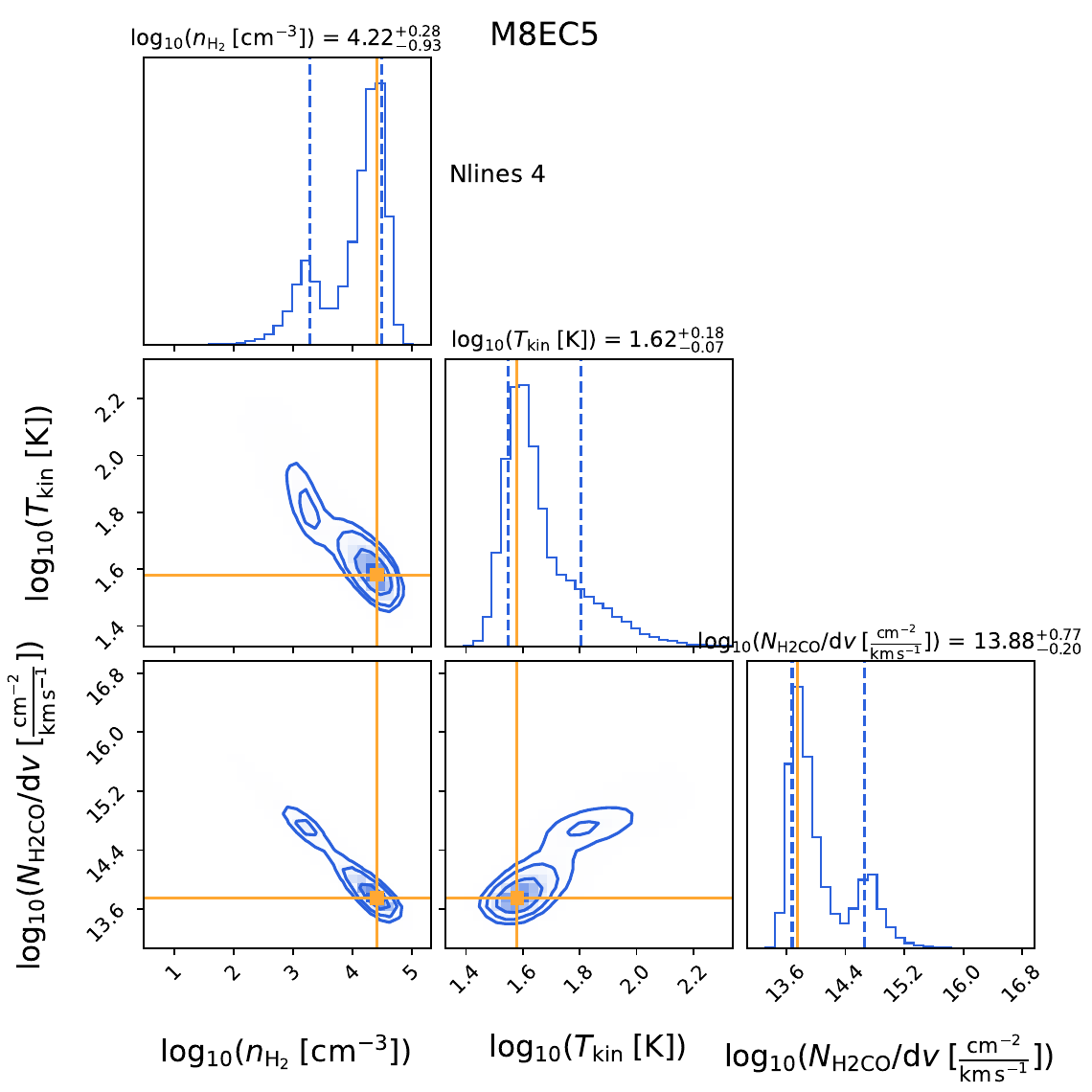}
                        \caption{Posterior probability distributions of the H$_2$ volume density ($n_{\U{H}_2}$), kinetic temperature ($T_\U{kin}$), and H$_2$CO column density per velocity bin ($N_\U{H2CO}/\U{dv}$) at EC5.}
                        \label{fig:app:corner_M8EC5}
                \end{figure}

\begin{figure}[htbp]
                        \centering
                        \includegraphics[width=0.999\linewidth]{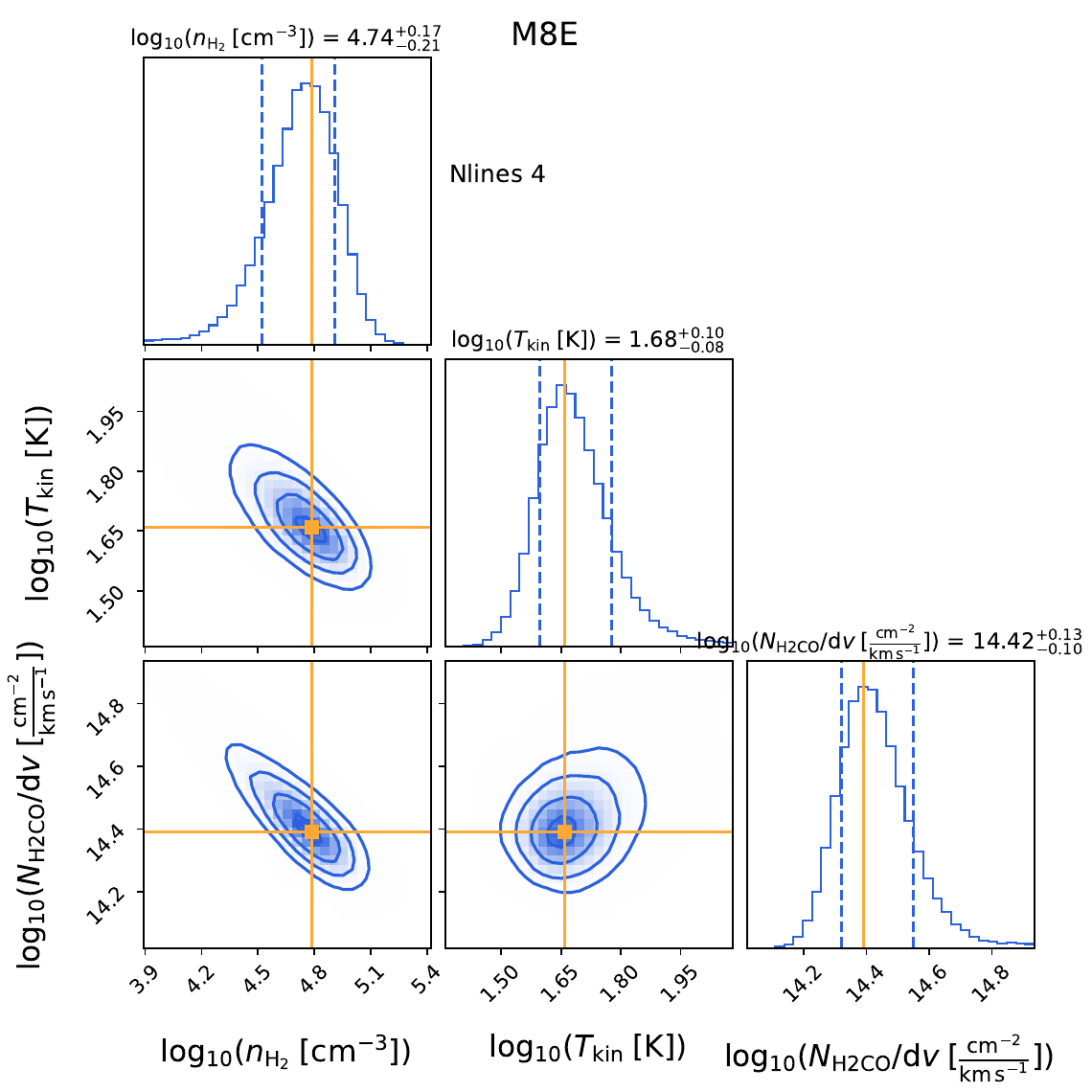}
                        \caption{Posterior probability distributions of the H$_2$ volume density ($n_{\U{H}_2}$), kinetic temperature ($T_\U{kin}$), and H$_2$CO column density per velocity bin ($N_\U{H2CO}/\U{dv}$) at E.}
                        \label{fig:app:corner_M8E}
                \end{figure}

\begin{figure}[htbp]
                        \centering
                        \includegraphics[width=0.999\linewidth]{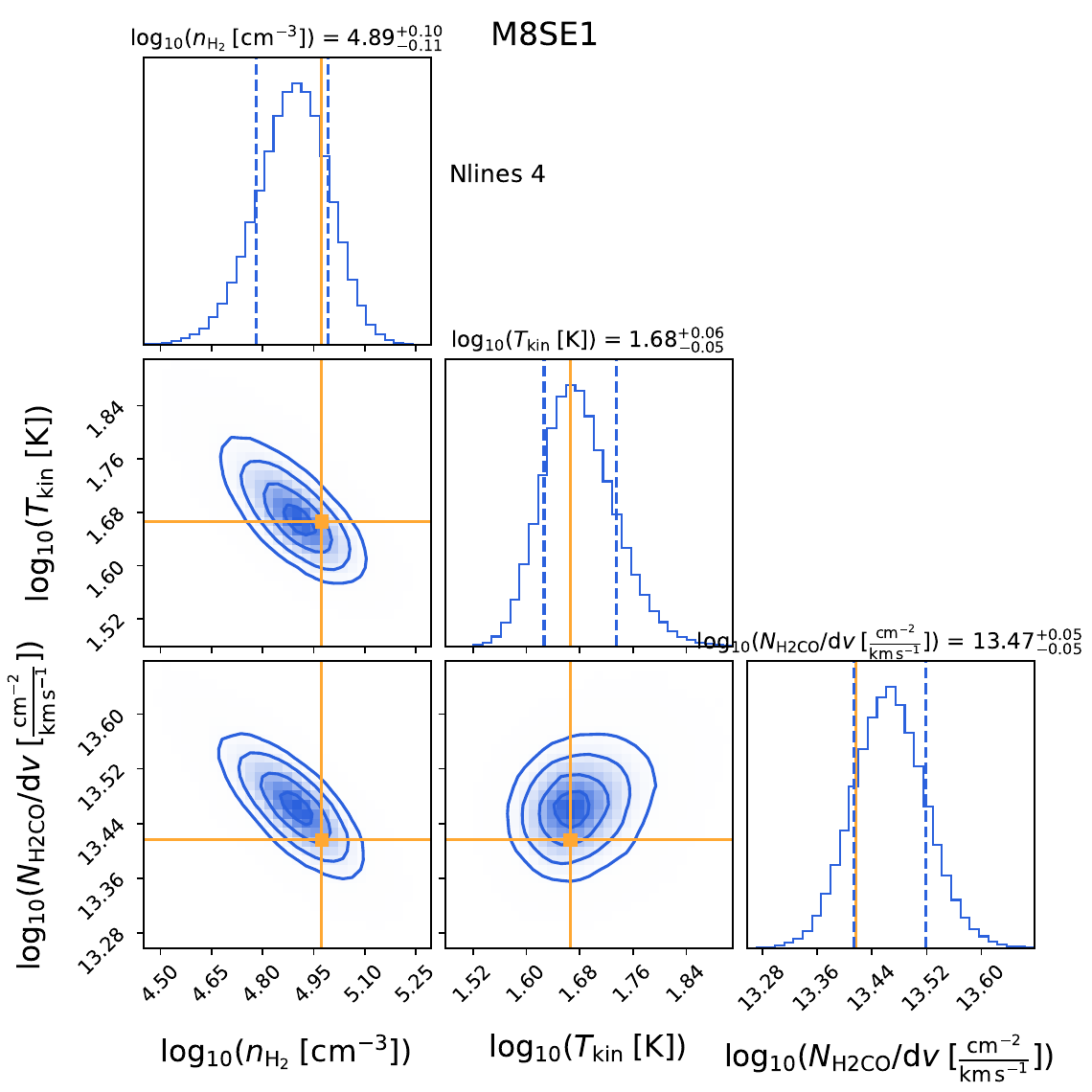}
                        \caption{Posterior probability distributions of the H$_2$ volume density ($n_{\U{H}_2}$), kinetic temperature ($T_\U{kin}$), and H$_2$CO column density per velocity bin ($N_\U{H2CO}/\U{dv}$) at SE1.}
                        \label{fig:app:corner_M8SE1}
                \end{figure}

\begin{figure}[htbp]
                        \centering
                        \includegraphics[width=0.999\linewidth]{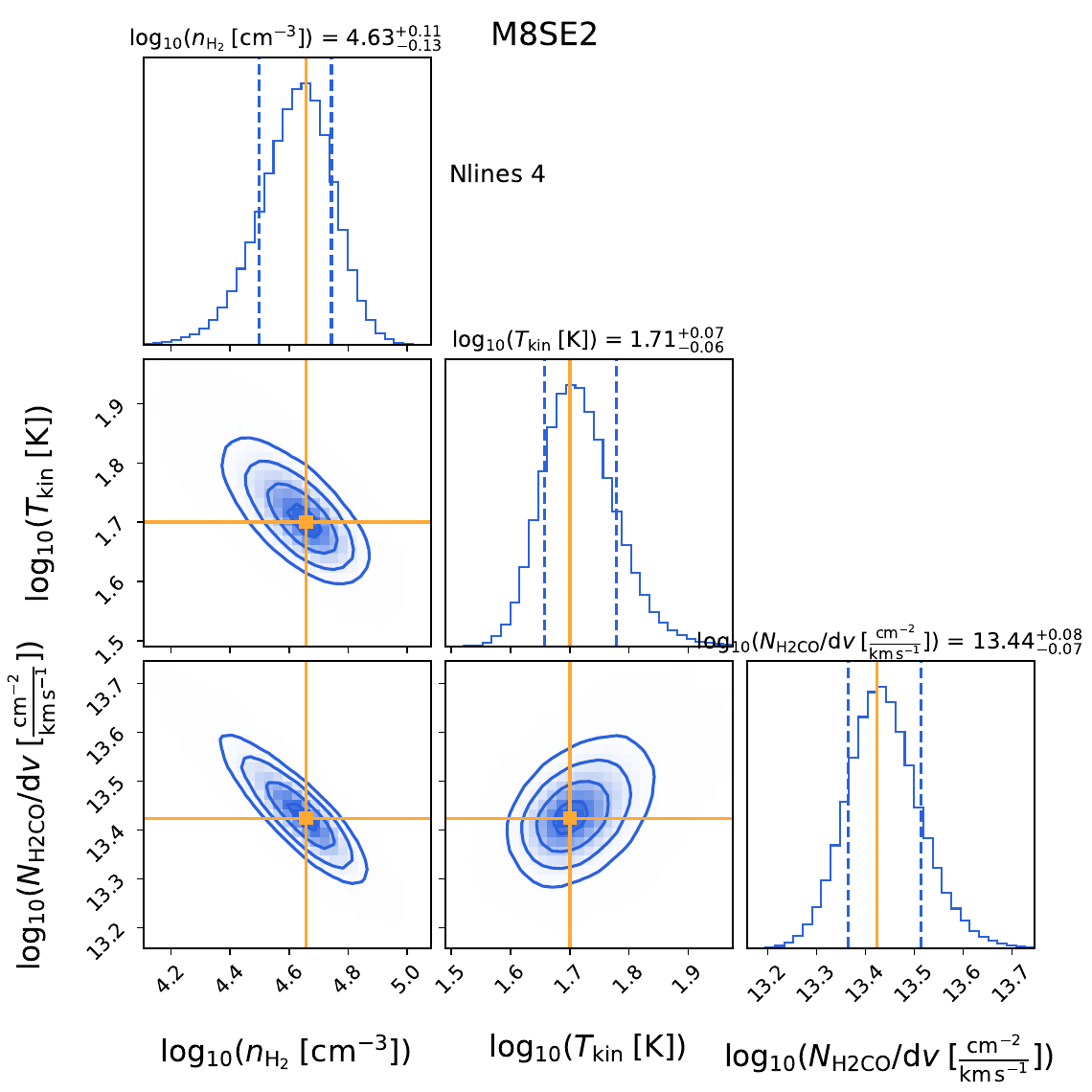}
                        \caption{Posterior probability distributions of the H$_2$ volume density ($n_{\U{H}_2}$), kinetic temperature ($T_\U{kin}$), and H$_2$CO column density per velocity bin ($N_\U{H2CO}/\U{dv}$) at SE2.}
                        \label{fig:app:corner_M8SE2}
                \end{figure}

\begin{figure}[htbp]
                        \centering
                        \includegraphics[width=0.999\linewidth]{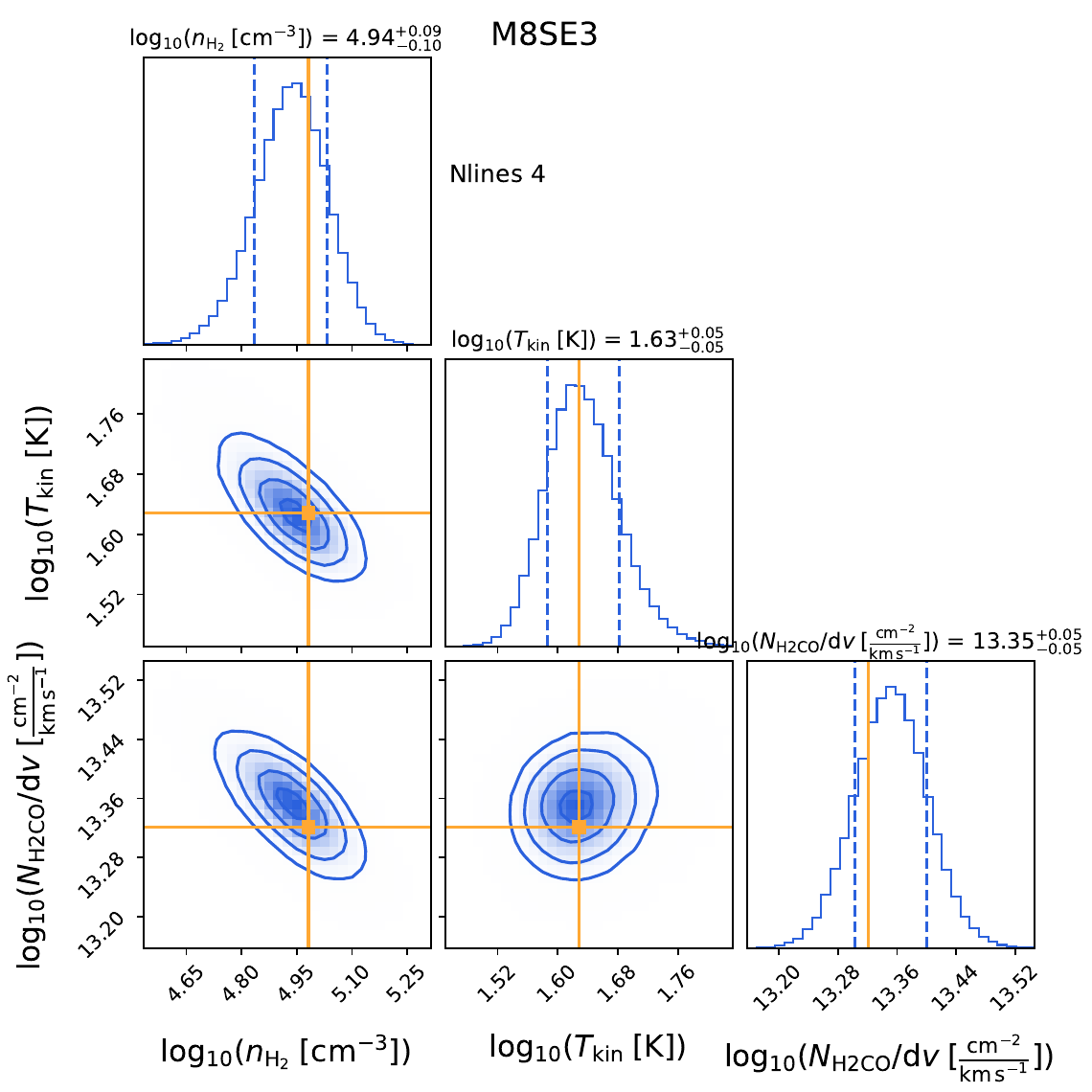}
                        \caption{Posterior probability distributions of the H$_2$ volume density ($n_{\U{H}_2}$), kinetic temperature ($T_\U{kin}$), and H$_2$CO column density per velocity bin ($N_\U{H2CO}/\U{dv}$) at SE3.}
                        \label{fig:app:corner_M8SE3}
                \end{figure}

\begin{figure}[htbp]
                        \centering
                        \includegraphics[width=0.999\linewidth]{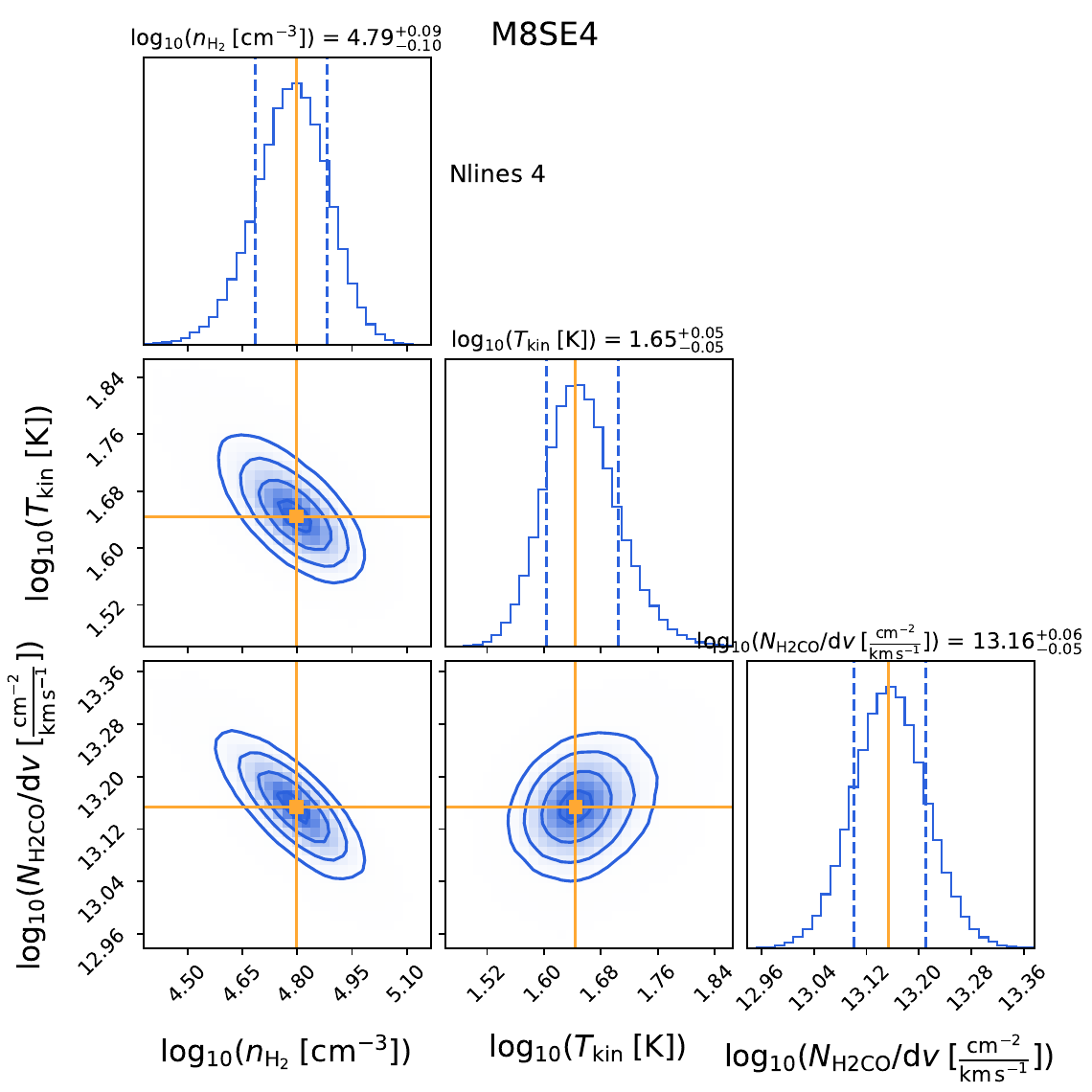}
                        \caption{Posterior probability distributions of the H$_2$ volume density ($n_{\U{H}_2}$), kinetic temperature ($T_\U{kin}$), and H$_2$CO column density per velocity bin ($N_\U{H2CO}/\U{dv}$) at SE4.}
                        \label{fig:app:corner_M8SE4}
                \end{figure}

\begin{figure}[htbp]
                        \centering
                        \includegraphics[width=0.999\linewidth]{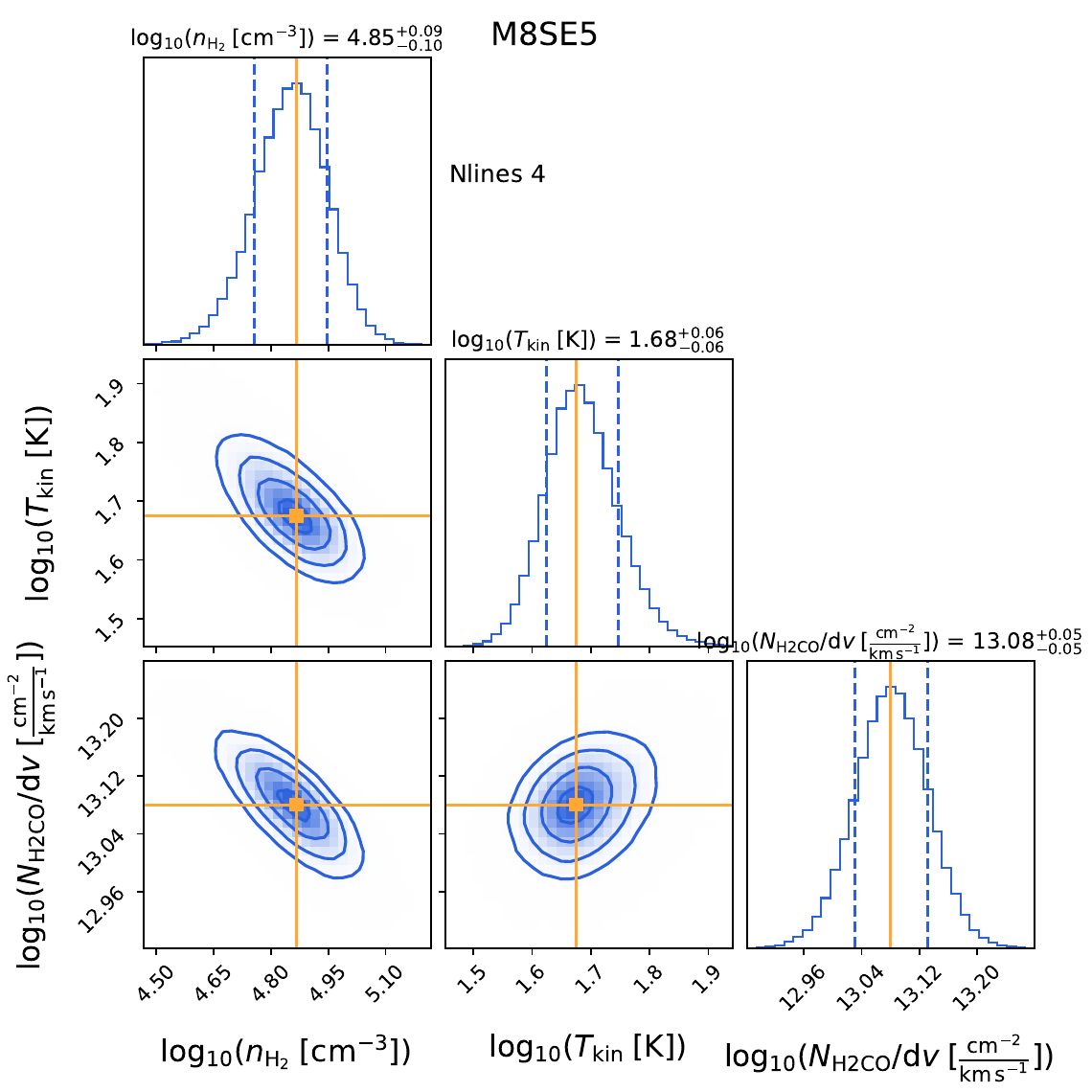}
                        \caption{Posterior probability distributions of the H$_2$ volume density ($n_{\U{H}_2}$), kinetic temperature ($T_\U{kin}$), and H$_2$CO column density per velocity bin ($N_\U{H2CO}/\U{dv}$) at SE5.}
                        \label{fig:app:corner_M8SE5}
                \end{figure}

\begin{figure}[htbp]
                        \centering
                        \includegraphics[width=0.999\linewidth]{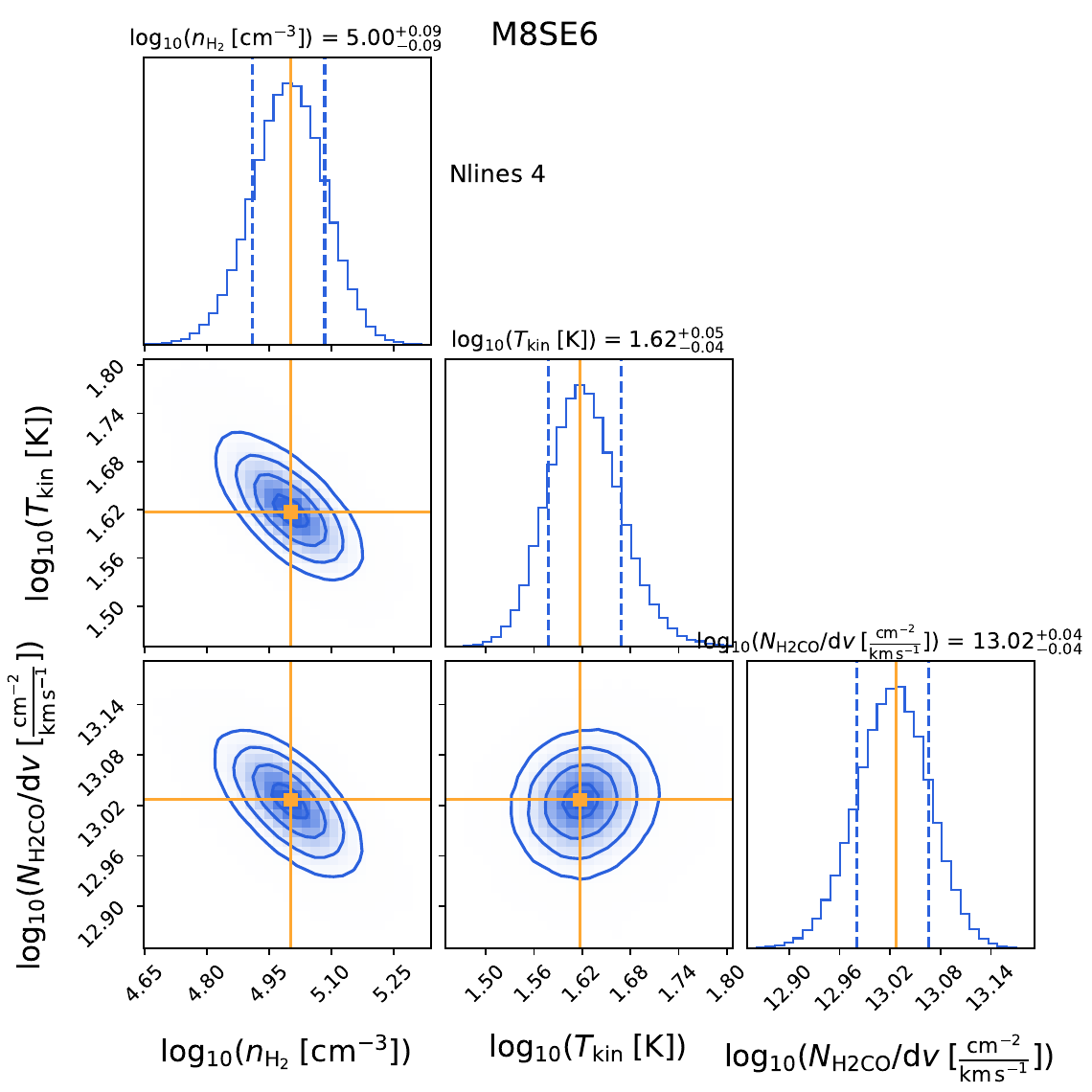}
                        \caption{Posterior probability distributions of the H$_2$ volume density ($n_{\U{H}_2}$), kinetic temperature ($T_\U{kin}$), and H$_2$CO column density per velocity bin ($N_\U{H2CO}/\U{dv}$) at SE6.}
                        \label{fig:app:corner_M8SE6}
                \end{figure}

\begin{figure}[htbp]
                        \centering
                        \includegraphics[width=0.999\linewidth]{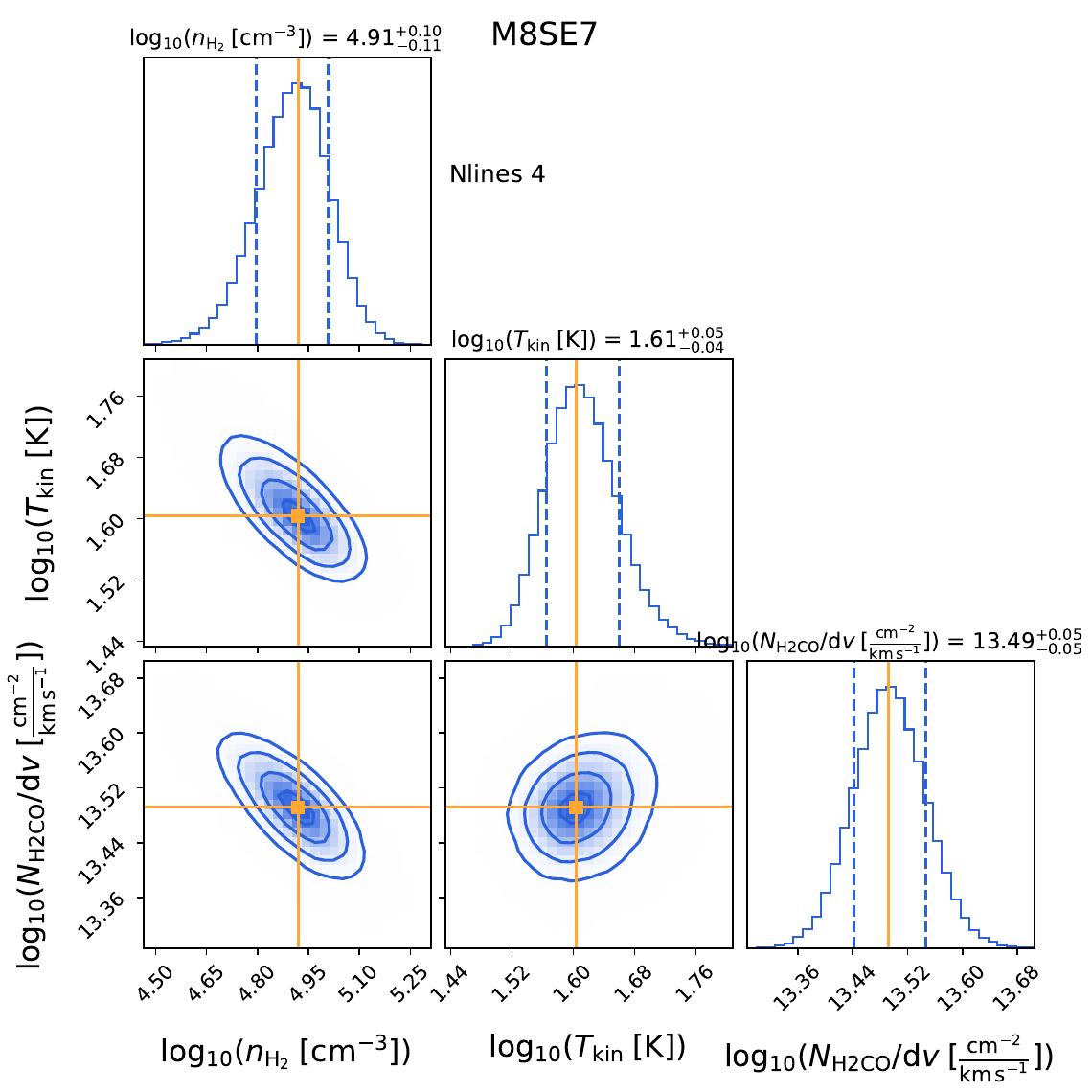}
                        \caption{Posterior probability distributions of the H$_2$ volume density ($n_{\U{H}_2}$), kinetic temperature ($T_\U{kin}$), and H$_2$CO column density per velocity bin ($N_\U{H2CO}/\U{dv}$) at SE7.}
                        \label{fig:app:corner_M8SE7}
                \end{figure}

\begin{figure}[htbp]
                        \centering
                        \includegraphics[width=0.999\linewidth]{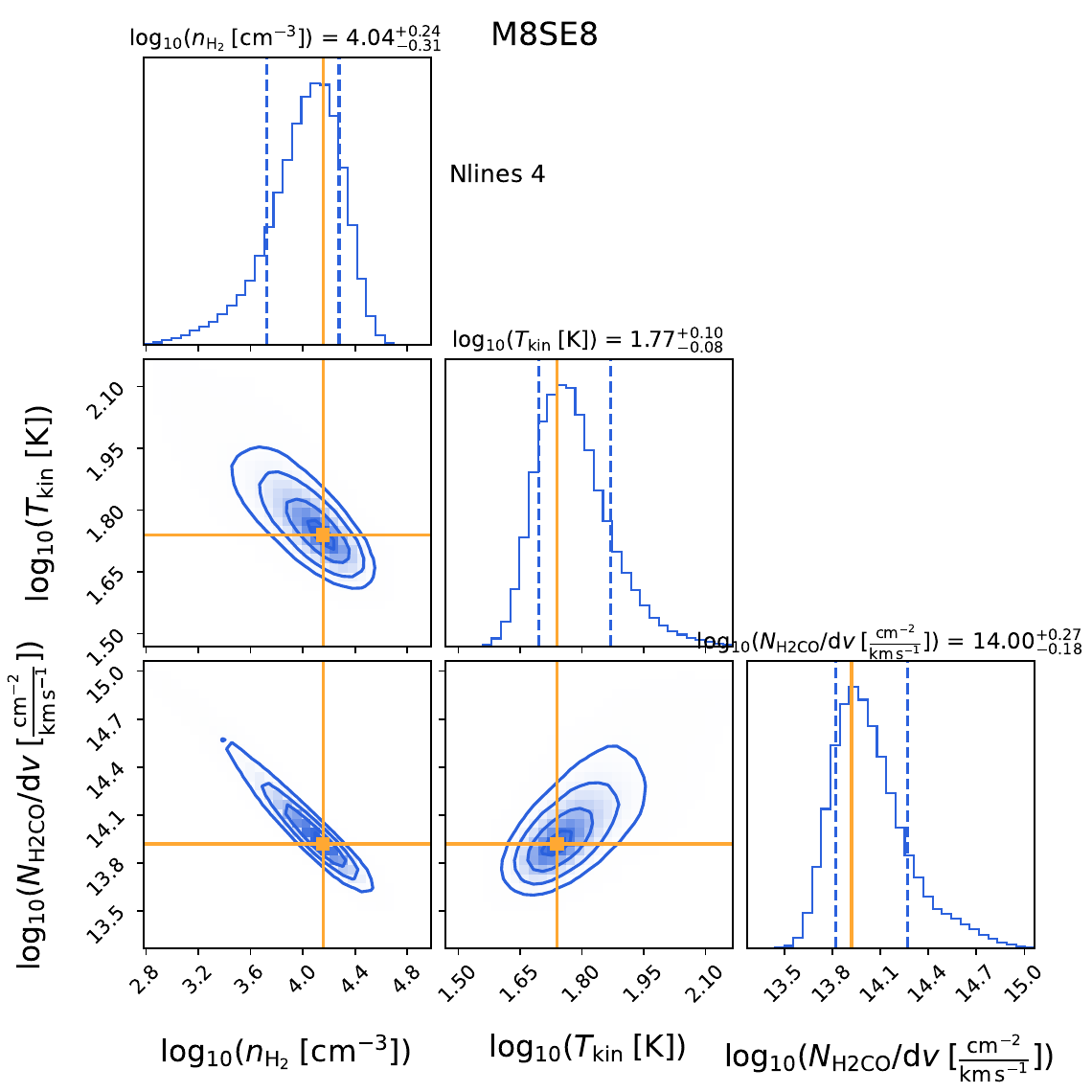}
                        \caption{Posterior probability distributions of the H$_2$ volume density ($n_{\U{H}_2}$), kinetic temperature ($T_\U{kin}$), and H$_2$CO column density per velocity bin ($N_\U{H2CO}/\U{dv}$) at SE8.}
                        \label{fig:app:corner_M8SE8}
                \end{figure}

\begin{figure}[htbp]
                        \centering
                        \includegraphics[width=0.999\linewidth]{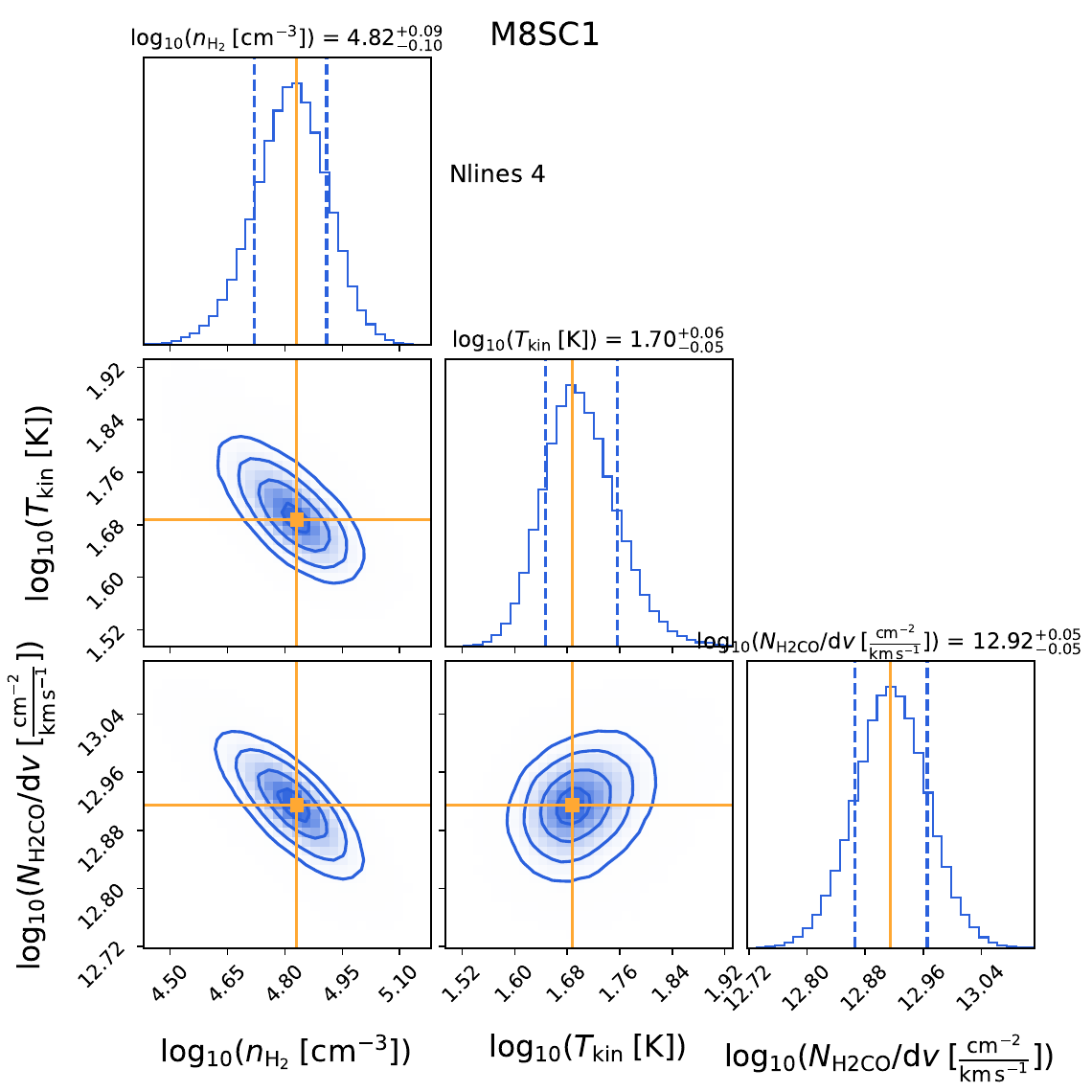}
                        \caption{Posterior probability distributions of the H$_2$ volume density ($n_{\U{H}_2}$), kinetic temperature ($T_\U{kin}$), and H$_2$CO column density per velocity bin ($N_\U{H2CO}/\U{dv}$) at SC1.}
                        \label{fig:app:corner_M8SC1}
                \end{figure}

\begin{figure}[htbp]
                        \centering
                        \includegraphics[width=0.999\linewidth]{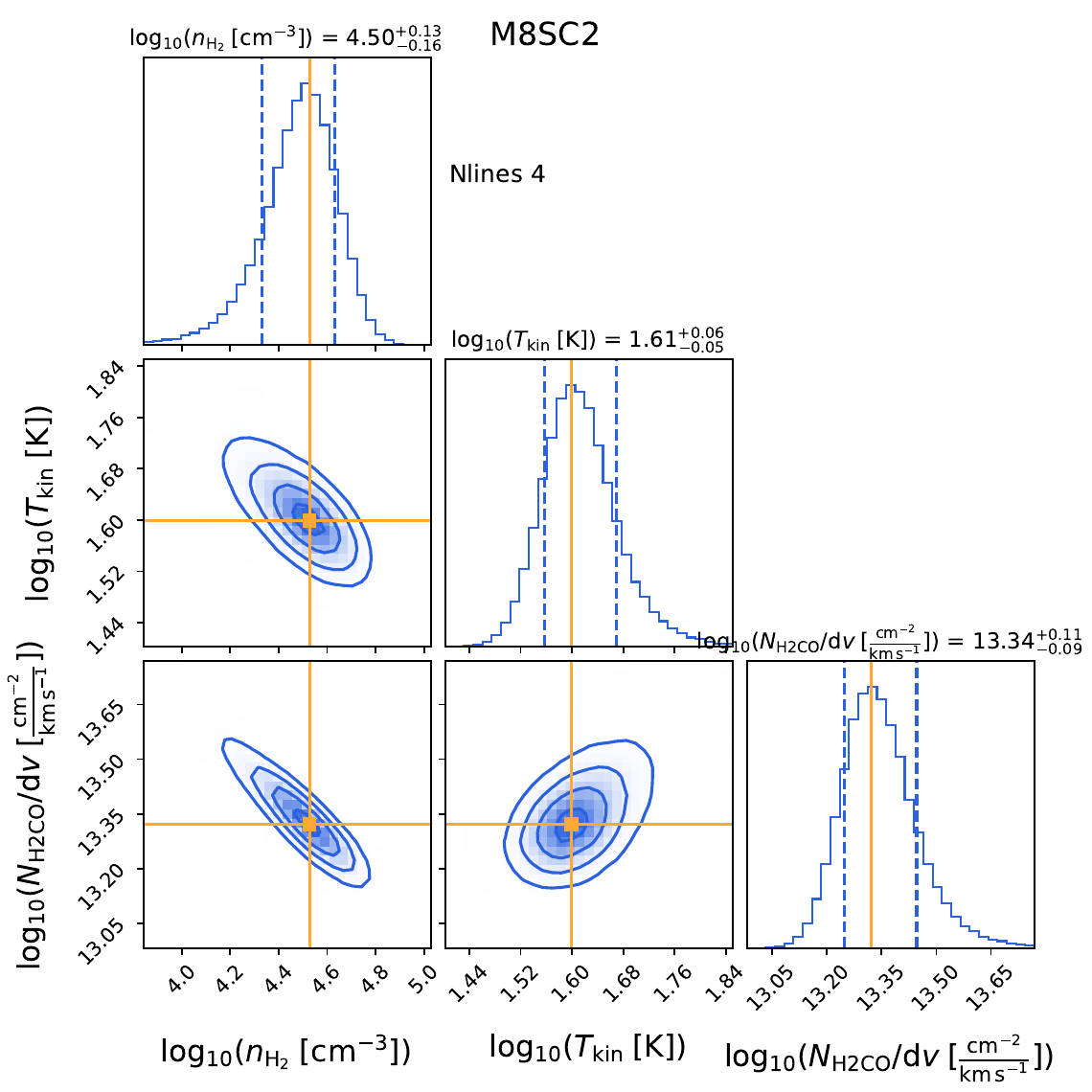}
                        \caption{Posterior probability distributions of the H$_2$ volume density ($n_{\U{H}_2}$), kinetic temperature ($T_\U{kin}$), and H$_2$CO column density per velocity bin ($N_\U{H2CO}/\U{dv}$) at SC2.}
                        \label{fig:app:corner_M8SC2}
                \end{figure}

\begin{figure}[htbp]
                        \centering
                        \includegraphics[width=0.999\linewidth]{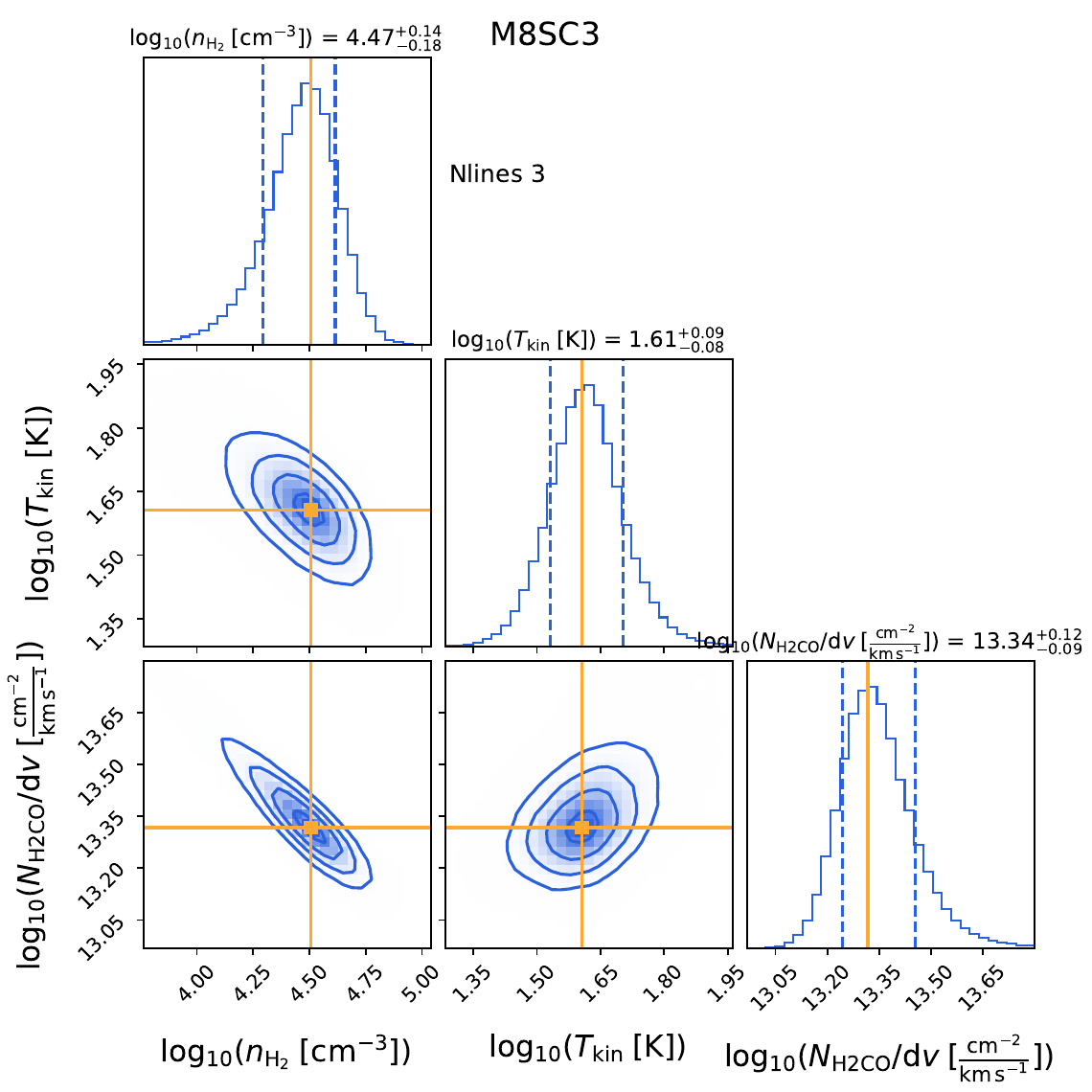}
                        \caption{Posterior probability distributions of the H$_2$ volume density ($n_{\U{H}_2}$), kinetic temperature ($T_\U{kin}$), and H$_2$CO column density per velocity bin ($N_\U{H2CO}/\U{dv}$) at SC3.}
                        \label{fig:app:corner_M8SC3}
                \end{figure}

\begin{figure}[htbp]
                        \centering
                        \includegraphics[width=0.999\linewidth]{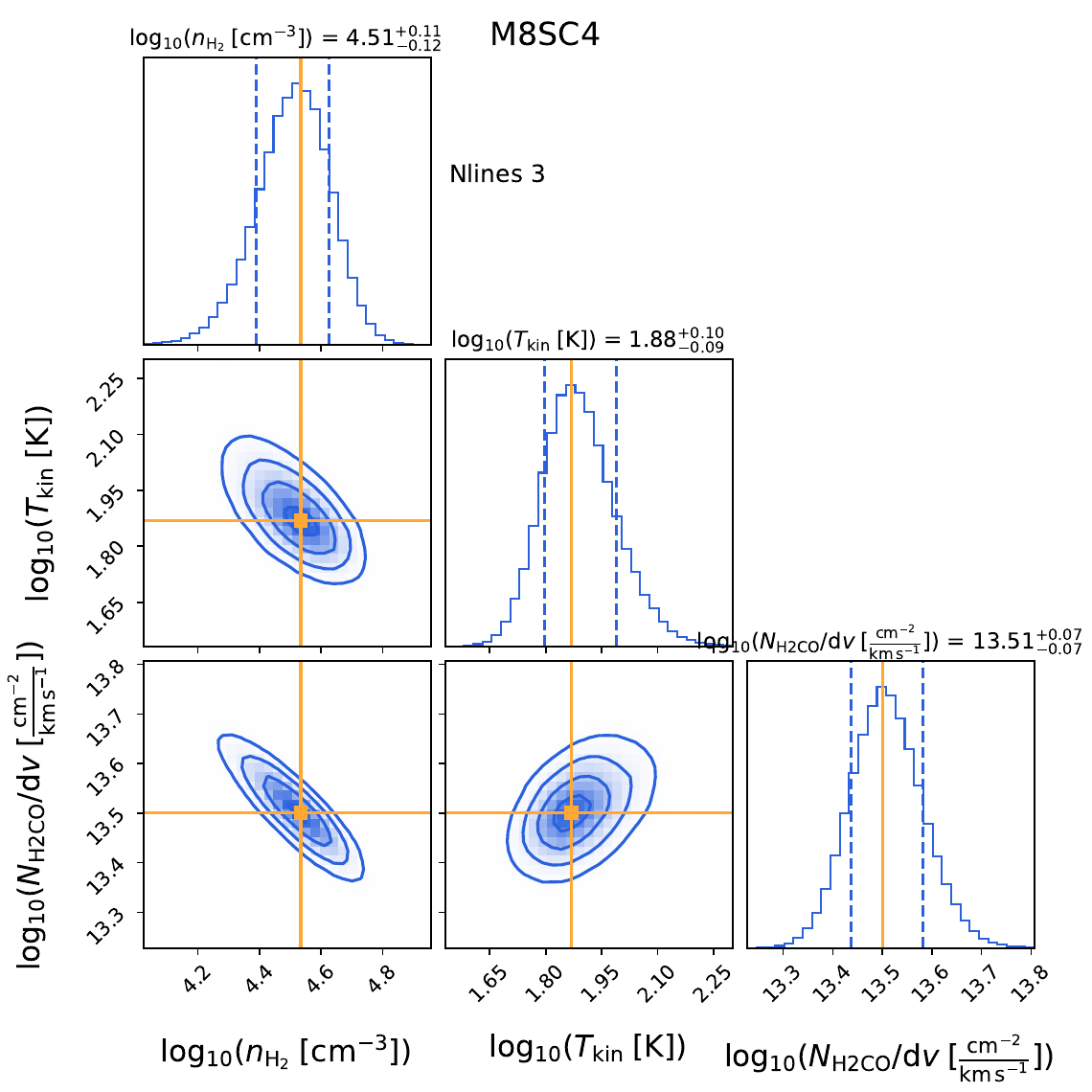}
                        \caption{Posterior probability distributions of the H$_2$ volume density ($n_{\U{H}_2}$), kinetic temperature ($T_\U{kin}$), and H$_2$CO column density per velocity bin ($N_\U{H2CO}/\U{dv}$) at SC4.}
                        \label{fig:app:corner_M8SC4}
                \end{figure}

\begin{figure}[htbp]
                        \centering
                        \includegraphics[width=0.999\linewidth]{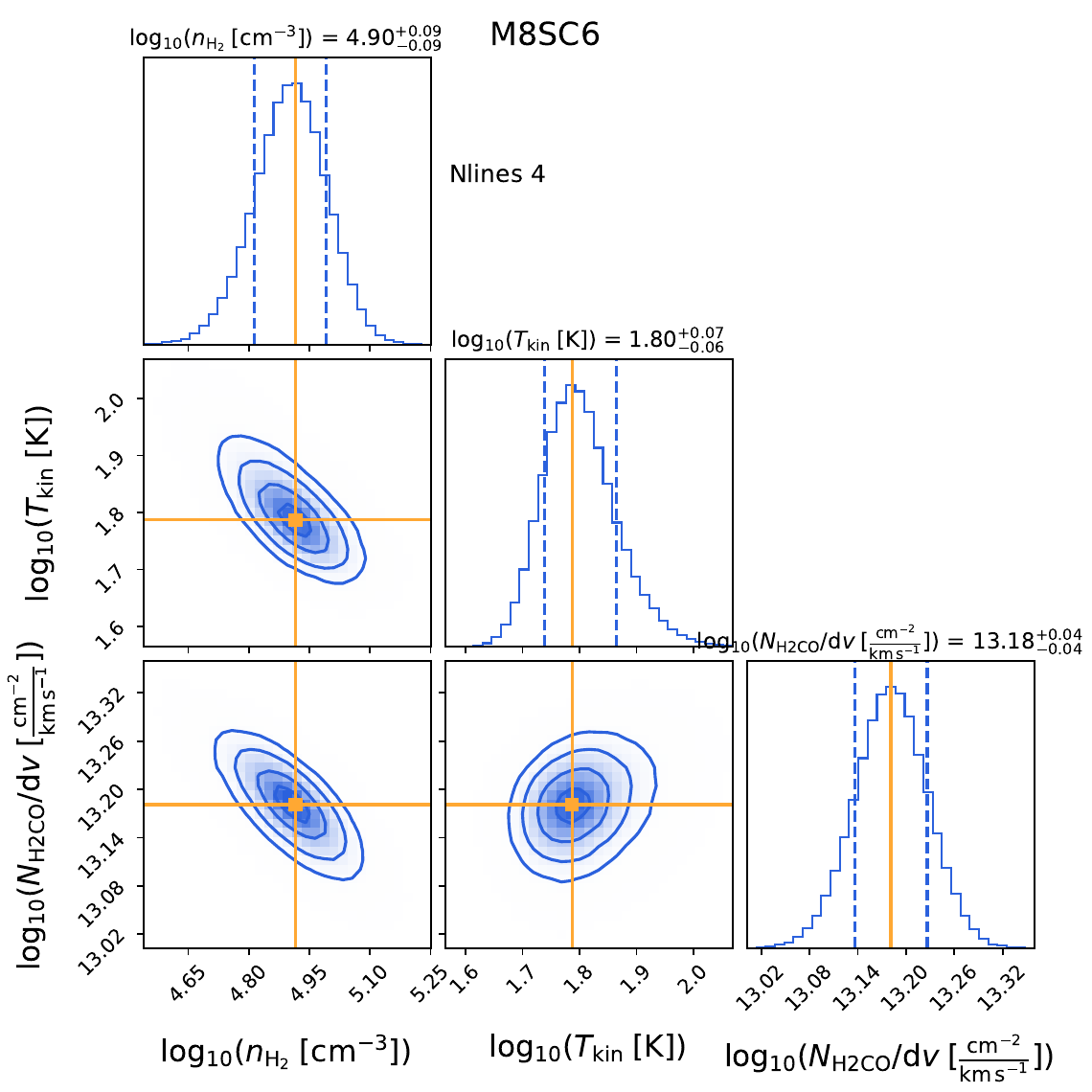}
                        \caption{Posterior probability distributions of the H$_2$ volume density ($n_{\U{H}_2}$), kinetic temperature ($T_\U{kin}$), and H$_2$CO column density per velocity bin ($N_\U{H2CO}/\U{dv}$) at SC6.}
                        \label{fig:app:corner_M8SC6}
                \end{figure}

\begin{figure}[htbp]
                        \centering
                        \includegraphics[width=0.999\linewidth]{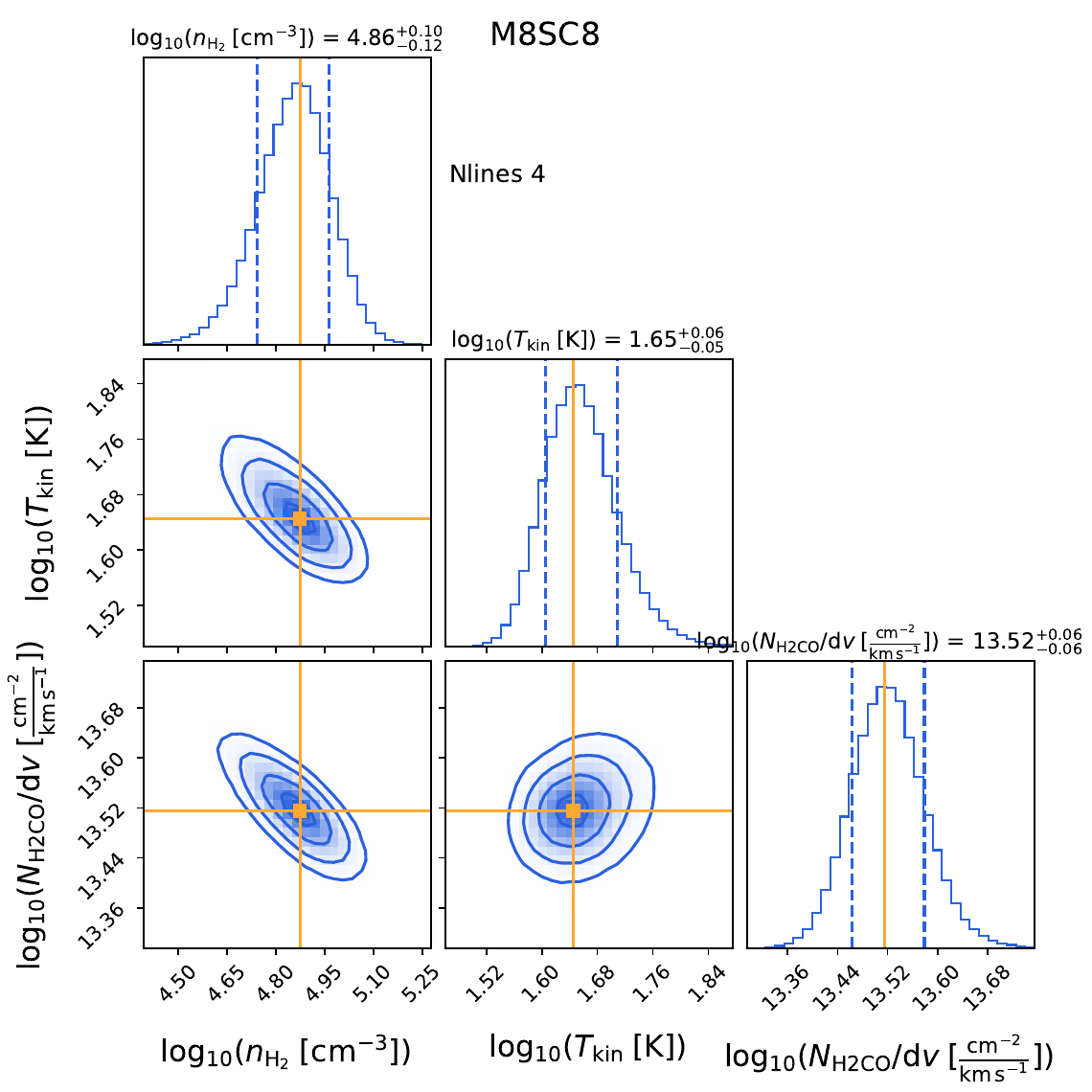}
                        \caption{Posterior probability distributions of the H$_2$ volume density ($n_{\U{H}_2}$), kinetic temperature ($T_\U{kin}$), and H$_2$CO column density per velocity bin ($N_\U{H2CO}/\U{dv}$) at SC8.}
                        \label{fig:app:corner_M8SC8}
                \end{figure}

                As described in Sects.~\ref{subsec:temps} and.~\ref{subsec:temps_ch3cn_ch3cch}, we derived the kinetic temperatures, H$_2$ volume densities and column densities of formaldehyde with an MCMC method, and the properties of acetonitrile and methyl acetylene using rotation diagrams. This appendix gives a short overview of the concepts used in rotation diagrams and provides the resulting physical quantities in Table~\ref{tab:app:temps_cd} alongside the corresponding rotation diagrams in Figs.~\ref{fig:app:rd_M8HG}--\ref{fig:app:rd_M8SC9} and MCMC posterior density distributions in Figs.~\ref{fig:app:corner_M8HG}--\ref{fig:app:corner_M8SC8}.
                
                When multiple transitions of the same species are detected, it is possible to derive rotation temperatures $T_\U{rot}$ and column densities $N_\U{total}$ of the respective species. In the frame of an energy level system, the Boltzmann distribution can be written as
                \begin{equation}\label{eq:boltzmann_full}
                        \frac{N_\U{u}}{N_\U{total}} = \frac{g_\U{u}}{Q(T_\U{rot})} \exp\left(-\frac{E_\U{u}}{k_\U{B}  T_\U{rot}}\right),
                \end{equation}
                with the Boltzmann constant, $k_\U{B}$, the column density of a respective upper energy level, $N_\U{u}$, and the degeneracy, $g_\U{u}$, and energy, $E_\U{u}$, of the same level. $Q(T_\U{rot})$ is the rotational partition function that describes the sum over all rotational energy levels in a molecule for a given rotation temperature~\citep{Mangum2015rotdiag}. $N_\U{u}$ can be computed in the optically thin limit from the measured intensity $I$ of a transition between the upper energy level (u) and the lower energy level (l) according to~\citep{Goldsmith1999population} 
                \begin{equation}\label{eq:cd_thin}
                        N_\U{u}^\U{thin} = \frac{8 \pi k_\U{B} \nu^2 }{h c^3 A_\U{ul}} I,
                \end{equation}
                with the speed of light, $c$, the Planck constant, $h$, the frequency, $\nu$, of a respective transition and the corresponding spontaneous emission Einstein (A) coefficient, $A_\U{ul}$.
                
                By rearranging Eq.~\ref{eq:boltzmann_full} as
                \begin{equation}\label{eq:boltzmann_lin}
                        \ln \left(\frac{N_\U{u}}{g_\U{u}}\right) = - \frac{1}{k_\U{B} T_\U{rot}} E_\U{u} + \ln\left(\frac{N_\U{total}}{Q(T_\U{rot})}\right),
                \end{equation}
                it can be seen that the logarithm of the column density of an upper energy level $N_\U{u}$ is proportional to the energy of the respective level. This relation can be used to perform a linear fit on the known values of $\ln (N_\U{u}/g_\U{u})$ and $E_\U{u}$, in order to obtain the rotation temperature $T_\U{rot}$ from the slope and the column density from the ordinate interception.
                
                For this analysis, transitions of the same acetonitrile or methyl acetylene multiplets were fit simultaneously with multiple Gaussian components of the same width and velocity shift to account for line blending. According to Equation~\ref{eq:cd_thin}, the values of $N_\U{u}$ and therefore of $\ln (N_\U{u}/g_\U{u})$ depend on the velocity-integrated line intensity, $I$. Their uncertainty was therefore derived by propagating the uncertainty of $I$, which is composed of the fit uncertainty for the line profiles with the \texttt{MINIMIZE} function of \texttt{CLASS}, and an additional 10\% uncertainty to account for the calibration uncertainty of the data.
                
                \begin{figure}[htbp]
                        \centering
                        \includegraphics[width=0.999\linewidth]{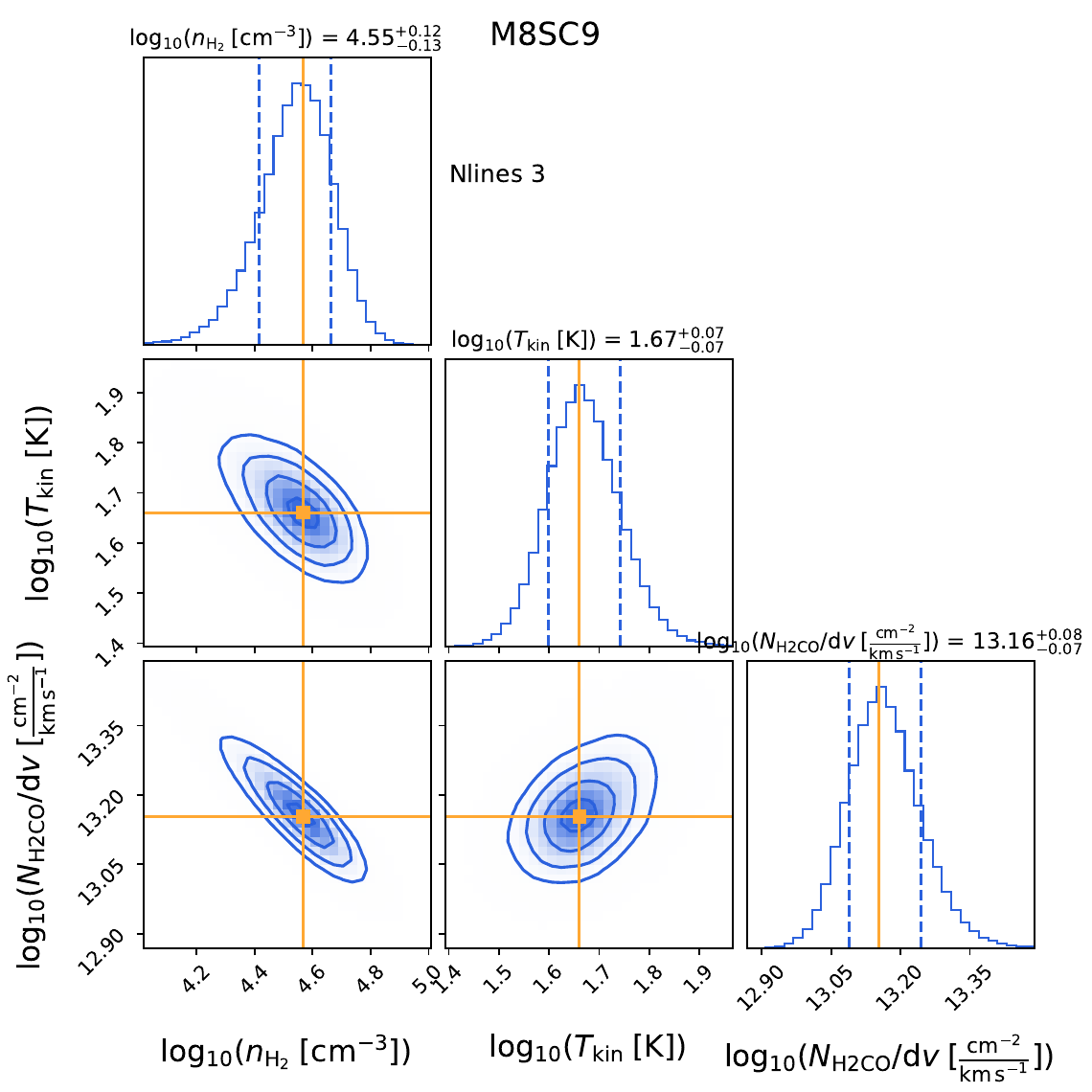}
                        \caption{Posterior probability distributions of the H$_2$ volume density ($n_{\U{H}_2}$), kinetic temperature ($T_\U{kin}$), and H$_2$CO column density per velocity bin ($N_\U{H2CO}/\U{dv}$) at SC9.}
                        \label{fig:app:corner_M8SC9}
                \end{figure}
                
                \clearpage
                \begin{table*}[htbp]
                        \caption{Rotational temperatures and column densities derived from para-formaldehyde (p-H$_2$CO), acetonitrile (CH$_3$CN), and methyl acetylene (CH$_3$C$_2$H). The fourth column contains the H$_2$ volume densities derived based on the observed H$_2$CO transitions.}
                        \label{tab:app:temps_cd}
                        \centering
                        \begin{tabular}{cccccccc}
                                \hline
                                \hline
                                Clump & $T$(H$_2$CO) & $N$(H$_2$CO) & $n_{\U{H}_2}$ & $T$(CH$_3$C$_2$H) & $N$(CH$_3$C$_2$H) & $T$(CH$_3$CN) & $N$(CH$_3$CN) \\
                                &  (K)       &  ($10^{12}\,\si{\per\centi\meter\squared}$) & ($10^{4}\,\si{\per\centi\meter\cubed}$)   &  (K)        & ($10^{12}\,\si{\per\centi\meter\squared}$)    &    (K)           &   ($10^{12}\,\si{\per\centi\meter\squared}$)  \\ \hline
                                HG & 55.0$^{+7.6}_{-8.9}$ & 53.7$^{+3.7}_{-3.7}$ & 21.9$^{+5.0}_{-5.0}$ & 32.5 $\pm$ 3.5 & 38.2 $\pm$ 7.8 & 20.0 $\pm$ 8.2 & 1.1 $\pm$ 0.8\\
                                WC1 & 38.9$^{+4.5}_{-4.5}$ & 29.5$^{+3.4}_{-4.1}$ & 7.4$^{+1.9}_{-1.9}$ & 35.0 $\pm$ 7.3 & 61.7 $\pm$ 22.3 & 20.8 $\pm$ 3.5 & 1.4 $\pm$ 0.4\\
                                WC2 & - & - & - & 43.6 $\pm$ 14.2 & 72.5 $\pm$ 39.5 & - & -\\
                                WC3 & - & - & - & - & - & - & -\\
                                WC4 & 85.1$^{+23.5}_{-27.4}$ & 5.6$^{+1.0}_{-0.9}$ & 8.5$^{+2.5}_{-2.5}$ & - & - & - & -\\
                                WC5 & - & - & - & - & - & - & -\\
                                WC6 & - & - & - & - & - & - & -\\
                                WC7 & - & - & - & - & - & - & -\\
                                WC8 & - & - & - & - & - & - & -\\
                                WC9 & - & - & - & - & - & - & -\\
                                SW1 & - & - & - & - & - & - & -\\
                                EC1 & 37.2$^{+4.3}_{-4.3}$ & 20.9$^{+3.4}_{-3.4}$ & 5.4$^{+1.4}_{-1.6}$ & - & - & 48.5 $\pm$ 10.9 & 2.5 $\pm$ 1.0\\
                                EC2 & 60.3$^{+9.7}_{-11.1}$ & 8.7$^{+1.2}_{-1.2}$ & 5.2$^{+1.2}_{-1.3}$ & - & - & - & -\\
                                EC3 & 49.0$^{+7.9}_{-7.9}$ & 10.5$^{+1.4}_{-1.7}$ & 4.4$^{+1.1}_{-1.2}$ & - & - & 46.4 $\pm$ 20.2 & 2.2 $\pm$ 1.5\\
                                EC4 & 38.0$^{+7.0}_{-19.3}$ & 112.2$^{+62.0}_{-214.4}$ & 1.8$^{+1.4}_{-4.3}$ & 27.7 $\pm$ 3.8 & 57.2 $\pm$ 14.3 & 27.7 $\pm$ 7.3 & 1.6 $\pm$ 0.7\\
                                EC5 & 41.7$^{+6.7}_{-17.3}$ & 75.9$^{+34.9}_{-134.5}$ & 1.7$^{+1.1}_{-3.6}$ & - & - & - & -\\
                                E & 47.9$^{+8.8}_{-11.0}$ & 263.0$^{+60.6}_{-78.7}$ & 5.5$^{+2.2}_{-2.7}$ & 30.0 $\pm$ 0.5 & 213.1 $\pm$ 11.7 & 33.3 $\pm$ 1.6 & 7.6 $\pm$ 0.7\\
                                SE1 & 47.9$^{+5.5}_{-6.6}$ & 29.5$^{+3.4}_{-3.4}$ & 7.8$^{+1.8}_{-2.0}$ & 22.3 $\pm$ 0.6 & 54.7 $\pm$ 4.5 & 33.2 $\pm$ 2.6 & 2.9 $\pm$ 0.4\\
                                SE2 & 51.3$^{+7.1}_{-8.3}$ & 27.5$^{+4.4}_{-5.1}$ & 4.3$^{+1.1}_{-1.3}$ & - & - & - & -\\
                                SE3 & 42.7$^{+4.9}_{-4.9}$ & 22.4$^{+2.6}_{-2.6}$ & 8.7$^{+1.8}_{-2.0}$ & 20.3 $\pm$ 1.5 & 31.4 $\pm$ 5.1 & 46.9 $\pm$ 17.2 & 3.3 $\pm$ 1.9\\
                                SE4 & 44.7$^{+5.1}_{-5.1}$ & 14.5$^{+1.7}_{-2.0}$ & 6.2$^{+1.3}_{-1.4}$ & - & - & - & -\\
                                SE5 & 47.9$^{+6.6}_{-6.6}$ & 12.0$^{+1.4}_{-1.4}$ & 7.1$^{+1.5}_{-1.6}$ & - & - & - & -\\
                                SE6 & 41.7$^{+3.8}_{-4.8}$ & 10.5$^{+1.0}_{-1.0}$ & 10.0$^{+2.1}_{-2.1}$ & - & - & - & -\\
                                SE7 & 40.7$^{+3.8}_{-4.7}$ & 30.9$^{+3.6}_{-3.6}$ & 8.1$^{+1.9}_{-2.1}$ & 14.9 $\pm$ 4.4 & 19.9 $\pm$ 11.7 & 37.5 $\pm$ 7.2 & 2.5 $\pm$ 0.8\\
                                SE8 & 58.9$^{+10.8}_{-13.6}$ & 100.0$^{+41.4}_{-62.2}$ & 1.1$^{+0.6}_{-0.8}$ & 36.6 $\pm$ 9.4 & 37.7 $\pm$ 16.7 & 39.0 $\pm$ 5.9 & 2.9 $\pm$ 0.8\\
                                SC1 & 50.1$^{+5.8}_{-6.9}$ & 8.3$^{+1.0}_{-1.0}$ & 6.6$^{+1.4}_{-1.5}$ & - & - & 17.9 $\pm$ 7.6 & 0.9 $\pm$ 0.7\\
                                SC2 & 40.7$^{+4.7}_{-5.6}$ & 21.9$^{+4.5}_{-5.5}$ & 3.2$^{+0.9}_{-1.2}$ & 34.1 $\pm$ 13.9 & 22.6 $\pm$ 16.1 & - & -\\
                                SC3 & 40.7$^{+7.5}_{-8.4}$ & 21.9$^{+4.5}_{-6.0}$ & 3.0$^{+1.0}_{-1.2}$ & - & - & - & -\\
                                SC4 & 75.9$^{+15.7}_{-17.5}$ & 32.4$^{+5.2}_{-5.2}$ & 3.2$^{+0.8}_{-0.9}$ & - & - & - & -\\
                                SC5 & - & - & - & - & - & - & -\\
                                SC6 & 63.1$^{+8.7}_{-10.2}$ & 15.1$^{+1.4}_{-1.4}$ & 7.9$^{+1.6}_{-1.6}$ & - & - & - & -\\
                                SC7 & - & - & - & - & - & - & -\\
                                SC8 & 44.7$^{+5.1}_{-6.2}$ & 33.1$^{+4.6}_{-4.6}$ & 7.2$^{+1.7}_{-2.0}$ & - & - & 42.8 $\pm$ 13.3 & 3.0 $\pm$ 1.5\\
                                SC9 & 46.8$^{+7.5}_{-7.5}$ & 14.5$^{+2.3}_{-2.7}$ & 3.5$^{+1.0}_{-1.1}$ & - & - & 11.3 $\pm$ 3.5 & 0.6 $\pm$ 0.4\\
                                C1 & - & - & - & - & - & - & -\\
                                C2 & - & - & - & - & - & - & -\\
                                C3 & - & - & - & - & - & - & -\\
                                \hline
                        \end{tabular}
                \end{table*}
                
                \clearpage
                \section{Column densities of detected species at all clumps in M8}
                \label{app:cd}
                Table~\ref{tab:app:cd} provides column density values or upper limits for all species detected in M8 for all clumps, as described in Sect.~\ref{subsec:cd}. The full version of this table is available at the CDS.
                
                \begin{table}[htbp]
                        \caption{Column density values and upper limits for 50 species at HG (extract).}
                        \label{tab:app:cd}
                        \centering
                        \begin{tabular}{cccc}
                                \hline
                                \hline
                                Clump & Species & $N_1$(Species) & $N_2$(Species) \\
                                &  & (cm\,$^{-2}$)     & (cm\,$^{-2}$)  \\ \hline
                                HG & $^{13}$CN & $<\num{2.6e+13}$ & - \\
                                HG & c-C$_3$H & $<\num{1.7e+13}$ & - \\
                                HG & C$^{13}$CH & $<\num{8.2e+13}$ & - \\
                                HG & C$_2$D & $<\num{1.4e+14}$ & - \\
                                HG & C$_2$H & $\num{1.8e+15}$ & \\
                                HG & CN & $\num{1.0e+15}$ & $\num{1.2e+14}$ \\
                                HG & HCO & $\num{1.1e+14}$ & \\
                                HG & NO & $<\num{1.1e+15}$ & - \\
                                HG & NS & $<\num{3.7e+13}$ & - \\
                                HG & $^{13}$CH$_3$OH-A$^+$ & $<\num{5.8e+13}$ & - \\
                                HG & $^{13}$CO & $\num{1.8e+17}$ & $\num{1.3e+16}$ \\
                                HG & $^{13}$C$^{18}$O & $<\num{2.0e+15}$ & - \\
                                HG & $^{13}$CS & $\num{1.1e+13}$ & \\
                                HG & $^{13}$C$^{34}$S & $<\num{1.7e+12}$ & - \\
                                HG & C$_3$H$^+$ & $<\num{1.1e+12}$ & - \\
                                HG & C$_4$H & $\num{2.6e+13}$ & \\
                                HG & C$_2$S & $\num{7.1e+12}$ & \\
                                HG & CF$^+$ & $\num{8.1e+12}$ & \\
                                HG & CH$_3$C$_2$H & $\num{3.8e+13}$ & \\
                                HG & CH$_3$CHO-A & $<\num{2.4e+13}$ & - \\
                                HG & CH$_3$CN & $\num{1.1e+12}$ & \\
                                HG & CH$_3$OH-A$^+$ & $\num{1.9e+14}$ & \\
                                HG & CH$_3$SH-E & $<\num{4.5e+13}$ & - \\
                                HG & CO & $\num{4.5e+17}$ & $\num{1.3e+17}$ \\
                                HG & C$^{17}$O & $\num{5.6e+16}$ & \\
                                HG & C$^{18}$O & $\num{2.9e+16}$ & $\num{5.1e+14}$ \\
                                HG & CS & $\num{2.4e+14}$ & $\num{4.7e+12}$ \\
                                HG & C$^{33}$S & $\num{9.3e+12}$ & \\
                                HG & $^{34}$CS & $\num{3.3e+13}$ & \\
                                HG & DC$_3$N & $<\num{8.8e+11}$ & - \\
                                HG & DCN & $\num{6.5e+12}$ & \\
                                HG & DCO$^+$ & $\num{6.1e+11}$ & \\
                                HG & DNC & $<\num{2.9e+12}$ & - \\
                                HG & H$_2^{13}$CO & $<\num{2.6e+12}$ & - \\
                                HG & H$_2$C$_2$O & $<\num{1.2e+13}$ & - \\
                                HG & H$_2$CO & $\num{5.4e+13}$ & $\num{1.4e+13}$ \\
                                HG & H$_2$CS & $\num{2.6e+13}$ & \\
                                HG & H$_2$C$^{34}$S & $<\num{1.1e+13}$ & - \\
                                HG & H$_2$S & $<\num{1.1e+14}$ & - \\
                                HG & H$^{13}$CC$_2$H & $<\num{1.1e+12}$ & - \\
                                HG & H$^{13}$CN & $\num{1.2e+13}$ & \\
                                HG & H$^{13}$CO$^+$ & $\num{3.7e+12}$ & $\num{2.9e+11}$ \\
                                HG & HC$_3$N & $\num{1.6e+13}$ & \\
                                HG & HC$_5$N & $<\num{2.6e+12}$ & - \\
                                HG & HC$^{13}$CCN & $<\num{9.9e+11}$ & - \\
                                HG & HC$_2^{13}$CN & $<\num{9.9e+11}$ & - \\
                                HG & HC$_3^{15}$N & $<\num{8.2e+11}$ & - \\
                                HG & HCN & $\num{2.8e+14}$ & \\
                                HG & HC$^{15}$N & $\num{2.3e+12}$ & \\
                                HG & HCNO & $<\num{7.4e+11}$ & - \\
                                \hline
                        \end{tabular}
                        \tablefoot{Uncertainties are estimated to be of order 10\% based on the calibration uncertainty of the telescopes. If two velocity components are detected, the column densities $N_1$ and $N_2$ are calculated separately. For undetected species, an upper limit of the respective column density is calculated based on the baseline RMS. Corresponding values are marked with a $<$. The complete table containing all detected species at all M8 clumps is available at the CDS.} 
                \end{table}
        \end{appendix}
\end{document}